\documentclass[12pt]{mypaper}
\pdfoutput=1
\usepackage{multirow}
\usepackage{graphicx}
\usepackage{rotate}
\usepackage{rotating}
\usepackage{relsize}
\usepackage{lineno}
\usepackage{slashed}
\usepackage{url}
\usepackage{amsmath}
\usepackage{longtable}
\usepackage[font=small]{caption}
\usepackage[bottom]{footmisc}

\usepackage{latexsym}
\usepackage{amssymb}
\usepackage{epsfig,amsmath,graphics}
\usepackage{color}
\usepackage{dcolumn}
\usepackage{slashed}
\usepackage{comment,latexsym} 
\usepackage{bm}
\usepackage{verbatim}
\usepackage{tabularx}
\usepackage{dcolumn}
\definecolor{colorLink}{rgb}{0.7,0,0}
\definecolor{colorCite}{rgb}{0,.7,0}
\definecolor{colorURL}{rgb}{0,0,0.7}
\usepackage[colorlinks=true,linktocpage=true,linkcolor=colorLink,citecolor=colorCite,urlcolor=colorURL]{hyperref}
\usepackage{setspace}
\usepackage{xspace}
\usepackage{bigints}
\usepackage{multirow}
\usepackage{cite}

\newcommand{\atlas}{{ATLAS}}
\newcommand{\cms}{{CMS}}

\newcommand{\pt}{\ensuremath{p_T}}
\newcommand{\HT}{\ensuremath{H_T}}
\newcommand{\meff}{\ensuremath{m_\text{eff}}}

\newcommand{\OO}{\ensuremath{\mathcal{O}}}

\def\be{\begin{equation}}
\def\ee{\end{equation}}
\newcommand{\beq}{\begin{equation}}
\newcommand{\eeq}{\end{equation}}
\def\bea{\begin{eqnarray}}
\def\eea{\end{eqnarray}}

\newcommand{\GeV}{{\text{ GeV}}}
\newcommand{\TeV}{{\text{ TeV}}}

\newcommand{\MET}{\ensuremath{E_{T}^{\mathrm{miss}}}}

\newcommand{\NegSpace}{\vspace{-10pt}}

\newcommand{\minitab}[2][l]{\begin{tabular}{#1}#2\end{tabular}}

\newcommand{\hiddensubsection}[1]{
    \stepcounter{subsection}
    \subsection*{\arabic{section}.\arabic{subsection}\hspace{1em}{#1}}
}

\def\ack{\section*{Acknowledgments}}

\begin{document}

%
\begin{titlepage}
\flushright{SLAC-PUB-15857}
\vspace{-25pt}
\title{\Large SUSY Simplified Models at\\
 14, 33, and 100 TeV Proton Colliders}
\vspace{-30pt}
 \begin{Authlist}
Timothy Cohen
\vspace{-5pt}
\Instfoot{slac}{SLAC National Accelerator Laboratory, Menlo Park, USA}
Tobias Golling
\vspace{-5pt}
\Instfoot{yale}{ Yale University, New Haven, USA}
Mike Hance
\vspace{-5pt}
\Instfoot{lbl}{Lawrence Berkeley National Laboratory, Berkeley, USA}
Anna Henrichs
\vspace{-5pt}
\Instfoot{yale}{ Yale University, New Haven, USA}
Kiel Howe
\vspace{-5pt}
\Instfoot{sitp}{Stanford Institute for Theoretical Physics, Stanford University, Stanford, USA}
\vspace{-10pt}
\Instfoot{slac}{SLAC National Accelerator Laboratory, Menlo Park, USA}
Joshua Loyal
\vspace{-5pt}
\Instfoot{yale}{ Yale University, New Haven, USA}
Sanjay Padhi
\vspace{-5pt}
 \Instfoot{ucsd}{University of California, San Diego, USA}
Jay G. Wacker
\vspace{-5pt}
\Instfoot{slac}{SLAC National Accelerator Laboratory, Menlo Park, USA}
\end{Authlist}

\begin{abstract}
$\quad$ Results are presented for a variety of SUSY Simplified Models at the $14$ TeV LHC as well as a $33$ and $100$ TeV proton collider.  Our focus is on models whose signals are driven by colored production.  We present projections of the upper limit and discovery reach in the gluino-neutralino (for both light and heavy flavor decays), squark-neutralino, and gluino-squark Simplified Model planes.  Depending on the model a jets + \MET, mono-jet, or same-sign di-lepton search is applied.  The impact of pileup is explored.  This study utilizes the Snowmass backgrounds and combined detector.  Assuming $3000 \text{ fb}^{-1}$ of integrated luminosity, a gluino that decays to light flavor quarks can be discovered below $2.3$ TeV at the $14$ TeV LHC and below $11$ TeV at a $100$ TeV machine.
\end{abstract}
\end{titlepage}

\clearpage
\setcounter{page}{2}

\begin{spacing}{0}
\tableofcontents
\end{spacing}

\setstretch{1.1}
\section{Introduction}
The Large Hadron Collider (LHC) has completed its $8$ TeV run.   Searches for a wide variety of beyond the Standard Model (SM) states, both in the context of Supersymmetry (SUSY) and otherwise, have been (and are being) performed.  The absence so far for any signatures of new particles lurking in the $8\TeV$ data does not deter the expectation that new physics will be accessible at colliders.  The next stage of the energy frontier collider effort will begin once the LHC has completed its upgrade to a center-of-mass energy approaching $14$ TeV.  In addition, discussions of collider physics beyond the LHC have begun; of particular relevance here are plans for a proton collider with $\sqrt{s}\sim100\TeV$.  In light of all this activity, it is interesting to develop a quantitative picture for the physics potential of the next phase of the LHC and beyond.

This work provides a study of the reach of the LHC upgrade and future proton colliders in the context of SUSY Simplified Models \cite{Alwall:2008va, Alwall:2008ag, Alves:2011wf}. Supersymmetry is one of the best-motivated possibilities for new physics within the reach of future machines.  Supersymmetric models provide a framework for constructing collider searches that generically cover additional motivated extensions of the SM; most importantly signals that involve missing energy and/or heavy flavor production.  Furthermore, there is a cornucopia of results on SUSY extensions to the SM using $8 \TeV$ LHC data that provide a useful reference point for comparing the reach of future colliders.  Clearly, assessing the ability to search for new physics in the context of SUSY is a convenient benchmark for understanding the general physics potential of future proton colliders.

Given the vast possibilities for signatures that can be realized within the SUSY framework, we choose to work with the signature driven approach of Simplified Models.  The philosophy underlying Simplified Models is simple: isolate the minimal field content required to produce a specific SUSY signature --- it then becomes tractable to optimize a search such that it provides the maximal reach in both mass and $\sigma \times \text{BR}$.  In practice, Simplified Models are IR-defined Lagrangian based theories that consist of a minimal number of particles and couplings; by keeping the number of free parameters to $\OO(\text{a few})$, it is possible to understand the consequences of a given experiment for the entire parameter space.  

Note that Simplified Models do not capture certain features of ``complete" SUSY models; this approach remains agnostic about complimentary phenomenology, \emph{e.g.}~dark matter.  For contrast consider the UV-motivated simplified parameter space of the CMSSM; many of these models do contain multiple collider accessible particles, but it is only possible to explore the full parameter space with a non-trivial combination of experiments including proton colliders and dark matter detection \cite{Cohen:2013kna}.  Another fruitful approach for understanding complementarity between experiments is the reduced IR parameter space of the pMSSM \cite{Cahill-Rowley:2013vfa}.  However, it can be challenging to interpret and generalize the results of CMSSM or pMSSM specific collider searches to more generic settings.

The parameter space of SUSY Simplified Models has been explored in great detail at the 8 TeV LHC by both the ATLAS and CMS collaborations (for a recent overview of the experimental exclusions and the implications for SUSY models, see \cite{Craig:2013cxa}). In this work we focus on minimal SUSY extensions of the SM with colored initial states.  These models are expected to provide the greatest sparticle mass reach at hadron colliders.

In particular, motivated by expectations for the ``first signatures" of SUSY, we study the following Simplified Models:
\vspace{-10pt}
\renewcommand{\arraystretch}{1.4}
\setlength{\tabcolsep}{8pt}
\begin{center}
\begin{tabular}{c|c|c}
Section & Simplified Model & Decay Channel\\
\hline
\ref{sec:GNLightFlavor} and \ref{sec:GNCompressed}& Gluino-neutralino with light flavor decays &  $\widetilde{g} \rightarrow q\,\overline{q}\,\widetilde{\chi}_1^0$ \\
\ref{sec:QN} and \ref{sec:QNCompressed} &  Squark-neutralino & $\widetilde{q} \rightarrow q\,\widetilde{\chi}_1^0$ \\
\ref{sec:GluinoSquarkNeutralino}& Gluino-squark with a massless neutralino & $\widetilde{g} \rightarrow \big (q\,\overline{q}\,\widetilde{\chi}_1^0 / q\, \widetilde{q}^*\big)$; $\widetilde{q} \rightarrow \big( q\,\widetilde{\chi}_1^0 / q\,\widetilde{g} \big) $ \\
 \ref{sec:GNHeavyFlavor}& Gluino-neutralino with heavy flavor decays &  $\widetilde{g} \rightarrow t\,\overline{t}\,\widetilde{\chi}_1^0$ 
\end{tabular}
\end{center}
Our analyses loosely follow existing public 8 TeV search strategies from the ATLAS and CMS collaborations with optimizations performed to account for the higher luminosity and energy.  We study the impact of pileup conditions to estimate how our conclusions could be altered by the harsh environments of running proton colliders at high instantaneous luminosity.  Additional studies on the impact of systematic uncertainties are provided for a few models.

Discovery reach and exclusions limits are given for the following collider scenarios:
\vspace{-10pt}
\begin{center}
\renewcommand{\arraystretch}{1.5}
\setlength{\tabcolsep}{12pt}
\begin{tabular}{c|c|c}
Machine & $\sqrt{s}$ & Final Integrated Luminosity\\
\hline
LHC Phase I & $14 \,\TeV$ & $300$ fb$^{-1}$ \\
HL-LHC or LHC Phase II &  $14 \,\TeV$ & $3000$ fb$^{-1}$ \\
HE-LHC & $33 \,\TeV$ & $3000$ fb$^{-1}$ \\
VLHC & $100 \,\TeV$ & $3000$ fb$^{-1}$
\end{tabular}
\end{center}

The results presented in this work use the common Snowmass backgrounds \cite{Avetisyan:2013onh}, which were generated using the Open Science Grid \cite{Avetisyan:2013dta}.  The Snowmass detector framework \cite{Anderson:2013kxz} was used for signal and background event reconstruction. QCD backgrounds were not simulated as the preselection cuts have been demonstrated to effectively eliminate any QCD contamination.  Note that all results presented here are based on existing Monte Carlo and detector simulation tools extended to $33 \TeV$ and $100 \TeV$.   We do not investigate the uncertainties related to the extrapolation of parton distribution functions or the modeling of electroweak contributions to the parton shower at high collision energies (for some discussion of these issues, see the Snowmass report from the energy frontier QCD working group \cite{Campbell:2013qaa}).

While studies assuming center-of mass energies beyond $14 \TeV$ do exist, for example the famous EHLQ paper on SSC collider physics \cite{Eichten:1984eu}, the results presented here represent some of the first computations that have been done using modern Monte Carlo and detector simulation tools. This work is a broad first step in the realistic assessment of the capabilities of future proton colliders for new particle searches, This work, along with other Snowmass 2013 studies of new physics searches at $33\TeV$ \cite{Agashe:2013fda,Avetisyan:2013rca,Bhattacharya:2013iea,Varnes:2013pxa} and $100\TeV$ \cite{Andeen:2013zca,Apanasevich:2013cta,Degrande:2013yda,Stolarski:2013msa,Yu:2013wta,Zhou:2013raa} colliders, provides a useful reference for evaluating future experimental options and a launching point for further detailed investigation. 

The rest of the paper is organized as follows. Section \ref{sec:Validation} compares our results with an official $14$ TeV ATLAS study.  The remaining sections are divided by the particular choice of Simplified Model, with separate sections for the searches targeted in the compressed regions of parameter space.  Sections \ref{sec:GNLightFlavor}-\ref{sec:GluinoSquarkNeutralino} describe the searches and results for Simplified Models with hadronic final states, with the details of the common analyses and the impact of pile-up and systematics discussed in the context of the gluino-neutralino model in Sec.~\ref{sec:GNLightFlavor} and for compressed spectra in Sec.~\ref{sec:GNCompressed}. Section \ref{sec:GNHeavyFlavor} presents the analysis and sensitivity of a leptonic search for the gluino-neutralino model with heavy flavor decays. An appendix provides the details of the Monte Carlo framework employed for this study.

A companion paper \cite{Cohen:2013zla} provides a summary of the results and lessons learned.   Its purpose is to emphasize the compelling case for future proton colliders.

\section{Validation}
\label{sec:Validation}
In order to validate our event generation and weighting procedures, we have made a comparison with an ATLAS study on the capabilities of the high luminosity 14 TeV LHC \cite{ATL-PHYS-PUB-2013-002}.  Specifically, ATLAS provides distributions for benchmark points in the gluino-squark plane with 
\NegSpace
\begin{itemize}
\item $m_{\widetilde{g}} = 3200 \GeV$; $m_{\widetilde{q}} = 3200 \GeV$ 
\item $m_{\widetilde{g}} = 2800 \GeV$; $m_{\widetilde{q}} = 2400 \GeV$ 
\end{itemize}
\NegSpace
where the following requirements are enforced: $\MET/\sqrt{H_T} > 15 \GeV^{1/2}$, no leptons, and four jets with $p_T > 60 \GeV$.

\begin{figure}[!htb]
\begin{center}
\includegraphics[width=0.45\textwidth]{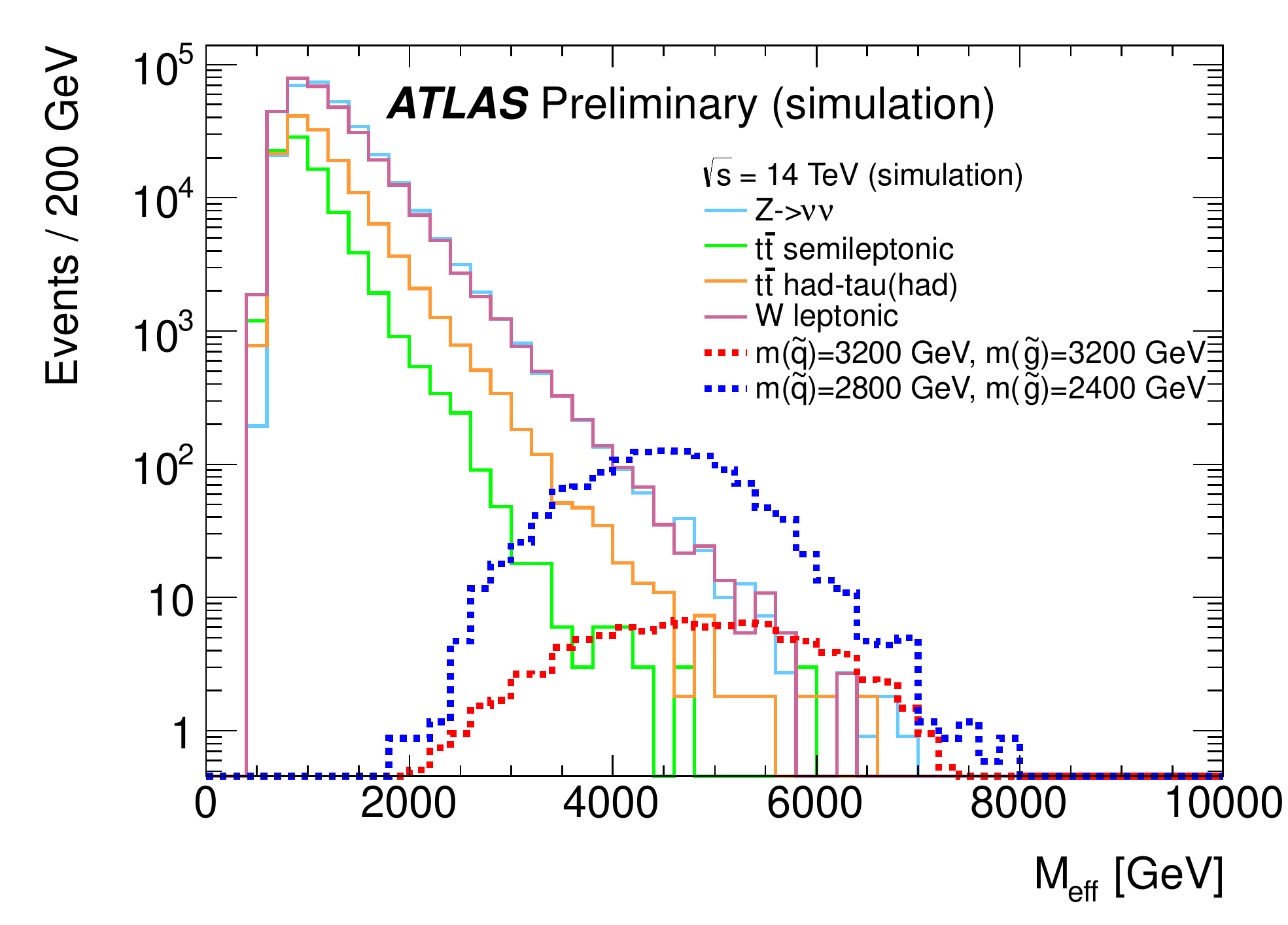}
\includegraphics[width=0.45\textwidth]{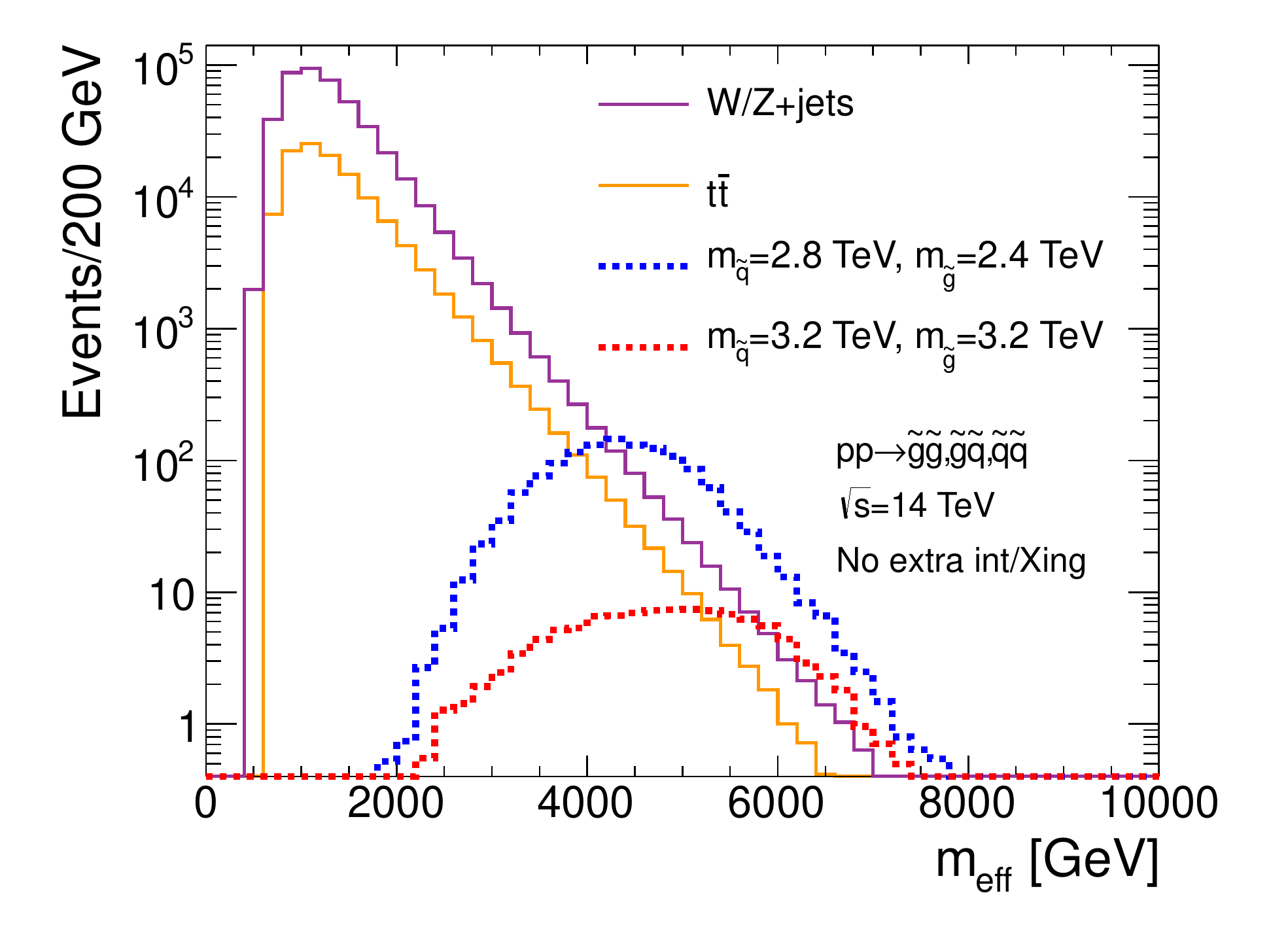} 
\caption{Plotted on the left [right] is the histogram of \meff~for the validation model as determined by ATLAS [this study].  Note that an exact comparison between the background not possible since we use the Snowmass particle containers, \emph{i.e.}~$W$ + jets and $Z$ + jets are plotted together. The background histograms are not stacked.  }
\label{fig:validation}
\end{center}
\end{figure}

In the left panel of Fig.~\ref{fig:validation} we show the \meff~distribution for signal and backgrounds from \cite{ATL-PHYS-PUB-2013-002} and on the right we show our analogous distribution.  The signal distributions are the same within the tolerance of the systematic uncertainty assumed below.   We find that the parts of the distributions within the cut regions for our analyses appear to be consistent to within 10\%.  The $\MET/\sqrt{H_T}$ distribution, also provided in \cite{ATL-PHYS-PUB-2013-002}, is also  consistent with our results.  Finally, we note that while we use slightly different search strategies (and have used different detector simulations) we obtain similar results for the gluino-squark plane presented below in Sec.~\ref{sec:GluinoSquarkNeutralino}.

\pagebreak
\section{The Gluino-Neutralino Model with Light Flavor Decays}
\label{sec:GNLightFlavor}
In the ``gluino-neutralino model with light flavor decays", the gluino $\widetilde{g}$ is the only kinematically accessible colored particle. The squarks are completely decoupled and do not contribute to gluino production diagrams. The gluino undergoes a prompt three-body decay through off-shell squarks,  $\widetilde{g} \rightarrow q\,\overline{q}\,\widetilde{\chi}^0_1$, where $q = u,d,c,s$ are the first and second generation quarks and $\widetilde{\chi}^0_1$ is a neutralino LSP. The branching ratios to all four flavors of light quark are taken to be equal. The only two relevant parameters are the gluino mass $m_{\widetilde{g}}$ and the neutralino mass $m_{\widetilde{\chi}^0_1}$.  This model can be summarized by:
\begin{center}
\NegSpace
\renewcommand{\arraystretch}{1.3}
\setlength{\tabcolsep}{12pt}
\begin{tabular}{c|c|c}
BSM particles & production & decays \\
\hline
$\widetilde{g},\,\widetilde{\chi}^0_1$ & $p\,p\rightarrow \widetilde{g}\, \widetilde{g}$ & $\widetilde{g} \rightarrow q\,\overline{q}\,\widetilde{\chi}^0_1 $ 
\end{tabular}
\end{center}

This model is motivated by (mini-)split supersymmetry scenarios, where the scalar superpartners are heavier than the gauginos \cite{Wells:2004di, ArkaniHamed:2004fb, Giudice:2004tc, Arvanitaki:2012ps, ArkaniHamed:2012gw}.  The final state is four (or more) hard jets and missing energy.  Therefore, this signature provides a good proxy with which to investigate the power of a traditional jets plus $\MET$ style hadron collider search strategy to discriminate against background.  The current preliminary limits on this model using $20$ fb$^{-1}$ of $8$ TeV data are $m_{\widetilde{g}} = 1350 \text{ GeV}$ (ATLAS \cite{ATLAS-CONF-2013-047}) and $m_{\widetilde{g}} = 1200 \text{ GeV}$ (CMS \cite{CMS-PAS-SUS-13-012}) assuming a massless neutralino.

We simulated matched \texttt{MadGraph} samples for $\widetilde{g}\,\widetilde{g}$ with up to 2 additional generator level jets for the following points in parameter space:\footnote{We include $1\GeV$ for an example where the neutralino is effectively massless; the second line of neutralino masses is chosen to cover the bulk of the gluino-neutralino plane; the final line is chosen to ensure coverage in the ``compressed" region.}  
\begin{center}
\NegSpace
\renewcommand{\arraystretch}{1.3}
\setlength{\tabcolsep}{12pt}
\begin{tabular}{c|l}
BSM particles & masses \\
\hline
\hline
$m_{\widetilde{g}}$ $\big[14 \TeV\big]$ & $ (315, 397, 500, 629, 792, 997, 1255, 1580, 1989, 2489, 2989, 3489) \GeV$   \\
$m_{\widetilde{g}}$ $\big[33 \TeV\big]$ & $ (500, 629, 792, 997, 1255, 1580, 1989, 2504, 3152, $ \\
 & $3968, 4968, 5968, 6968) \GeV $\\
$m_{\widetilde{g}}$ $\big[100 \TeV\big]$ & $ (1000, 1259, 1585, 1995, 2512, 3162, 3981, 5012, 6310,$   \\
 &  $7944, 9944, 11944, 13944, 15944) \GeV$ \\
\hline
					    & $1\GeV$ \\
$m_{\widetilde{\chi}^0_1}$  & $(0.2, 0.4, 0.6, 0.7, 0.8 , 0.9)\times m_{\widetilde{g}}$ \\
					    & $m_{\widetilde{g}}-(100\GeV, 50\GeV,15\GeV,5\GeV)$
\end{tabular}
\end{center}
We find that including pileup does not significantly change the results of this study and present results below for only the no-pileup case. We discuss the effect of pile-up in more detail in Sec.~\ref{sec:pileupGOGO}.

\subsection{Dominant Backgrounds}
\label{sec:GOGO_Backgrounds}
The background is dominated by $W/Z + \text{jets}$, with subdominant contributions from $t\,\overline{t}$ production.  Single top events and $W/Z$ events from vector boson fusion processes are also illustrated in several figures, and are negligible.  In all cases, there are decay modes which lead to multi-jet signatures. The \MET~can come from a variety of sources, such as neutrinos, jets/leptons which are lost down the beam pipe, and energy smearing effects. 

\subsection{Analysis Strategy}
\label{sec:GOGO_Strategy}
The gluino-neutralino model with light flavor decays can be probed with an analysis inspired by  the ATLAS analysis in~\cite{ATL-PHYS-PUB-2013-002}.  After an event preselection, rectangular cuts on one or more variables are optimized at each point in parameter space to yield maximum signal significance.  Specifically, we simultaneously scan a two-dimensional set of cuts on \MET\ and \HT, where \MET\ is the magnitude of the missing transverse momentum and \HT\ is defined as the scalar sum of jet \pt\,. In contrast, the discriminating variable used by ATLAS is \meff, the scalar sum of \HT\ and \MET.  We require jets to have $p_T>30$ GeV and $|\eta|<3.5$.  Electrons and muons are selected by requiring $p_T > 10$ GeV and $|\eta |<2.6$.  

In detail, our analysis strategy proceeds as follows:

\textbf{\textsc{Preselection}}
\NegSpace
\begin{itemize}
\item zero selected electrons or muons
\item $\MET > 100$ GeV
\item at least 4 jets with $p_T > 60$ GeV
\end{itemize}
\NegSpace

After preselection, a requirement is placed on $\MET/\sqrt{\HT}$ to further reduce the QCD background, where the dominant source of \MET\,is from jet mismeasurement.  A cut on the leading jet $p_T$ is applied to reduce backgrounds from hard ISR jets.  Finally, a two dimensional scan over cuts on \HT~and \MET is performed to determine the maximum significance.

\textbf{\textsc{Search Strategy}: Simultaneous optimization over \HT~and \MET}
\NegSpace
\begin{itemize}
\item $\MET/\sqrt{\HT} > 15$ GeV$^{1/2}$
\item The leading jet $p_T$ must satisfy $p_T^\text{leading} < 0.4\, \HT$
\item $\MET > (\MET)_\text{optimal}$
\item $\HT > (\HT)_\text{optimal}$
\end{itemize}

The $14$, $33$, and $100$ TeV analyses all use the same set of fixed cuts, differing only in the optimization over $(\MET)_\text{optimal}$ and $(\HT)_\text{optimal}$. In practice, scaling the $\MET/\sqrt{\HT}$ cut with CM energy may be desirable to reduce QCD background, and we have verified that this has no effect on the efficiencies for the signal and dominant backgrounds for the models under study.  

Note that this search yields some power to discover these models in the difficult region of parameter space where the neutralino is degenerate with the gluino.  Section~\ref{sec:GNCompressed} will provide the results of a search that is specifically targeted for this region of parameter space.

\begin{figure}[!tb]
\begin{center}
    \includegraphics[width=0.45\textwidth]{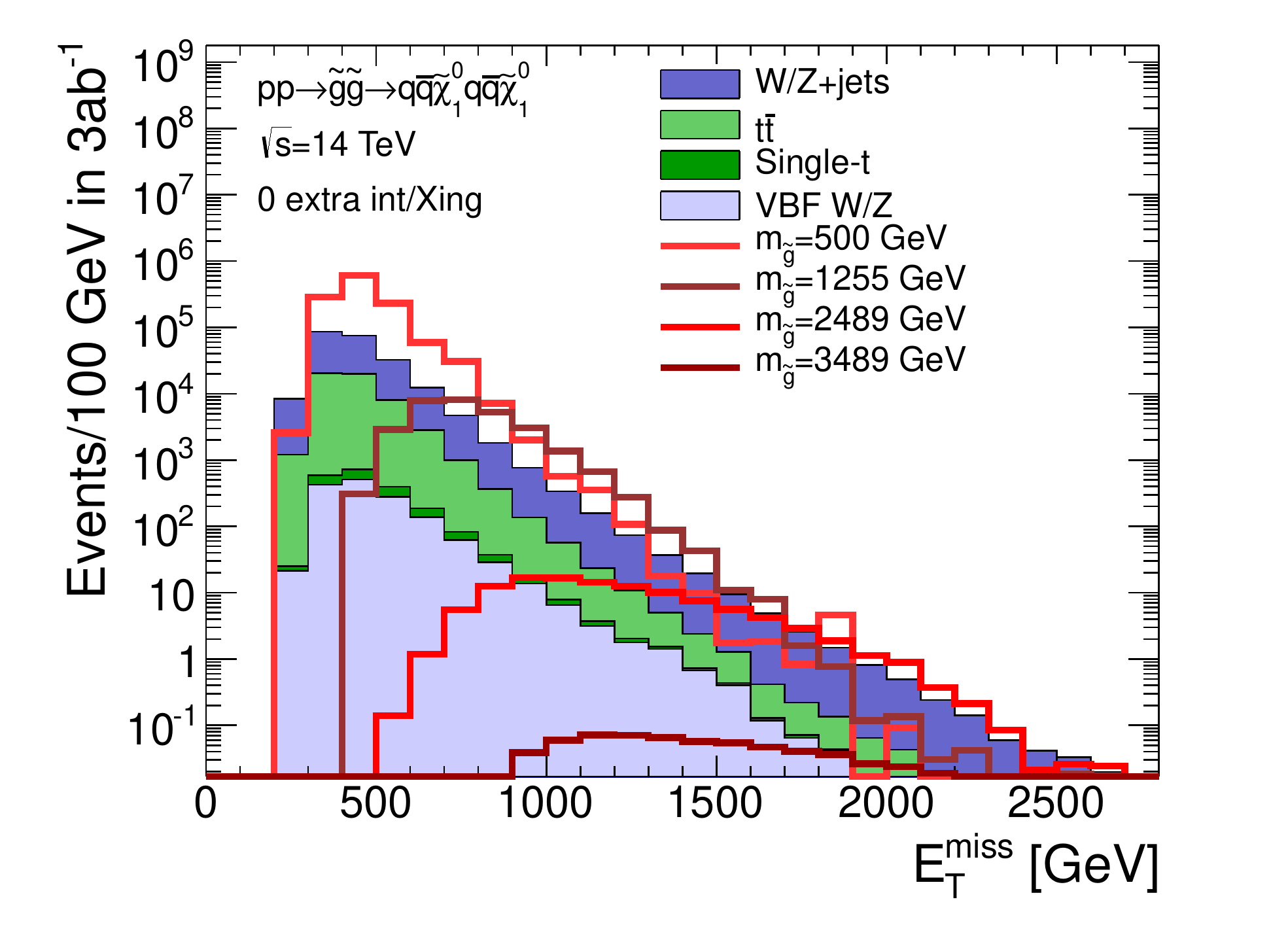}
    \includegraphics[width=0.45\textwidth]{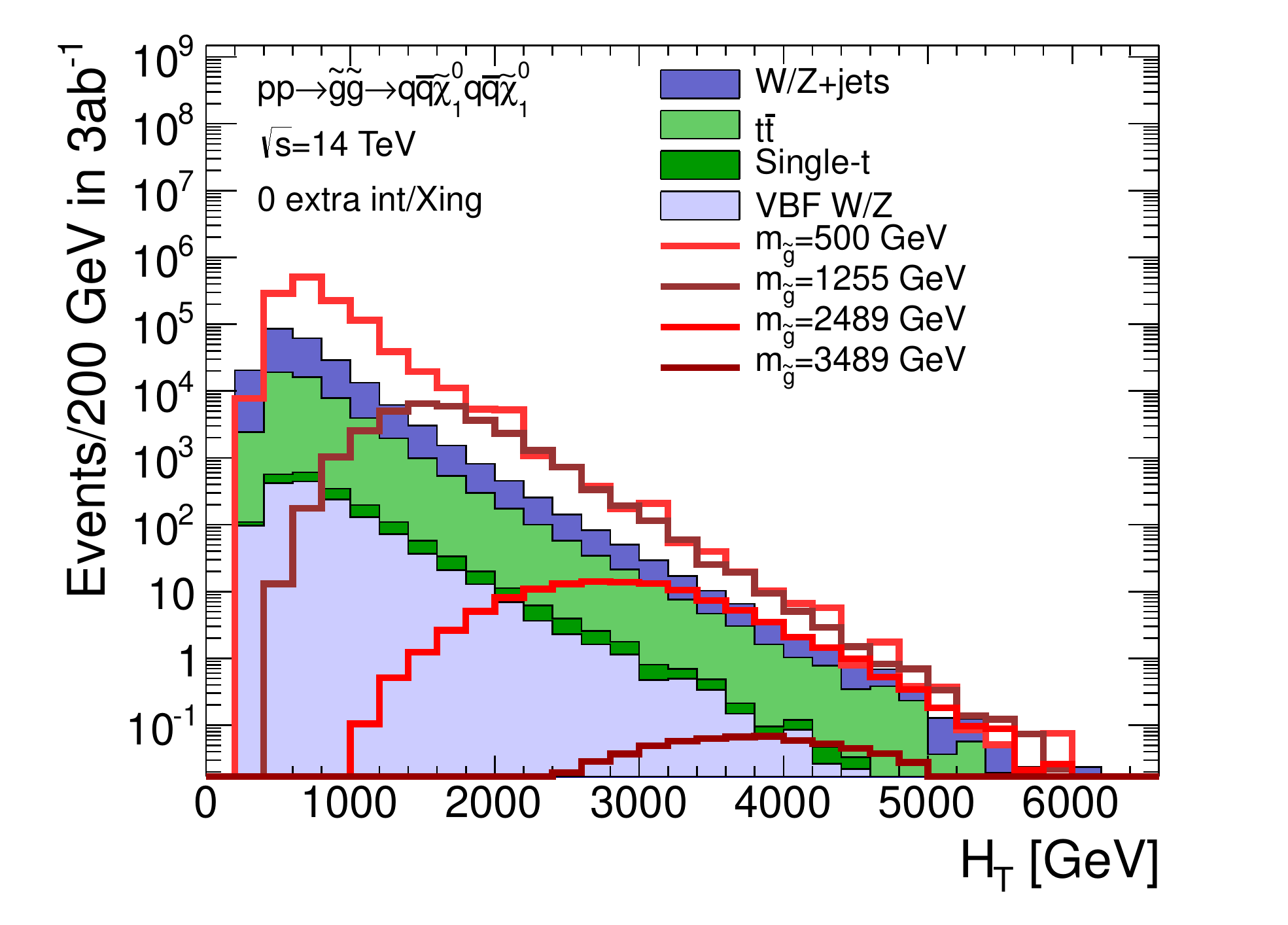}
\caption{Histogram of \MET~[left] and \HT~[right] after preselection cuts for background and a range of gluino-neutralino models with light flavor decays at the $14$ TeV LHC.  The neutralino mass is $1\GeV$ for all signal models.}
\label{fig:GOGO_alljet_presel_distributions}
\end{center}
\end{figure}

\hiddensubsection{Analysis: 14 TeV}
\addcontentsline{toc}{subsection}{3.3-3.9 $\,\,\,\,$ Analysis and Results: $14$, $33$, and $100$ TeV}
In Fig.~\ref{fig:GOGO_alljet_presel_distributions} we show histograms of $\MET$ [left] and \HT~[right] for signal and background at $\sqrt{s}=14\TeV$ after applying the preselection cuts listed in Sec.~\ref{sec:GOGO_Strategy}. Because the tails of the signal and background distributions have a similar slope, the optimization procedure generally leads to  cuts near the bulk of the signal distribution. We find that the searches are systematics limited when the optimal cuts are applied (see Sec.~\ref{sec:GOGO_systematics} below for a detailed discussion).

Using the search strategy discussed above in Sec.~\ref{sec:GOGO_Strategy}, it is possible to explore the potential reach for the gluino-neutralino model at the 14 TeV LHC.  Table~\ref{tab:GNbulkCounts} gives a few example of the number of events that result from this cut flow for background and three example signal points: $m_{\widetilde{g}} = 500,\,1255,\text{ and } 2489 \GeV$ with $m_{\widetilde{\chi}^0_1} = 1 \GeV$.  Each choice of the \MET~and \HT~cuts given in Table~\ref{tab:GNbulkCounts} corresponds to the optimal cut for one of the given signal points.  The hardness of the cut increases with the mass of the gluino.  Note that for $14$ TeV proton collisions, the $V$+jets background dominates over the events from $t\,\overline{t}$.  From this table, it is possible to infer that $500 \GeV$ and $1255 \GeV$ gluinos could be easily discovered while the $2489 \GeV$ would only yield a few $\sigma$ hint using the full power of the high luminosity LHC. 

\begin{table}[h!]
\renewcommand{\arraystretch}{1.4}
\setlength{\tabcolsep}{7pt}
\footnotesize
\vskip 10pt
  \begin{centering}
    \begin{tabular}{| r | r r | r | r  r  r |}
      \hline
      &&&&\multicolumn{3}{c|}{$m_{\widetilde{g}}$ [GeV]}\\
      Cut                                                      &$V$+jets                 &$t\bar{t}$           &Total BG&                     500&                     1255&                     2489  \\
      \hline
      \hline
      $\mathrm{Preselection}$                                  &$2.07 \times 10^{7}$     &$2.47 \times 10^{7}$  &$4.54 \times 10^{7}$&$3.08 \times 10^{7}$&$1.03 \times 10^{5}$&                      173  \\
      \hline
      $\MET/\sqrt{H_{T}} > 15 \text{ GeV}^{1/2}$  &$4.45 \times 10^{5}$     &$1.20 \times 10^{5}$  &$5.65 \times 10^{5}$&$1.34 \times 10^{6}$&$3.14 \times 10^{4}$&                       95            \\
      $p_{\mathrm{T}}^{\mathrm{leading}} < 0.4\times \HT$&$1.69 \times 10^{5}$     &$5.16 \times 10^{4}$   &$2.21 \times 10^{5}$&$7.62 \times 10^{5}$&$1.68 \times 10^{4}$&                     52.9 \\
      \hline
      \hline
$\MET > 450$ GeV&\multirow{2}{*}{$4.73 \times 10^{4}$}&\multirow{2}{*}{$1.84 \times 10^{4}$}&\multirow{2}{*}{$6.57 \times 10^{4}$}&\multirow{2}{*}{\color{red}$5.57 \times 10^{5}$}&\multirow{2}{*}{$2.98 \times 10^{4}$}&\multirow{2}{*}{                      115}\\
$\HT > 800$ GeV          &&&&&&\\
\hline
$\MET > 800$ GeV&\multirow{2}{*}{$1.22 \times 10^{3}$}&\multirow{2}{*}{                      554}&\multirow{2}{*}{$1.78 \times 10^{3}$}&\multirow{2}{*}{$1.14 \times 10^{4}$}&\multirow{2}{*}{\color{red}$9.36 \times 10^{3}$}&\multirow{2}{*}{                      110}\\
$\HT > 1650$ GeV         &&&&&&\\
\hline
$\MET > 1050$ GeV&\multirow{2}{*}{                     55.5}&\multirow{2}{*}{                     30.1}&\multirow{2}{*}{                     85.6}&\multirow{2}{*}{                      297}&\multirow{2}{*}{                      288}&\multirow{2}{*}{                   \color{red}  57.2}\\
$\HT > 2600$ GeV         &&&&&&\\
\hline
    \end{tabular}
    \caption{Number of expected events for $\sqrt{s} = 14$ TeV with 3000 fb$^{-1}$ integrated luminosity for the background processes and selected gluino masses for the gluino-neutralino model with light flavor decays.  The neutralino mass is $1 \GeV$.  Three choices of cuts on \MET~and \HT~are provided, and for each gluino mass column the entry in the row corresponding to the ``optimal" cuts is marked in red.
    }
    \label{tab:GNbulkCounts}
  \end{centering}
\end{table}

\begin{figure}[h!]
  \centering
    \includegraphics[width=.48\columnwidth]{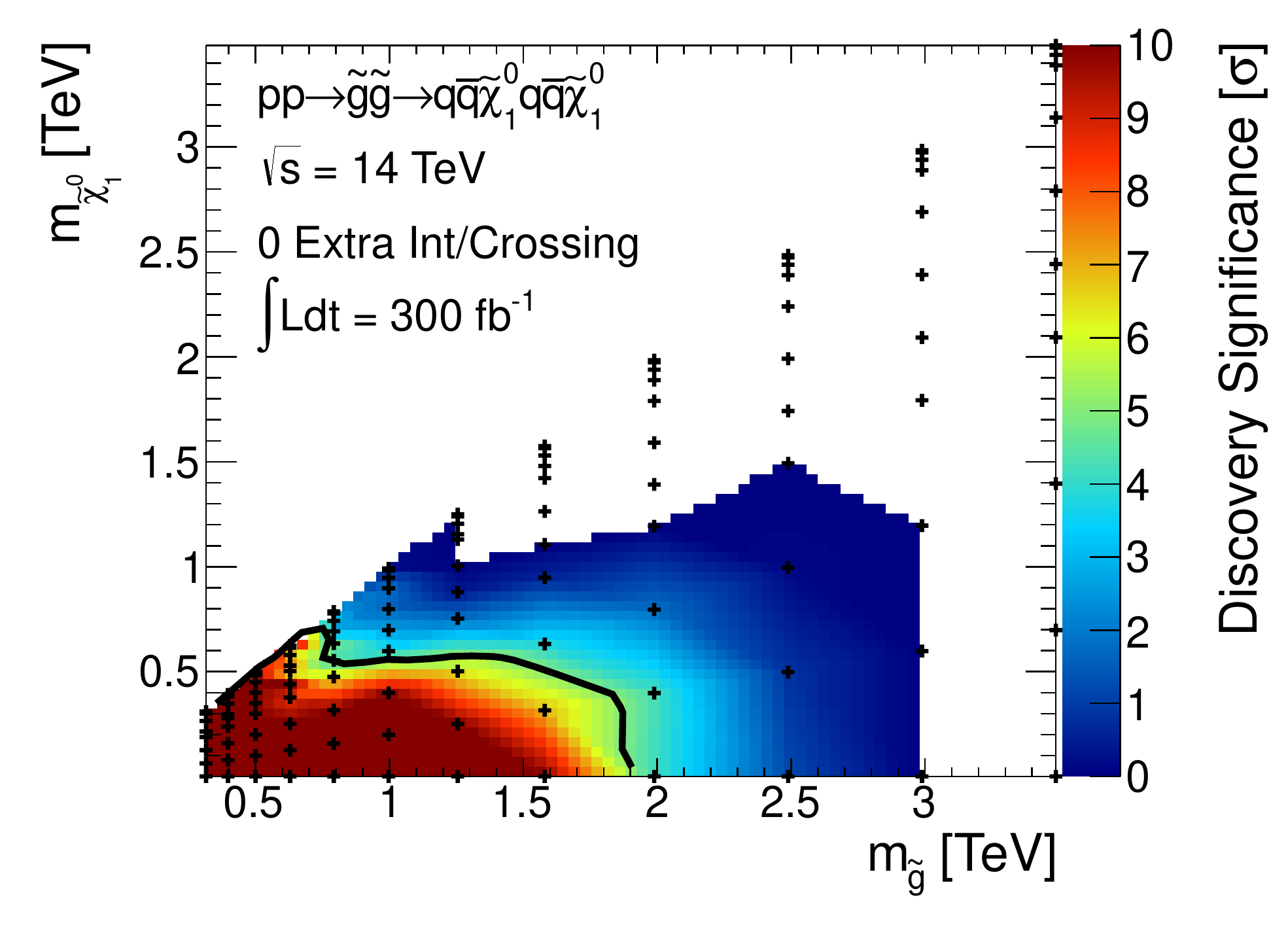}
  \includegraphics[width=.48\columnwidth]{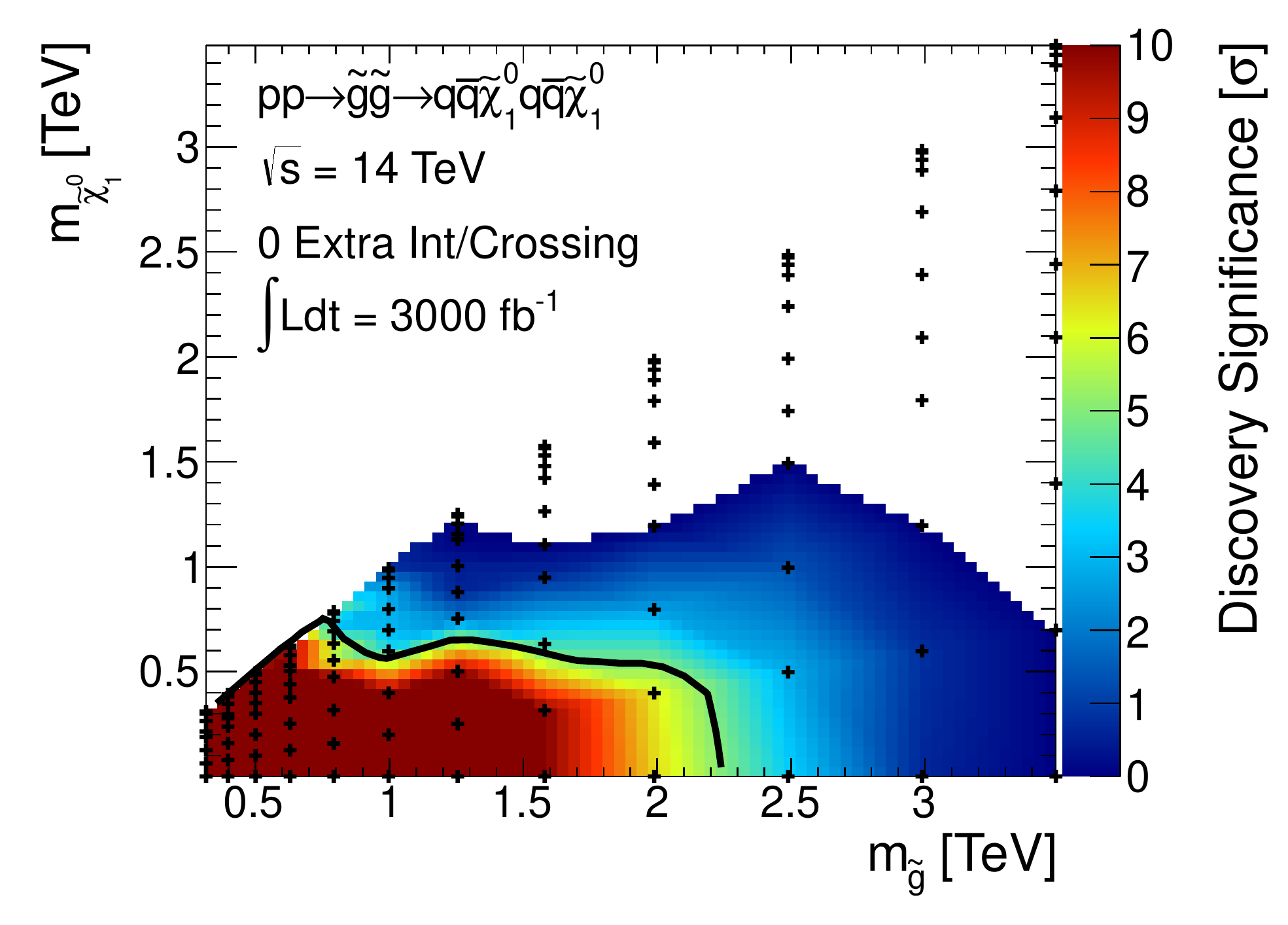}
    \includegraphics[width=.48\columnwidth]{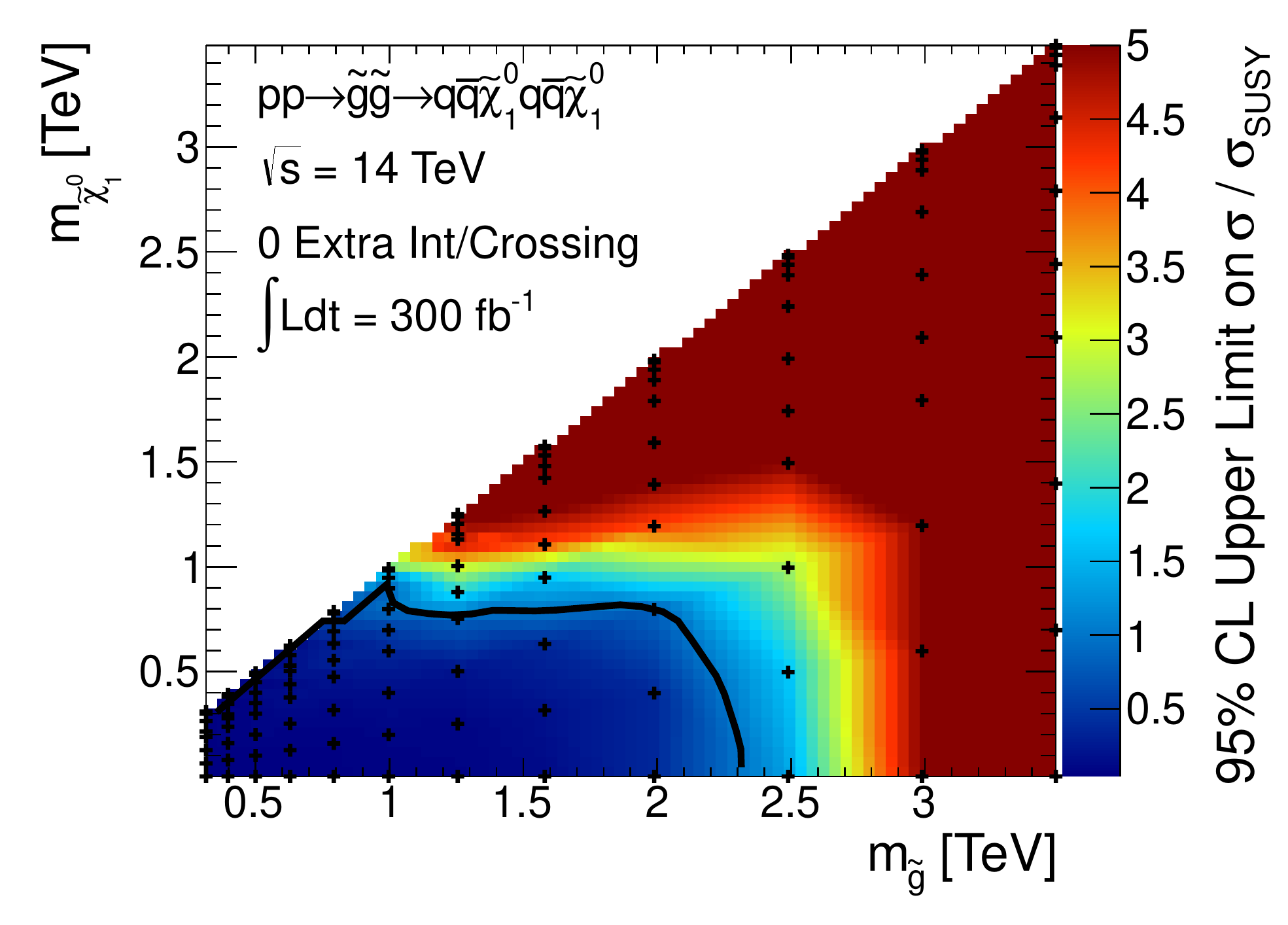}
  \includegraphics[width=.48\columnwidth]{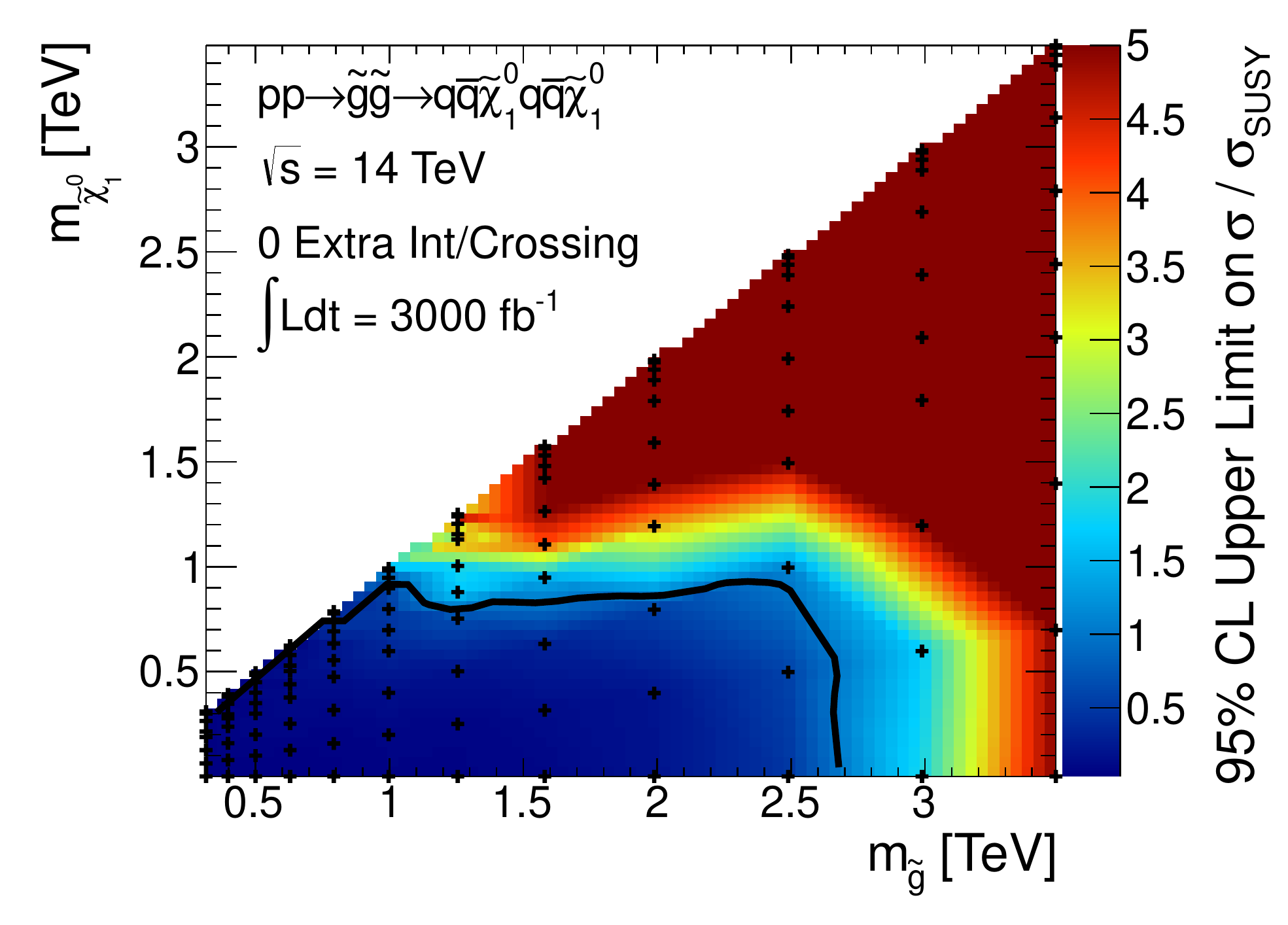}
  \caption{Results at $14 \TeV$ for the gluino-neutralino model with light flavor decays are given in the $m_{\widetilde{\chi}^0_1}$ versus $m_{\widetilde{g}}$ plane.  The top [bottom] row shows the expected $5\sigma$ discovery reach [95\% confidence level upper limits] for gluino pair production.  Mass points to the left/below the contours are expected to be probed at $300 \text{ fb}^{-1}$ [left] and $3000 \text{ fb}^{-1}$ [right].  A 20\% systematic uncertainty is assumed and pileup is not included.  The black crosses mark the simulated models.}
\label{fig:GOGO_14_NoPileUp_results}
\end{figure}

\pagebreak
\hiddensubsection{Results: 14 TeV}
\label{sec:GOGO_Results14TeV}
The $5\sigma$ discovery [top] and $95\%$ CL~limits [bottom] for the gluino-neutralino model are shown in Fig.~\ref{fig:GOGO_14_NoPileUp_results}.  The left [right] panel assumes $300 \text{ fb}^{-1}$ $\left[3000 \text{ fb}^{-1}\right]$ of integrated luminosity.  The signal and background yields after optimized cuts are provided as inputs to a \verb.RooStats. routine to calculate 95\% CL exclusion intervals using the CL$_s$ method along with the expected signal $p_0$ values.  A $20\%$ systematic uncertainty is applied to the backgrounds.  The assumed signal systematics are outlined in the appendix.  

Using the NLO gluino pair production cross section one can make a very naive estimate for the reach of a given collider.  For example, we find that the choice of gluino mass which would yield $10$ events at $300 \text{ fb}^{-1}$ $\big(3000 \text{ fb}^{-1}\big)$ is $2.8\,(3.3) \TeV$.  This roughly corresponds to the maximal possible reach one could expect for a given luminosity using $14 \TeV$ proton collisions.  

Using a realistic simulation framework along with the search strategy employed here, the $14 \TeV$ $300 \text{ fb}^{-1}$ limit with massless neutralinos is projected to be at a gluino mass of $2.3 \TeV$ (corresponding to 110 events), while the $14 \TeV$ $3000 \text{ fb}^{-1}$ limit is projected to be at $2.7 \TeV$ (corresponding to 175 events).  Furthermore, the $14 \TeV$ LHC with $3000 \text{ fb}^{-1}$ could discover a gluino as heavy as $2.3\TeV$ if the neutralino is massless, while for $m_{\widetilde{\chi}^0_1}\gtrsim 500 \GeV$ the gluino mass reach rapidly diminishes.

\begin{figure}[!h]
\begin{center}
    \includegraphics[width=0.45\textwidth]{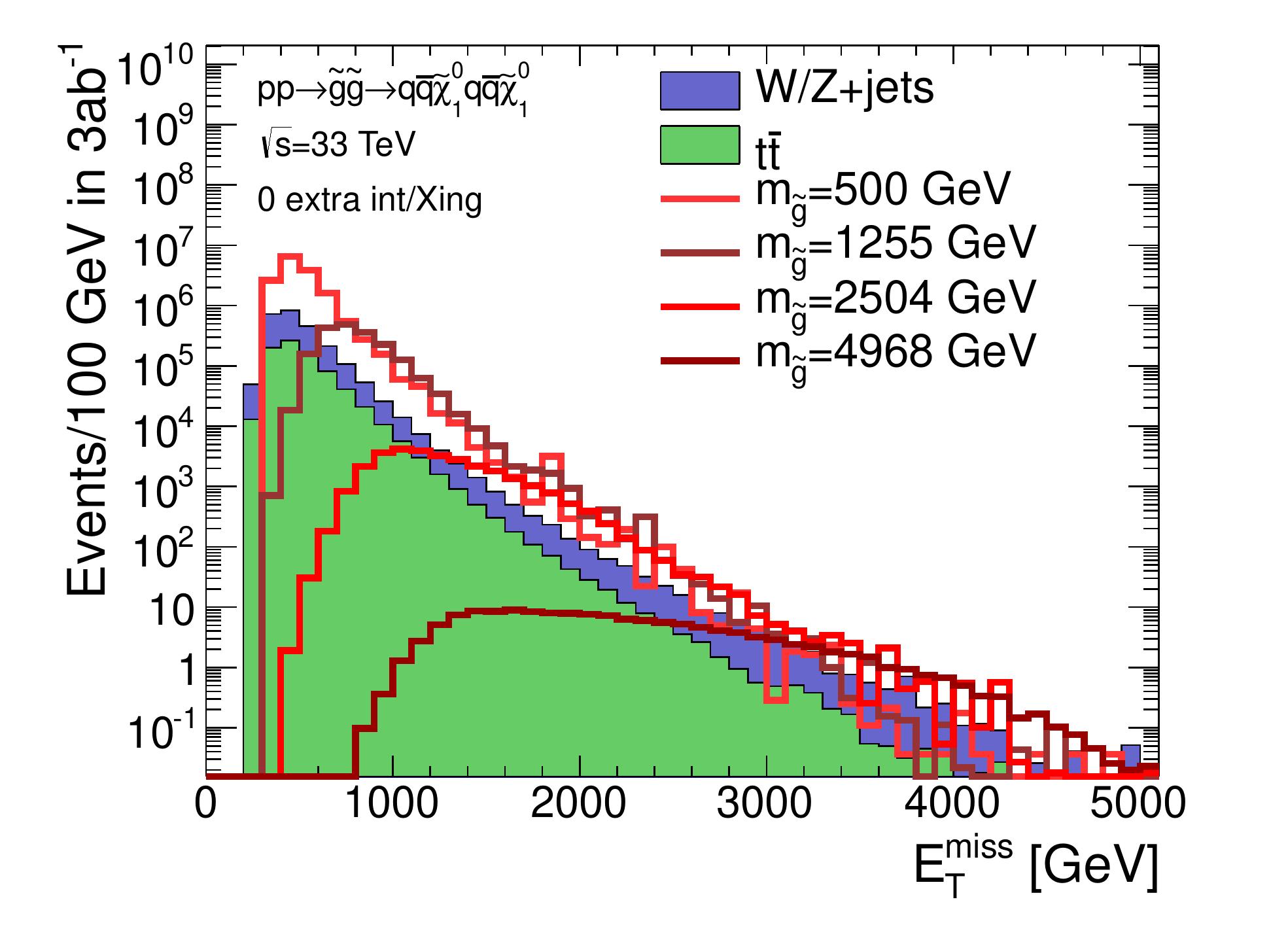}
    \includegraphics[width=0.45\textwidth]{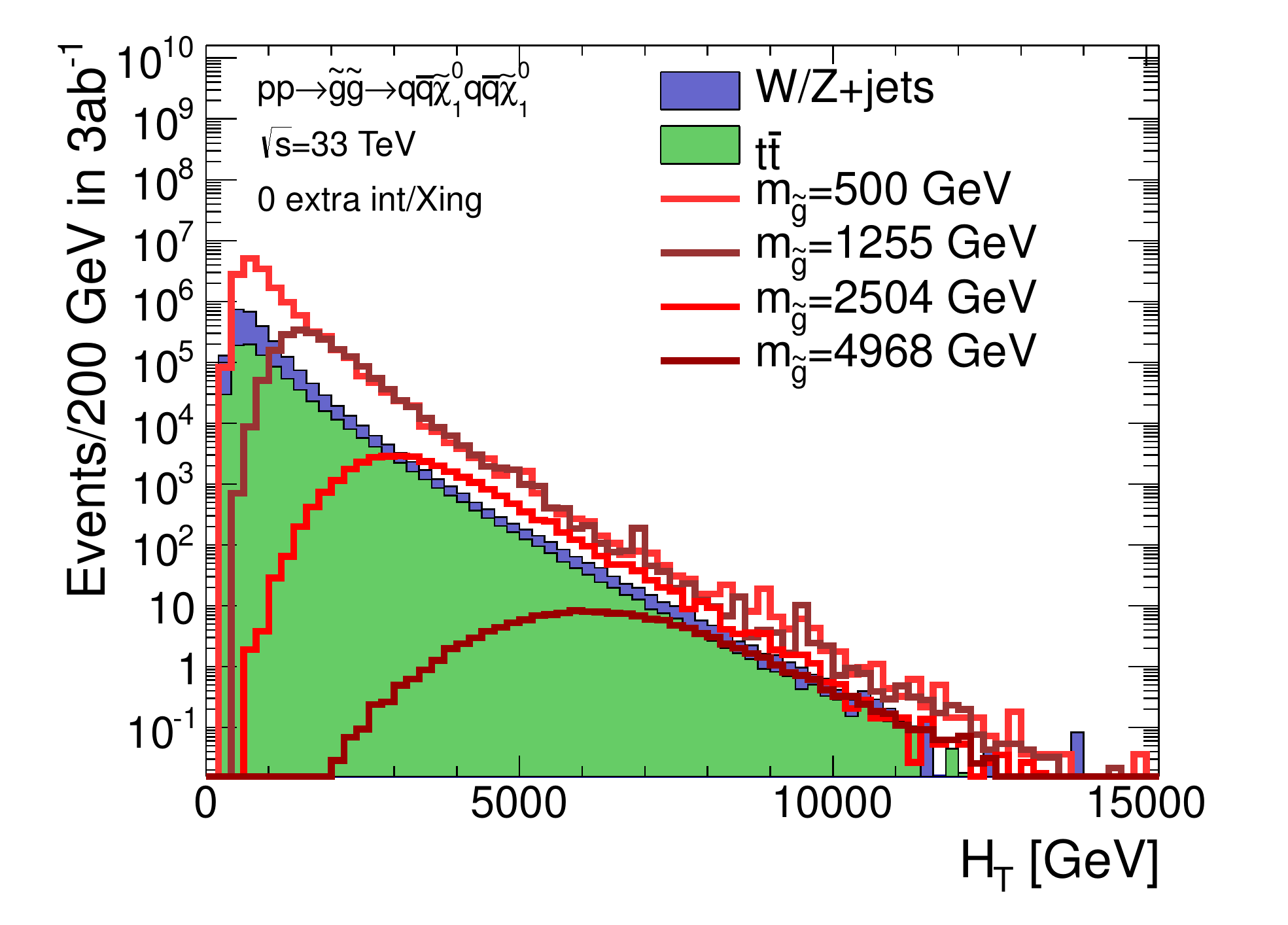}
\caption{Histogram of \MET~[left] and \HT~[right] after preselection cuts for background and a range of gluino-neutralino models with light flavor decays at a $33$ TeV proton collider.  The neutralino mass is $1\GeV$ for all signal models.}
\label{fig:GOGO_alljet_presel_distributions_33TeV}
\end{center}
\end{figure}

\pagebreak
\hiddensubsection{Analysis: 33 TeV}
In Fig.~\ref{fig:GOGO_alljet_presel_distributions_33TeV} we show histograms of $\MET$ [left] and \HT~[right] for signal and background at $\sqrt{s}=33\TeV$ after applying the preselection cuts listed  in Sec.~\ref{sec:GOGO_Strategy}. Because the tails of the signal and background distributions have a similar slope, the optimization procedure generally leads to cuts near the bulk of the signal distribution. Moving to a higher center-of-mass energy allows for harder cuts to be placed, which in turn implies fewer background events survive the requirements.   At $33$ TeV, we find that the searches are systematics limited when the optimal cuts are applied (see Sec.~\ref{sec:GOGO_systematics} below for a detailed discussion).  

Using the search strategy discussed above in Sec.~\ref{sec:GOGO_Strategy}, it is possible to explore the potential reach for the gluino-neutralino model at a $33$ TeV proton collider.  Table~\ref{tab:GNbulkCounts33TeV} gives a few example of the number of events that result from this cut flow for background and three example signal points: $m_{\widetilde{g}} = 1255,\, 3152,\text{ and } 4968 \GeV$ with $m_{\widetilde{\chi}^0_1} = 1 \GeV$.  Each choice of the \MET~and \HT~cuts given in Table~\ref{tab:GNbulkCounts33TeV} corresponds to the optimal cuts for one of the given signal points.    The hardness of the cut increases with the mass of the gluino.  It is clear that the ratio of $t\,\overline{t}$ to $V+$jets background is growing with regards to the $14$ TeV search; this is due to the higher probability for gluon scattering as $\sqrt{s}$ increases.  It would likely be advantageous to veto $b$-tagged jets to further reduce the background from top quarks, and similarly to veto $\tau$-tagged jets to further reduce $W/Z$+jets. From this table, it is possible to infer that gluinos as heavy as $\sim 5$ TeV could be discovered at a $33$ TeV collider.  \HT\ cuts as hard as $5$ TeV are required to extract the most information from this data set.

\begin{table}[h!]
  \renewcommand{\arraystretch}{1.4}
  \setlength{\tabcolsep}{5pt}
  \footnotesize
  \begin{centering}
    \begin{tabular}{| r | r r | r | r  r  r |}
\hline
&&&&\multicolumn{3}{c|}{$m_{\widetilde{g}}$ [GeV]}\\
                                               Cut&                 $V$+jets&               $t\overline{t}$&                 Total BG&                     $1255$&                     $3152$&                     $4968$\\
\hline
\hline
$                             \mathrm{Preselection}$&{$1.55 \times 10^{8}$}&{$2.86 \times 10^{8}$}&{$4.42 \times 10^{8}$}&$1.22 \times 10^{7}$&$1.89 \times 10^{4}$&                      $316$\\
\hline
$\MET/\sqrt{\HT} > 15 \GeV^{1/2}$&{$4.50 \times 10^{6}$}&{$1.93 \times 10^{6}$}&{$6.44 \times 10^{6}$}&$3.50 \times 10^{6}$&$1.13 \times 10^{4}$&                      229\\
$p_T^{\mathrm{leading}} < 0.4\times \HT$&{$1.70 \times 10^{6}$}&{$8.02 \times 10^{5}$}&{$2.50 \times 10^{6}$}&$1.94 \times 10^{6}$&$6.94 \times 10^{3}$&                      139\\
\hline
\hline
$          \MET > 900 \GeV$&\multirow{2}{*}{$4.06 \times 10^{4}$}&\multirow{2}{*}{$5.65 \times 10^{4}$}&\multirow{2}{*}{$9.71 \times 10^{4}$}&\multirow{2}{*}{\color{red}$8.36 \times 10^{5}$}&\multirow{2}{*}{$6.93 \times 10^{3}$}&\multirow{2}{*}{                      $139$}\\
$                         \HT> 1900 \GeV$&&&&&&\\
\hline
$         \MET > 2100 \GeV$&\multirow{2}{*}{                      $127$}&\multirow{2}{*}{                      $103$}&\multirow{2}{*}{                      $230$}&\multirow{2}{*}{$1.44 \times 10^{3}$}&\multirow{2}{*}{\color{red} $1.10 \times 10^{3}$}&\multirow{2}{*}{                     $78.5$}\\
$                         \HT > 3800 \GeV$&&&&&&\\
\hline
$         \MET > 2750 \GeV$&\multirow{2}{*}{                       $10$}&\multirow{2}{*}{                      $7.8$}&\multirow{2}{*}{                     $17.8$}&\multirow{2}{*}{                     $53.1$}&\multirow{2}{*}{                     $91.2$}&\multirow{2}{*}{            \color{red}         $33.6$}\\
$                        \HT > 5150 \GeV$&&&&&&\\
\hline
    \end{tabular}
    \caption{Number of expected events for $\sqrt{s} = 33$ TeV with 3000 fb$^{-1}$ integrated luminosity for the background processes and selected gluino masses for the gluino-neutralino model with light flavor decays.  The neutralino mass is $1 \GeV$.  Three choices of cuts on \MET~and \HT~are provided, and for each gluino mass column the entry in the row corresponding to the ``optimal" cuts is marked in red.
    }
    \label{tab:GNbulkCounts33TeV}
  \end{centering}
\end{table}

\pagebreak

\begin{figure}[h!]
  \centering
  \includegraphics[width=.48\columnwidth]{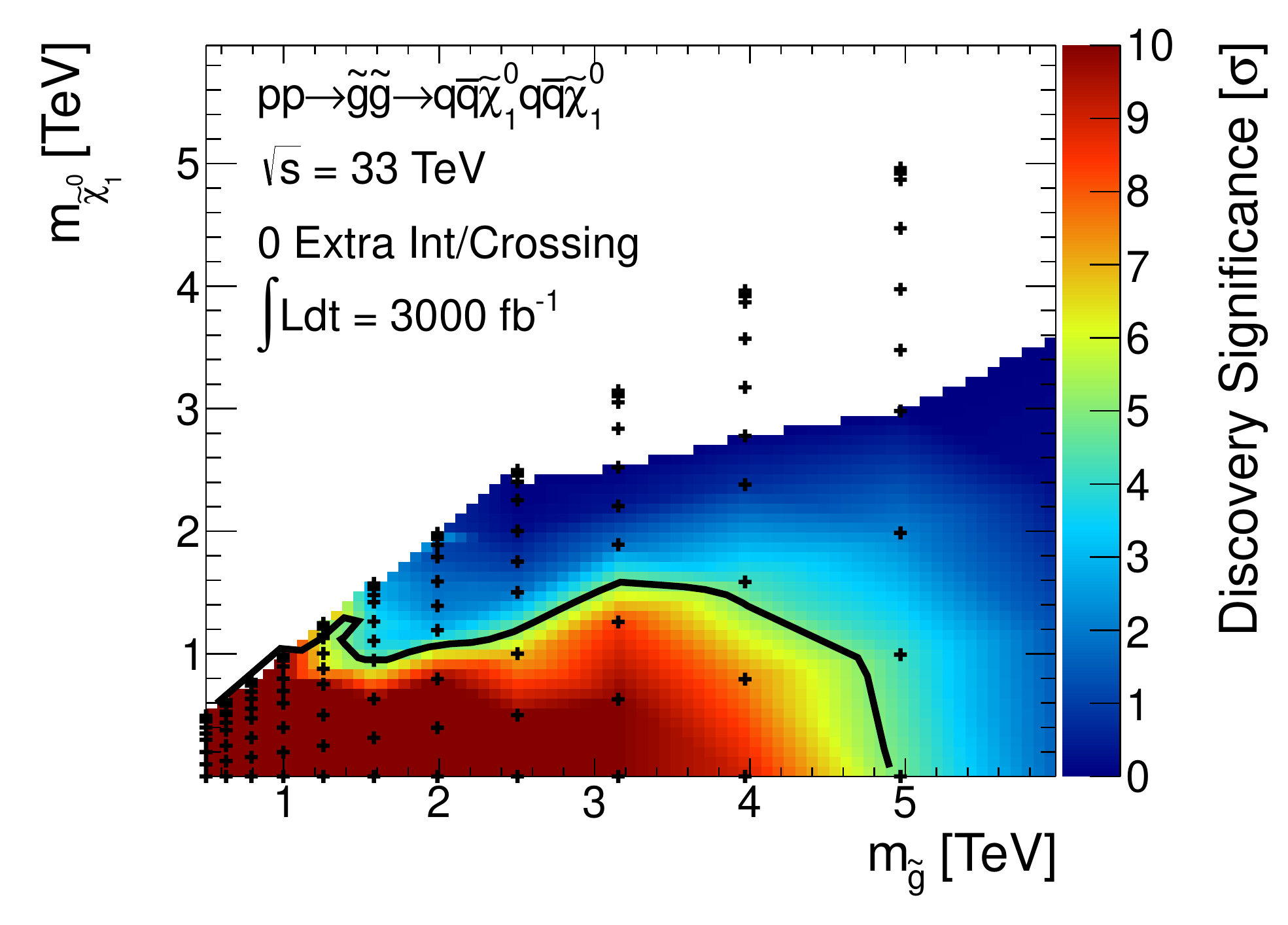}
  \includegraphics[width=.48\columnwidth]{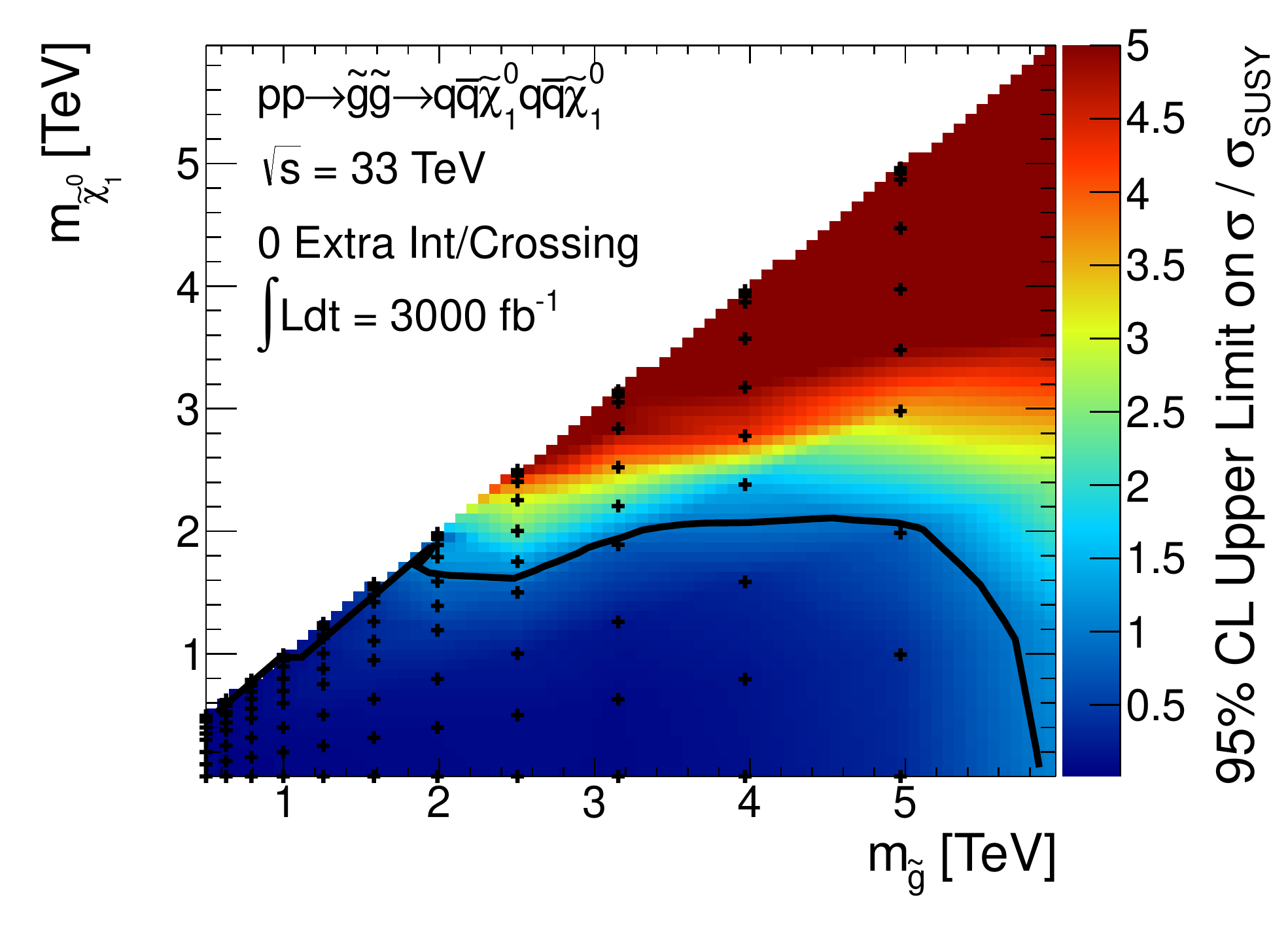}
  \caption{Results at $33 \TeV$ for the gluino-neutralino model with light flavor decays are given in the $m_{\widetilde{\chi}^0_1}$ versus $m_{\widetilde{g}}$ plane.  The left [right] plot shows the expected $5\sigma$ discovery reach [95\% confidence level upper limits] for gluino pair production.  Mass points to the left/below the contours are expected to be probed at $3000 \text{ fb}^{-1}$.  A 20\% systematic uncertainty is assumed and pileup is not included.  The black crosses mark the simulated models.}
\label{fig:GOGO_33_NoPileUp_results}
\end{figure}

\hiddensubsection{Results: 33 TeV}
The $5\sigma$ discovery [left] and $95\%$ C.L.~limits [right] for the gluino-neutralino model are shown in Fig.~\ref{fig:GOGO_33_NoPileUp_results}, assuming $3000 \text{ fb}^{-1}$ of integrated luminosity.  $20\%$ systematic uncertainty is applied to the backgrounds.  The assumed signal systematic are outlined in the appendix.  Pileup is not included; a demonstration that pileup will not significantly change these results is given in Sec.~\ref{sec:pileupGOGO} below.

Using the NLO gluino pair production cross section one can make a very naive estimate for the reach of a given collider.  For example, we find that the choice of gluino mass which would yield $10$ events at $3000 \text{ fb}^{-1}$ is $6.7 \TeV$.  This roughly corresponds to the maximal possible reach one could expect for a given luminosity using $33 \TeV$ proton collisions.  

Using a realistic simulation framework along with the search strategy employed here the $33 \TeV$ $3000 \text{ fb}^{-1}$ limit with massless neutralinos is projected to be $5.8 \TeV$ (corresponding to 61 events).  Furthermore, the $33 \TeV$ proton collider with $3000 \text{ fb}^{-1}$ could discover a gluino as heavy as $4.8\TeV$ if the neutralino is massless, while for $m_{\widetilde{\chi}^0_1}\gtrsim 1 \TeV$ the gluino mass reach rapidly diminishes. 

\begin{figure}[!h]
\begin{center}
    \includegraphics[width=0.45\textwidth]{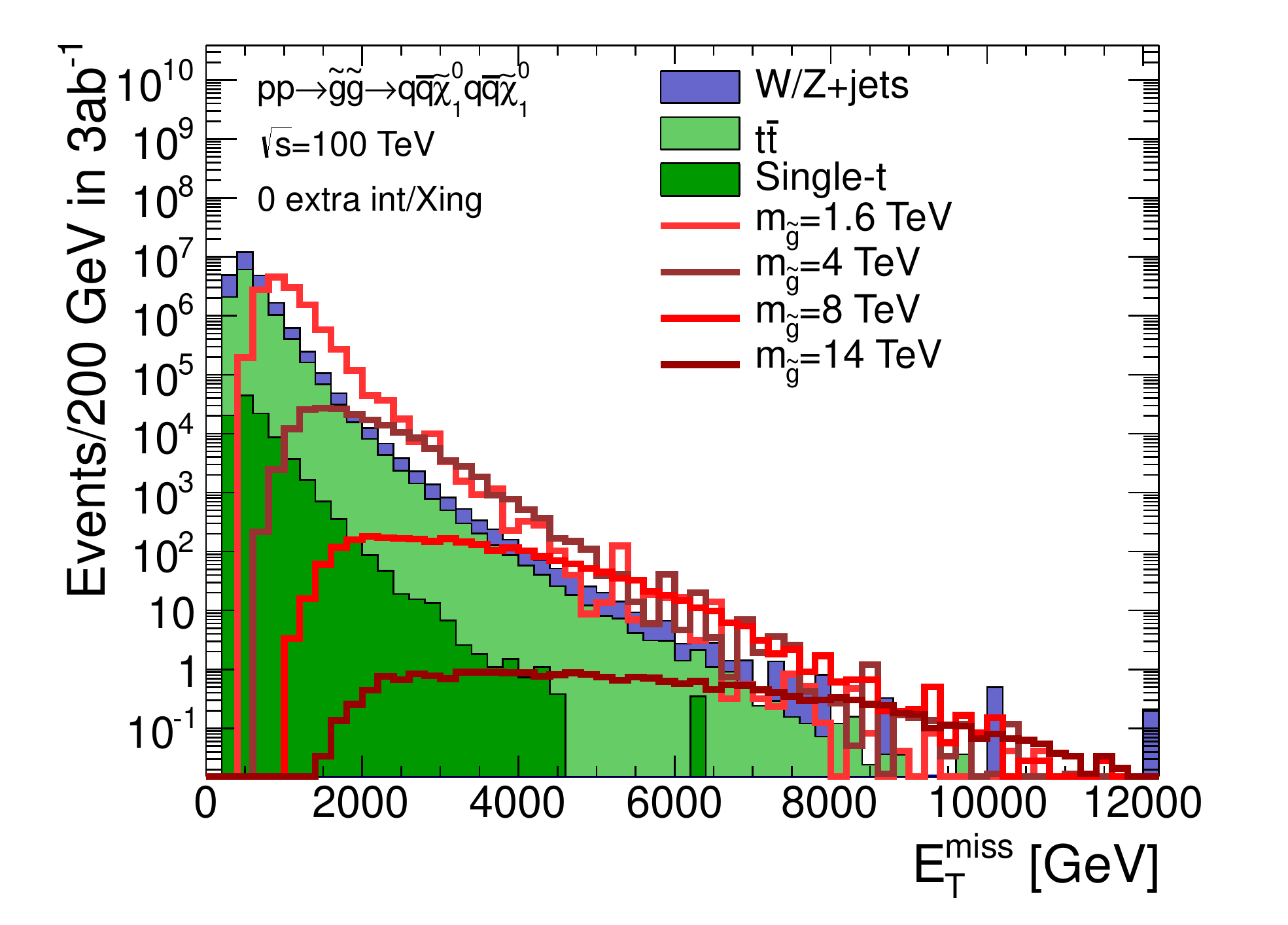}
    \includegraphics[width=0.45\textwidth]{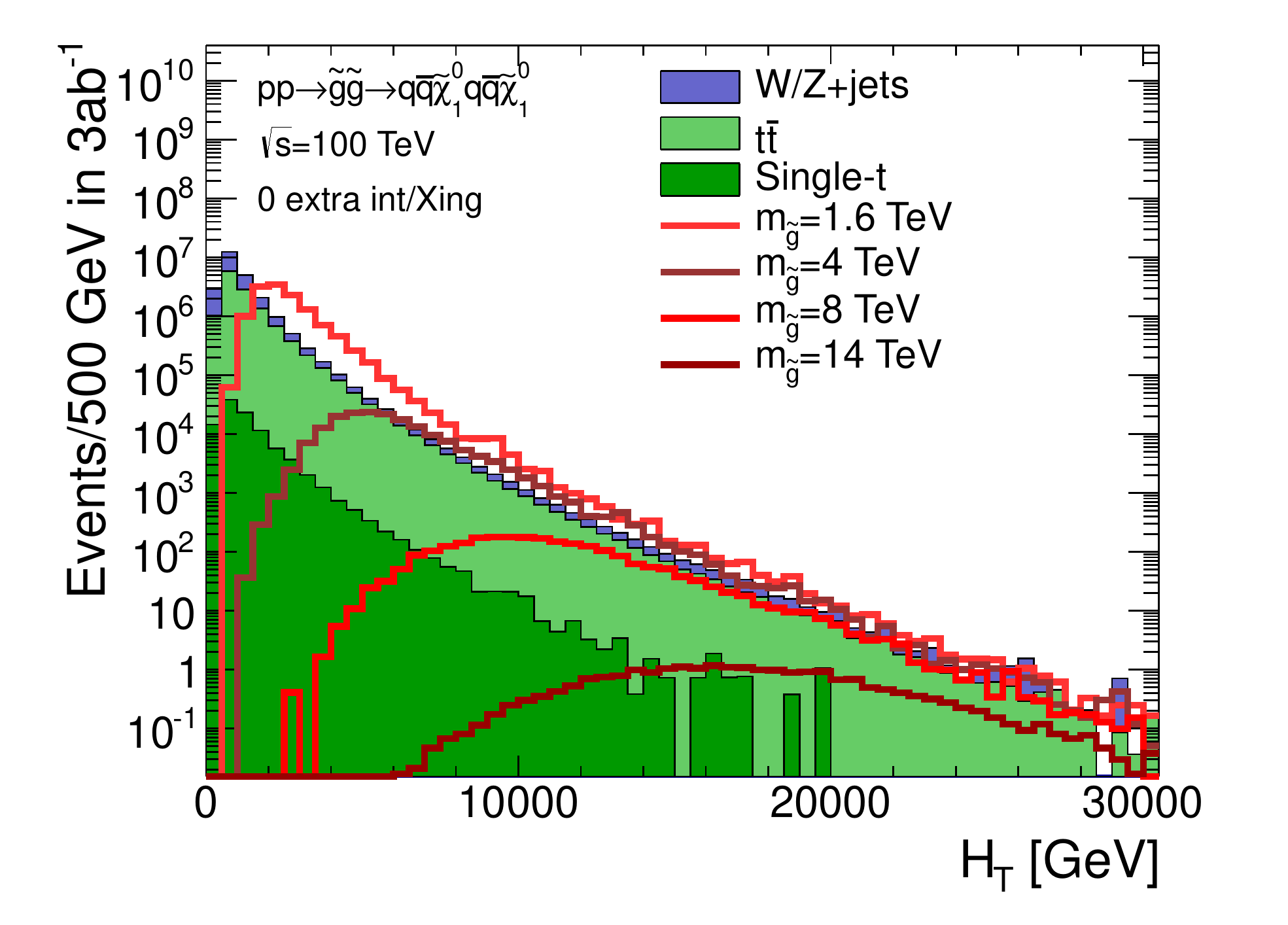}
\caption{Histogram of \MET~[left] and \HT~[right] after preselection cuts for background and a range of gluino-neutralino models with light flavor decays at a $100$ TeV proton collider.  The neutralino mass is $1\GeV$ for all signal models.}
\label{fig:GOGO_alljet_presel_distributions_100TeV}
\end{center}
\end{figure}

\pagebreak
\hiddensubsection{Analysis: 100 TeV}
In Fig.~\ref{fig:GOGO_alljet_presel_distributions_100TeV} we show histograms of $\MET$ [left] and \HT~[right] for signal and background at $\sqrt{s}=100\TeV$ after applying the preselection cuts listed in Sec.~\ref{sec:GOGO_Strategy}. Because the tails of the signal and background distributions have a similar slope, the optimization procedure generally leads to cuts near the bulk of the signal distribution. Moving to a higher center-of-mass energy allows for harder cuts to be placed, which in turn implies fewer background events survive the requirements. For example, we find that the signal efficiencies at the high gluino mass edge of our limits are several times larger at $100 \TeV$ than at $14 \TeV$.   We find that the searches are only barely systematics limited when the optimal cuts are applied (see Sec.~\ref{sec:GOGO_systematics} below for a detailed discussion).

Using the search strategy discussed above in Sec.~\ref{sec:GOGO_Strategy}, it is possible to explore the potential reach for the gluino-neutralino model at a $100$ TeV proton collider.  Table~\ref{tab:GNbulkCounts100TeV} gives a few example of the number of events that result from this cut flow for background and three example signal points: $m_{\widetilde{g}} = 5012,\, 9944,\text{ and } 13944 \GeV$ with $m_{\widetilde{\chi}^0_1} = 1 \GeV$.  Each choice of the \MET~and \HT~cuts given in Table~\ref{tab:GNbulkCounts100TeV} corresponds to the optimal cuts for one of the given signal points.   The hardness of the cut increases with the mass of the gluino.   It is clear that the ratio of $t\,\overline{t}$ to $V+$jets background is growing with regards to the $14$ and $33$ TeV searches; this is due to the higher probability for gluon scattering as $\sqrt{s}$ increases.  In this analysis it would certainly be advantageous to veto $b$-tagged jets to further reduce the background from top quarks.  From this table, it is possible to infer that gluinos as heavy as $\sim 10$ TeV could be discovered at a $100$ TeV collider.  \HT\ cuts as hard as $12$ TeV are required to extract the most information from this data set.

\begin{table}[tbp]
  \renewcommand{\arraystretch}{1.4}
  \setlength{\tabcolsep}{5pt}
  \footnotesize
  \begin{centering}
    \begin{tabular}{| r | r r | r | r  r  r  |}
\hline
&&&&\multicolumn{3}{c|}{$m_{\widetilde{g}}$ [GeV]}\\
                                               Cut&                 $V$+jets&               $t\overline{t}$&                 Total BG&                     
$5012$&                     $9944$&                    $13944$\\
\hline
\hline
$ \mathrm{Preselection}$&{$1.64 \times 10^{9}$}&{$3.33 \times 10^{9}$}&{$4.97 \times 10^{9}$}&$1.12 \times 10^{5}$&                     $ 876$&                     $43.4$\\
\hline
$\MET/\sqrt{\HT} > 15 \GeV^{1/2}$&{$3.59 \times 10^{7}$}&{$3.31 \times 10^{7}$}&{$6.90 \times 10^{7}$}&$7.99 \times 10^{4}$&                     $ 740$&                   $  38.8$\\
$p_T^{\mathrm{leading}} < 0.4\times \HT $&{$1.19 \times 10^{7}$}&{$1.25 \times 10^{7}$}&{$2.44 \times 10^{7}$}&$4.87 \times 10^{4}$&                      $443$ &                     $22.7$ \\
\hline
\hline
$         \MET > 5150 \GeV$&\multirow{2}{*}{                    $ 21.6$}&\multirow{2}{*}{                    $ 33.1$}&\multirow{2}{*}{                    $ 54.8$ }&\multirow{2}{*}{                    \color{red}  $216$}&\multirow{2}{*}{                     $91.6$}&\multirow{2}{*}{                     $10.7$}\\
$                         \HT > 9550 \GeV$&&&&&&\\
\hline
$         \MET > 5530 \GeV$&\multirow{2}{*}{                       $12$}&\multirow{2}{*}{                     $18.9$}&\multirow{2}{*}{                     $30.9$}&\multirow{2}{*}{                      $136$}&\multirow{2}{*}{                 \color{red}    $67.4$}&\multirow{2}{*}{                     $ 9.2$}\\
$                         \HT > 9750 \GeV$&&&&&&\\
\hline
$        \MET > 6150 \GeV$&\multirow{2}{*}{                      $4.1$}&\multirow{2}{*}{                     $ 6.3$}&\multirow{2}{*}{                     $10.4$}&\multirow{2}{*}{                     $33.6$}&\multirow{2}{*}{                     $29.6$}&\multirow{2}{*}{           \color{red}           $6.8$}\\
$                       \HT > 11700 \GeV$&&&&&&\\                                               
\hline
    \end{tabular}
    \caption{ Number of expected events for $\sqrt{s} = 100$ TeV with 3000 fb$^{-1}$ integrated luminosity for the background processes and selected gluino masses for the gluino-neutralino model with light flavor decays.  The neutralino mass is $1 \GeV$.  Three choices of cuts on \MET~and \HT~are provided, and for each gluino mass column the entry in the row corresponding to the ``optimal" cuts is marked in red.
    }
    \label{tab:GNbulkCounts100TeV}
  \end{centering}
\end{table}

\begin{figure}[h!]
  \centering
  \includegraphics[width=.48\columnwidth]{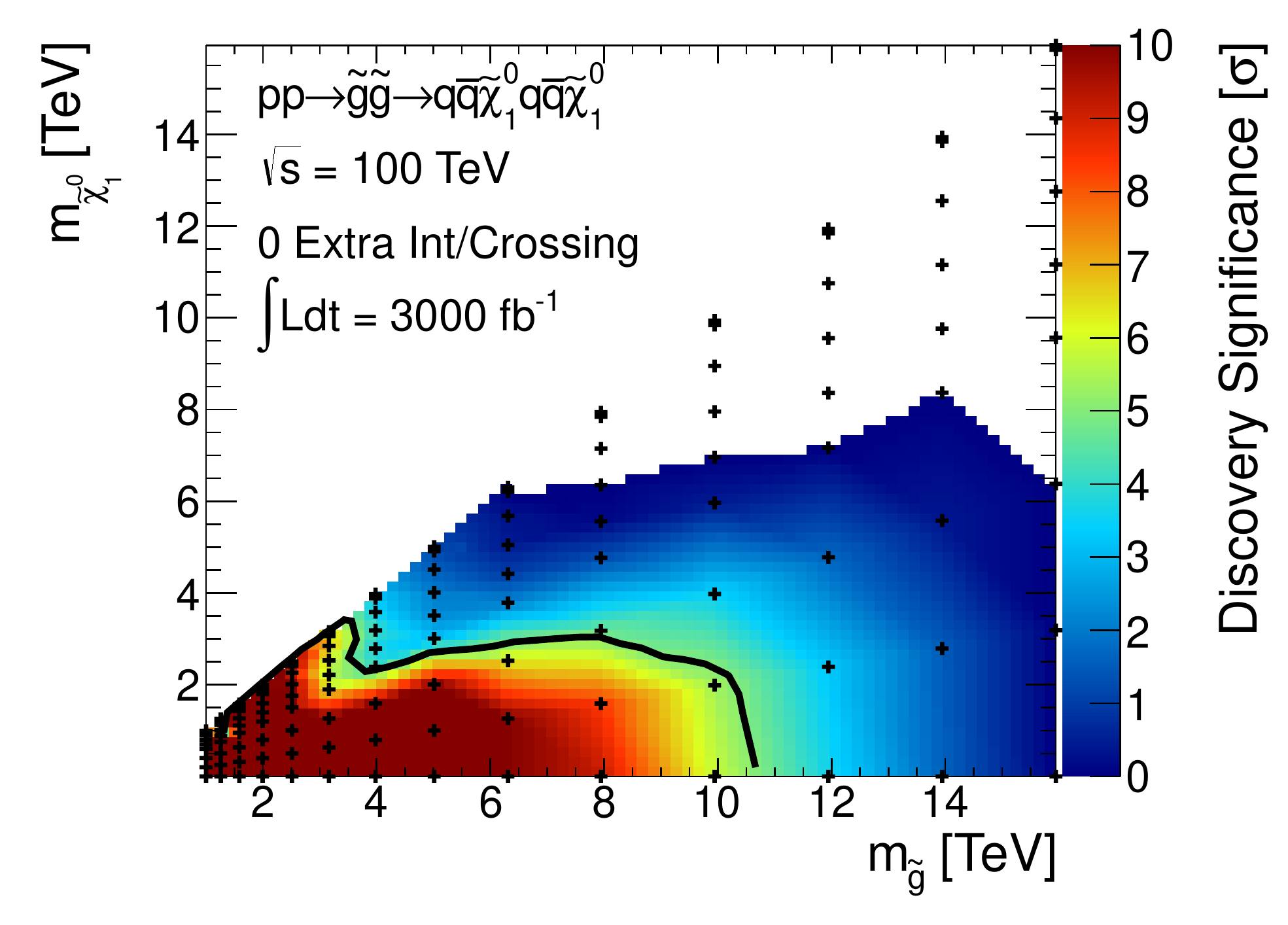}
  \includegraphics[width=.48\columnwidth]{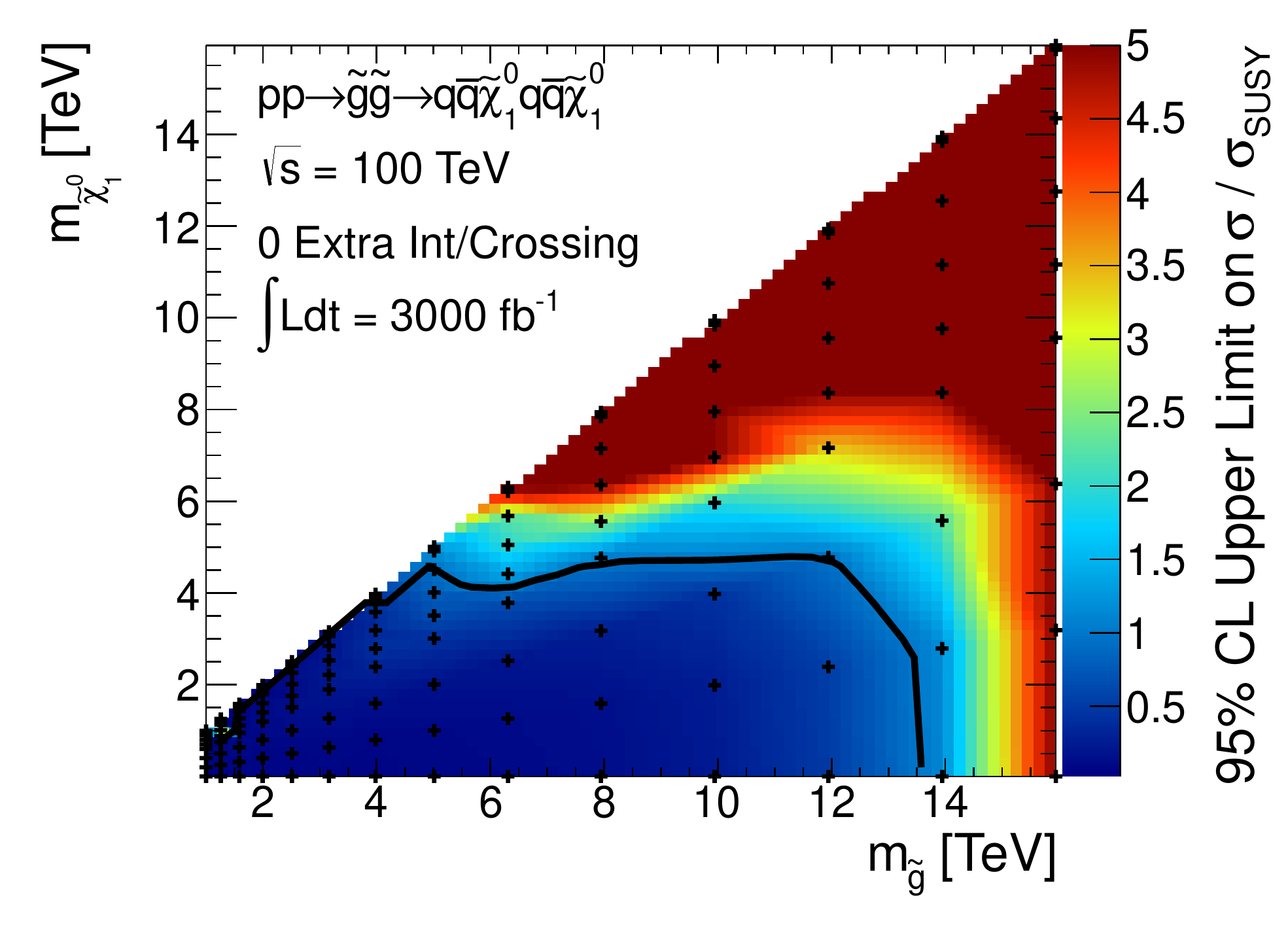}
  \caption{Results at $100 \TeV$ for the gluino-neutralino model with light flavor decays are given in the $m_{\widetilde{\chi}^0_1}$ versus $m_{\widetilde{g}}$ plane.  The left [right] plot shows the expected $5\sigma$ discovery reach [95\% confidence level upper limits] for gluino pair production.  Mass points to the left/below the contours are expected to be probed at $3000 \text{ fb}^{-1}$.  A 20\% systematic uncertainty is assumed and pileup is not included.  The black crosses mark the simulated models.}
\label{fig:GOGO_100_NoPileUp_results}
\end{figure}

\pagebreak
\hiddensubsection{Results: 100 TeV}
The $5\sigma$ discovery [left] and $95\%$ CL~limits [right] for the gluino-neutralino model are shown in Fig.~\ref{fig:GOGO_100_NoPileUp_results}, assuming $3000 \text{ fb}^{-1}$ of integrated luminosity.  $20\%$ systematic uncertainty is applied to the backgrounds.  The assumed signal systematic are outlined in the appendix.  Pileup is not included; a demonstration that pileup will not significantly change these results is given in Sec.~\ref{sec:pileupGOGO} below.

Using the NLO gluino pair production cross section one can make a very naive estimate for the reach of a given collider.  For example, we find that the choice of gluino mass which would yield $10$ events at $3000 \text{ fb}^{-1}$ is $16.1 \TeV$.  This roughly corresponds to the maximal possible reach one could expect for a given luminosity using $100 \TeV$ proton collisions.  

Using a realistic simulation framework along with the search strategy employed here the $100 \TeV$ $3000 \text{ fb}^{-1}$ limit with massless neutralinos is projected to be $13.5 \TeV$ (corresponding to 60 events).  Furthermore, the $100 \TeV$ proton collider with $3000 \text{ fb}^{-1}$ could discover a gluino as heavy as $11\TeV$ if the neutralino is massless, while for $m_{\widetilde{\chi}^0_1}\gtrsim 1 \TeV$ the gluino mass reach rapidly diminishes. 

The next section provides a comparison of the impact that the four Snowmass collider scenarios can have on the parameter space of this model.

\begin{figure}[h!]
  \centering
  \includegraphics[width=.48\columnwidth]{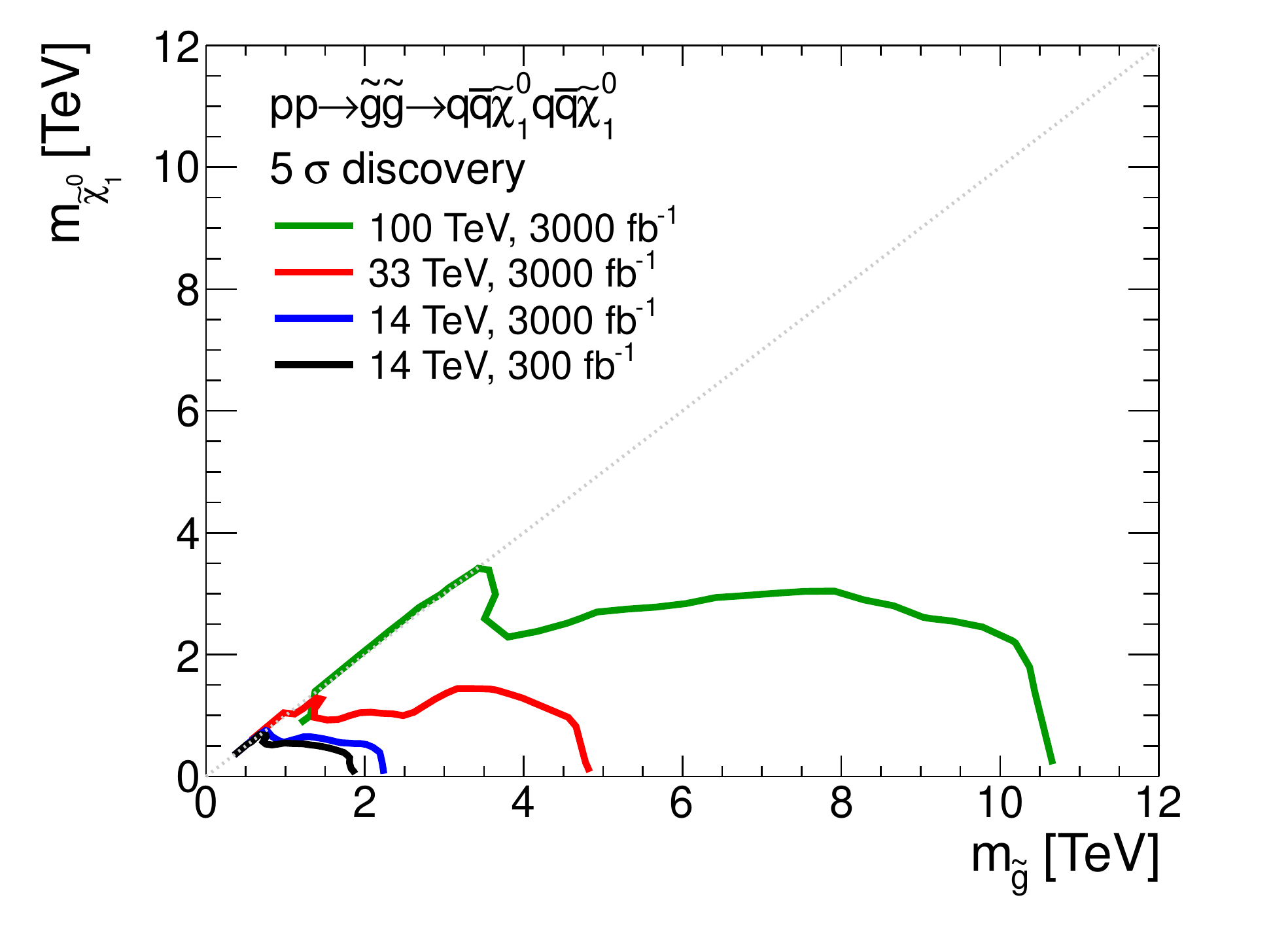}
  \includegraphics[width=.48\columnwidth]{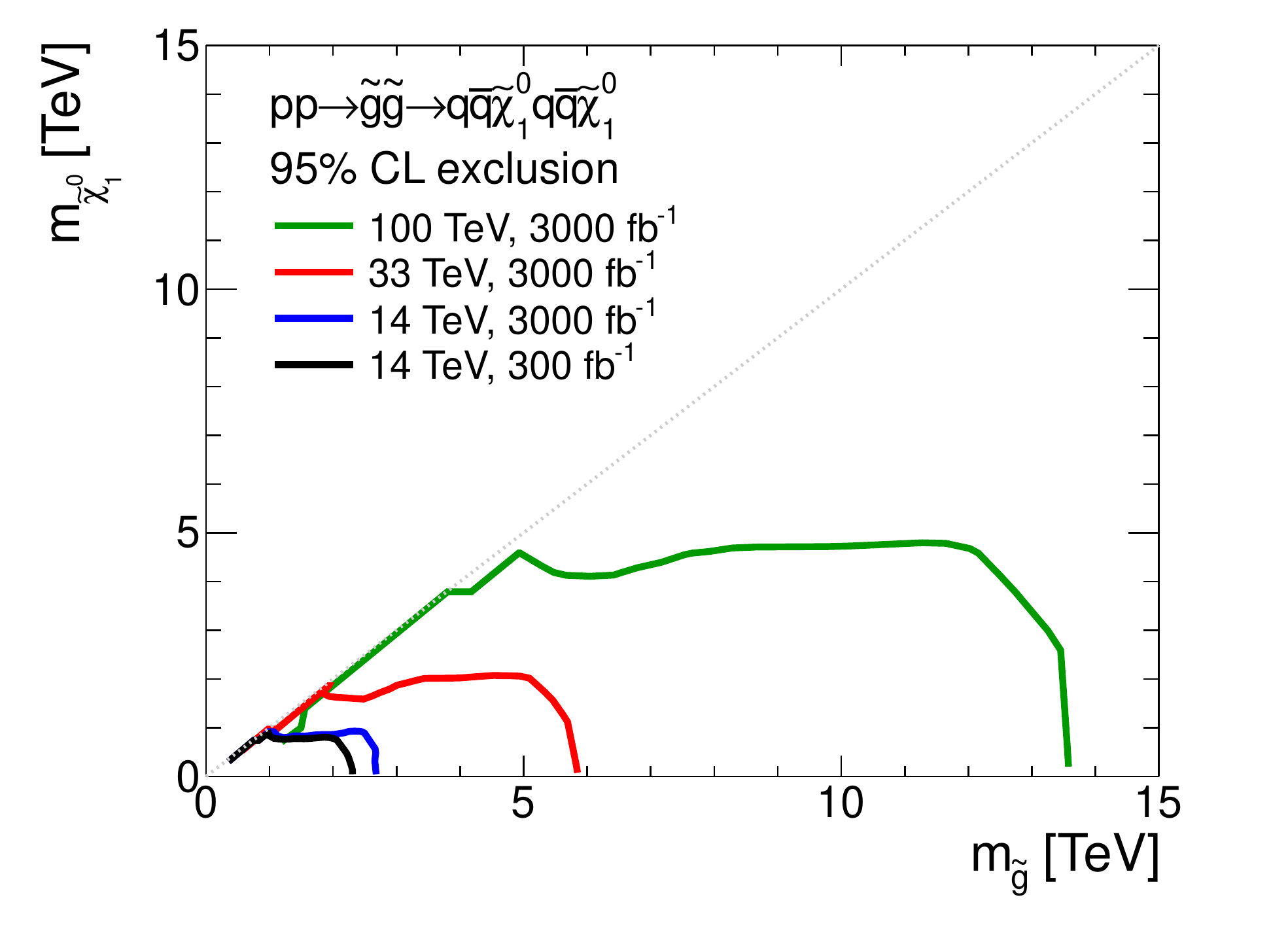}
  \caption{Results for the gluino-neutralino model with light flavor decays.  The left [right] panel shows the $5\,\sigma$ discovery reach [$95\%$ CL exclusion] for the four collider scenarios studied here.  A $20\%$ systematic uncertainty is assumed and pileup is not included.}
    \label{fig:GOGO_Comparison}
\end{figure}

\pagebreak
\hiddensubsection{Comparing Colliders}
The multi-jet plus \MET~signature of the gluino-neutralino model with light flavor decays provides a useful study with which to compare the potential impact of different proton colliders.  Figure \ref{fig:GOGO_Comparison} shows the $5\sigma$ discovery reach [$95\%$ CL exclusion] for two choices of integrated luminosity at $14$ TeV, along with the full data set assumed for $33$ and $100$ TeV.  At $14$ TeV, the factor of $10$ increase in luminosity leads to a modest increase by $350\GeV$ in the gluino limits. The smallness of this increase is due to the rapidly falling cross section. Furthermore, because the signal regions are not background-free, the improvement in cross section-limit does not match the factor of $10$ increase in luminosity; the shift in mass reach corresponds to only roughly a factor of five in the gluino production cross-section. For lighter gluinos, there is no improvement to the range of accessible neutralino masses. This is because the systematic uncertainty dominates in the signal regions for these models except in the high gluino mass tail.

 In contrast, increasing the center-of-mass energy has a tremendous impact on the experimentally available parameter space, since now much heavier gluinos can be produced without relying on the tails of parton distributions to supply the necessary energy.  Figure \ref{fig:GOGO_Comparison} makes a compelling case for investing in future proton colliders which can operate at these high energies.

\begin{figure}[h!]
  \centering
  \includegraphics[width=.48\columnwidth]{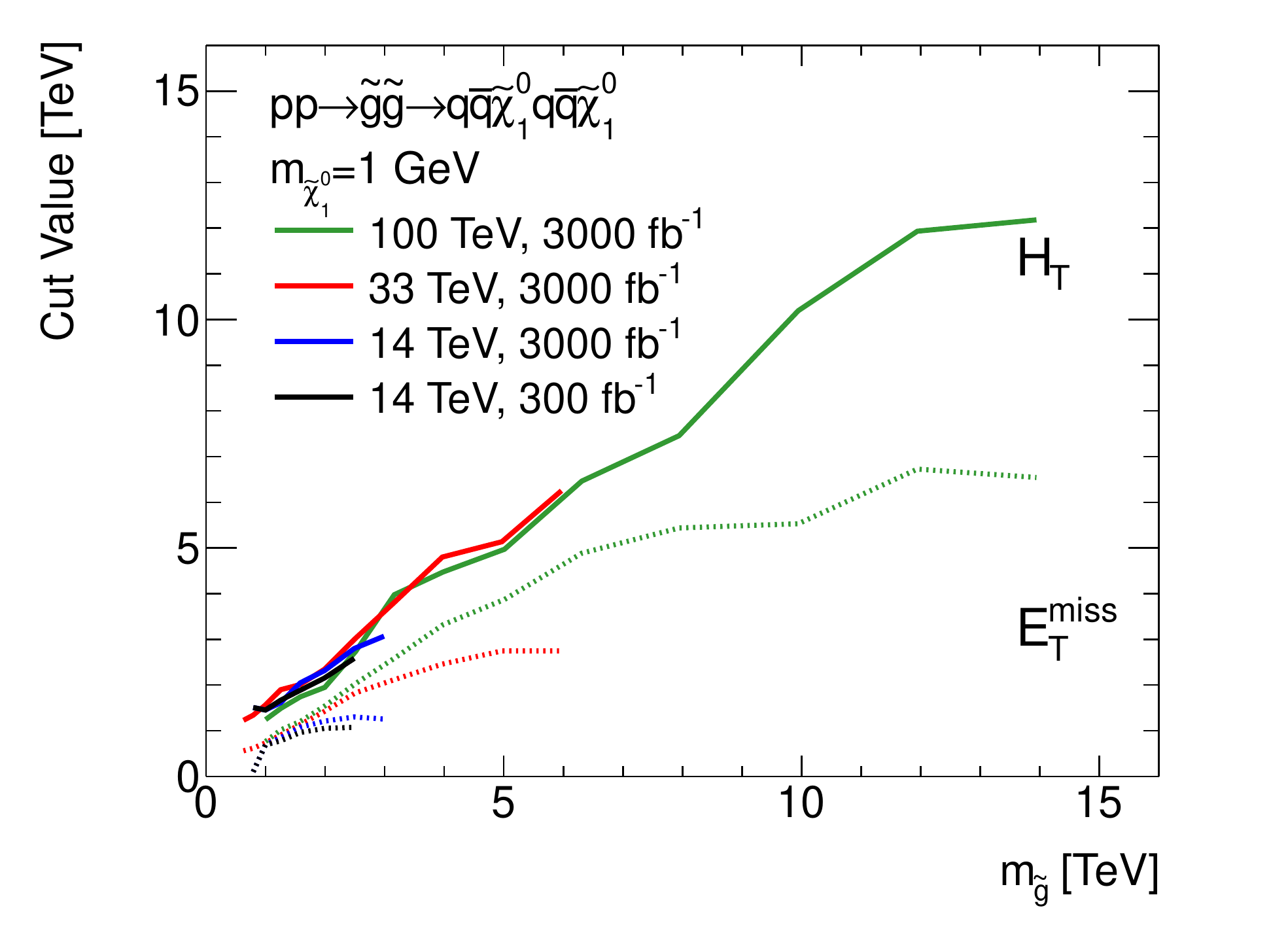}
  \caption{The optimal $\HT$ [solid] and $\MET$ [dashed] cuts for the gluino-neutralino model with light flavor decays as a function of gluino mass (assuming a $1$ GeV neutralino) for the four collider scenarios studied here.}
    \label{fig:GOGO_OptimalCut}
\end{figure}

Figure~\ref{fig:GOGO_OptimalCut} provides a comparison of the optimal cut at the different colliders that results from applying the analysis discussed in Sec.~\ref{sec:GOGO_Strategy} as a function of gluino mass (assuming a $1$ GeV neutralino).  It is interesting to note that the slope of the $\HT$ cut is larger than that for the $\MET$ cut.  The search is taking advantage of the tremendous energy that is imparted to jets when these heavy gluinos decay.  Furthermore, it is also interesting that the $\HT$ cuts track very closely between machines (until mass of the gluino becomes so heavy that a given collider can no longer produce them in appreciable quantities), while the $\MET$ cuts begin to flatten out for very high mass gluinos.  This can be understood by inspecting the histograms provided in Figs.~\ref{fig:GOGO_alljet_presel_distributions},  \ref{fig:GOGO_alljet_presel_distributions_33TeV}, and \ref{fig:GOGO_alljet_presel_distributions_100TeV}.  The signal and background distributions have different shapes for $\MET$, while the $\HT$ of signal and background tend to fall off with a similar slope in the tails.  The cut on on $\HT$ therefore simply scales with the gluino mass, while the optimization for $\MET$ is more subtle.  Finally, it is worth noting that due to the increase of the ratio of $t\,\overline{t}$ to $V+$jets events as $\sqrt{s}$ increases, it is likely worth exploring the addition of a veto on $b$-tagged jets for the higher energy colliders.

It is clear from these results that all four collider scenarios can have tremendous impact on our understanding of the gluino-neutralino parameter space.  The next sections are devoted to exploring various details related to these conclusions.

\pagebreak
\subsection{Comparing Optimization Strategies}
\label{sec:GOGO_Optimization}
\begin{figure}[bp]
  \centering
  \includegraphics[width=.48\columnwidth]{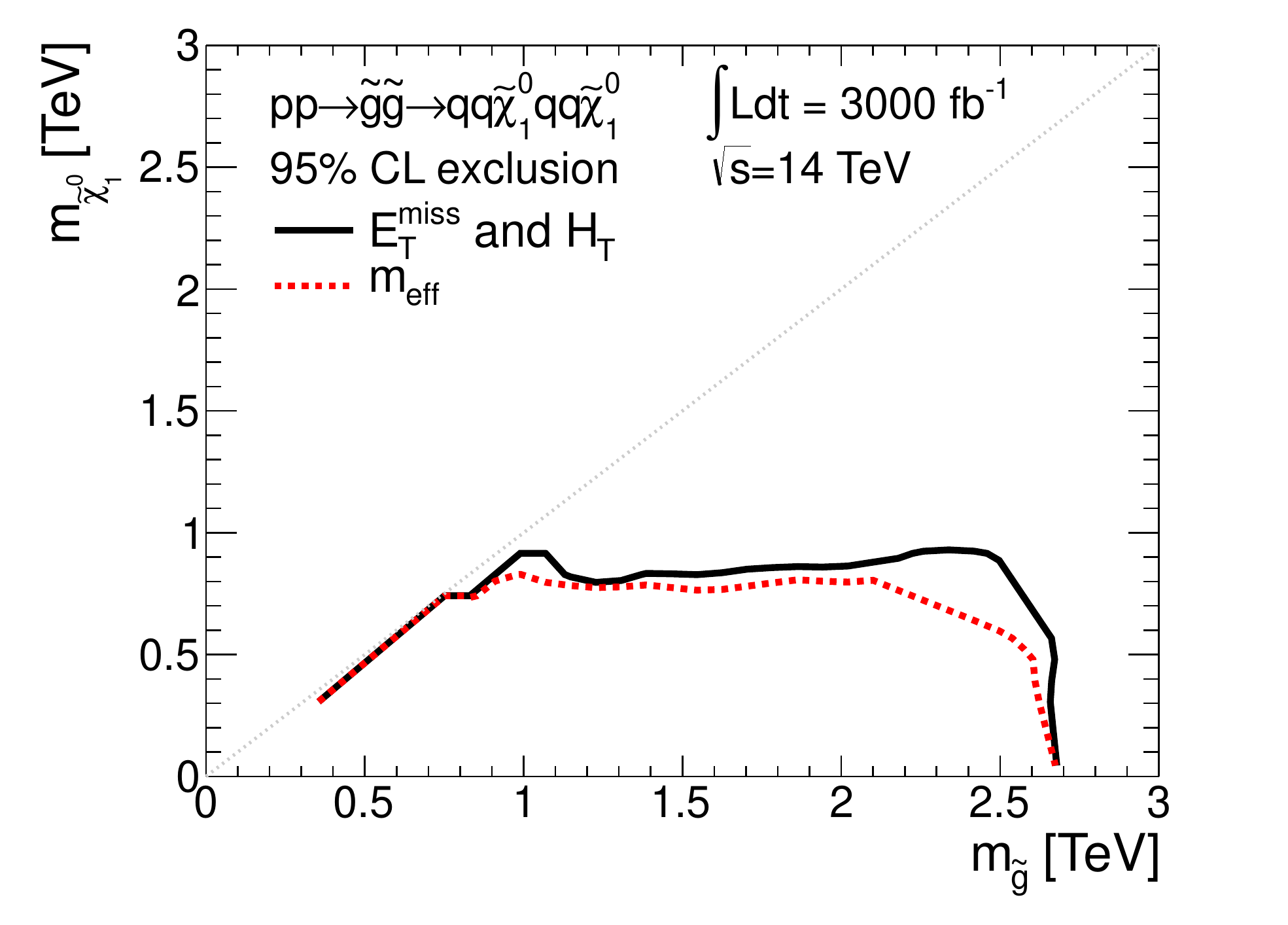}
  \includegraphics[width=.48\columnwidth]{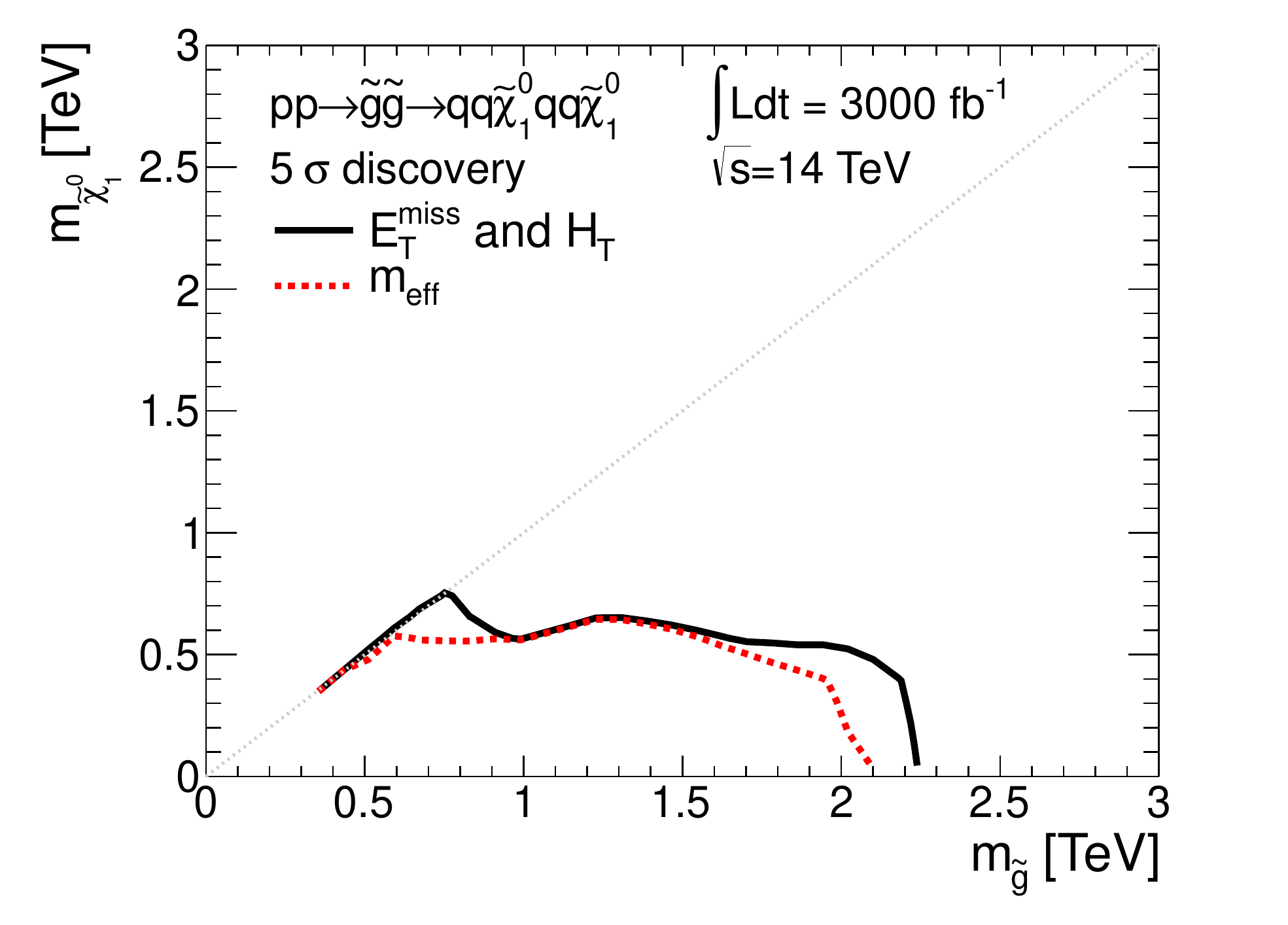}
  \caption{The $5\sigma$ discovery contours [right] and expected $95\%$ CL exclusion limits [left] for the one-dimensional \meff~[red, dotted],~and two-dimensional \MET~and \HT~[black, solid] optimization strategies.}
  \label{fig:1Dvs2D}
\end{figure}

A two-dimensional optimization over cuts on $\MET$~and $\HT$ is employed here.  This is the most significant difference between our strategy and the cuts used in the ATLAS analysis~\cite{ATL-PHYS-PUB-2013-002}, which only optimizes over $\meff\equiv \HT + \MET$.  The purpose of this section is to quantify the gain in significance from performing the two-dimensional optimization.  In Fig.~\ref{fig:1Dvs2D}, we plot the results of our mockup of the ATLAS one-dimensional scan along with the contours derived in this study by optimizing cuts over both \MET~and \HT.  The two-dimensional strategy improves the reach for several regions of the signal grid.  Therefore, we also use the two-dimensional strategy to study the squark-neutralino and gluino-squark signal models in the following sections.

\begin{figure}[h!]
  \centering
  \includegraphics[width=.48\columnwidth]{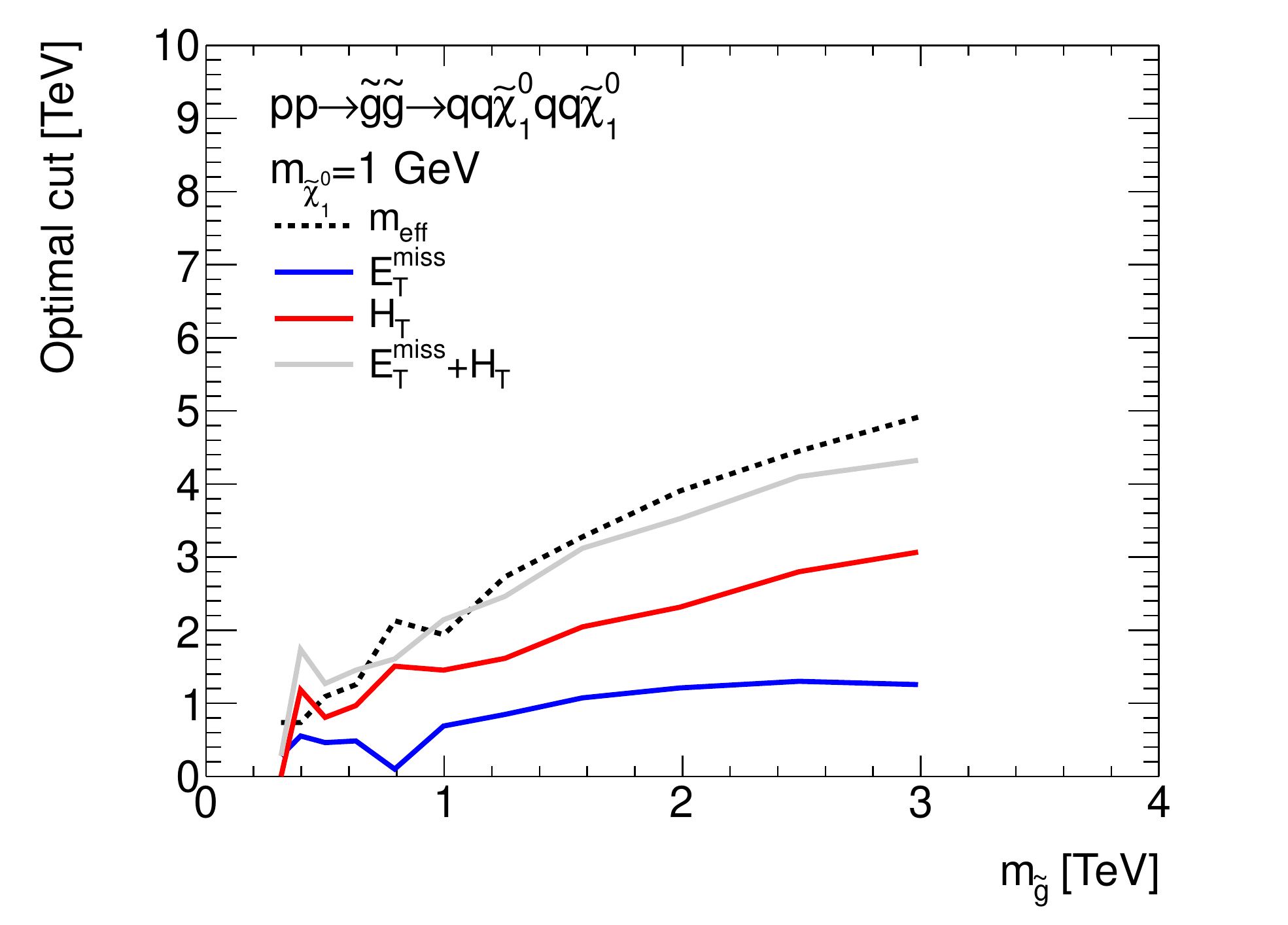}
  \caption{Optimal cuts on \meff~[black, dotted],~\MET~[blue, solid], and \HT~[red, solid]\ for the $14$ TeV LHC with $3000 \text{ fb}^{-1}$.  The mass of the neutralino is $1 \GeV$.  Also plotted is \MET+\HT\ [grey, solid] for the two-dimensional optimization, which allows direct comparison with \meff~from the one-dimensional strategy.}
  \label{fig:optimalcuts}
\end{figure}

It is interesting to compare the scaling of the optimal cut as a function of the gluino mass for the two optimization strategies.  The optimal cuts which result for the \meff~only [black, dotted] strategy along with the \MET~[blue, solid] and \HT~[red, solid] are plotted in Fig.~\ref{fig:optimalcuts}.  Also shown is $\MET+\HT$ [grey, solid], which allows a direct comparison between the one-dimensional and two-dimensional optimizations.  Above $\sim 1\TeV$, the cuts all increase monotonically as a function of $m_{\widetilde{g}}$.  The single cut on \meff~tends to be harder than the sum of $\MET$ and $\HT$.  The two-dimensional optimization can cover a wider parameter space of cuts, which allows it to take advantage of more complete information about the shape of the distribution.  This allows it to perform as well or better with a slightly softer effective \meff~cut.  

\pagebreak
\subsection{Impact of Systematic Uncertainties}
\label{sec:GOGO_systematics}
A systematic uncertainty of 20\% was assumed for the background normalization in the results we have presented. It is likely that the experiments will significantly reduce these uncertainties with larger datasets and an improved understanding of their detectors; it is also possible that this value is aggressive given our current knowledge (or lack thereof) of physics at higher $\sqrt{s}$.  It is therefore interesting to understand the impact of different systematic uncertainties on the discovery reach.

\begin{figure}[h!]
  \centering
  \includegraphics[width=.48\columnwidth]{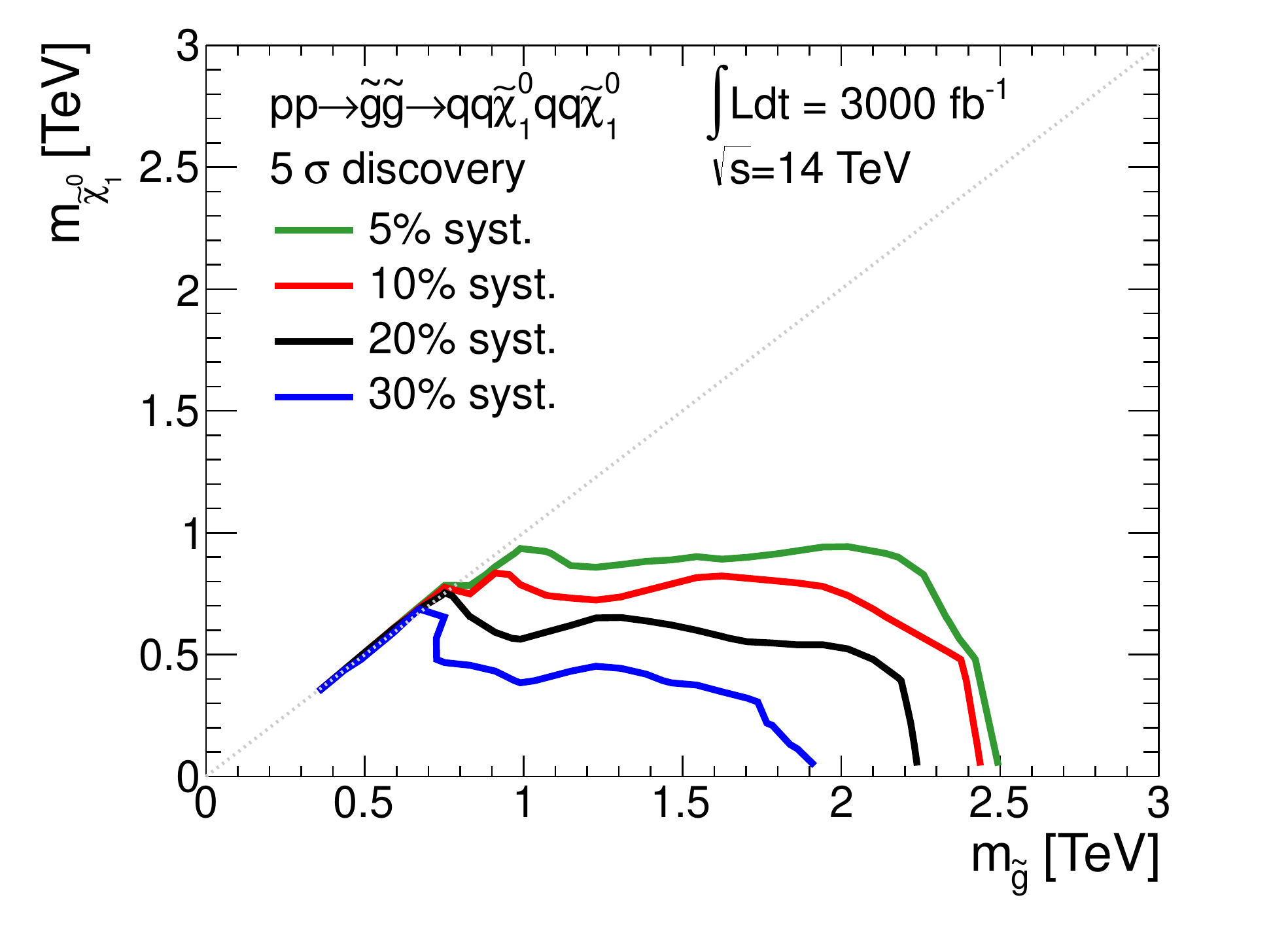}
    \includegraphics[width=.48\columnwidth]{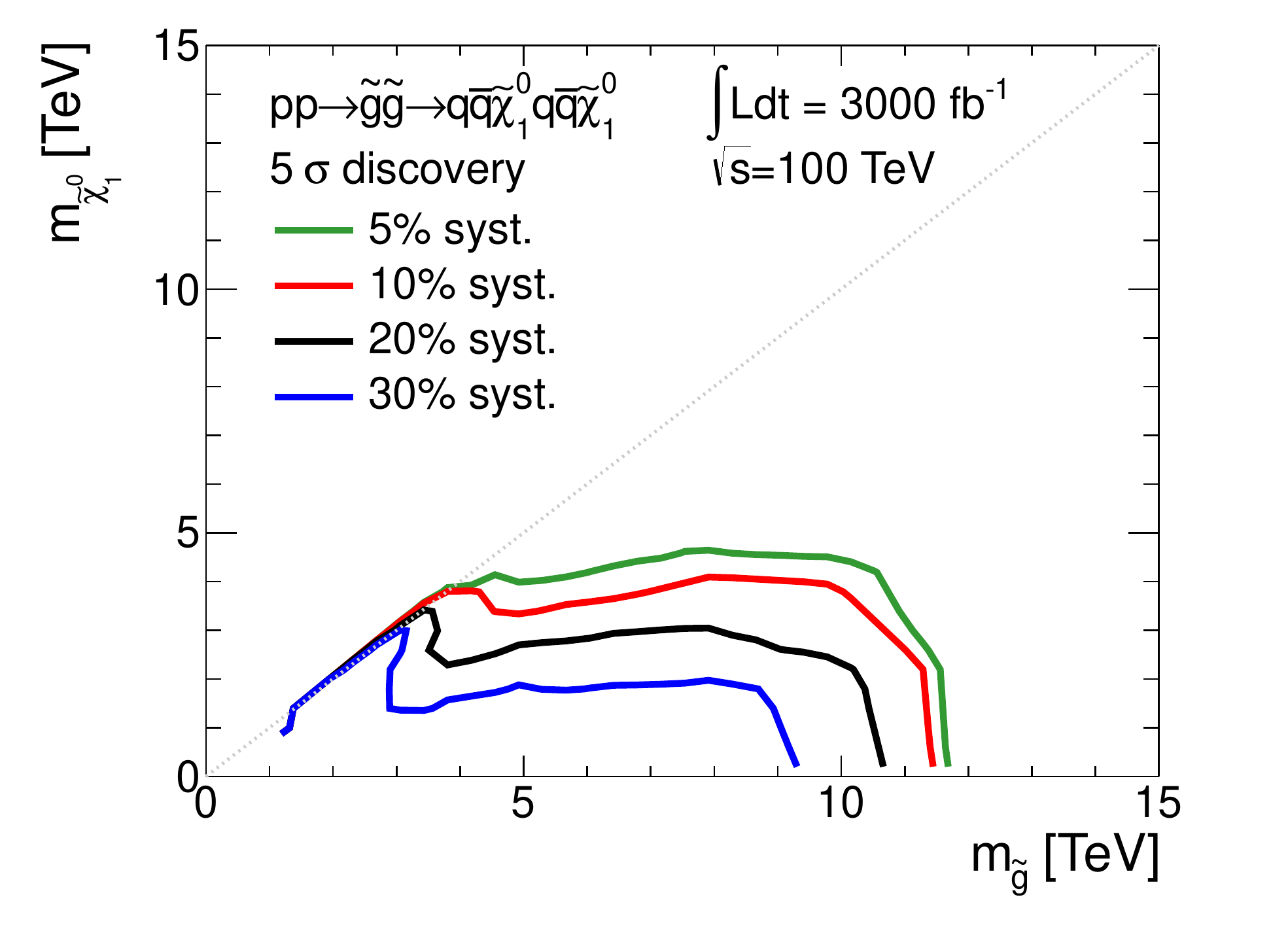}
  \caption{Expected $5\sigma$ discovery contours for the $\sqrt{s} = 14$ TeV LHC [left] and a $100$ TeV proton collider [right] with $3000$ fb$^{-1}$.  The different curves correspond to various assumptions for the systematic uncertainty on the background: $5\%$ [green], $10\%$ [red], $20\%$ [blue], and $30\%$ [black].}
  \label{fig:uncertainties}
\end{figure}

Figure~\ref{fig:uncertainties} shows the impact from a change in the systematic uncertainty for gluino discovery at $14\TeV$ and $100\TeV$ with $3000$fb$^{-1}$.  Varying the systematic background uncertainty from 30\% to 5\%, the discovery reach increases by roughly $600 \GeV$ ($3.4 \TeV$) in $m_{\widetilde{g}}$ at $14\TeV$ ($100\TeV$) and the coverage in $m_{\widetilde{\chi}_1^0}$ direction is roughly doubled.  The impact of systematic uncertainties on the $95\%$ exclusion limits is less dramatic.  Note that in this analysis we reoptimized the \MET\ and \HT\ cuts for each choice of systematic uncertainty.  As the LHC continues to run, the improvements in our understanding of the relevant backgrounds will be useful in extending the physics potential of the machine.

\subsection{Impact of Pileup}
\label{sec:pileupGOGO}
In order to reach an integrated luminosity of $3000 \text{ fb}^{-1}$, the instantaneous luminosity at the LHC will need to increase significantly with respect to previous runs.  There will be a corresponding increase in the number of pile-up events per bunch crossing.  It is crucial to understand the impact this environment will have on the expected reach using this data set.  

\begin{figure}[h!]
  \centering
  \includegraphics[width=.48\columnwidth]{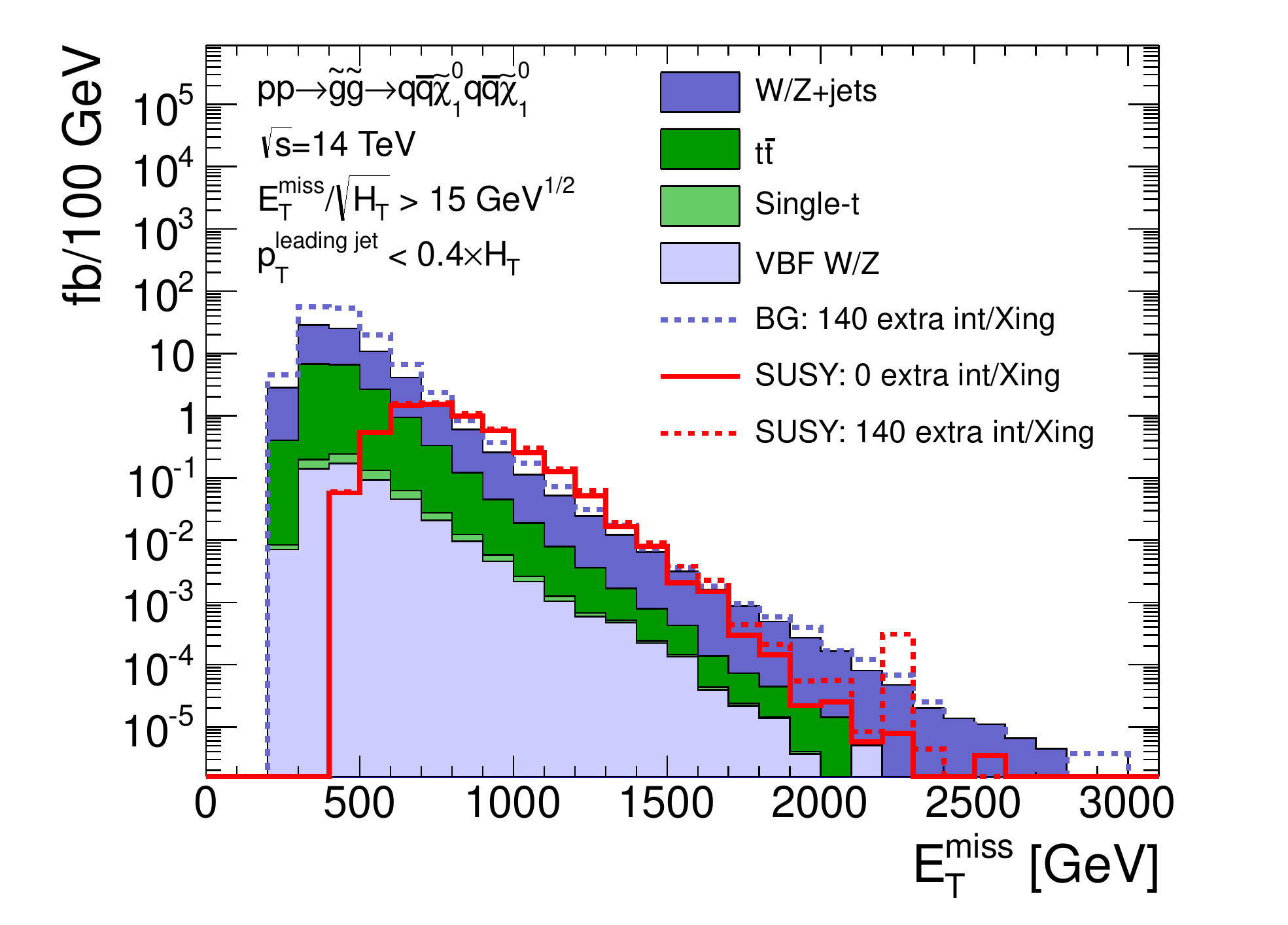}
  \includegraphics[width=.48\columnwidth]{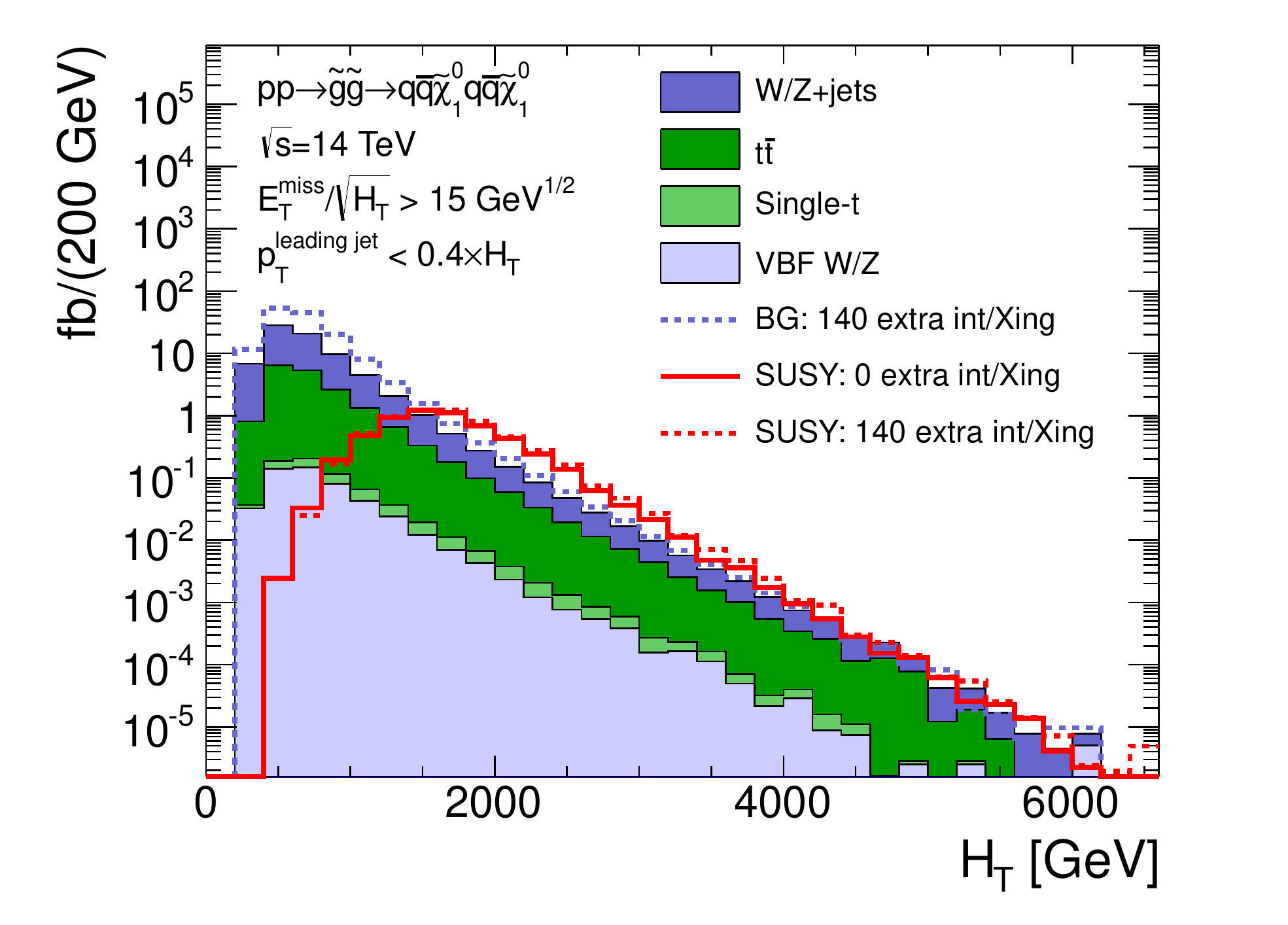}
  \caption{Signal and background \MET\ [left] and \HT\ [right] distributions at the $14$ TeV LHC for events with no pileup [solid] and the sum of backgrounds for events with 140 additional $pp$ interactions per bunch crossing [dashed].  Additional interactions increase the background rates at low \MET, but have little impact on the final analysis due to the tight \MET~cuts.} 
 \label{fig:GOGO_pileup_14_distributions}
\end{figure}

To study this in detail, we repeated the full analysis with signal and background samples that include 140 additional minimum-bias interactions.  The \texttt{Delphes} based Snowmass simulation includes a pileup suppression algorithm that primarily impacts the \MET\ resolution \cite{Anderson:2013kxz}.  Figure~\ref{fig:GOGO_pileup_14_distributions} shows the \MET~[left] and \HT~[right] distributions with and without pileup.  The samples with pileup follow the distributions without pileup closely, especially in the search regions.  We also observe that the the largest effects of pileup is at at low values of \MET-significance, and
are therefore suppressed by the requirement that $\MET/\sqrt{\HT}>15 \text{ GeV}^{1/2}$.  

\begin{figure}[h!]
  \centering
  \includegraphics[width=.48\columnwidth]{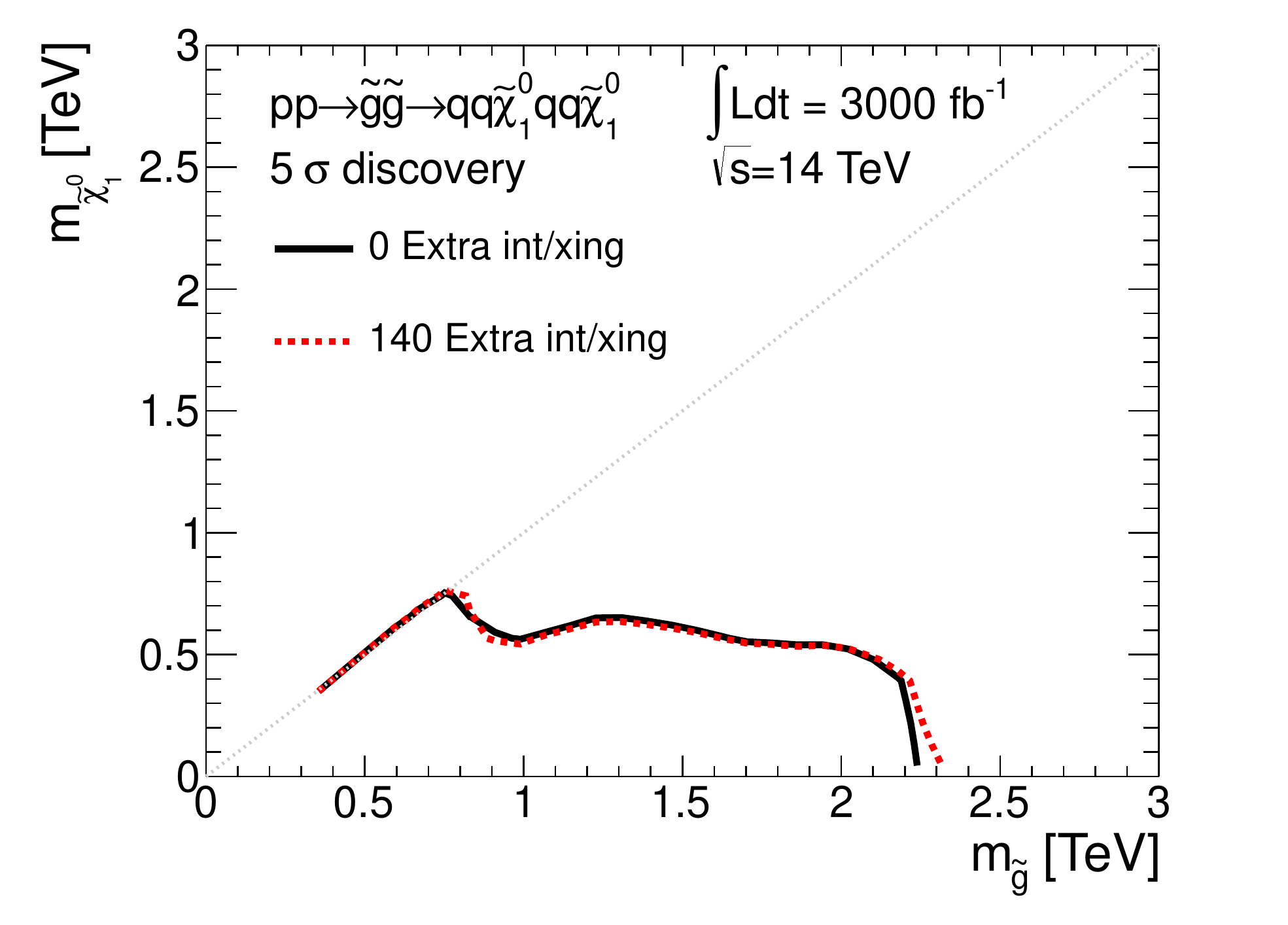}
  \includegraphics[width=.48\columnwidth]{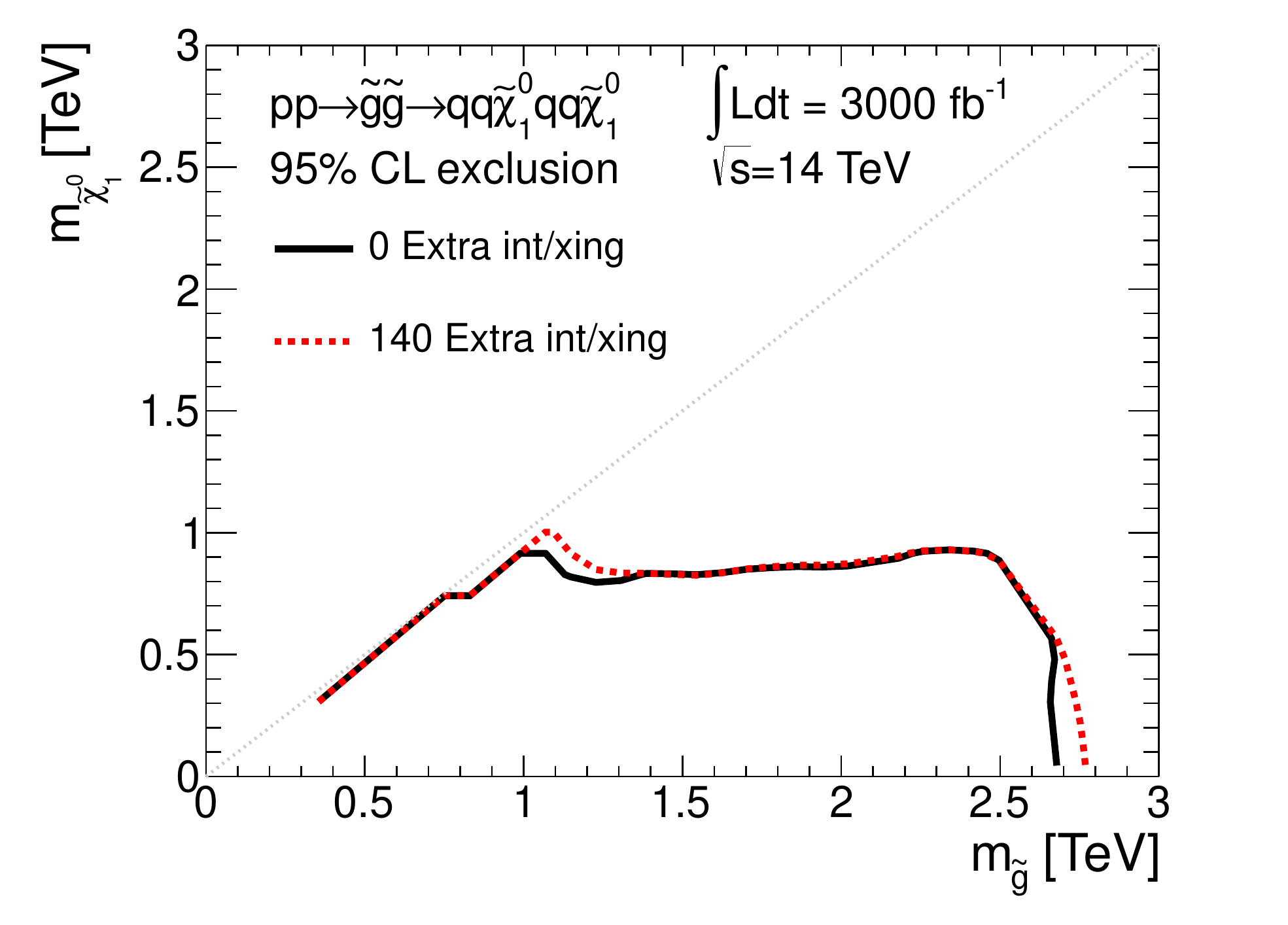}
  \caption{Discovery contours [right] and expected limits [left] for the analyses performed with [red, dotted] and without [black, solid] pileup at the $14$ TeV LHC with $3000 \text{ fb}^{-1}$ integrated luminosity.}
  \label{fig:pileup_14_results}
\end{figure}

The impact of pileup on the discovery significance [left] and limits [right] are shown in Fig.~\ref{fig:pileup_14_results}.  Given that the \HT\ and \MET\ distributions are effectively unchanged, it is not surprising that the results are very similar with and without pileup.  The contours with and without pileup each lie within the other's 1$\sigma$ confidence interval, and we find no evidence that this reflects anything other than statistical fluctuations for a few signal points.  We can safely assume that pileup has a small impact on this analysis.

\pagebreak
\section{The Compressed Gluino-Neutralino Model with Light Flavor Decays}
\label{sec:GNCompressed}
This section is devoted to analyses which target the compressed region of parameter space for the gluino-neutralino with light-flavor decays Simplified Model discussed in Sec.~\ref{sec:GNLightFlavor}, where
\be
m_{\widetilde{g}} - m_{\widetilde{\chi}_1^0} \equiv \Delta m \ll m_{\widetilde{g}}.
\ee
For models with this spectrum, the search strategy of Sec.~\ref{sec:GNLightFlavor} does not provide the optimal reach.  With compressed spectra the gluino decays only generate soft partons, thereby suppressing the \HT~signals and reducing the efficiency for passing the $4$ jet requirement.  A more effective strategy for compressed spectra searches relies instead on events with hard initial state radiation (ISR) jets to discriminate signal from background.  

In this study, we will apply two different search strategies that are optimized for this kinematic configuration and will choose the one which leads to the most stringent bound on the production cross section for each point in parameter space.  Some of the cuts chosen below are inspired by recent public results from \atlas~\cite{ATLAS-CONF-2012-147} and \cms~\cite{CMS-PAS-EXO-12-048} on monojet searches.  For recent work on the compressed region of parameter space see \cite{Bhattacherjee:2013wna}, and for a discussion of the theoretical uncertainties see \cite{Dreiner:2012sh}.

\subsection{Dominant Backgrounds}
\label{sec:Backgrounds_CompressedGOGO}
The dominant background is the production of a $Z$ boson in association with jets, where the $Z$ boson decays into a pair of neutrinos ($Z \rightarrow \nu \nu$), leading to events with jets and a significant amount of missing transverse energy. Subleading backgrounds are the production of a $W$ boson which decays leptonically $\big(W \rightarrow \ell\, \nu\big)$ in association with jets, where the charged lepton is not reconstructed properly.  Finally, when considering events with a significant number of jets, $t\bar{t}$ production in the fully hadronic decay channel $\big (t\rightarrow b\,q\,q'\big)$ can be relevant.

\hiddensubsection{Two Analysis Strategies: 14 TeV}
\addcontentsline{toc}{subsection}{4.2-4.8 $\,\,\,\,$ Analysis and Results: $14$, $33$, and $100$ TeV}
\label{sec:FourAnalysisStrategies}
This section is devoted to a description of the two analysis strategies employed to search for the compressed regions of the gluino-neutralino parameter space.  Applications to the $14$ TeV LHC will be presented for illustration; $33$ and $100$ TeV will be discussed below.  The following preselections are common to both approaches.

\textbf{\textsc{Preselection}}
\NegSpace
\begin{itemize}
\item lepton veto: any event with an electron or muon with $p_T > 10$ GeV and $|\eta| < 2.5$ is discarded 
\item jets are required to have $p_T > 30$ GeV and $|\eta| < 4.5$
\item the leading jet must be reconstructed within $|\eta| < 2.5$
\item $\MET > 100 \GeV$
\end{itemize}
\NegSpace
The first set of cuts implemented in this analysis is based on the public monojet search from the \atlas\ collaboration \cite{ATLAS-CONF-2012-147}. 

\begin{figure}[!htb]
\begin{center}
\includegraphics[width=0.6\textwidth]{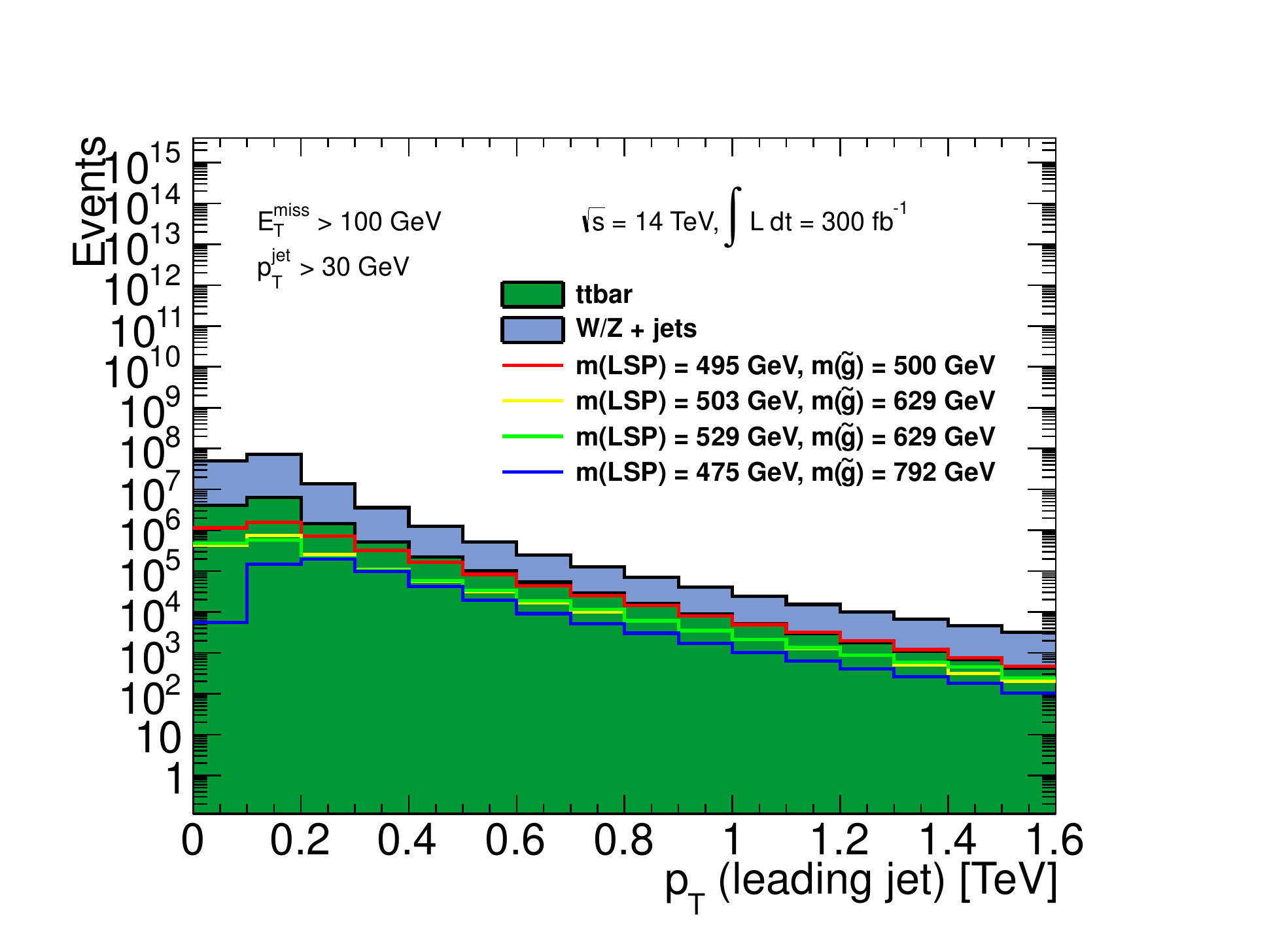}
\caption{Histogram of the leading jet \pt~for signal and background at the $14 \TeV$ LHC with $300 \text{ fb}^{-1}$ after the preselection for a range of gluino and neutralino masses in the compressed region.}
\label{fig:lj_pt_presel_distributions}
\end{center}
\end{figure}

\textbf{\textsc{Search Strategy 1}: Leading jet based selection}
\NegSpace
\begin{itemize}
\item at most 2 jets
\item leading jet must have $p_T > (\text{leading jet }\pt)_\text{optimal}$ and $|\eta| < 2.0$
\item second jet is allowed if $\Delta \varphi(j_{2},\MET) > 0.5$
\item $\MET > \big(\MET\big)_\text{optimal}$
\end{itemize}
\NegSpace
where both $\big(\MET\big)_\text{optimal}$ and $(\text{leading jet }\pt)_\text{optimal}$ are determined simultaneously by taking the values in the range $1 - 10$ TeV that yields the strongest exclusion.   Figure~\ref{fig:lj_pt_presel_distributions} shows the distribution of the leading jet $p_T$ and illustrates the ability to distinguish signal from background using this variable.

Table~\ref{tab:lj_pt_counts} shows the expected signal and background yields for the signal region with cuts on the leading jet $\pt > (500 \GeV, 1 \TeV, 2 \TeV)$ and $\MET > (500 \GeV, 1 \TeV, 2 \TeV)$ . This analysis is expected to be especially powerful for very small mass differences, when no jets except for a hard ISR jet can be reconstructed. 

\begin{table}[h!]
\begin{centering}
\renewcommand{\arraystretch}{1.6}
\setlength{\tabcolsep}{6pt}
\footnotesize
\vskip 10pt
\begin{tabular}{|r|rr|r|rrr|}
\hline
&&&&\multicolumn{3}{c|}{$\big(m_{\widetilde{g}},\,m_{\widetilde{\chi}^0_1}\big)\quad$ [GeV]} \\
Cut & $V$+jets & $t\,\overline{t}$ & Total BG & $(500,\,495)$ & $(792,\,787)$ & $(997,\,992)$\\
\hline
\hline
Preselection & $1.3 \times 10^9$  & $1.2 \times 10^{8}$ & $1.4 \times 10^{9} $ & $4.0 \times 10^{7}$  & $2.6 \times 10^6 $ & $6.0 \times 10^5$ \\
\hline
$\MET > 120 $ GeV, $\leq 2$ jets & $6.0 \times 10^8$ & $7.0 \times 10^6$ & $6.1 \times 10^8$& $1.9 \times 10^7$  & $1.4 \times 10^6$&  $2.8 \times 10^5$\\
$p_{T} (j_1) > 120 $ GeV,  $|\eta (j_1)| < 2.0$ & $3.2 \times 10^8$ & $3.2 \times 10^6$ & $3.6 \times 10^8$ & $1.2 \times 10^7$ & $9.2 \times 10^5$ & $1.9 \times 10^5$\\
$\Delta{\varphi(j_2,\MET}) > 0.5$ &  $2.4 \times 10^8$ & $1.6 \times 10^6$ &$2.4 \times 10^8$ & $8.0 \times 10^6$ & $6.4 \times 10^5 $ & $1.3 \times 10^5$\\
\hline
\hline
$\MET > 500 $ GeV  & \multirow{2}{*}{$4.5 \times 10^5$} & \multirow{2}{*}{$1.7 \times 10^3$} & \multirow{2}{*}{$4.5 \times 10^5$} & \multirow{2}{*}{\color{black}$4.5 \times 10^5$}  & \multirow{2}{*}{$5.9 \times 10^4$} & \multirow{2}{*}{$1.6 \times 10^4$}\\
$ p_{T} (j_1) > 500 $ GeV &&&&&&\\
\hline
$\MET > 1 $ TeV  & \multirow{2}{*}{$9.4 \times 10^3$} & \multirow{2}{*}{$13$} & \multirow{2}{*}{$9.4 \times 10^3$} & \multirow{2}{*}{$1.9 \times 10^4$} & \multirow{2}{*}{\color{black}$4.4 \times 10^3$} & \multirow{2}{*}{$1.6 \times 10^3$}\\
$ p_{T} (j_1)> 1$ TeV &&&&&&\\
\hline
$\MET > 2 $ TeV & \multirow{2}{*}{$49$} & \multirow{2}{*}{$0$} & \multirow{2}{*}{$49$} & \multirow{2}{*}{$87$} & \multirow{2}{*}{$38$} & \multirow{2}{*}{\color{black}$18$}\\
 $ p_{T} (j_1 )> 2$ TeV &&&&&&\\
\hline
\end{tabular}
\caption{Number of expected events for $\sqrt{s} = 14$ TeV and $3000$ fb$^{-1}$ for the background processes and three gluino-neutralino models in the compressed region. The leading jet $p_T$ based selection with various cuts is applied.  Three choices of cuts are provided for illustration.}
\label{tab:lj_pt_counts}
\end{centering}
\end{table}

\begin{figure}[!htb]
\begin{center}
\includegraphics[width=0.6\textwidth]{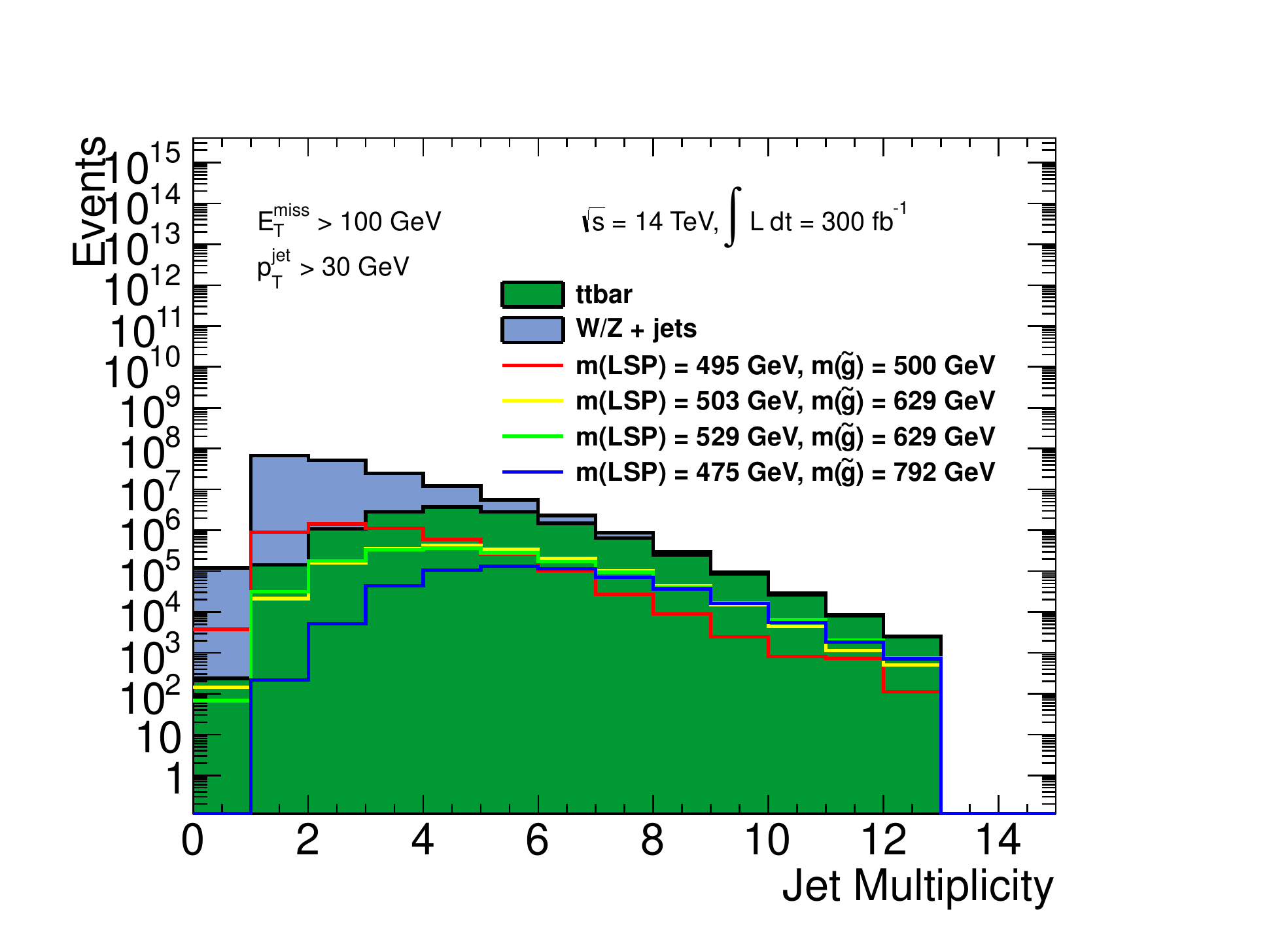}
\caption{Histogram of jet multiplicities for signal and background at the $14 \TeV$ LHC with $300 \text{ fb}^{-1}$ after the preselection for a range of gluino and neutralino masses in the compressed region.}
\label{fig:alljet_presel_distributions}
\end{center}
\end{figure}

\pagebreak
\textbf{\textsc{Search Strategy 2}: \MET~based selection without jet veto}
\NegSpace
\begin{itemize}
\item leading jet with $p_T > 110$ GeV and $|\eta| < 2.4$
\item $\MET > (\MET)_\text{optimal}$
\end{itemize}
\NegSpace
with $\MET$ varied in the range $(1,10) \TeV$.  No requirement is placed on a maximum number of jets.  Figure~\ref{fig:alljet_presel_distributions} shows that already for signal scenarios with small mass differences it is likely to reconstruct more than one jet in the event.  Note that for higher jet multiplicities the production of top quark pairs in the fully hadronic decay mode starts to dominate over $W/Z$ + jets production.

Table~\ref{tab:alljets_counts} shows the expected number of signal and background for three choices of the \MET~cut. Compared to the previous selection one can see that a significantly larger number of events are selected, especially for the larger mass differences. In addition, for this selection top pair production can make a non-negligible contribution to the total number of background events.

\begin{table}[h!]
\begin{centering}
\renewcommand{\arraystretch}{1.8}
\setlength{\tabcolsep}{5pt}
\footnotesize
\vskip 10pt
\begin{tabular}{|r|rr|r|rrr|}
\hline
&&&&\multicolumn{3}{c|}{$\big(m_{\widetilde{g}},\,m_{\widetilde{\chi}^0_1}\big)\quad$ [GeV]} \\
Cut & $V$+jets & $t\,\overline{t}$ & Total BG & $(500,\,495)$ & $(792,\,787)$ & $(997,\,992)$\\
\hline
\hline
Preselection & $1.3 \times 10^9$  & $1.2 \times 10^{8}$ & $1.4 \times 10^{9} $ & $4.0 \times 10^{7}$  & $2.6 \times 10^6 $ & $6.0 \times 10^5$ \\
\hline
$p_{T} (j_1) > 110 $ GeV, $  |\eta (j_1)| < 2.4$ & $7.9 \times 10^8$  & $8.1 \times 10^7$  & $8.7 \times 10^8$ & $3.0 \times 10^7$ & $2.1 \times 10^6$ & $4.8 \times 10^5$ \\
\hline
\hline
$\MET > 500 $ GeV & $2.1 \times 10^6$ & $1.5 \times 10^5$ & $2.3 \times 10^6$ & \color{black}$2.4 \times 10^6$ & $2.9 \times 10^5$  & $7.5 \times 10^4$\\
\hline
$\MET > 1 $ TeV & $4.9 \times 10^4$ & $2.1 \times 10^3$ & $5.2 \times 10^4$ & $1.3 \times 10^5$ & \color{black}$2.5 \times 10^4$ & $8.1 \times 10^3$\\
\hline
$\MET > 2$ TeV & $278$ & $3$ & $282$ & $900$ & $328$ &\color{black} $133$\\
\hline
\end{tabular}
\caption{Number of expected events for $\sqrt{s} = 14$ TeV and $3000$ fb$^{-1}$ for the background processes and selected signal processes. The selection without a veto on additional jets with cuts on \MET~is applied. Three choices of cuts are provided for illustration.}
\label{tab:alljets_counts}
\end{centering}
\end{table}

\begin{figure}[!htb]
\begin{center}
\includegraphics[width=0.48\textwidth]{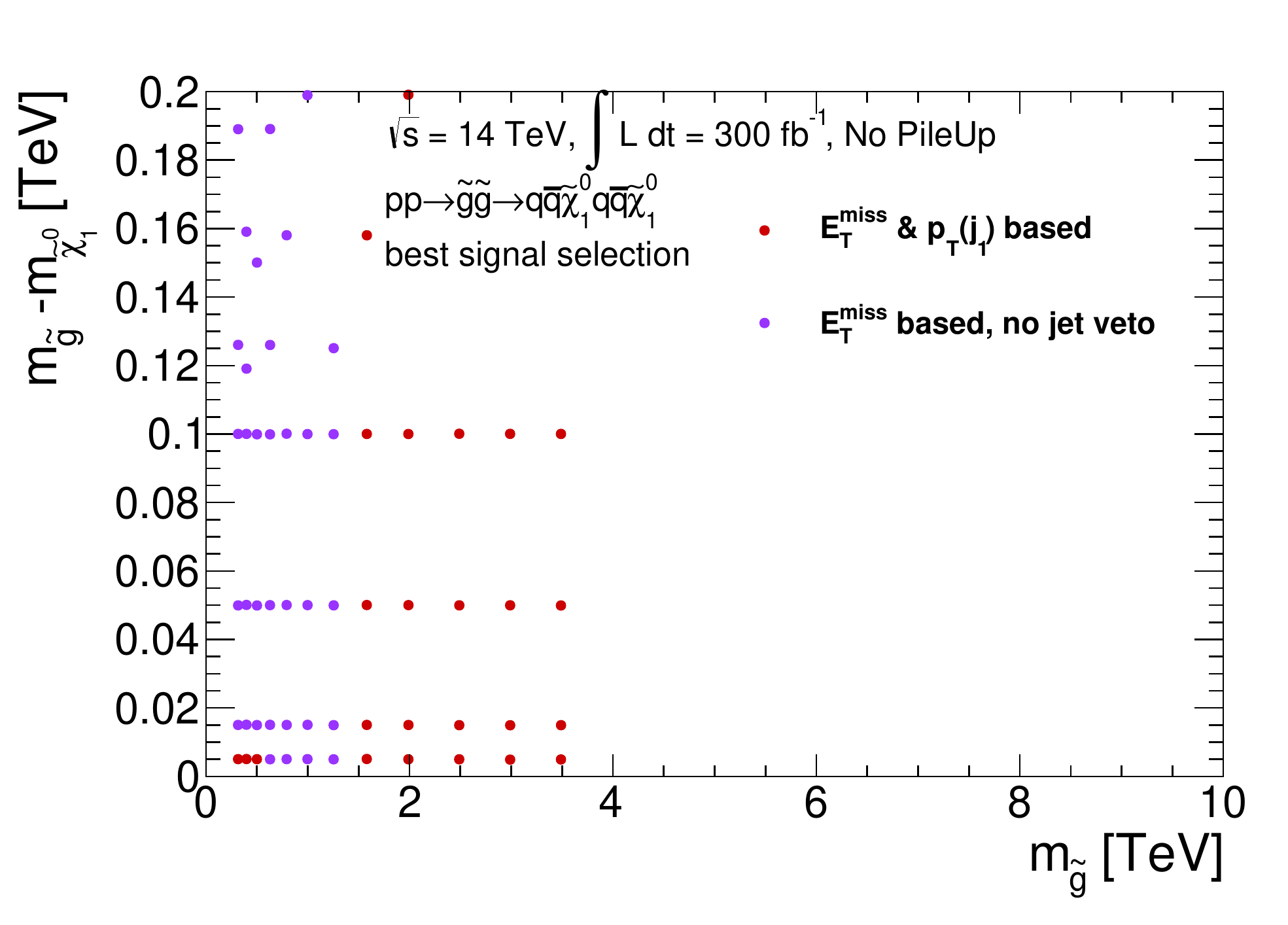}
\includegraphics[width=0.48\textwidth]{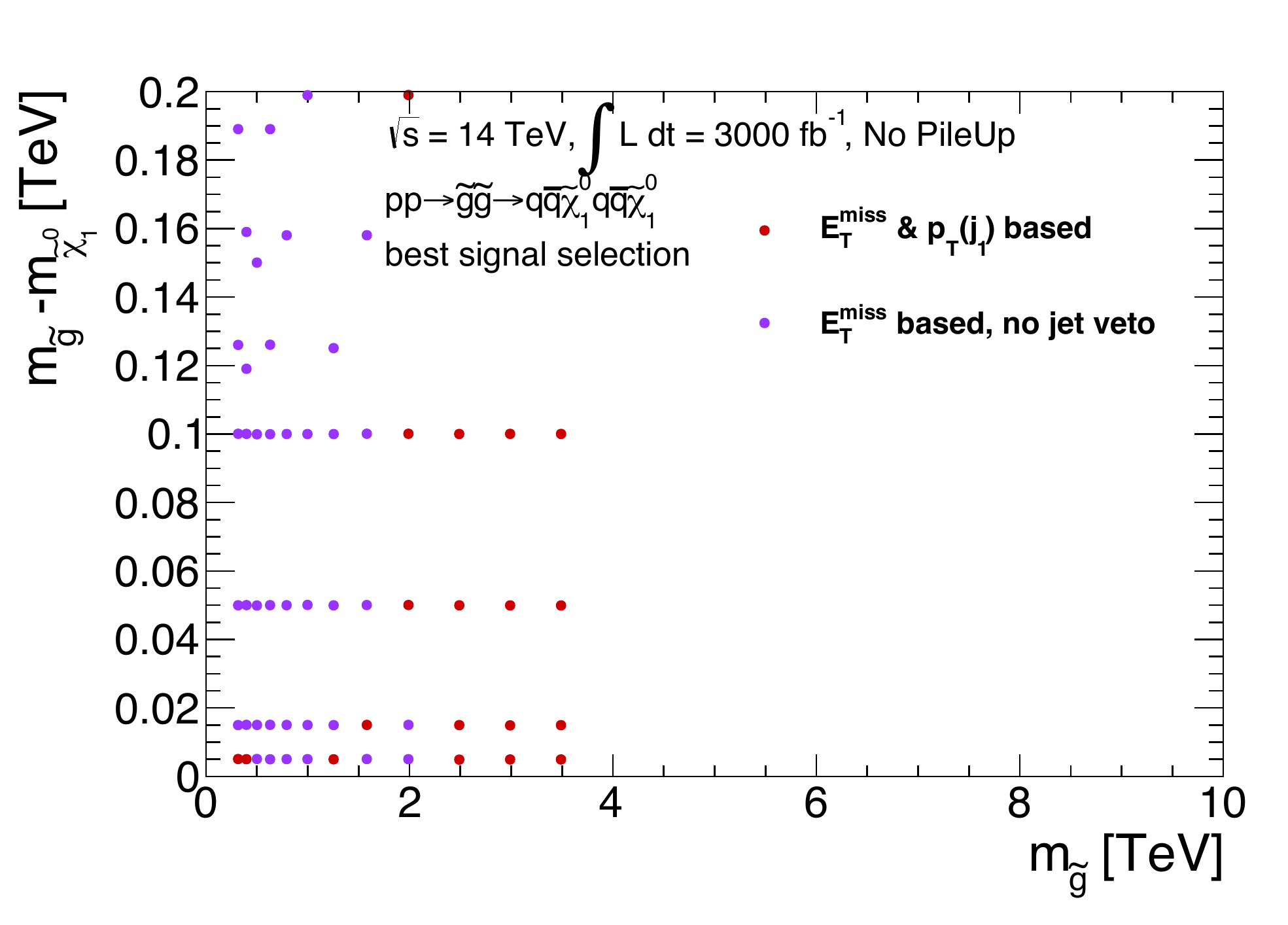}
\caption{The choice of analysis that lead to the best discovery reach for a given point in parameter space for an integrated luminosity of $300$ fb$^{-1}$ [left] and $3000$ fb$^{-1}$ [right] at the $14$ TeV LHC for the compressed region of the gluino-neutralino Simplified Model with light flavor decays. The colors refer to the analyses as presented above: red circle = leading jet based, purple circle = $\MET$-based.  For very high gluino masses, neither analysis can exclude the signal process.}
\label{fig:GNLightFlavorCompressedBestSearch}
\end{center}
\end{figure}

\pagebreak
\hiddensubsection{Results: 14 TeV}
We now apply the compressed analysis to the gluino-neutralino model.  Figure~\ref{fig:GNLightFlavorCompressedBestSearch} shows which of the two selection strategies lead to the best discovery reach in the $m_{\widetilde{\chi}_1^0}\text{ -- }m_{\widetilde{g}}$ plane.  For lighter gluinos and very small values of $\Delta m$ the leading jet based search dominates, while for higher masses and less compression the more inclusive $\MET$~based search leads to the strongest exclusion.  Note that for the points with $m_{\widetilde{g}} \gtrsim 2 \TeV$ neither analysis can exclude the model so that the choice is not particularly relevant.

\begin{figure}[!htb]
\begin{center}
\includegraphics[width=0.48\textwidth]{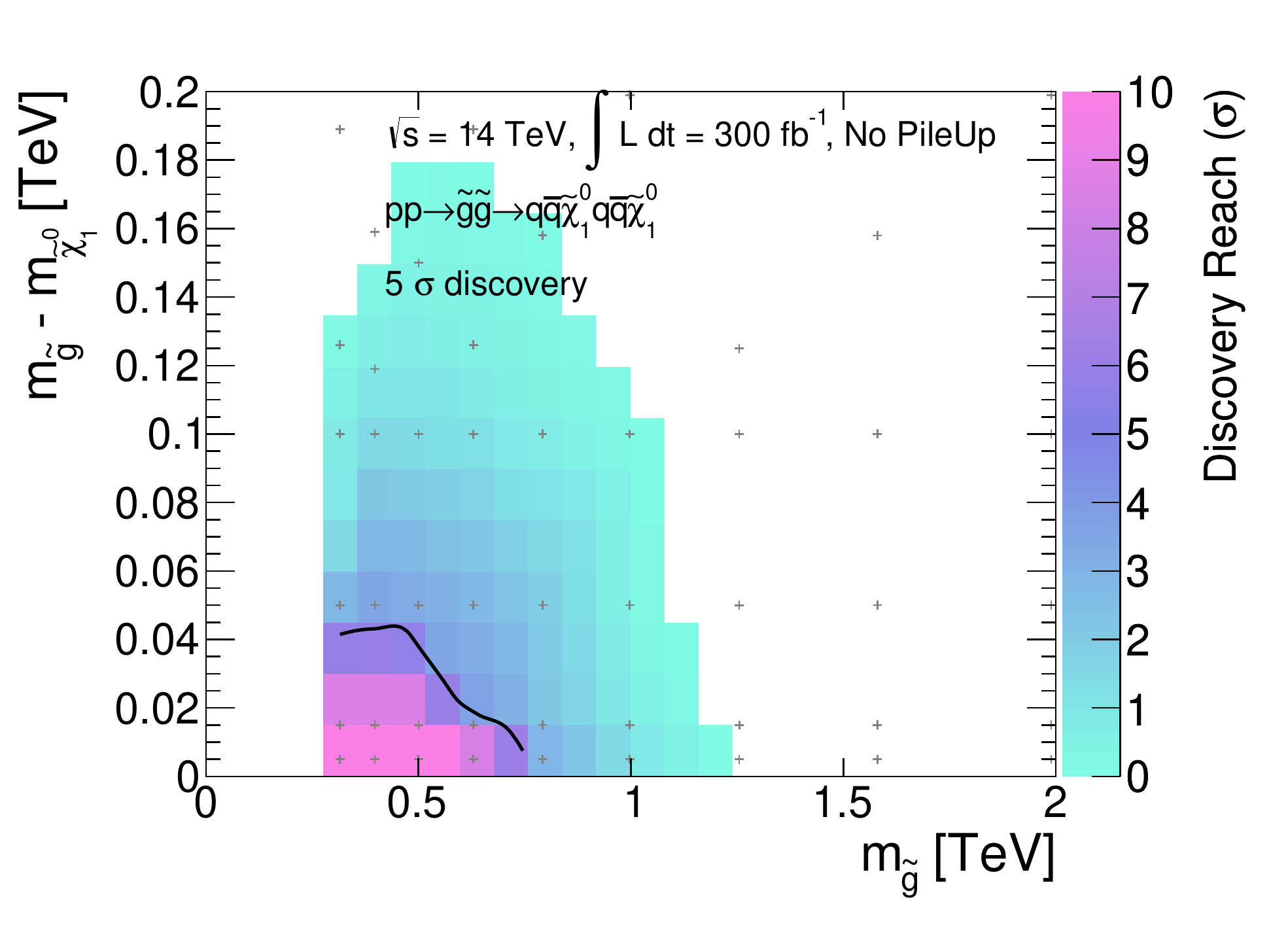}
\includegraphics[width=0.48\textwidth]{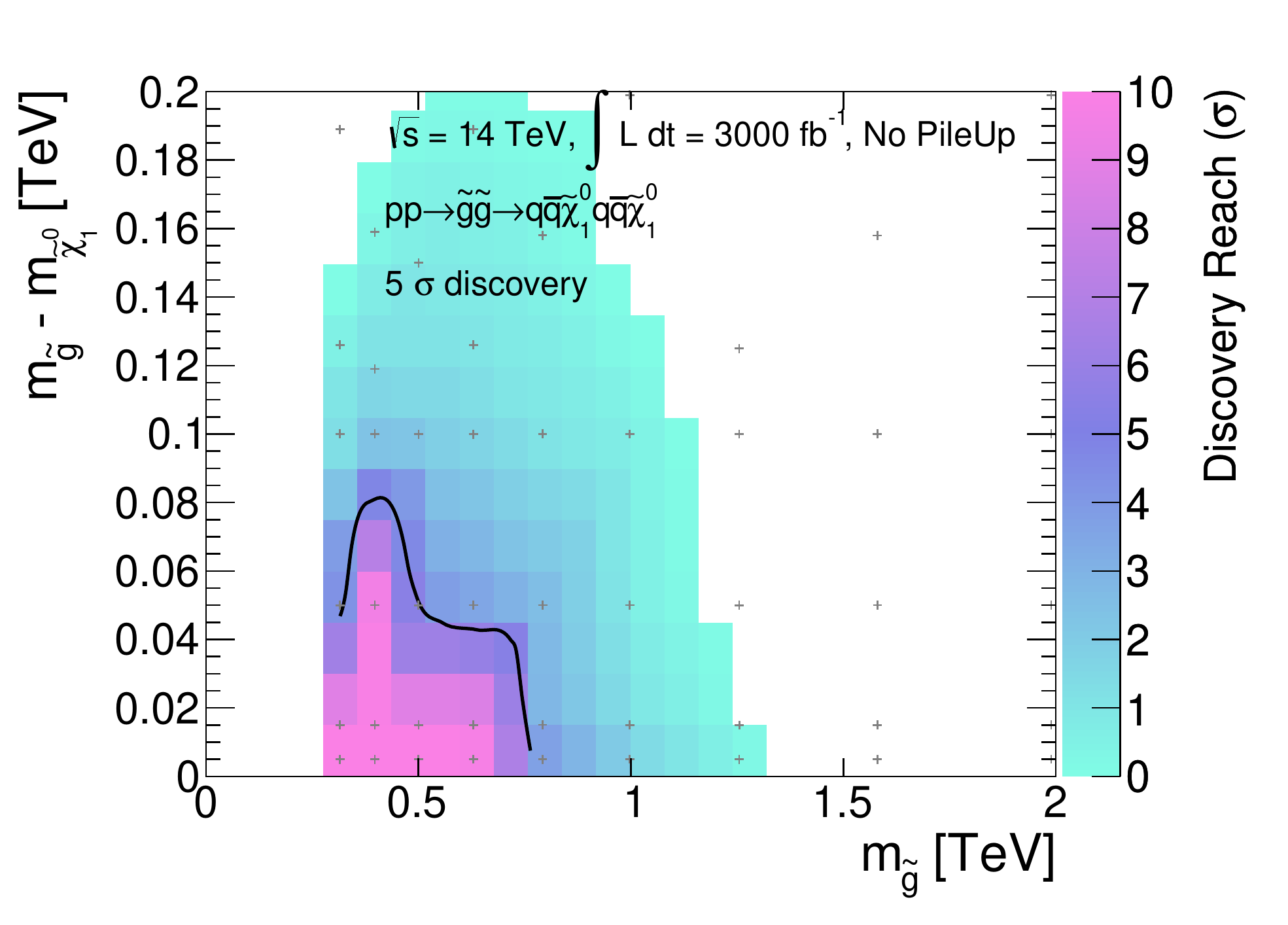}
\includegraphics[width=0.48\textwidth]{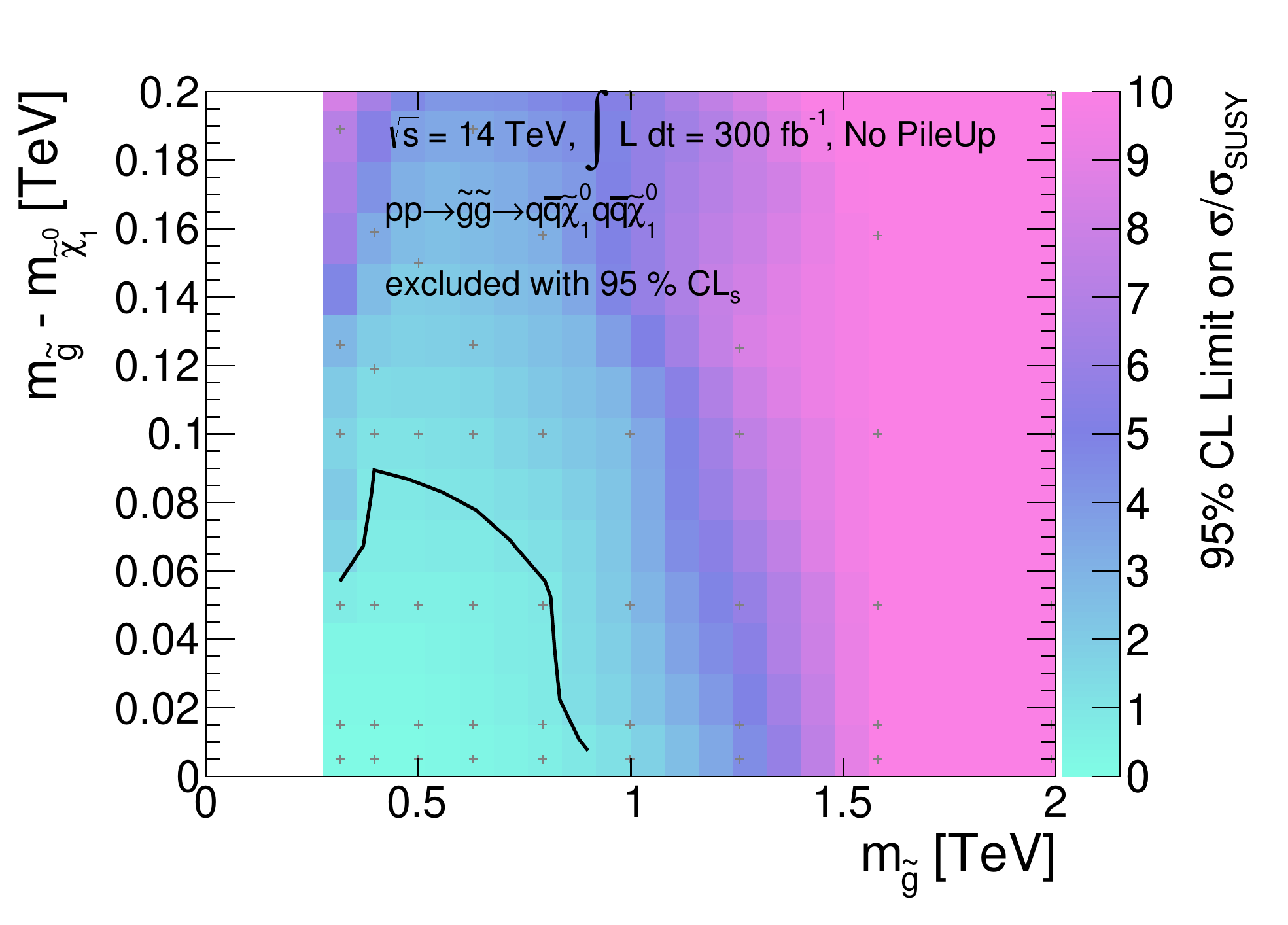}
\includegraphics[width=0.48\textwidth]{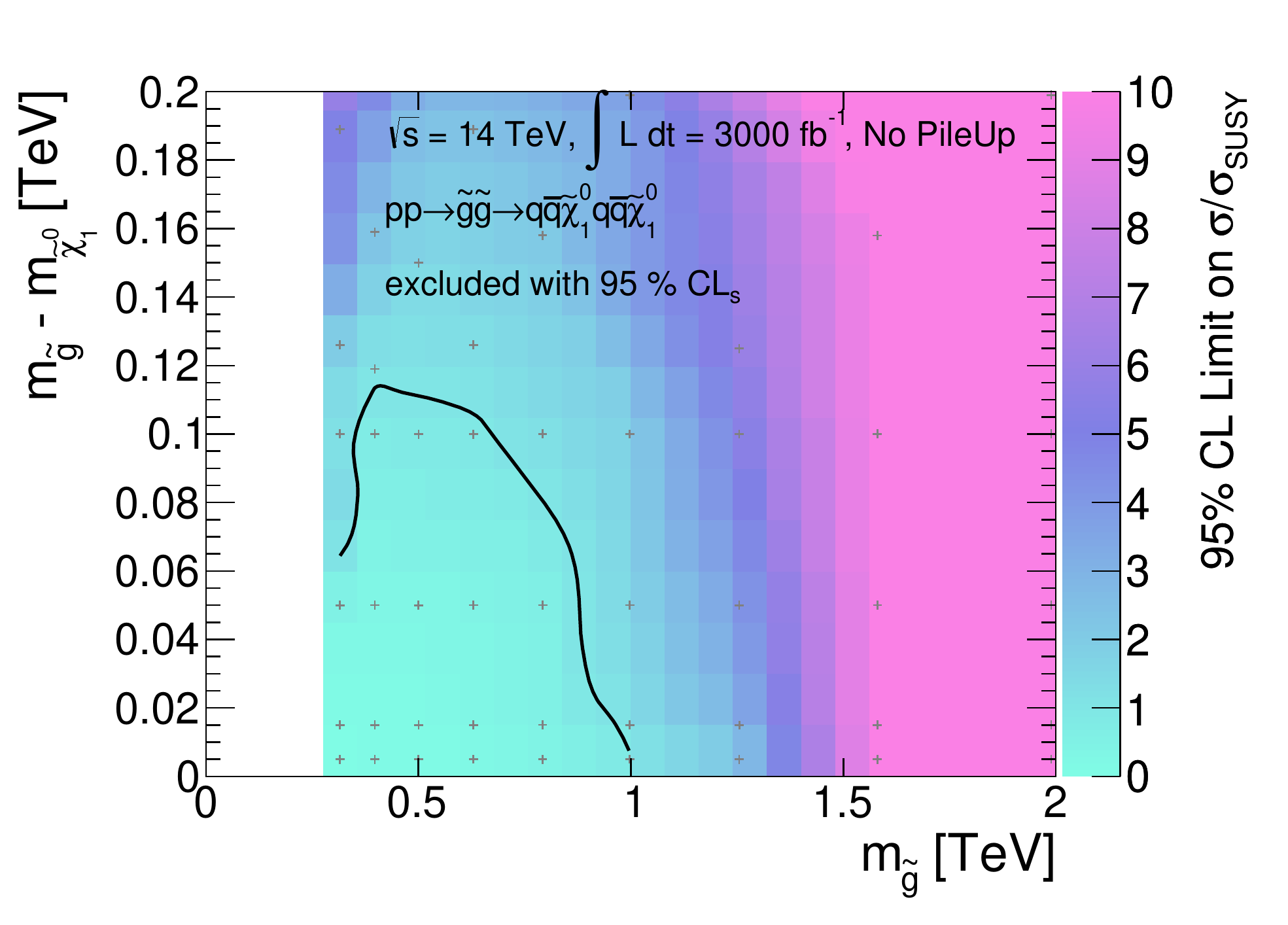}
\caption{Results for the compressed gluino-neutralino model with light flavor decays at the $14$ TeV LHC are given in the $m_{\widetilde{g}} - m_{\widetilde{\chi}_1^0}$ versus $m_{\widetilde{g}}$ plane.  The top [bottom] row shows the expected $5\sigma$ discovery reach [95\% confidence level upper limits] for gluino pair production.  Mass points to the left/below the contours are expected to be probed at $300$ fb$^{-1}$ [left] and $3000$ fb$^{-1}$ [right].  A $20\%$ systematic uncertainty is assumed for the background and pileup is not included.}
\label{fig:GNLightFlavorCompressedResults}
\end{center}
\end{figure}

The results for integrated luminosities of $300$ fb$^{-1}$ and $3000$ fb$^{-1}$ at the $14\TeV$ LHC are shown in Fig.~\ref{fig:GNLightFlavorCompressedResults}. A $20\%$ systematic uncertainty is applied to the backgrounds.  The assumed signal systematic are outlined in the appendix. Pile-up is neglected for these results; its impact is explored in Sec.~\ref{sec:GOGOpileup_compressed} below.  

With $300$ fb$^{-1}$ of data this search can exclude gluino masses of up to approximately $900$ GeV for a mass difference of $5$ GeV, with reduced reach for larger mass differences. The limits increase to around $1$ TeV with a factor of $10$ more data. This improves the reach near the degenerate limit by roughly $200\GeV$ compared to the $\HT$-based analysis described in Sec.~\ref{sec:GNLightFlavor}; the $\HT$-based searches do not begin to set stronger limits until $\Delta\gtrsim50\GeV$.  The combined discovery reach is shown in the bottom row of Fig.~\ref{fig:GNLightFlavorCompressedResults}. The discovery reach of this search is gluino masses up to 800 GeV near the degenerate limit. Unlike the exclusion reach, the discovery reach for this search is not a substantial improvement over the $\HT$-based analysis, even in the degenerate limit.  This occurs because the signal efficiency using these searches is such that there are not enough events to reach $5\sigma$ confidence. Overall, it is clear that the $14$ TeV LHC can have profound implications for models with compressed spectra.

\begin{figure}[!htb]
\begin{center}
\includegraphics[width=0.5\textwidth]{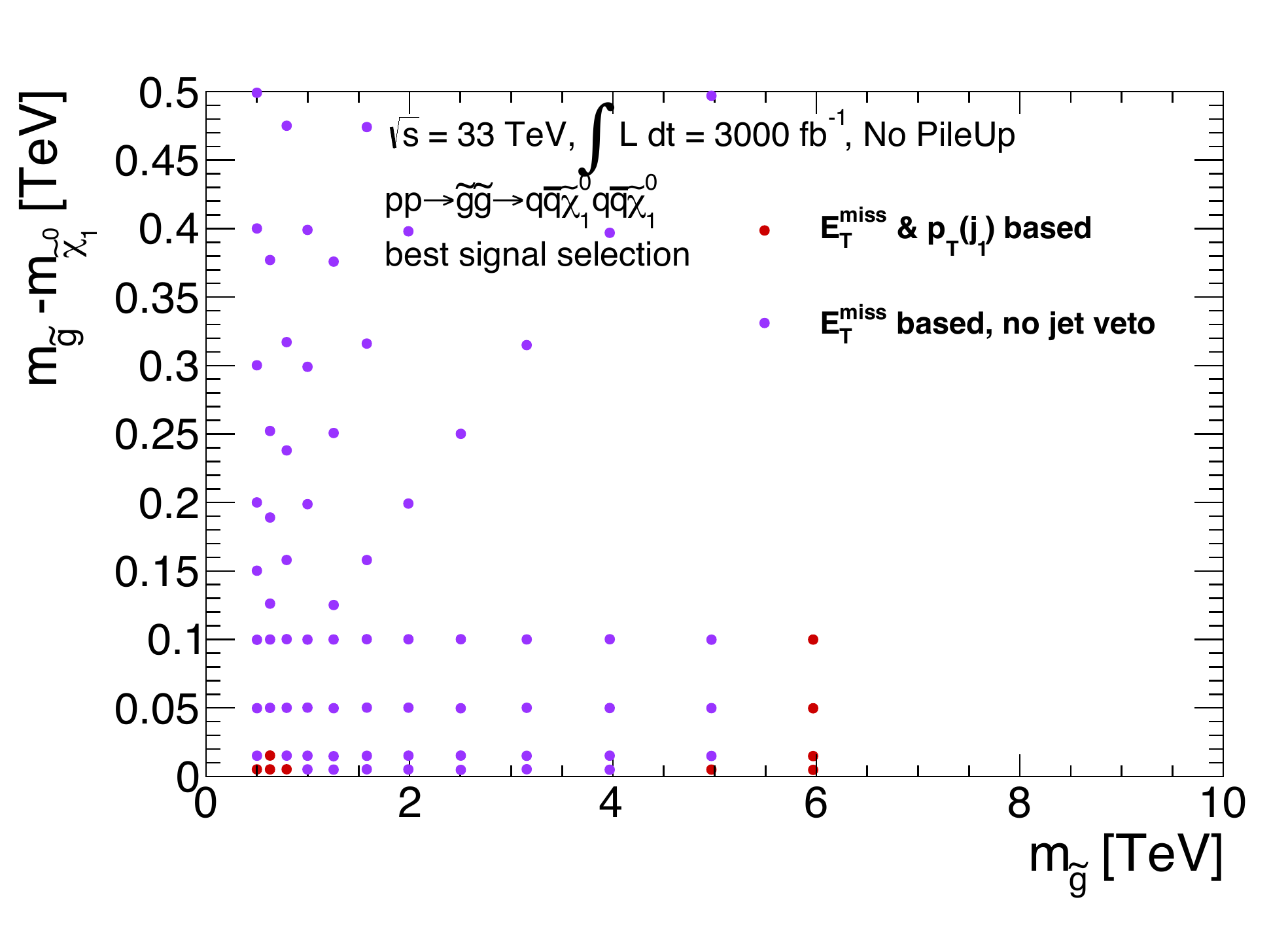}
\caption{The choice of analysis that lead to the best discovery reach for a given point in parameter space for an integrated luminosity of 3000 fb$^{-1}$ at a $33$ TeV proton collider for the compressed region of the gluino-neutralino Simplified Model with light flavor decays. The colors refer to the analyses as presented above: red circle = leading jet based, purple circle = $\MET$-based. For very high gluino masses, neither analyses can exclude the signal process.}
\label{fig:GNLightFlavorCompressedBestSearch_33TeV}
\end{center}
\end{figure}

\pagebreak
\hiddensubsection{Analysis: 33 TeV}
This section is devoted to the details of the $33$ TeV analysis in the compressed region of the gluino-neutralino model with light flavor decays.    As the center-of-mass energy increases, the average $p_T$ of an ISR jet would also increase.  This implies that the probability for more than one ISR jets to pass the preselection cuts will be correspondingly higher, causing the \MET-based search without additional jet veto to have the best acceptance of our search strategies.  From Fig.~\ref{fig:GNLightFlavorCompressedBestSearch_33TeV}, it is clear that this intuition holds; the $\MET$ based search gives the optimal significance for most of the parameter space studied here.

\begin{figure}[!htb]
\begin{center}
\includegraphics[width=0.6\textwidth]{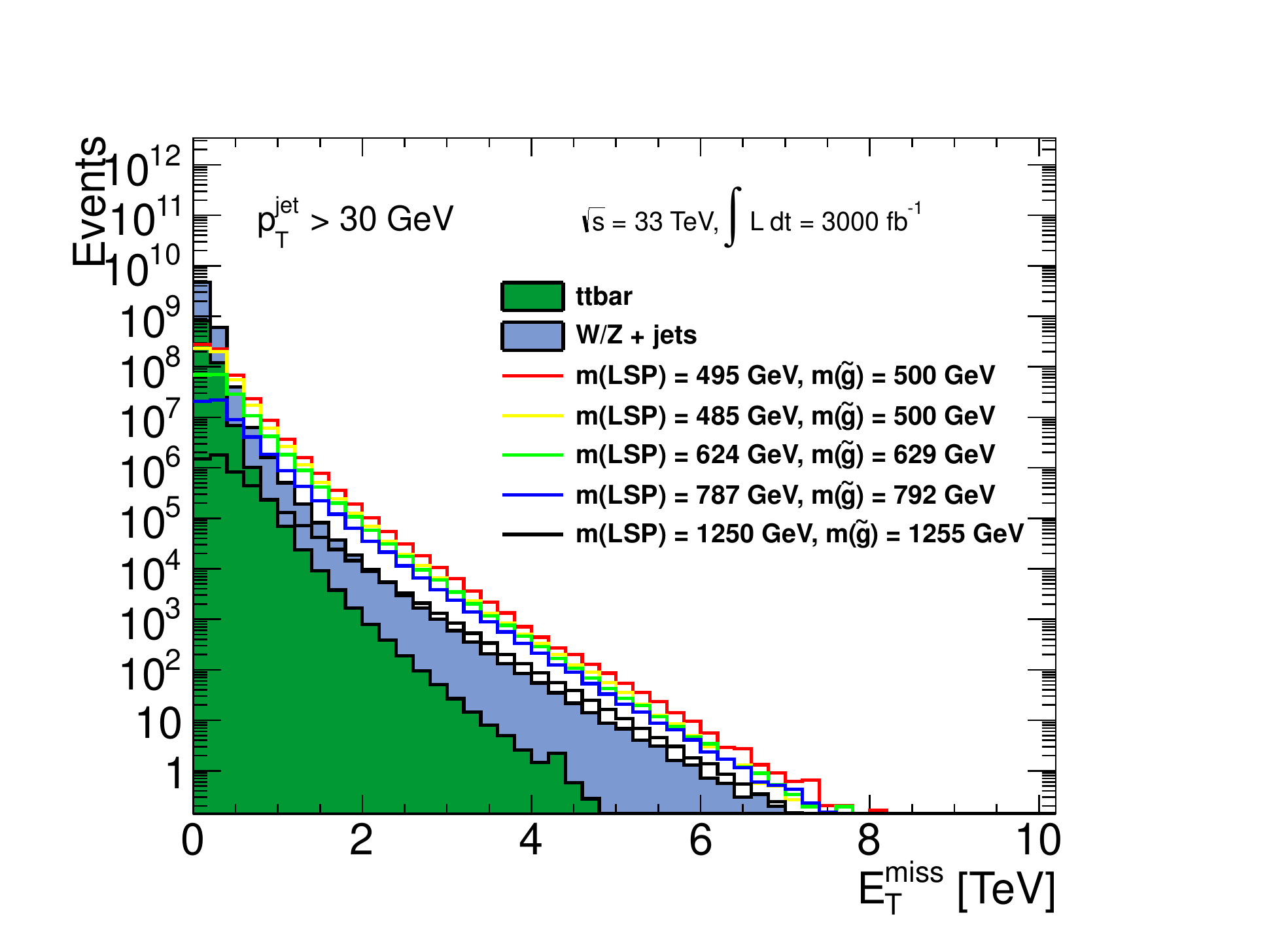}
\caption{Histogram of $\MET$ for signal and background at a $33$ TeV proton collider with $3000 \text{ fb}^{-1}$ after the preselection for a range of gluino and neutralino masses in the compressed region.}
\label{fig:met_presel_distributions_33TeV}
\end{center}
\end{figure}

Figure~\ref{fig:met_presel_distributions_33TeV} gives histograms of $\MET$ distribution for background and a variety of signal models.  It is clear that a cut on $\MET$ can be used to distinguish signal from background.  This can be seen quantitatively using Table \ref{tab:alljets_counts_33TeV}, where the cut flows are given for background and three signal models.

\begin{table}[h!]
\begin{centering}
\renewcommand{\arraystretch}{1.8}
\setlength{\tabcolsep}{5pt}
\footnotesize
\vskip 10pt
\begin{tabular}{|r|rr|r|rrr|}
\hline
&&&&\multicolumn{3}{c|}{$\big(m_{\widetilde{g}},\,m_{\widetilde{\chi}^0_1}\big)\quad$ [GeV]} \\
Cut & $V$+jets & $t\,\overline{t}$ & Total BG & $(797,\,792)$ & $(997,\,992)$ & $(1580,\,1575)$\\
\hline
\hline
Preselection & $4.2 \times 10^{9}$ & $8.6 \times 10^{8}$ & $5.1 \times 10^{9}$ & $5.5 \times 10^{7}$ & $ 1.7 \times 10^{7}$ & $1.2 \times 10^{6}$\\
\hline
$p_{T}^{\text{leadjet}} > 110 \text{ GeV}, |\eta^{\text{leadjet}}| < 2.4$ & $2.6 \times 10^{9}$ & $6.5 \times 10^{8}$ & $3.3 \times 10^{9}$ & $4.8 \times 10^{7}$ & $1.5 \times 10^{7}$ & $1.1 \times 10^{6}$\\
\hline
\hline
$\MET > 1 \text{ TeV}$ & $7.5 \times 10^{5}$ & $1.1 \times 10^{5}$ & $8.6 \times 10^{5}$ & $1.8 \times 10^{6}$ & $\color{black}8.0 \times 10^{5}$ & $1.0 \times 10^{5}$\\
\hline
$\MET > 3 \text{ TeV}$ & $1.5 \times 10^{3}$ & 62 & $1.5 \times 10^{3}$ & $6.1 \times 10^{3}$ & $4.0 \times 10^{3}$ & $\color{black}1.1 \times 10^{3}$\\
\hline
$\MET > 5 \text{ TeV}$ & $19$ & $0$ & $19$ & $62$ & $50$ & $19$ \\
\hline
\end{tabular}
\caption{Number of expected events for $\sqrt{s} = 33$ TeV and $3000$ fb$^{-1}$ for the background processes and selected signal processes. The selection without a veto on additional jets with cuts on \MET~is applied.  Three choices of cuts are provided for illustration.}
\label{tab:alljets_counts_33TeV}
\end{centering}
\end{table}

\begin{figure}[!htb]
\begin{center}
\includegraphics[width=0.48\textwidth]{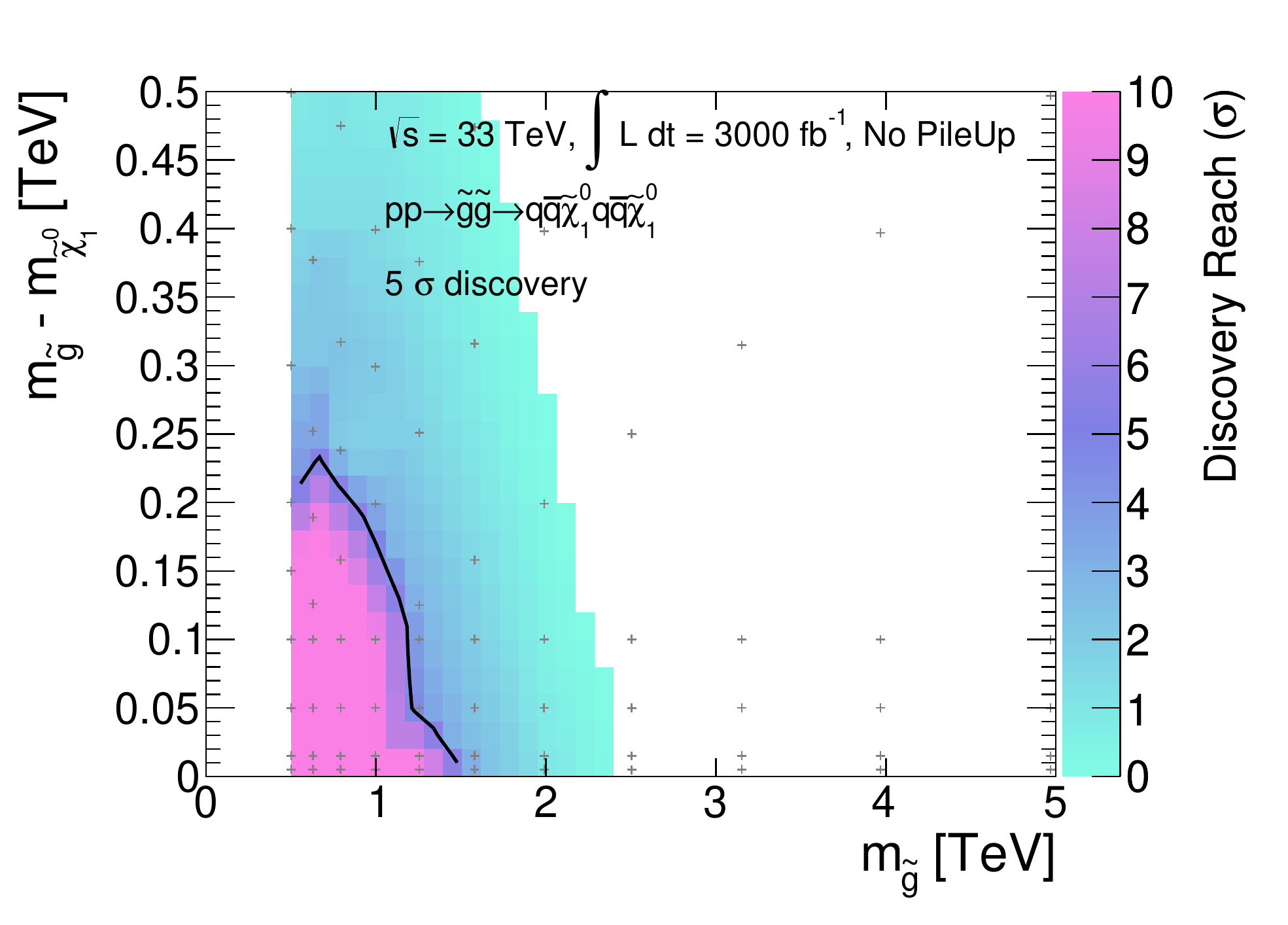}
\includegraphics[width=0.48\textwidth]{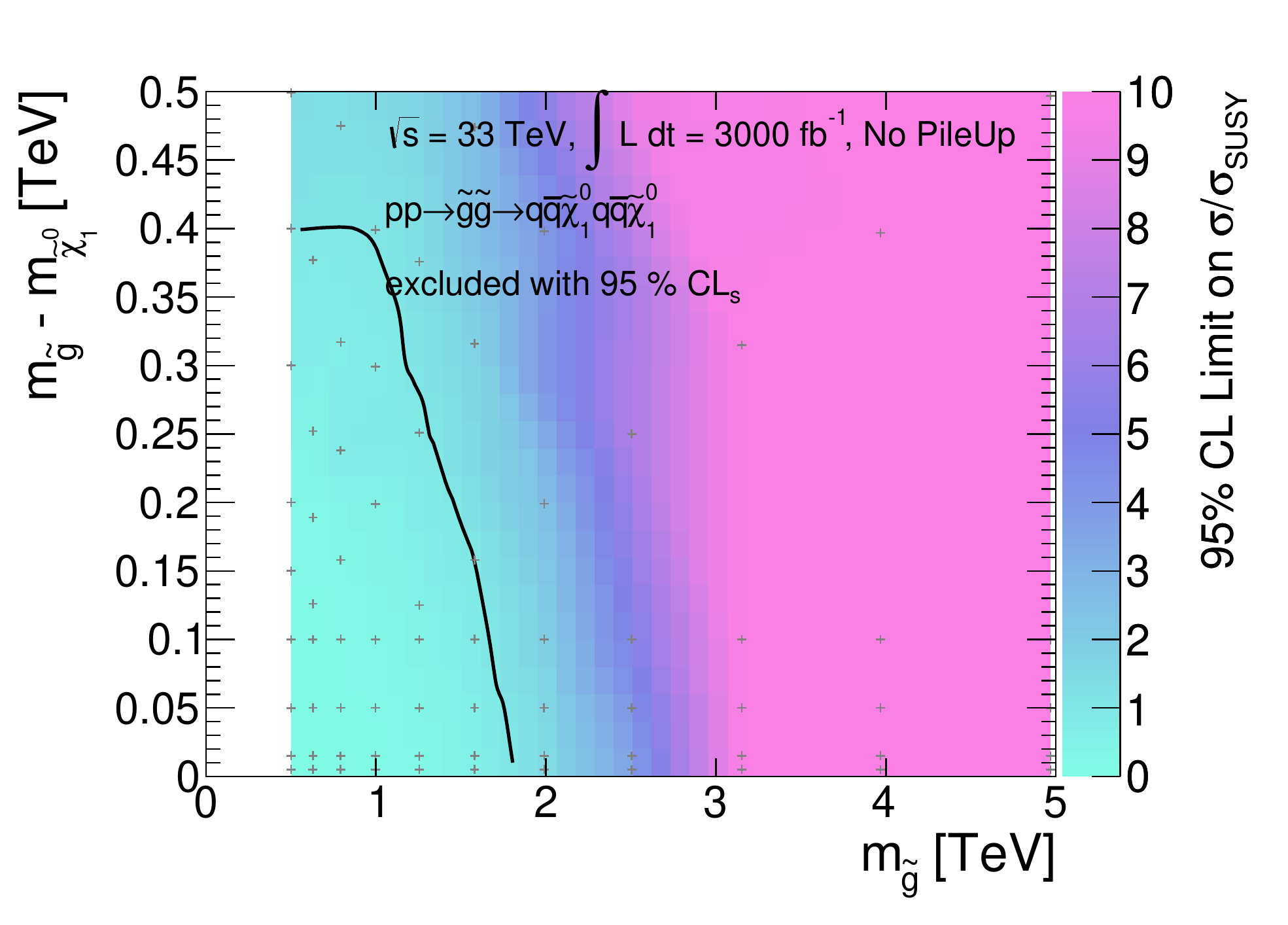}
\caption{Results for the compressed gluino-neutralino model with light flavor decays at a $33$ TeV proton collider with $3000 \text{ fb}^{-1}$ are given in the $m_{\widetilde{g}} - m_{\widetilde{\chi}_1^0}$ versus $m_{\widetilde{g}}$ plane.  The left [right] plot shows the expected $5\sigma$ discovery reach [95\% confidence level upper limits] for gluino pair production.  Mass points to the left/below the contours are expected to be probed at $3000$ fb$^{-1}$ [right].  A 20\% systematic uncertainty is assumed for the background and pileup is not included.}
\label{fig:GNLightFlavorCompressedResults_33TeV}
\end{center}
\end{figure}

\pagebreak
$ $
\pagebreak
\hiddensubsection{Results: 33 TeV}
The $5\sigma$ discovery [left] and $95\%$ C.L.~limits [right] for the gluino-neutralino model are shown in Fig.~\ref{fig:GNLightFlavorCompressedResults_33TeV}, assuming $3000 \text{ fb}^{-1}$ of integrated luminosity.  $20\%$ systematic uncertainty is applied to the backgrounds.  The assumed signal systematic are outlined in the appendix.  Pileup is not included; a demonstration that pileup will not significantly change these results is given in Sec.~\ref{sec:GOGOpileup_compressed} below.

For a $33$ TeV proton collider with $3000$ fb$^{-1}$ of data, the exclusion reach for a mass difference of $5$ GeV covers gluino masses of up to approximately $1.8$ TeV, with reduced reach for larger mass differences.   For very small mass differences in the range of $5$ to $50$ GeV discoveries could be made for gluino masses up to $1.4$ TeV.  This search improves the exclusion (discovery) reach near the degenerate limit by roughly $800\GeV$ ($400\GeV$) compared to the $\HT$-based analysis described in Sec.~\ref{sec:GNLightFlavor}; the $\HT$-based searches do not begin to set stronger limits until $\Delta\gtrsim50\GeV$. Overall, it is clear that a $33$ TeV proton collider can have profound implications for models with compressed spectra.

\begin{figure}[!htb]
\begin{center}
\includegraphics[width=0.5\textwidth]{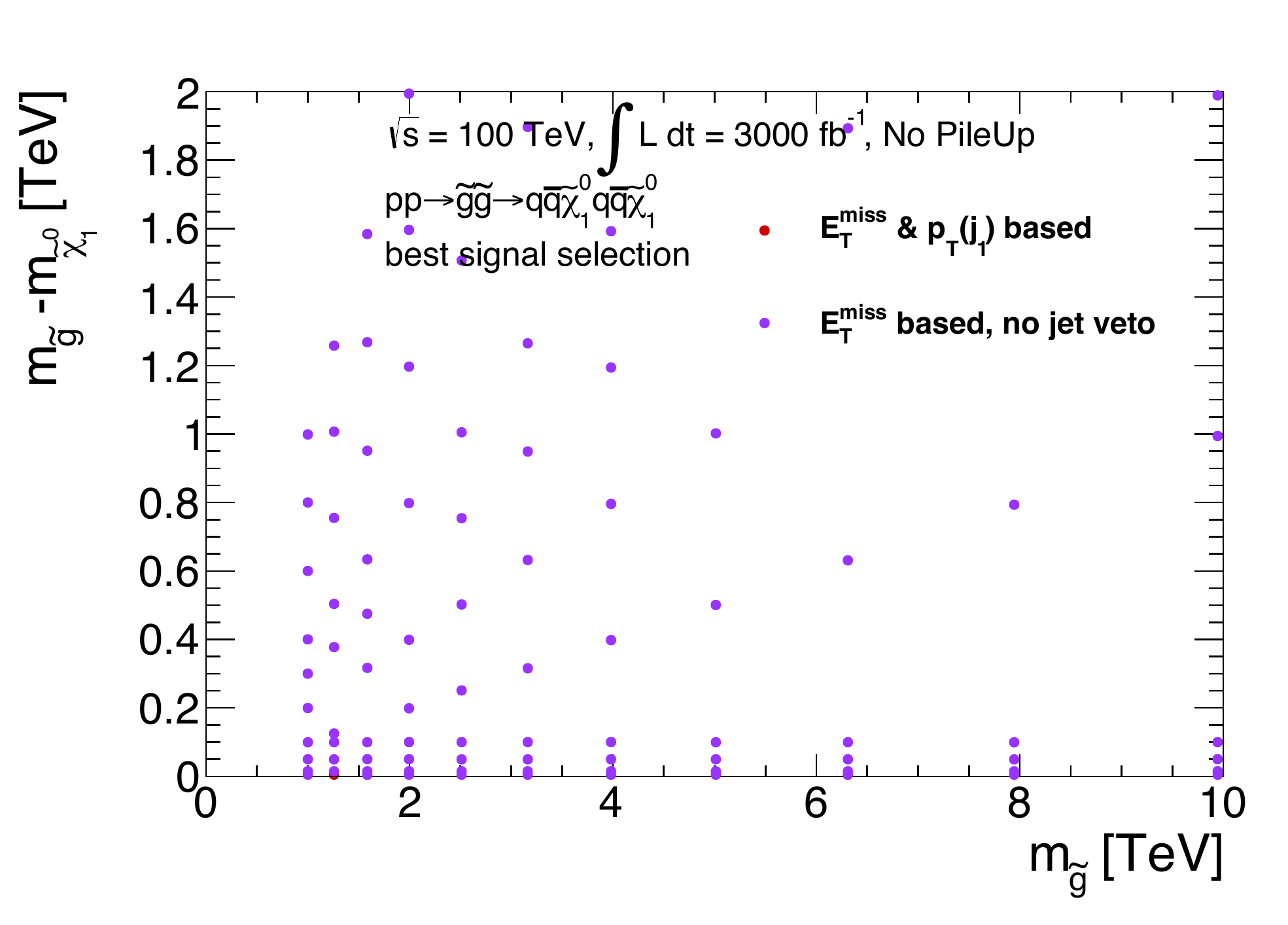}
\caption{The choice of analysis that lead to the best discovery reach for a given point in parameter space for an integrated luminosity of 3000 fb$^{-1}$ at a $100$ TeV proton collider for the compressed region of the gluino-neutralino Simplified Model with light flavor decays. The colors refer to the analyses as presented above: red circle = leading jet based, purple circle = $\MET$-based.}
\label{fig:GNLightFlavorCompressedBestSearch_100TeV}
\end{center}
\end{figure}

\begin{figure}[!h]
\begin{center}
\includegraphics[width=0.6\textwidth]{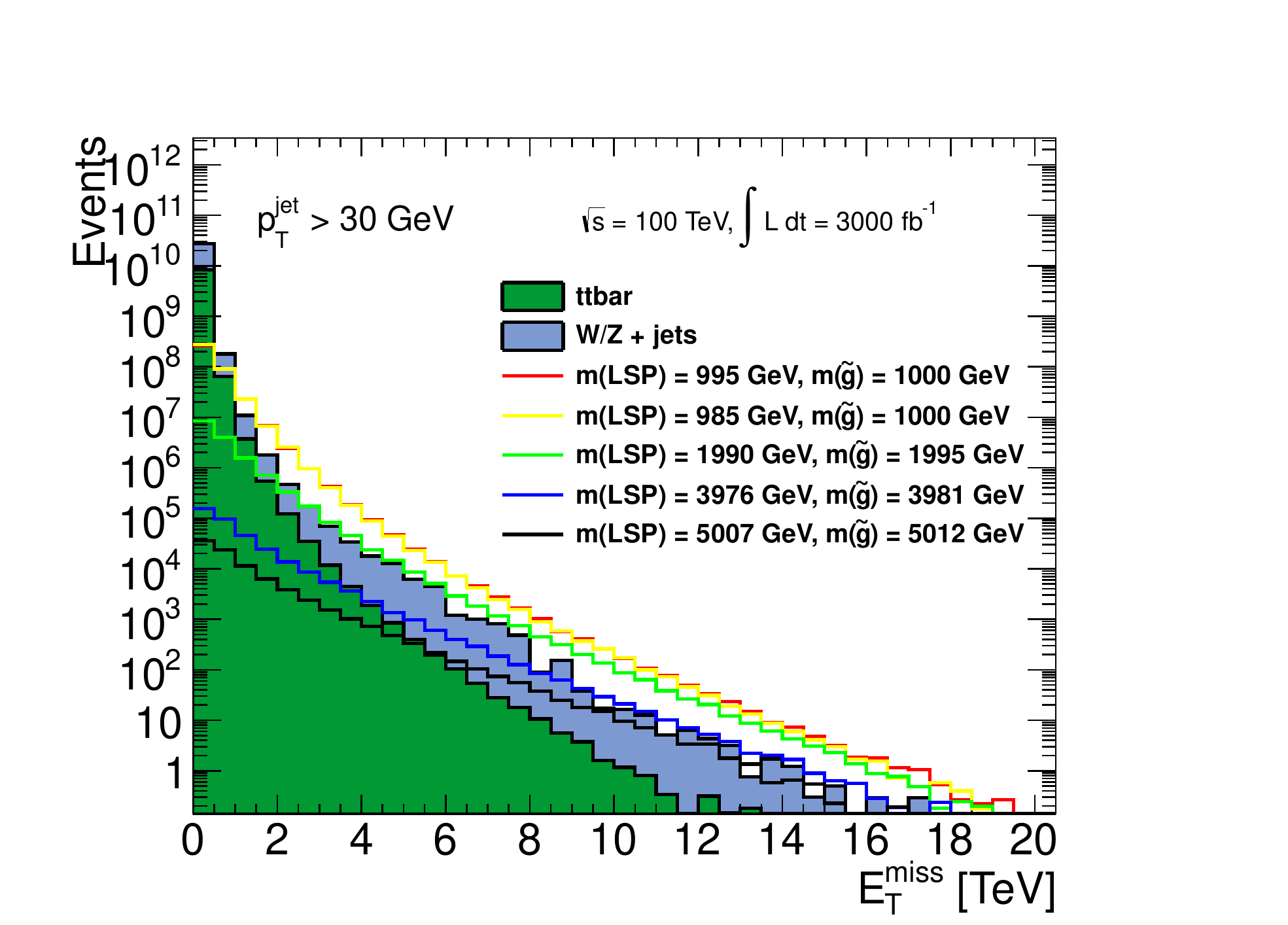}
\caption{Histogram of $\MET$ for signal and background at a $100$ TeV proton collider with $3000 \text{ fb}^{-1}$ after the preselection for a range of gluino and neutralino masses in the compressed region.}
\label{fig:met_presel_distributions_100TeV}
\end{center}
\end{figure}

\pagebreak
\hiddensubsection{Analysis: 100 TeV}
This section is devoted to the details of the $100$ TeV analysis in the compressed region of the gluino-neutralino model with light flavor decays.    As the center-of-mass energy increases, the average $p_T$ of an ISR jet would also increase.  This implies that the probability for more than one ISR jets to pass the preselection cuts will be correspondingly higher, causing the \MET-based search to have the best acceptance of our search strategies.  From Fig.~\ref{fig:GNLightFlavorCompressedBestSearch_100TeV}, it is clear that this intuition holds; the $\MET$ based search gives the optimal significance for all of the probable parameter space.  Figure~\ref{fig:met_presel_distributions_100TeV} gives histograms of $\MET$ for background and a variety of signal models.   It is clear that a cut on $\MET$ can be used to distinguish signal from background.  This can be seen in Table \ref{tab:alljets_counts_100TeV} where the cut flows are given for background and three signal models.

\begin{table}[h!]
\begin{centering}
\renewcommand{\arraystretch}{1.8}
\setlength{\tabcolsep}{3.pt}
\footnotesize
\vskip 10pt
\begin{tabular}{|r|rr|r|rrr|}
\hline
&&&&\multicolumn{3}{c|}{$\big(m_{\widetilde{g}},\,m_{\widetilde{\chi}^0_1}\big)\quad$ [GeV]} \\
Cut & $V$+jets & $t\,\overline{t}$ & Total BG & $(1995,\,1990)$ & $(2512,\,2507)$ & $(5012,\,5007)$\\
\hline
\hline
$\text{Preselection}$ & $1.7 \times 10^{10}$ & $7.0 \times 10^{9}$ & $2.4 \times 10^{10}$ & $1.3 \times 10^{7}$ & $4.1 \times 10^{6}$ & $7.9 \times 10^{4}$\\
\hline
$p_{T}^{\text{leadjet}} > 110 \text{ GeV}, |\eta^{\text{leadjet}}| < 2.4$ & $1.2 \times 10^{10}$ & $6.1 \times 10^{9}$ & $1.9 \times 10^{10}$ & $1.3 \times 10^{7}$ & $4.1 \times 10^{6}$ & $7.9 \times 10^{4}$\\
\hline
\hline
$\MET > 3 \text{ TeV}$ & $1.3 \times 10^{5}$ & $2.0 \times 10^{4}$ & $1.5 \times 10^{5}$ & $1.9 \times 10^{5}$ & $9.4 \times 10^{4}$ & $4.8 \times 10^{3}$\\
\hline
$\MET > 6 \text{ TeV}$ & $3.6 \times 10^{3}$ & 229 & $3.8 \times 10^{3}$ & $8.0 \times 10^{3}$ & $5.1 \times 10^{3}$ & $509$ \\
\hline
$\MET > 9 \text{ TeV}$ & $100$ & $9$ & $109$ & $612$ & $410$ & $67$ \\
\hline
\end{tabular}
\caption{Number of expected events for $\sqrt{s} = 100$ TeV and $3000$ fb$^{-1}$ for the background processes and selected signal processes. The selection without a veto on additional jets with cuts on \MET~is applied.  Three choices of cuts are provided for illustration.}
\label{tab:alljets_counts_100TeV}
\end{centering}
\end{table}

\begin{figure}[!htb]
\begin{center}
\includegraphics[width=0.48\textwidth]{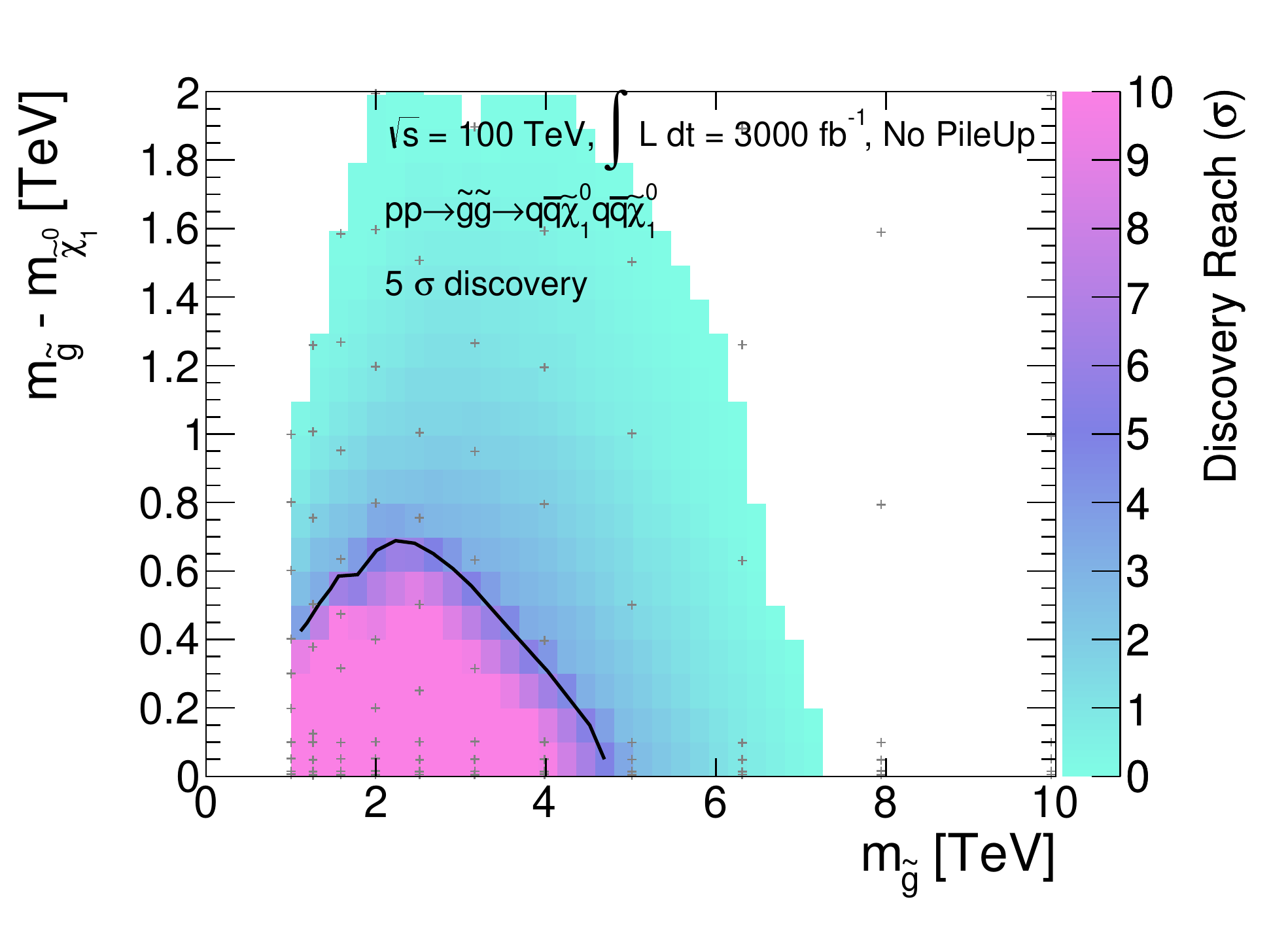}
\includegraphics[width=0.48\textwidth]{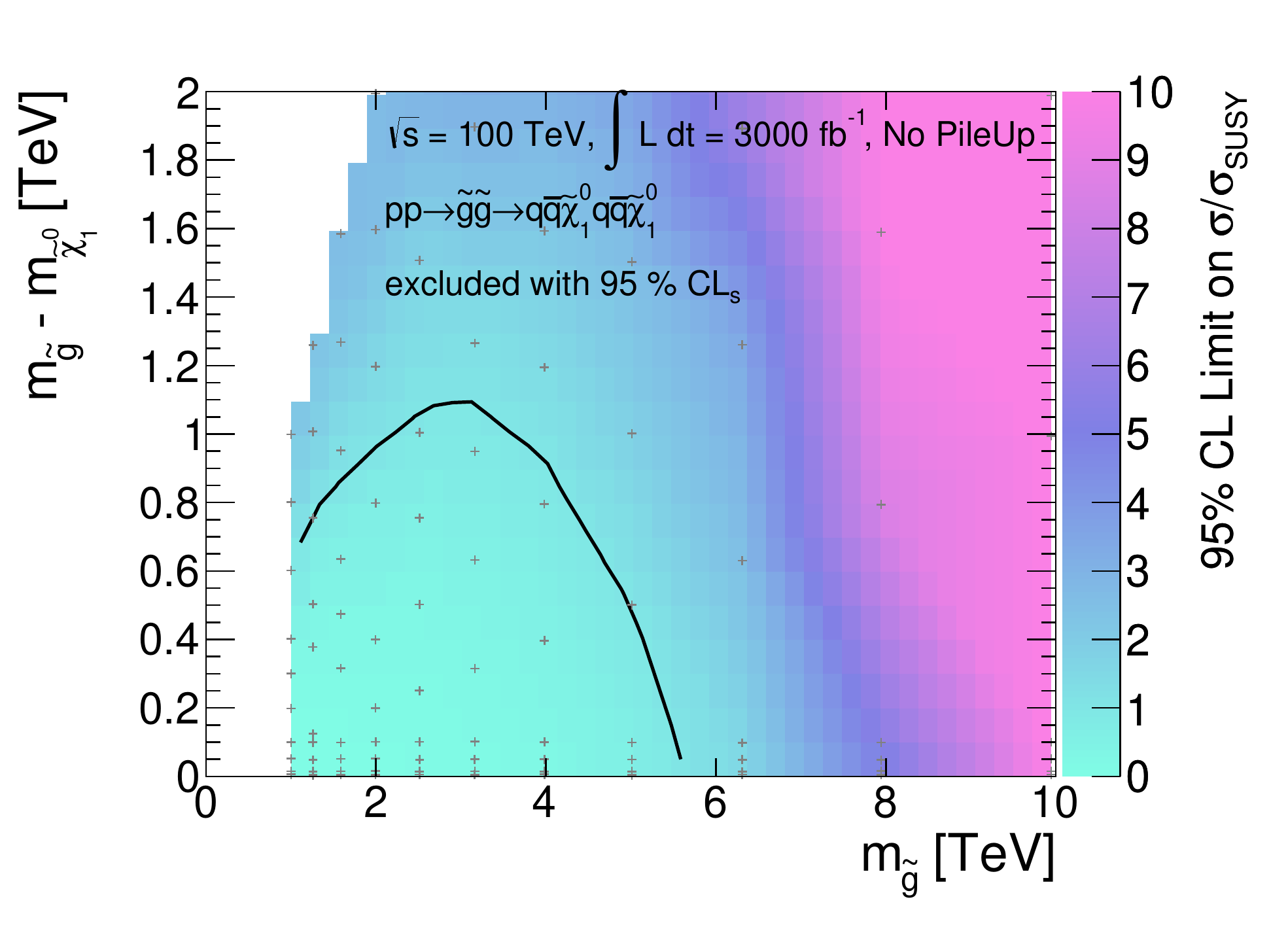}
\caption{Results for the compressed gluino-neutralino model with light flavor decays at a $100$ TeV proton collider are given in the $m_{\widetilde{g}} - m_{\widetilde{\chi}_1^0}$ versus $m_{\widetilde{g}}$ plane.  The left [right] plot shows the expected $5\sigma$ discovery reach [95\% confidence level upper limits] for gluino pair production.  Mass points to the left/below the contours are expected to be probed at $3000$ fb$^{-1}$ [right].  A 20\% systematic uncertainty is assumed for the background and pileup is not included.}
\label{fig:GNLightFlavorCompressedResults_100TeV}
\end{center}
\end{figure}

\pagebreak
$ $
\pagebreak
\hiddensubsection{Results: 100 TeV}
The $5\sigma$ discovery [left] and $95\%$ C.L.~limits [right] for the gluino-neutralino model are shown in Fig.~\ref{fig:GNLightFlavorCompressedResults_100TeV}, assuming $3000 \text{ fb}^{-1}$ of integrated luminosity.  $20\%$ systematic uncertainty is applied to the backgrounds.  The assumed signal systematic are outlined in the appendix.  Pileup is not included; a demonstration that pileup will not significantly change these results is given in Sec.~\ref{sec:GOGOpileup_compressed} below.

For a $100$ TeV proton collider with $3000$ fb$^{-1}$ of data, the exclusion reach for a mass difference of $5$ GeV covers gluino masses of up to approximately $5.7$ TeV, with reduced reach for larger mass differences.   For very small mass differences discoveries could be made for gluino masses up to $4.8$ TeV. This search improves the exclusion (discovery) reach near the degenerate limit by roughly $1.7\TeV$ ($1.3\TeV$) compared to the $\HT$-based analysis described in Sec.~\ref{sec:GNLightFlavor}; the $\HT$-based searches do not begin to set stronger limits until $\Delta\gtrsim500\GeV$.  Overall, it is clear that a $100$ TeV proton collider can have profound implications for models with compressed spectra.

\begin{figure}[h!]
  \centering
  \includegraphics[width=.48\columnwidth]{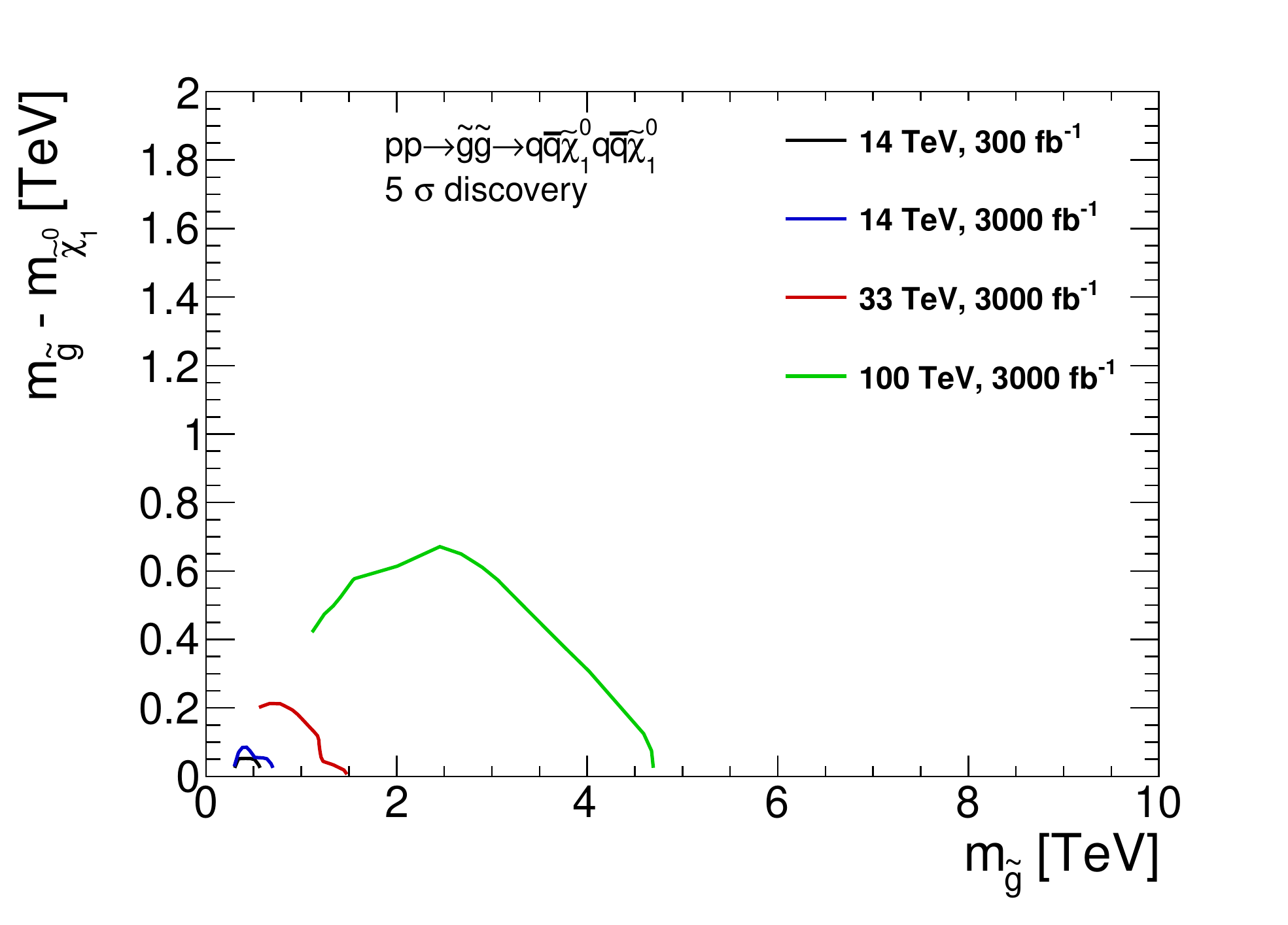}
  \includegraphics[width=.48\columnwidth]{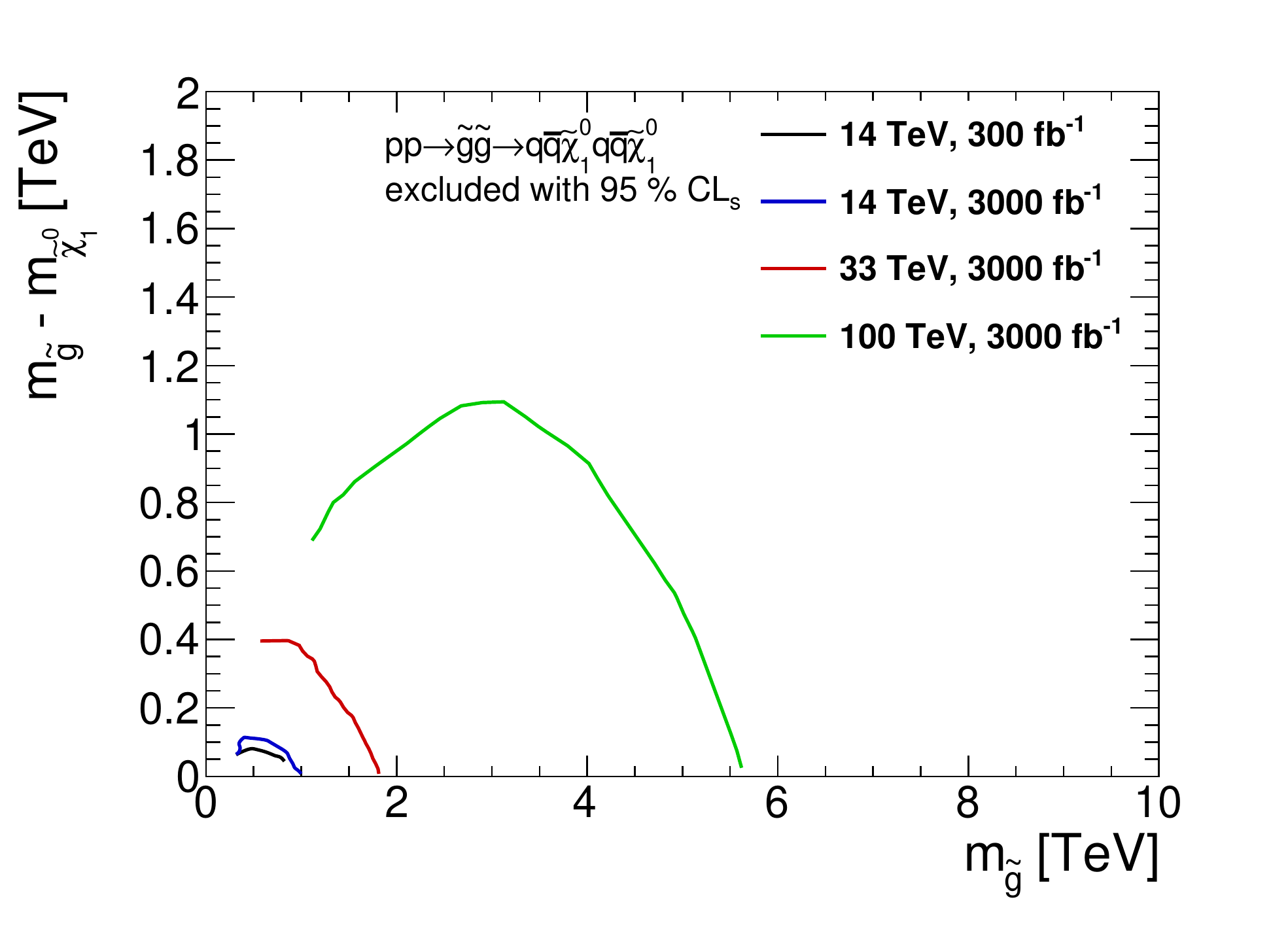}
  \caption{Results for the gluino-neutralino model with light flavor decays for the analyses that target the compressed region of parameter space.  The left [right] panel shows the $5\,\sigma$ discovery reach [$95\%$ CL exclusion] for the four collider scenarios studied here.  A $20\%$ systematic uncertainty is assumed and pileup is not included.}
    \label{fig:GOGO_Compressed_Comparison}
\end{figure}

\pagebreak
\hiddensubsection{Comparing Colliders}
The search for the gluino-neutralino model with light flavor decays in the compressed region provides an interesting case study with which to compare the potential impact of different proton colliders. Figure \ref{fig:GOGO_Compressed_Comparison} shows the $5\sigma$ discovery reach [$95\%$ CL exclusion] for two choices of integrated luminosity at $14$ TeV, along with the full data set assumed for $33$ and $100$ TeV. At $14$ TeV, the factor of $10$ increase in luminosity leads to a modest increase from $900$ GeV to $1000$ GeV in the gluino limits.  These limits have a strong dependence on the assumed systematic uncertainties.  Therefore, the increase in luminosity does not have a tremendous impact on the ability to probe higher mass gluinos.

In contrast, increasing the center-of-mass energy has a tremendous impact on the experimentally available parameter space.  For these machines, significantly heavier gluinos can be produced and more hard ISR jets are expected. For higher center-of-mass-energy, these searches specially targeted at the compressed region also become more and more important to fill in the gap in the reach of the untargeted search described in Sec.~\ref{sec:GNLightFlavor}. Figure \ref{fig:GOGO_Compressed_Comparison}  makes a compelling case for investing in future proton colliders which can operate at these high energies.

\pagebreak
\subsection{Impact of Systematic Uncertainties}
\label{sec:CompressedSystematic}
In the previous studies the systematic uncertainties on background are assumed to account for $20\%$ on the overall background normalization.  In the event of a discovery, it is likely that this error will be reduced dramatically as tremendous effort will be devoted to understanding these backgrounds in detail.  It is therefore interesting to study the impact of this assumption.

\begin{figure}[!htb]
\begin{center}
\includegraphics[width=0.48\textwidth]{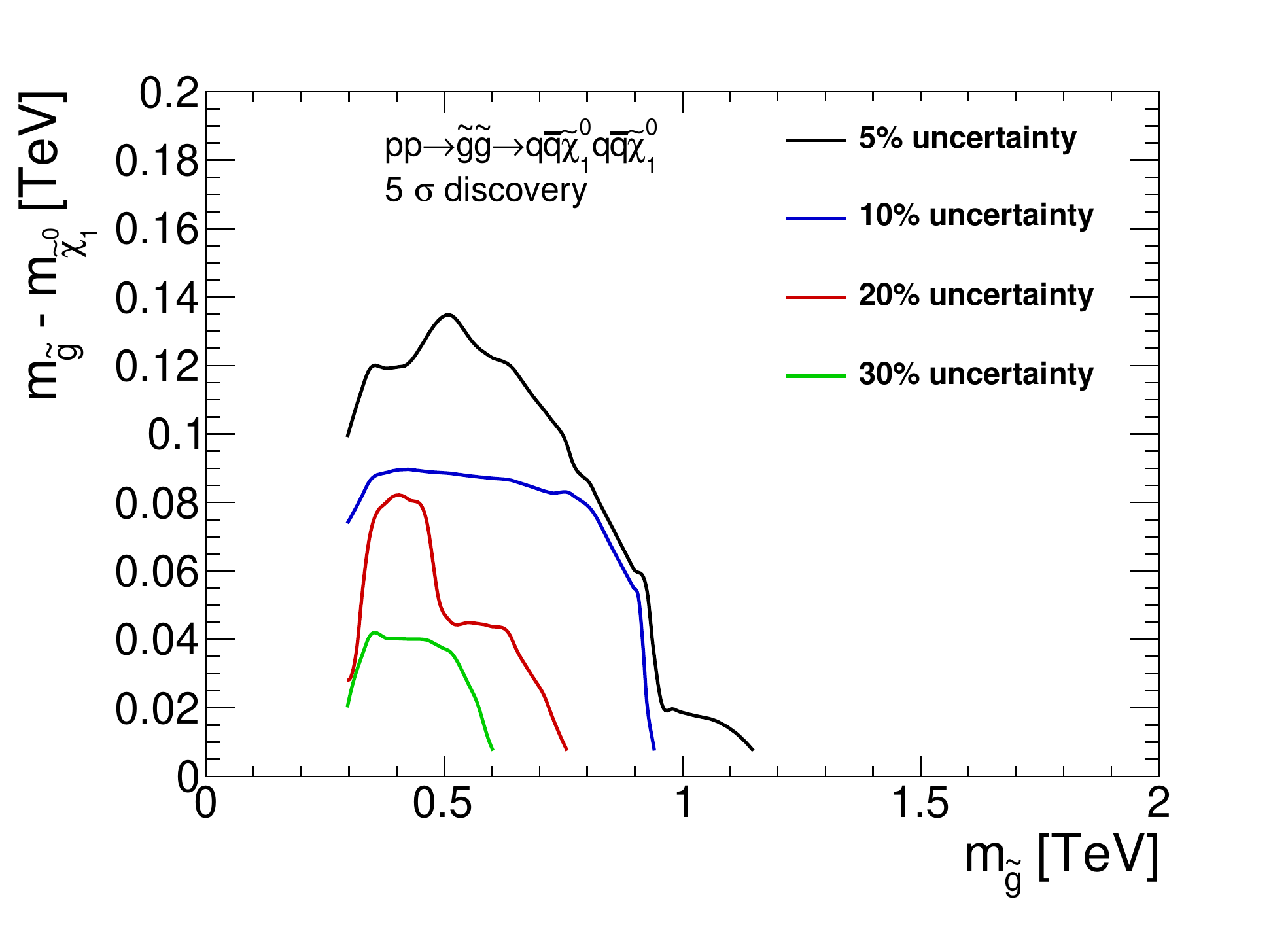}
\caption{Expected $5\sigma$ discovery contours for the $\sqrt{s} = 14$ TeV LHC with $3000$ fb$^{-1}$ using the \MET~search strategy using a fixed cut of $\MET > 2 \TeV$.  The different curves correspond to various assumptions for the systematic uncertainty on the background: $30\%$ [green], $20\%$ [red], $10\%$ [blue], and $5\%$ [black].}
\label{fig:GluinoNeutralino_Compressed_14TeV_uncertainty}
\end{center}
\end{figure}

Since the \MET-based search is most relevant in the region where the $5\sigma$ contour lies (see Figs.~\ref{fig:GNLightFlavorCompressedBestSearch} and \ref{fig:GNLightFlavorCompressedResults}), we demonstrate the impact of varying the systematic uncertainty for this search strategy for fixed cuts.  The results for 3000 fb$^{-1}$ of integrated luminosity at $14\TeV$ are shown in Fig.~\ref{fig:GluinoNeutralino_Compressed_14TeV_uncertainty}, where we fix $\MET > 2 \TeV$ and plot the $5\sigma$ discovery reach for $30\%$ [green], $20\%$ [red], $10\%$ [blue], and $5\%$ [black].  We see that a model with a degenerate gluino and neutralino could be discovered up to $\sim600$ GeV ($\sim1.1 \TeV$) for $30\%$ ($5\%$) systematic uncertainty.  The leading jet based search also has a comparable sensitivity to systematic uncertainties.  Improving our understanding of the background, which could be in principle achieved by studying this large data set carefully, could improve the gluino reach by more than $400 \GeV$.

\pagebreak
\subsection{Impact of Pileup}
\label{sec:GOGOpileup_compressed}
This section is devoted to an investigation of how the results for compressed spectra presented above would be affected by the presence of pileup.  As discussed in Sec.~\ref{sec:CompressedSystematic}, the strategy which yields the highest significance in the region where the $5\sigma$ contour lies is the \MET-based~search.  Therefore, we use this search to demonstrate the impact of different pileup conditions.

\begin{figure}[!htb]
\begin{center}
\includegraphics[width=0.48\textwidth]{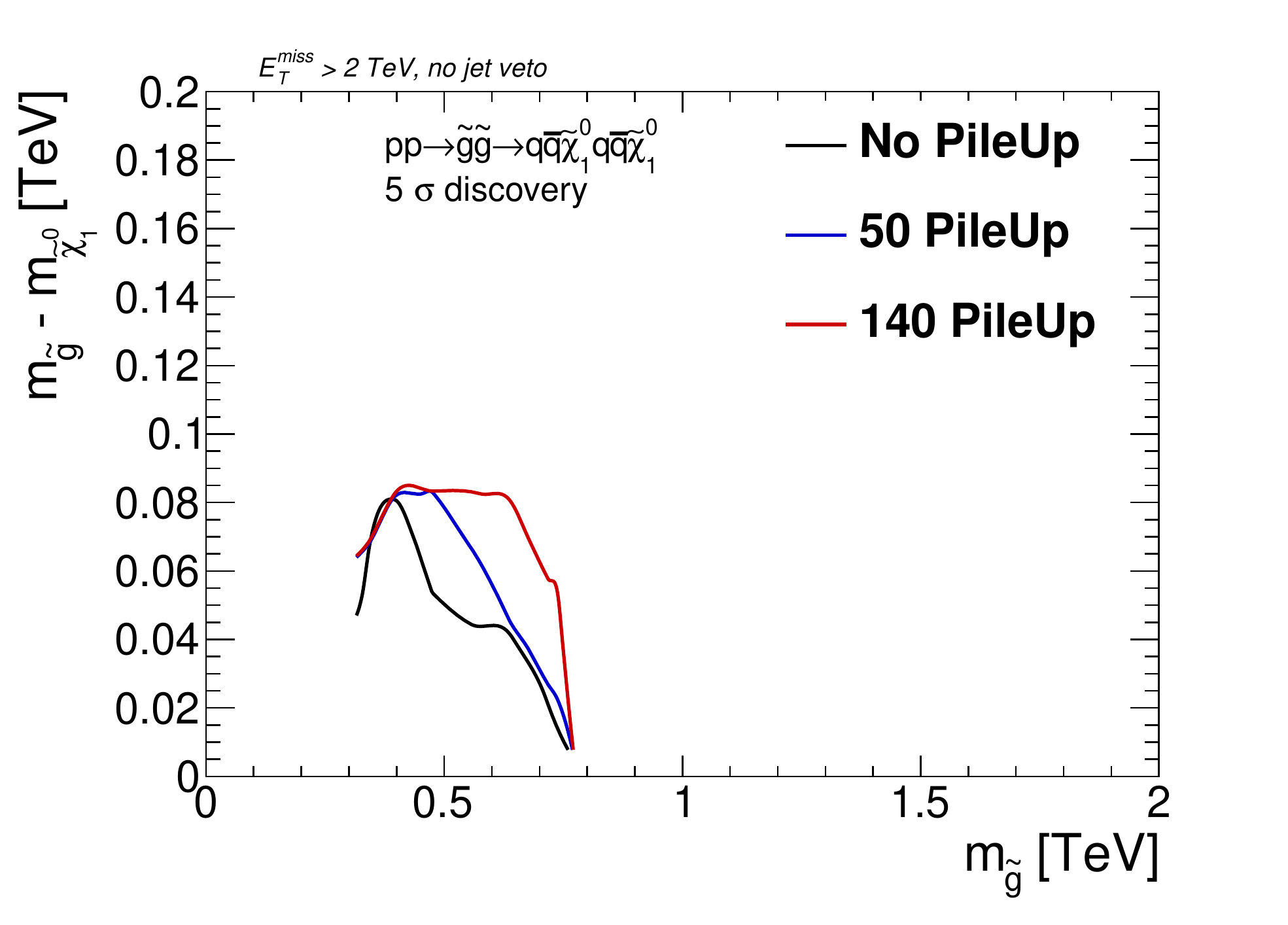}
\includegraphics[width=0.48\textwidth]{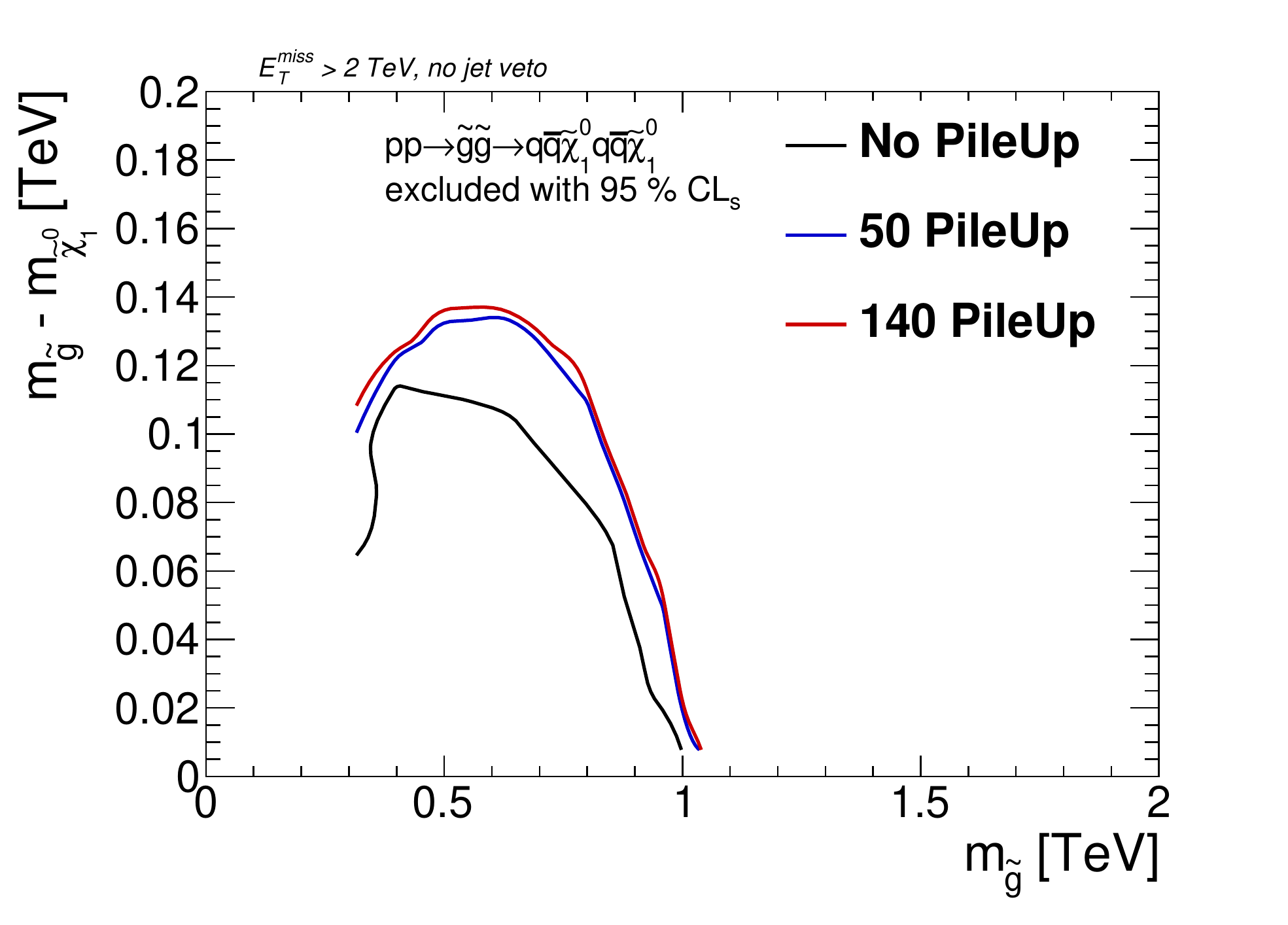}
\caption{Expected $5\sigma$ discovery contours [left] and $95\%$ CL limits [right] for the $\sqrt{s} = 14$ TeV LHC with $3000$ fb$^{-1}$ using the \MET~search strategy for a fixed cut of $\MET > 2 \TeV$.  The different curves correspond to the following average pileup: $0$ [black], $50$ [blue], $140$ [red].}
\label{fig:GluinoNeutralino_Compressed_14TeV_pileup}
\end{center}
\end{figure}

Figure~\ref{fig:GluinoNeutralino_Compressed_14TeV_pileup} gives the $5\sigma$ discovery contour [$95\%$ CL exclusion] on the left [right] for no pileup [black], 50 average events per bunch crossing [blue], and 140 average events per bunch crossing [red].  Surprisingly, we see that including pileup appears to increase the reach of this search.  One possibility the search is picking up more otherwise ``invisible" final states with soft ISR that become visible because there are pileup jets to push these events above the cut thresholds.  In other words, this apparent improvement is due to the fact that we have a fixed grid of cuts for our optimization scans. Note that the limits for $m_{\widetilde{g}} \simeq m_{\widetilde{\chi_1^0}}$ remain unchanged; the presence of pileup only impacts somewhat larger mass differences.

Note that all the curves in Fig.~\ref{fig:GluinoNeutralino_Compressed_14TeV_pileup} are computed for a fixed set of cuts, instead of reoptimizing for each pileup scenario which would obscure the comparison of the different limits.  It is clear that including pileup only makes the reach stronger.  The fact that we neglected pileup for the main results using these search strategies will imply that the limits we present are conservative.

\pagebreak

\section{The Squark-Neutralino Model}
\label{sec:QN}
In the ``squark-neutralino model", the first and second generation squarks
 $\widetilde{q} = \widetilde{u}_L, \widetilde{u}_R$, $\widetilde{d}_L, \widetilde{d}_R,$ $\widetilde{c}_L, \widetilde{c}_R,$ $\widetilde{s}_L, \widetilde{s}_R$ 
are the only kinematically accessible colored states. The gluino is completely decoupled from the squark production diagrams --- the squark production cross section is significantly reduced compared to models where the gluino is just above the kinematic limit. The squarks decay directly to the LSP and the corresponding quark, $\widetilde{q}_i \rightarrow q_i\,\widetilde{\chi}_1^0$. The only two relevant parameters are the squark mass $m_{\widetilde{q}}$, which is taken to be universal for the first two generations, and the neutralino mass $m_{\widetilde{\chi}_1^0}$.   The model is summarized as:
\NegSpace
\begin{center}
\renewcommand{\arraystretch}{1.3}
\setlength{\tabcolsep}{12pt}
\begin{tabular}{c|c|c}
BSM particles & production & decays \\
\hline
$\widetilde{q},\,\widetilde{\chi}_1^0$ & $p\,p \rightarrow \widetilde{q} \, \widetilde{q}^*$ & $\widetilde{q} \rightarrow q\,\widetilde{\chi}^0_1 $ 
\end{tabular}
\end{center}

Due to the structure of the renormalization group equations in the Minimal SUSY SM, a heavy gluino would tend to raise the squark masses; some tuning is required to achieve light squarks.  However, a class of theories with Dirac gluinos can be well approximated by this Simplified Model~\cite{Kribs:2013oda}.  The current preliminary limits on this model using $20$ fb$^{-1}$ of $8$ TeV data are $m_{\widetilde{q}} = 740 \text{ GeV}$ (ATLAS \cite{ATLAS-CONF-2013-047}) and $m_{\widetilde{q}} = 840 \text{ GeV}$ (CMS \cite{CMS-PAS-SUS-13-012}) assuming a massless neutralino.

Since the final state is two (or more) hard jets and missing energy, this model also serves to test the power of jets+\MET~style analyses.   The mass reach is not be nearly as high as in the gluino-neutralino light flavor decay model for two reasons: neglecting ISR and FSR, the final state has only two hard jets from the squark decays as opposed to four hard jets from the gluino decays, and cross section for producing squark pairs with the gluino completely decoupled is substantially lower than that for producing gluino pairs of the same mass.  Note that we checked that the $4$ jet requirement included in the jets+\MET~preselection does not have a detrimental impact on the squark results presented below.  

We simulated matched \texttt{MadGraph} samples for $\widetilde{q}\,\widetilde{q}^*$ with up to 2 additional generator level jets for the following points in parameter space:\footnote{We include $1\GeV$ for an example where the neutralino is effectively massless; the second line of neutralino masses is chosen to cover the bulk of the squark-neutralino plane; the final line is chosen to ensure coverage in the ``compressed" region.}  

\begin{center}
\renewcommand{\arraystretch}{1.3}
\setlength{\tabcolsep}{12pt}
\begin{tabular}{c|l}
BSM particles & masses \\
\hline
\hline
$m_{\widetilde{q}}$ $\big[14 \TeV\big]$ & $ (315, 397, 500, 629, 792, 997, 1255, 1580, 1989, 2489, 2989) \GeV$   \\
$m_{\widetilde{q}}$ $\big[33 \TeV\big]$ & $ (500, 629, 792, 997, 1255, 1580, 1989, 2489, 2989, 3489,$\\
 &  $3989, 4489, 4989, 5489, 5989) \GeV$   \\
$m_{\widetilde{q}}$ $\big[100 \TeV\big]$ & $ (1000, 1259, 1585, 1995, 2512, 3162, 3981, 5012, 6310, 7944, $   \\
 & $9944, 11944, 13944, 15944, 17944, 19944) \GeV$\\

\hline
					    & $1\GeV$ \\
$m_{\widetilde{\chi}^0_1}$  & $(0.2, 0.4, 0.6, 0.7, 0.8 , 0.9)\times m_{\widetilde{q}}$ \\
					    & $m_{\widetilde{q}}-(100\GeV, 50\GeV,15\GeV,5\GeV)$
\end{tabular}
\end{center}

The signature of this model is multi-jets and $\MET$.  Therefore, the dominant backgrounds are identical to the ones relevant for the gluino-neutralino model with light flavor decays and are discussed in Sec~\ref{sec:GOGO_Backgrounds}. We use the same analysis strategy as for the gluino-neutralino model, described in Sec.~\ref{sec:GOGO_Strategy}, to project discovery reach and limits for this model. Given that pileup had no impact on the results using this search strategy as demonstrated in Sec.~\ref{sec:pileupGOGO} above, we present here results only for the no pile-up scenario and expect little change when pileup is included.

Note that this search yields some power to discover these models in the difficult region of parameter space where the squark is degenerate with the neutralino, but Sec.~\ref{sec:QNCompressed} will provide the results of a search which is specifically targeted for this region of parameter space.

The next two sections give the details of the $14$ TeV LHC analysis and results.

\begin{figure}[tbp]
  \begin{center}
    \includegraphics[width=0.45\textwidth]{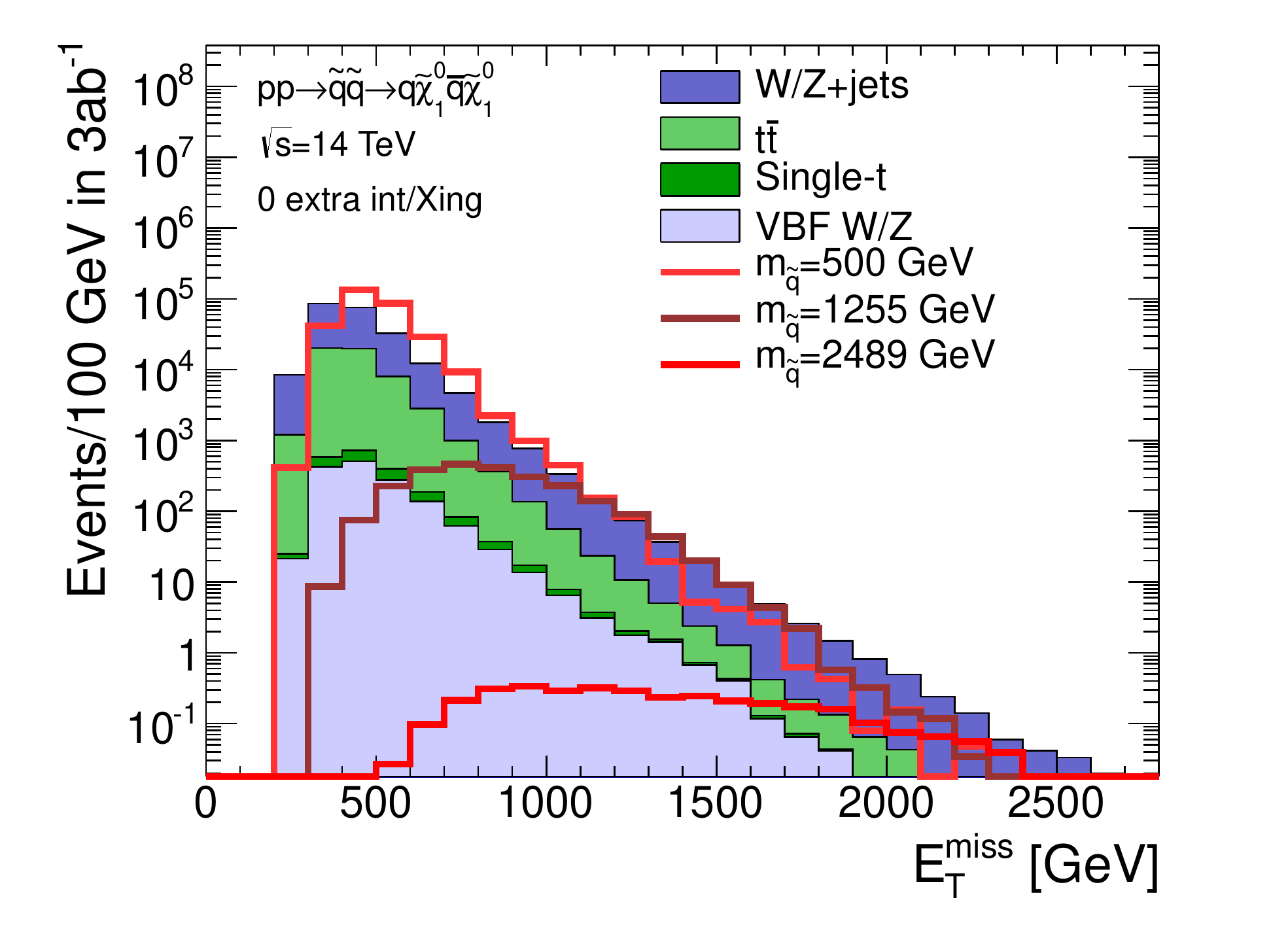}
    \includegraphics[width=0.45\textwidth]{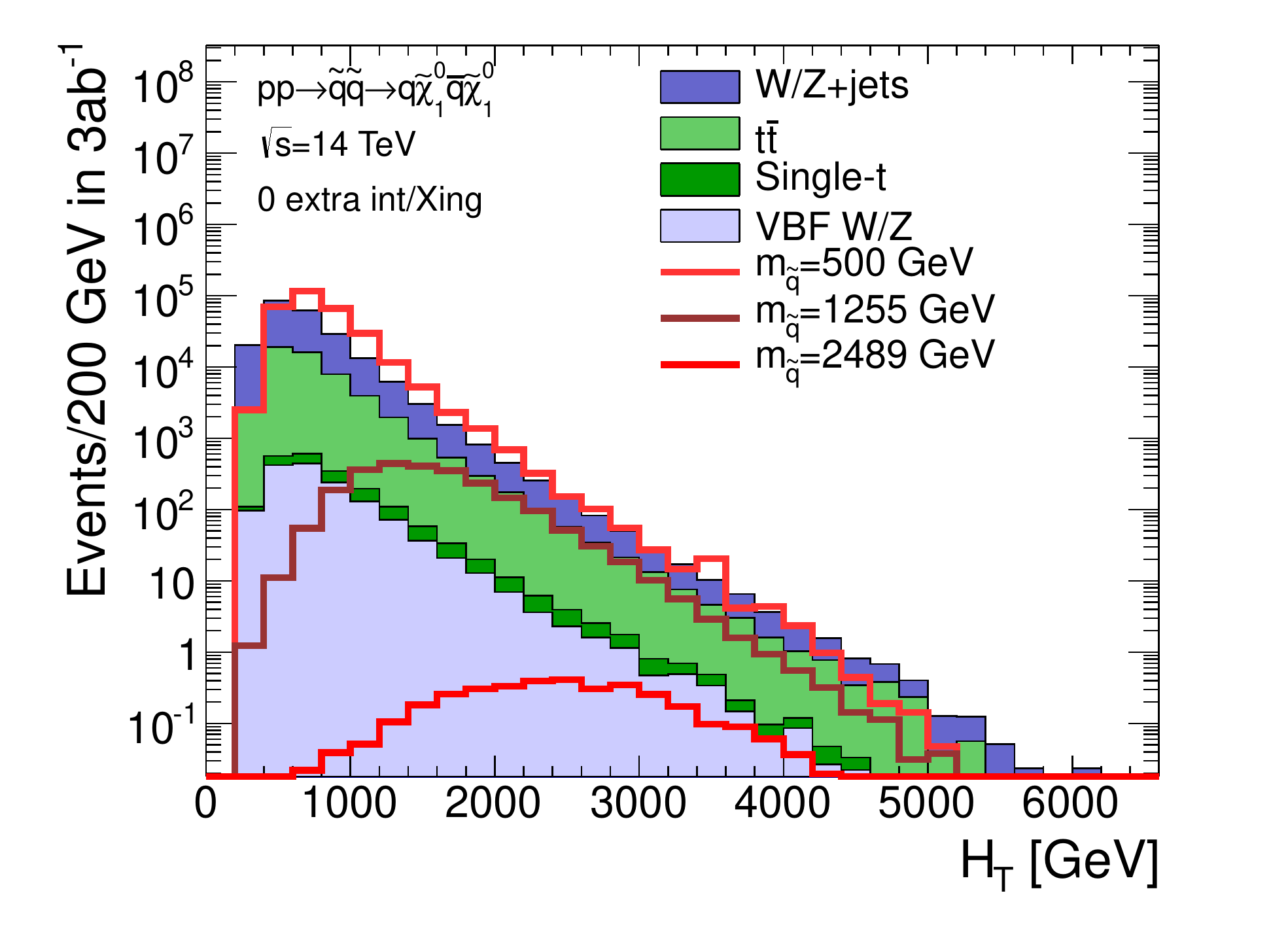}
    \caption{Histogram of \MET~[left] and \HT~[right] after preselection cuts for background and a range of squark-neutralino models at the $14$ TeV LHC.  The neutralino mass is $1\GeV$ for all signal models.}
    \label{fig:alljet_presel_distributions_SQSQB}
  \end{center}
\end{figure}

\hiddensubsection{Analysis: 14 TeV}
 Figure~\ref{fig:alljet_presel_distributions_SQSQB} shows the background and three signal distributions for the $14$ TeV LHC in the two kinematic variables which are scanned in this analysis: \MET~[left] and \HT~[right].  In Table~\ref{tab:SNbulkCounts} we give the number of events after each stage of cuts for the dominant backgrounds and two signal models.  From this table, it is clear that the $14$ TeV LHC with $3000 \text{ fb}^{-1}$ would be able to exclude (but not discover) squarks with mass of $1255 \GeV$.

\begin{table}[h!]
\renewcommand{\arraystretch}{1.4}
\setlength{\tabcolsep}{6pt}
\footnotesize
\vskip 10pt
  \begin{centering}
    \begin{tabular}{| r | r r | r | r  r |}
\hline
&&&&\multicolumn{2}{c|}{$\quad m_{\widetilde{q}}\quad$ [GeV]}\\
                                               Cut&                 $V$+jets&               $t\bar{t}$&                 Total BG&                     500&                     1255\\
\hline
\hline
$                             \text{Preselection}$&{$2.07 \times 10^{7}$}&{$2.47 \times 10^{7}$}&{$4.54 \times 10^{7}$}&$2.94 \times 10^{6}$&$1.41 \times 10^{4}$\\
$\MET/\sqrt{\HT} > 15 \text{ GeV}^{1/2}$&{$4.45 \times 10^{5}$}&{$1.20 \times 10^{5}$}&{$5.65 \times 10^{5}$}&$6.31 \times 10^{5}$&$8.48 \times 10^{3}$\\
$p_T^{\mathrm{leading}} < 0.4\times \HT$&{$1.69 \times 10^{5}$}&{$5.16 \times 10^{4}$}&{$2.21 \times 10^{5}$}&$3.05 \times 10^{5}$&$2.42 \times 10^{3}$\\
\hline
\hline
$\MET > 550$ GeV&\multirow{2}{*}{$3.98 \times 10^{4}$}&\multirow{2}{*}{$1.16 \times 10^{4}$}&\multirow{2}{*}{$5.15 \times 10^{4}$}&\multirow{2}{*}{\color{red}$1.27 \times 10^{5}$}&\multirow{2}{*}{$2.33 \times 10^{3}$}\\
$\HT > 600$ GeV          &&&&&\\
\hline
$\MET > 1200$ GeV&\multirow{2}{*}{                      174}&\multirow{2}{*}{                       23}&\multirow{2}{*}{                      197}&\multirow{2}{*}{                      185}&\multirow{2}{*}{\color{red}                      222}\\
$\HT > 1400$ GeV          &&&&&\\
\hline
    \end{tabular}
    \caption{Number of expected events for $\sqrt{s} = 14$ TeV with 3000 fb$^{-1}$ integrated luminosity for the background processes and selected squark masses for the squark-neutralino model.  The neutralino mass is $1 \GeV$.  Two choices of cuts on \MET~and \HT~are provided, and for each squark mass column the entry in the row corresponding to the ``optimal" cuts is marked in red. }
    \label{tab:SNbulkCounts}
  \end{centering}
\end{table}

\begin{figure}[bp]
  \centering
  \includegraphics[width=.48\columnwidth]{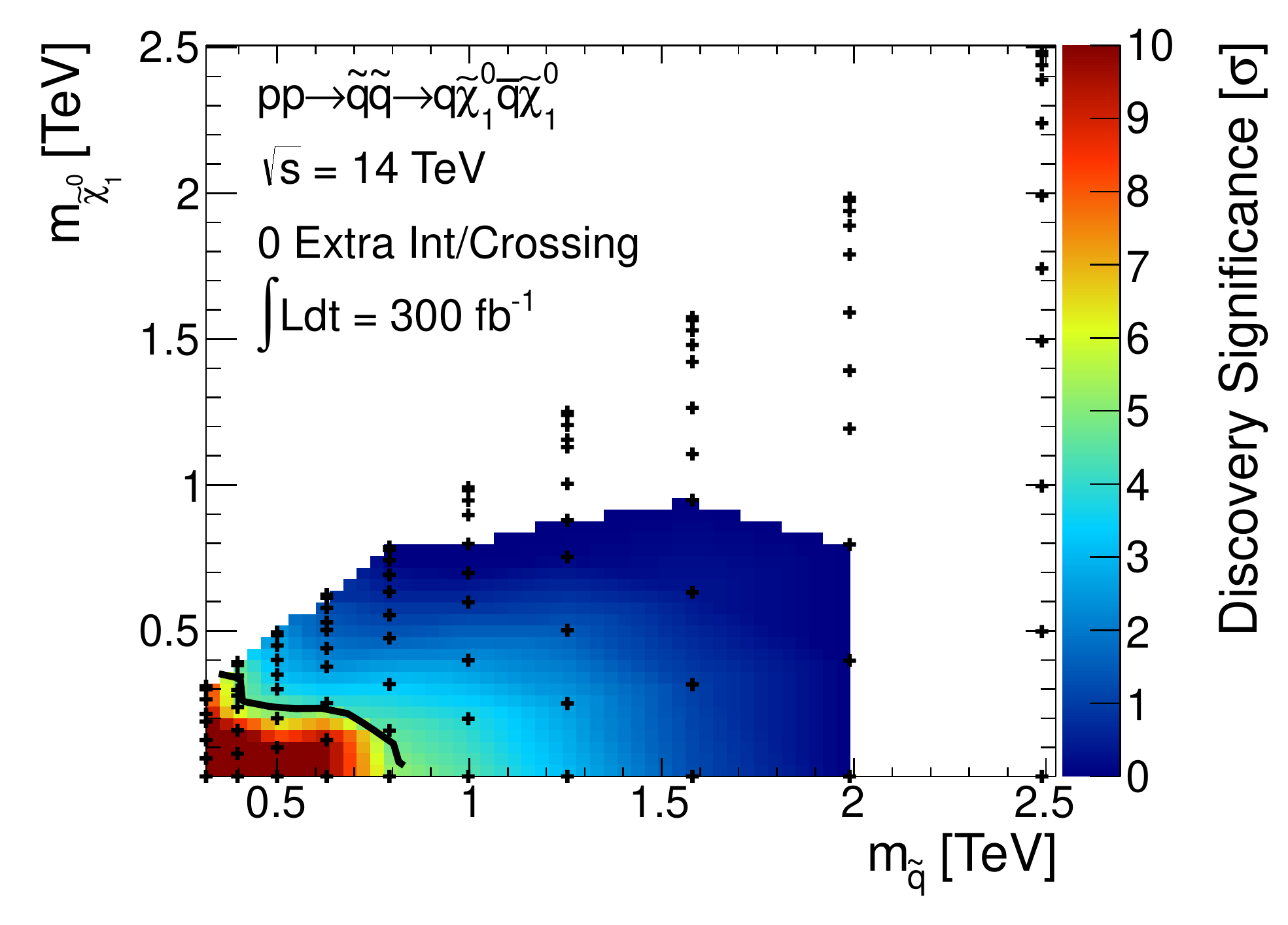}
  \includegraphics[width=.48\columnwidth]{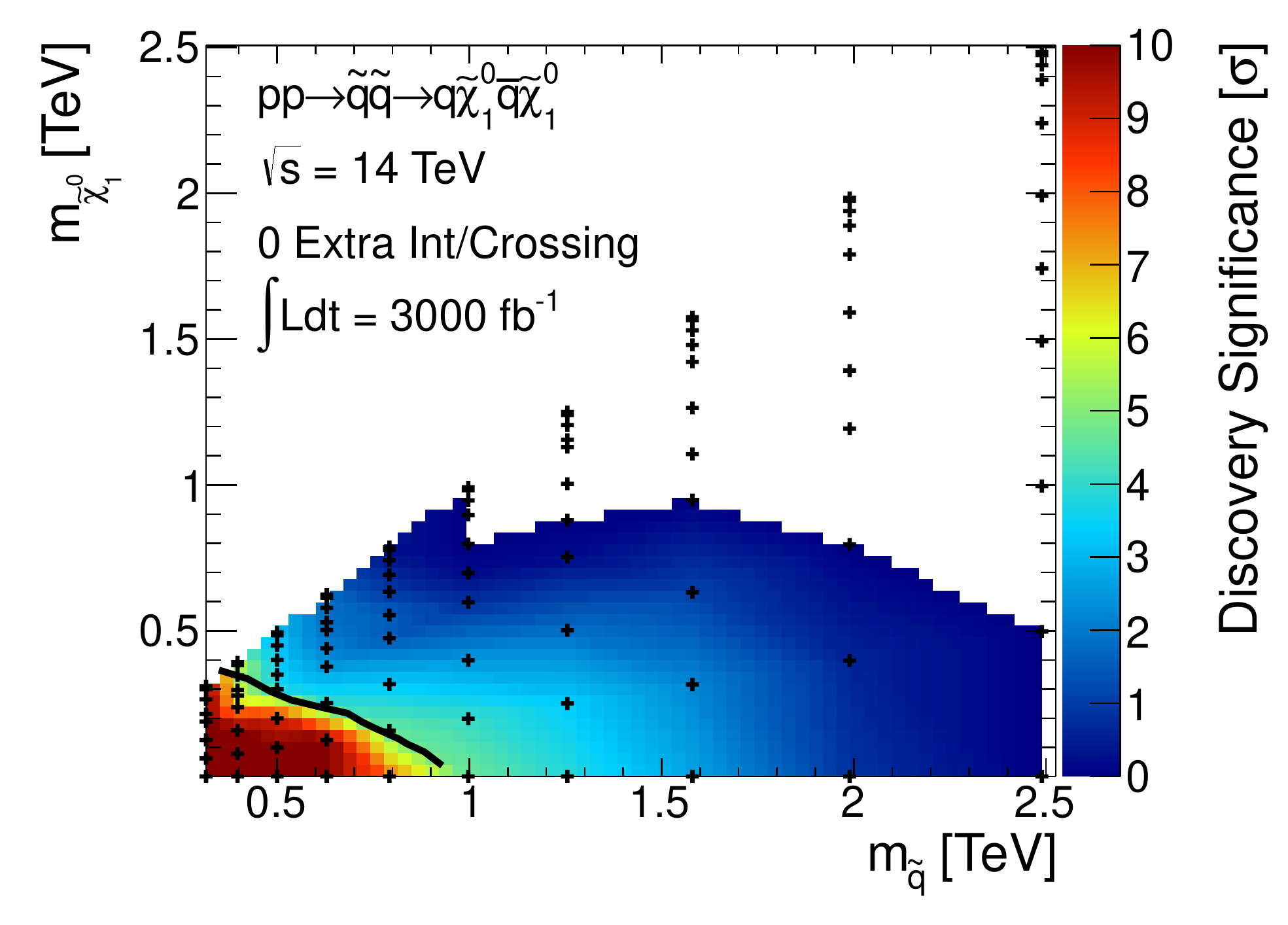}
  \includegraphics[width=.48\columnwidth]{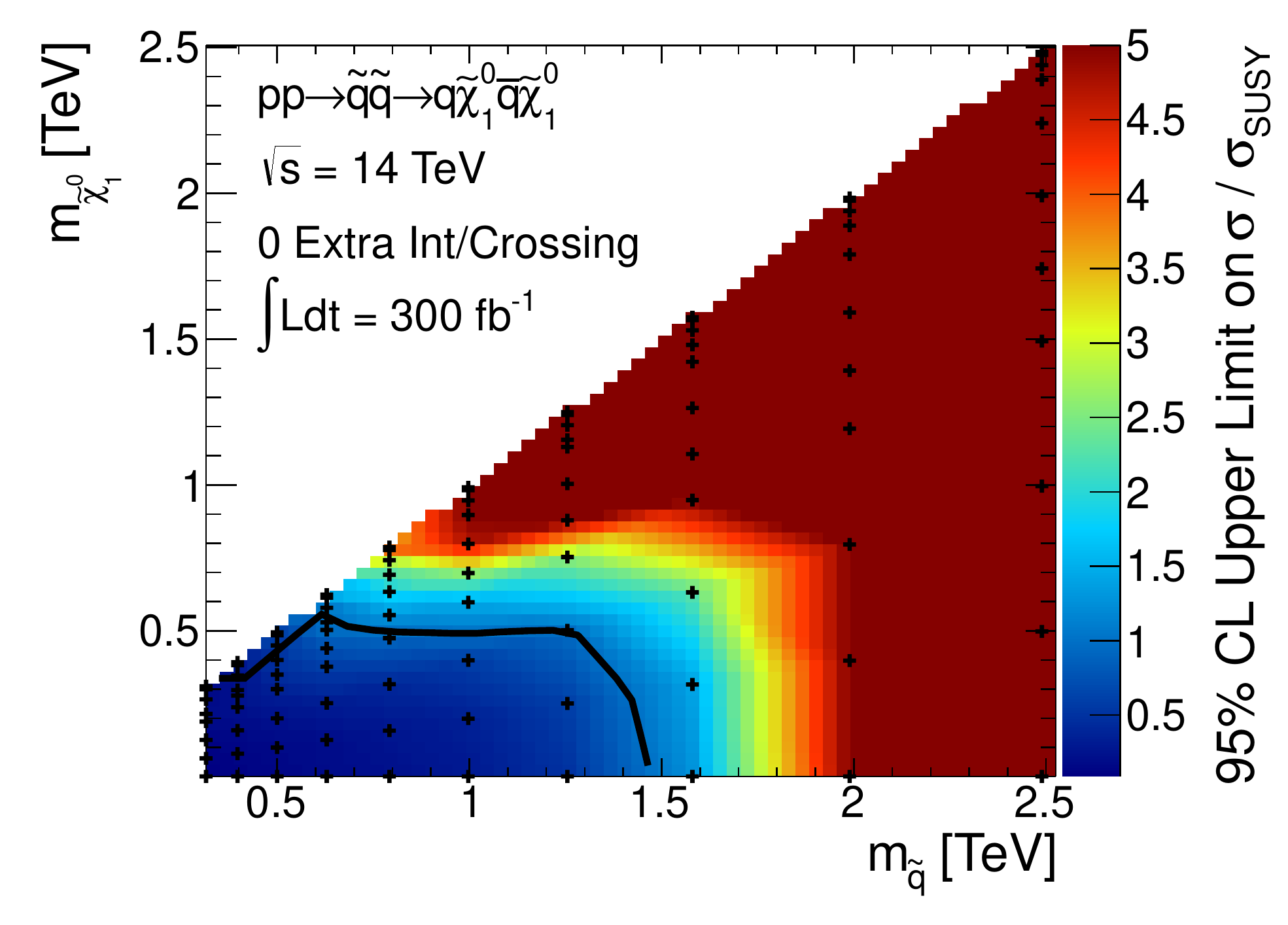}
  \includegraphics[width=.48\columnwidth]{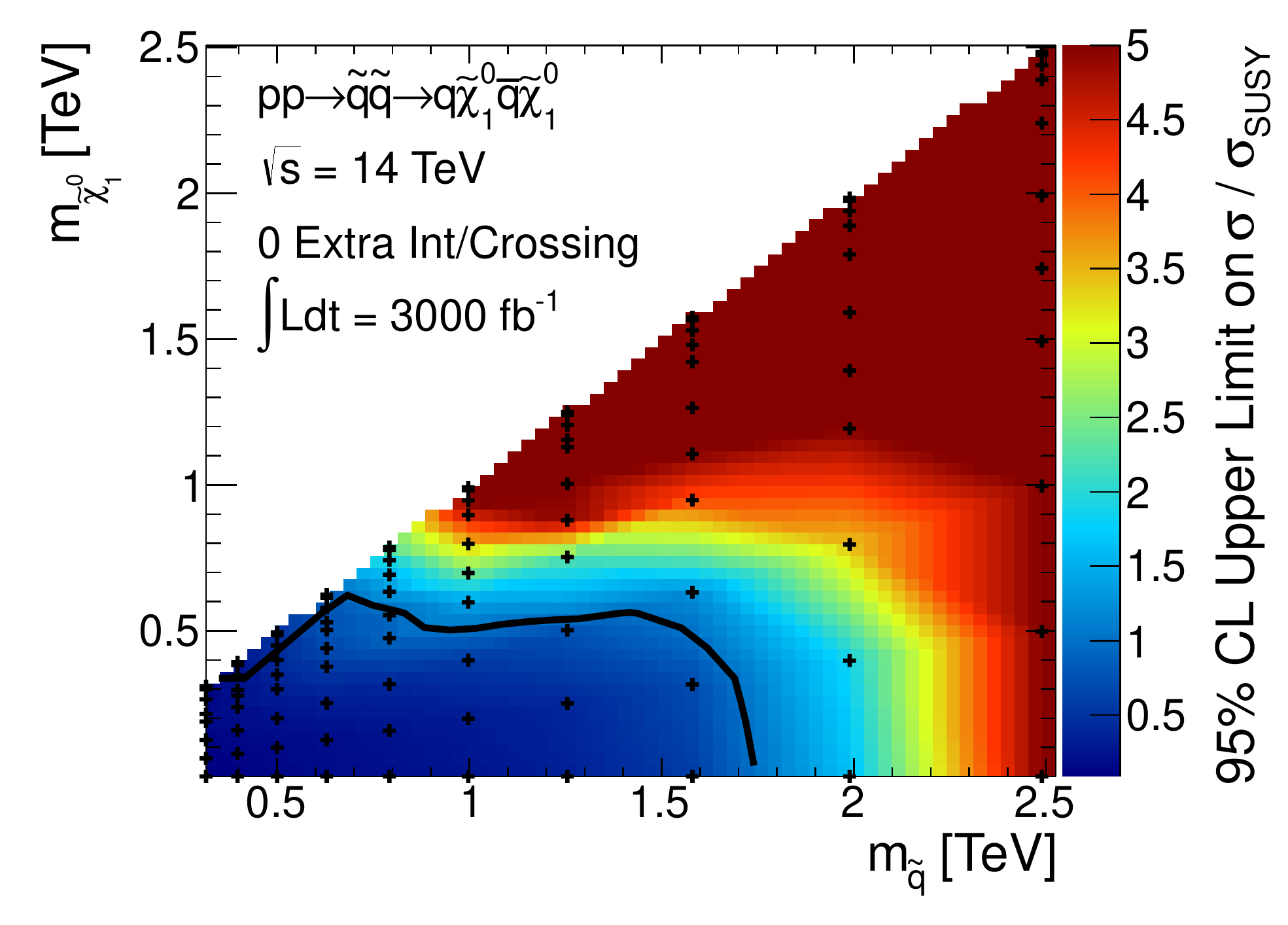}
  \caption{Results for the squark-neutralino model with light flavor decays are given in the $m_{\widetilde{\chi}_1^0}$ versus $m_{\widetilde{q}}$ plane.  The top [bottom] row shows the expected $5\sigma$ discovery reach [$95\%$ confidence level upper limits] for squark-anti-squark production at the $14$ TeV LHC.  Mass points to the left/below the contours are expected to be probed with $300$ fb$^{-1}$ [left] and $3000$ fb$^{-1}$ [right] of data.  A $20\%$ systematic uncertainty is assumed for the backgrounds. Pileup is not included.}
    \label{fig:SQSQ_14_NoPileUp_results}
\end{figure}

\pagebreak
\hiddensubsection{Results: 14 TeV}
The results for the squark-neutralino model are shown in Fig.~\ref{fig:SQSQ_14_NoPileUp_results} for the $14$ TeV LHC.  The left [right] panels give discovery significance [$95\%$ CL exclusion] contours in the $m_{\widetilde{\chi}_1^0}$ versus $m_{\widetilde{q}}$ plane.  The top [bottom], results assume $300\text{ fb}^{-1}$ [$3000\text{ fb}^{-1}$] of data.  As expected, the reach is significantly smaller than for the gluino-neutralino model with light flavor decays.  

Using the NLO squark pair production cross section one can make a very naive estimate for the reach of a given collider.  For example, we find that the choice of squark mass which would yield $10$ events at $300 \text{ fb}^{-1}$ $\big(3000 \text{ fb}^{-1}\big)$ is $2.4\,(2.9) \TeV$.  This roughly corresponds to the maximal possible reach one could expect for a given luminosity using $14 \TeV$ proton collisions.  

Using a realistic simulation framework along with the search strategy employed here the $14 \TeV$ $300 \text{ fb}^{-1}$ limit with massless neutralinos is projected to be $1.5 \TeV$ (corresponding to 1022 events), while the $14 \TeV$ $3000 \text{ fb}^{-1}$ limit is projected to be $1.7 \TeV$ (corresponding to 3482 events).  Given these huge numbers of events, it is possible that a different (or more sophisticated) search strategy would allow for greater sensitivity to these models --- this is outside the purview of the current study.   Finally, we note that the $14 \TeV$ LHC with $3000 \text{ fb}^{-1}$ could discover a squark as heavy as $800 \GeV$ if the neutralino is massless.  Unlike in the gluino-neutralino model, the squark mass discovery reach immediately begins to weaken significantly as soon as the neutralino mass is increased from the massless limit.

\begin{figure}[h!]
  \begin{center}
    \includegraphics[width=0.45\textwidth]{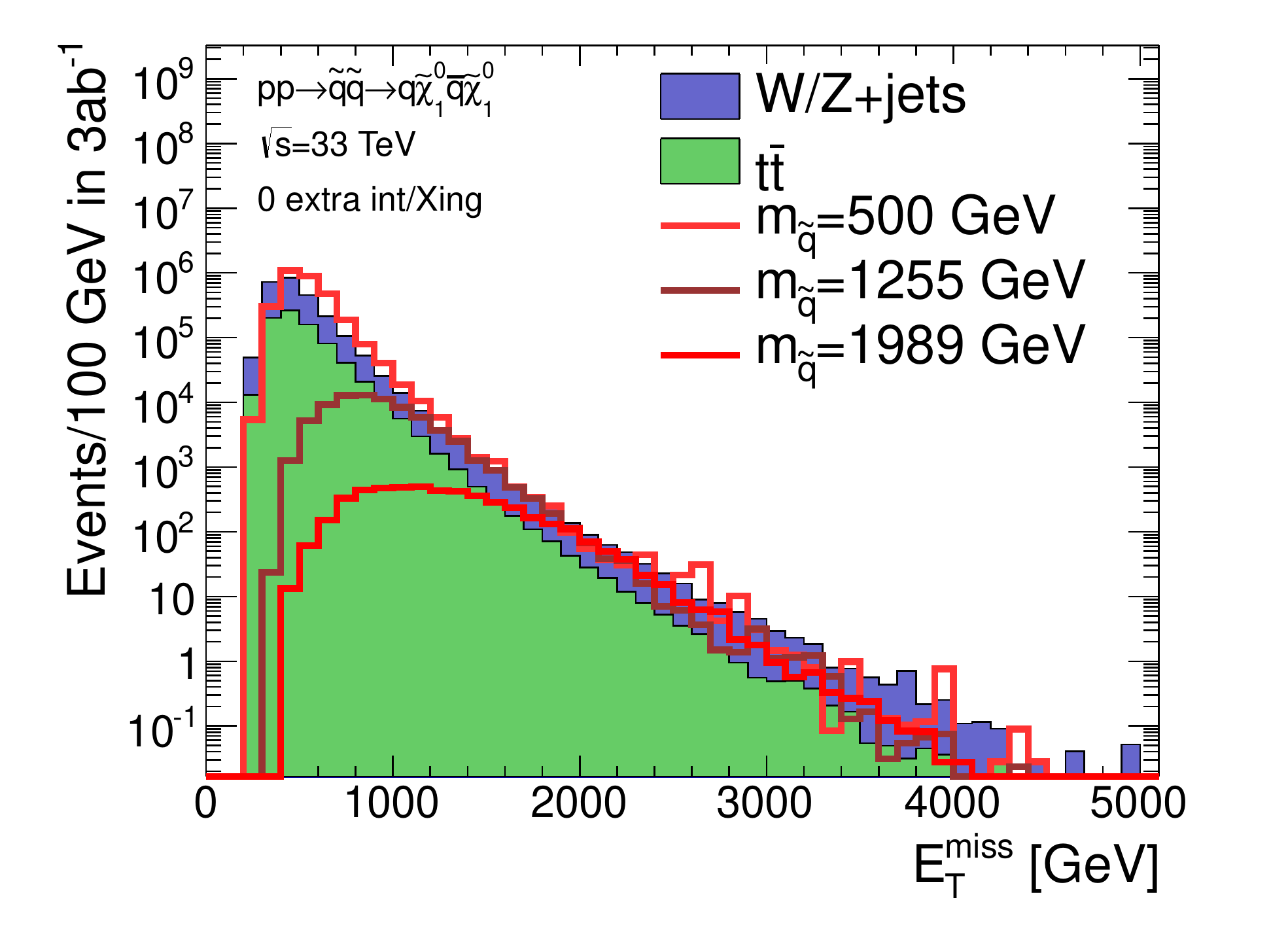}
    \includegraphics[width=0.45\textwidth]{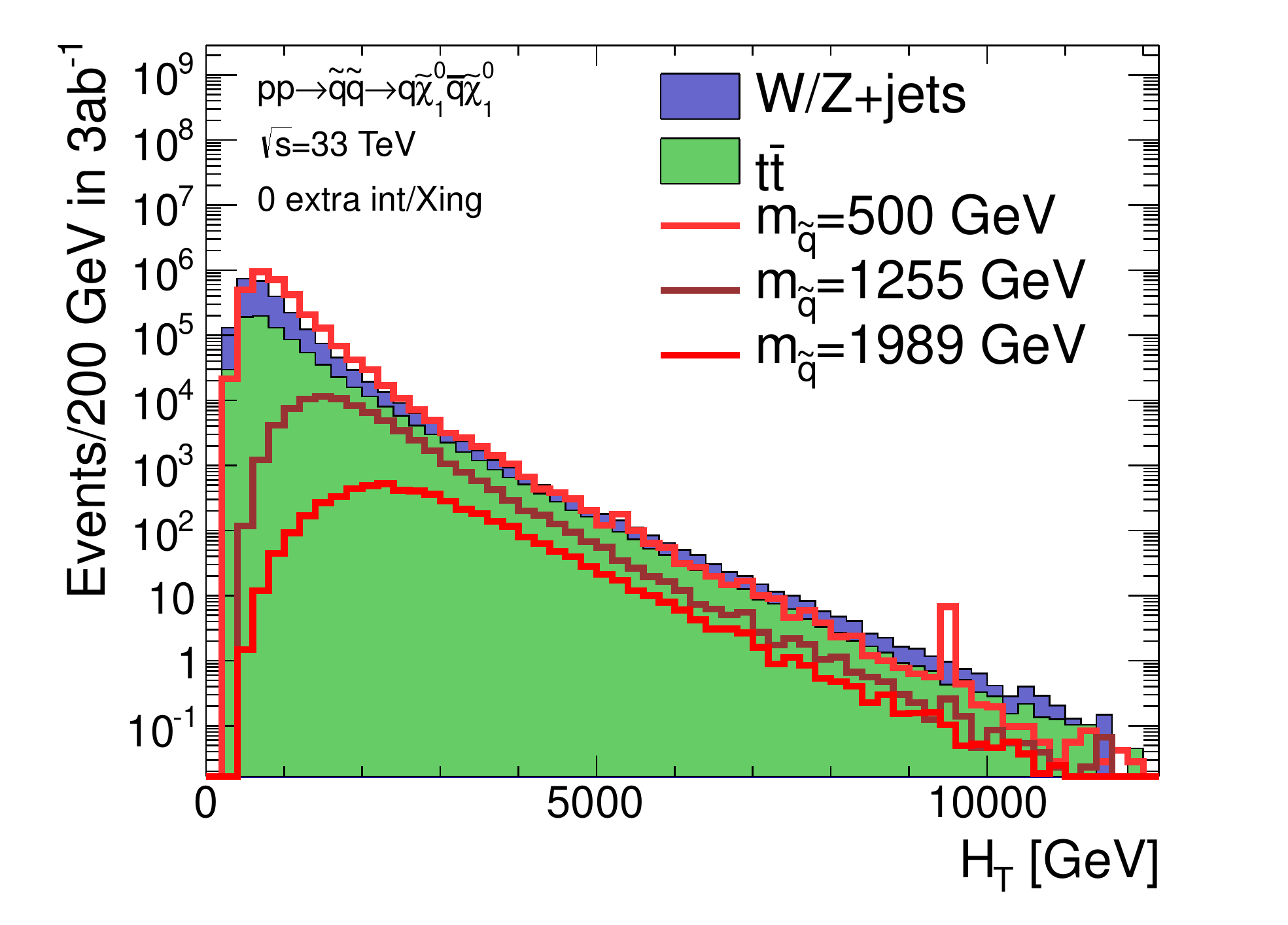}
    \caption{Histogram of \MET~[left] and \HT~[right] after preselection cuts for background and a range of squark-neutralino models at a $33$ TeV proton collider.  The neutralino mass is $1\GeV$ for all signal models.}
    \label{fig:alljet_presel_distributions_SQSQB_33TeV}
  \end{center}
\end{figure}

\pagebreak
\hiddensubsection{Analysis: 33 TeV}
Figure~\ref{fig:alljet_presel_distributions_SQSQB_33TeV} shows the background and three signal distributions for a $33$ TeV proton collider in the two kinematic variables which are scanned in this analysis: \MET~[left] and \HT~[right].  In Table~\ref{tab:SNbulkCounts_33TeV} we give the number of events after each stage of cuts for the dominant backgrounds and three signal models.  From this table, it is clear that a $33$ TeV proton collider with $3000 \text{ fb}^{-1}$ would be able to exclude (but not discover) squarks with mass of $3152 \GeV$.

\begin{table}[h!]
\renewcommand{\arraystretch}{1.4}
\setlength{\tabcolsep}{4pt}
\footnotesize
\vskip 10pt
  \begin{centering}
    \begin{tabular}{| r | r r | r | r  r r |}
\hline
&&&&\multicolumn{3}{c|}{$m_{\tilde{q}}$ [GeV]}\\
                                               Cut&                 $V$+jets&               $t\overline{t}$&                 Total BG&                      $629$&                     $1255$&                     $3152$\\
\hline
\hline
$                             \mathrm{Preselection}$&{$1.55 \times 10^{8}$}&{$2.86 \times 10^{8}$}&{$4.42 \times 10^{8}$}&$1.20 \times 10^{7}$&$3.99 \times 10^{5}$&                      $926$\\
\hline
$\MET/\sqrt{\HT} > 15 \GeV^{1/2}$&{$4.50 \times 10^{6}$}&{$1.93 \times 10^{6}$}&{$6.44 \times 10^{6}$}&$3.66 \times 10^{6}$&$2.34 \times 10^{5}$&                      762\\
$p_T^{\mathrm{leading}} < 0.4\times \HT $&{$1.70 \times 10^{6}$}&{$8.02 \times 10^{5}$}&{$2.50 \times 10^{6}$}&$1.61 \times 10^{6}$&$7.67 \times 10^{4}$&                      $150$\\
\hline
\hline
$         \MET > 650 \GeV$&\multirow{2}{*}{$1.13 \times 10^{6}$}&\multirow{2}{*}{$5.87 \times 10^{5}$}&\multirow{2}{*}{$1.72 \times 10^{6}$}&\multirow{2}{*}{\color{red} $1.55 \times 10^{6}$}&\multirow{2}{*}{$7.67 \times 10^{4}$}&\multirow{2}{*}{                      $150$}\\
$                          \HT > 650 \GeV$&&&&&&\\
\hline
$       \MET > 1300 \GeV$&\multirow{2}{*}{$1.08 \times 10^{4}$}&\multirow{2}{*}{$6.79 \times 10^{3}$}&\multirow{2}{*}{$1.76 \times 10^{4}$}&\multirow{2}{*}{$1.96 \times 10^{4}$}&\multirow{2}{*}{\color{red} $1.54 \times 10^{4}$}&\multirow{2}{*}{                      $124$}\\
$                         \HT > 1350 \GeV$&&&&&&\\
\hline
$         \MET > 2650 \GeV$&\multirow{2}{*}{                     $51.5$}&\multirow{2}{*}{                     $16.3$}&\multirow{2}{*}{                     $67.8$}&\multirow{2}{*}{                     $50.8$}&\multirow{2}{*}{                     $26.9$}&\multirow{2}{*}{                   \color{red}  $22.3$}\\
$                         \HT > 3350 \GeV$&&&&&&\\
\hline
    \end{tabular}
    \caption{Number of expected events for $\sqrt{s} = 33$ TeV with 3000 fb$^{-1}$ integrated luminosity for the background processes and selected squark masses for the squark-neutralino model.  The neutralino mass is $1 \GeV$.  Three choices of cuts on \MET~and \HT~are provided, and for each squark mass column the entry in the row corresponding to the ``optimal" cuts is marked in red.
    }
    \label{tab:SNbulkCounts_33TeV}
  \end{centering}
\end{table}

\begin{figure}[h!]
  \centering
  \includegraphics[width=.48\columnwidth]{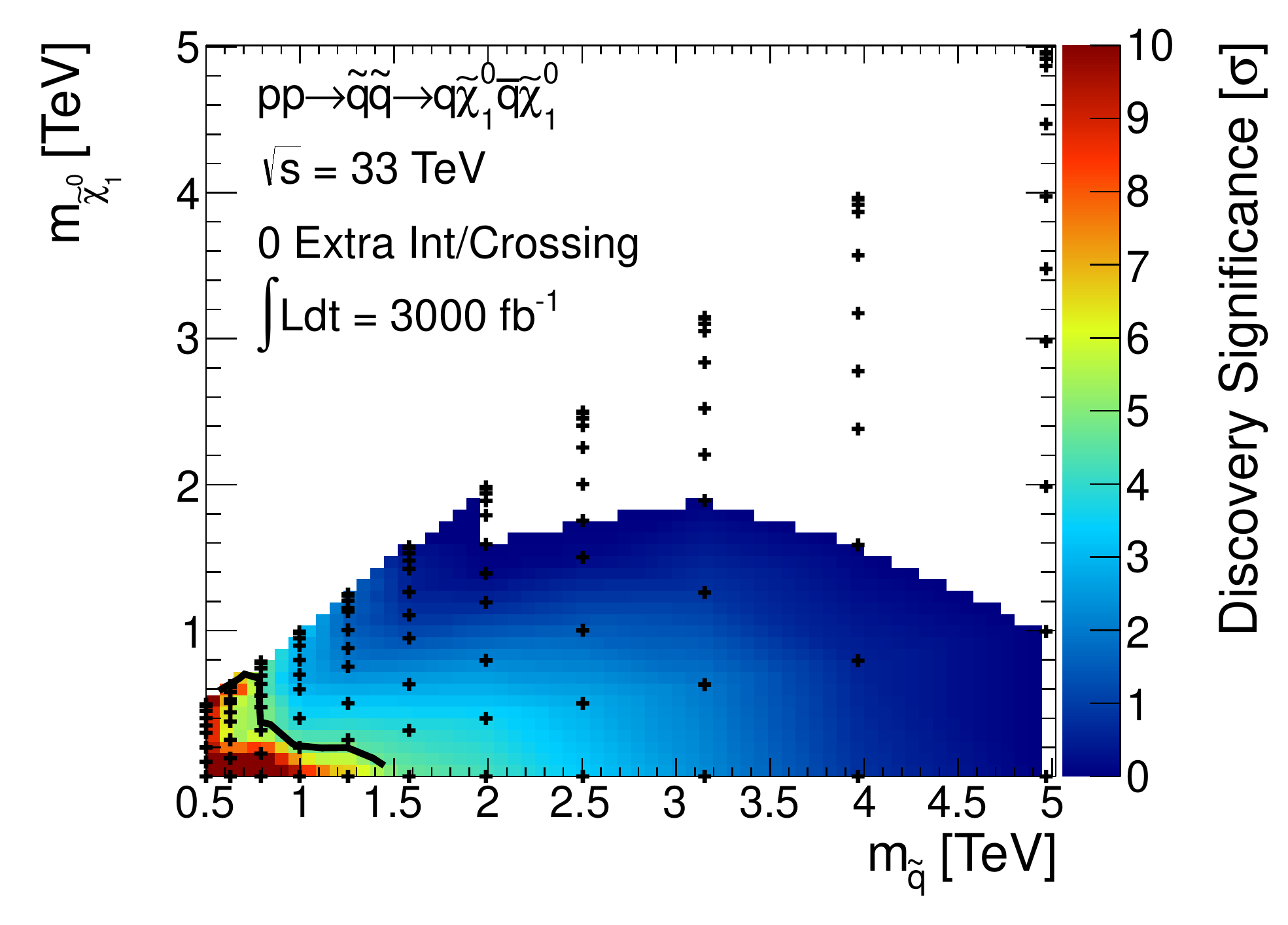}
  \includegraphics[width=.48\columnwidth]{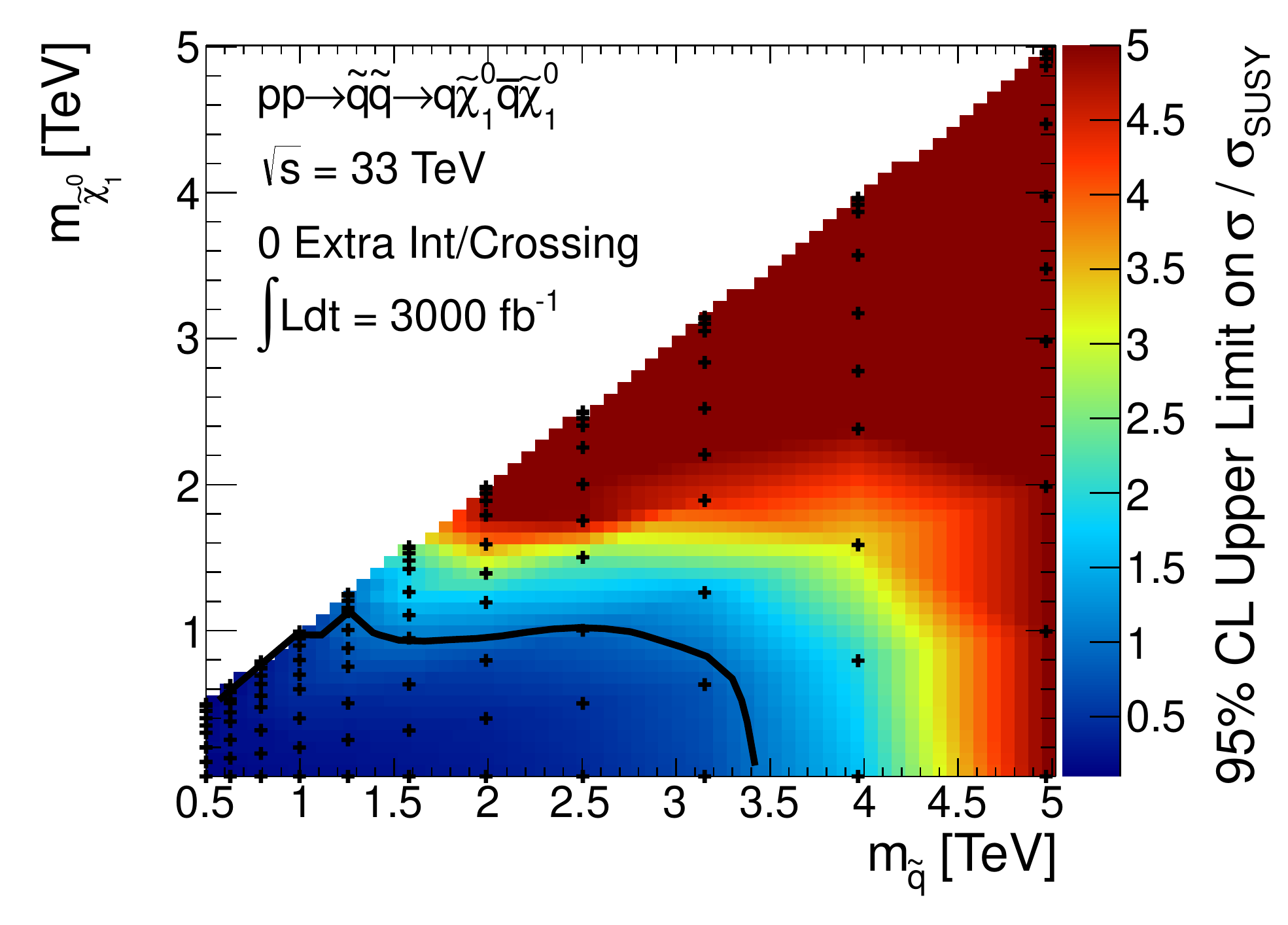}
  \caption{Results for the squark-neutralino model with light flavor decays are given in the $m_{\widetilde{\chi}_1^0}$ versus $m_{\widetilde{q}}$ plane.  The left [right] panels shows the expected $5\sigma$ discovery reach [95\% confidence level upper limits] for squark-anti-squark production at a $33$ TeV proton collider.  Mass points to the left/below the contours are expected to be probed with $3000$ fb$^{-1}$ of data [right].  A 20\% systematic uncertainty is assumed for the backgrounds. Pileup is not included.}
    \label{fig:SQSQ_33_NoPileUp_results}
\end{figure}

\hiddensubsection{Results: 33 TeV}
The results for the squark-neutralino model are shown in Fig.~\ref{fig:SQSQ_33_NoPileUp_results} for a $33$ TeV proton collider.  Discovery significance [$95\%$ CL exclusion] contours in the $m_{\widetilde{\chi}_1^0}$ versus $m_{\widetilde{q}}$ plane are shown on the left [right].  As expected, the reach is significantly smaller than for the gluino-neutralino model with light flavor decays.  

Using the NLO squark pair production cross section one can make a very naive estimate for the reach of a given collider.  For example, we find that the choice of squark mass which would yield $10$ events at $3000 \text{ fb}^{-1}$ is $5.8 \TeV$.  This roughly corresponds to the maximal possible reach one could expect for a given luminosity using $33 \TeV$ proton collisions.  

Using a realistic simulation framework along with the search strategy employed here the $33 \TeV$ $3000 \text{ fb}^{-1}$ limit with massless neutralinos is projected to be $3.4 \TeV$ (corresponding to 3482 events).  Given this huge number of events, it is possible that a different (or more sophisticated) search strategy would allow for greater sensitivity to these models --- this is outside the purview of the current study.  Finally, we note that the $33 \TeV$ proton collider with $3000 \text{ fb}^{-1}$ could discover a squark as heavy as $1.4\TeV$ if the neutralino is massless.  As in the $14$ TeV search, the squark mass discovery reach immediately begins to weaken significantly as the neutralino mass is increased from the massless limit. 

\pagebreak

\begin{figure}[h!]
  \begin{center}
    \includegraphics[width=0.45\textwidth]{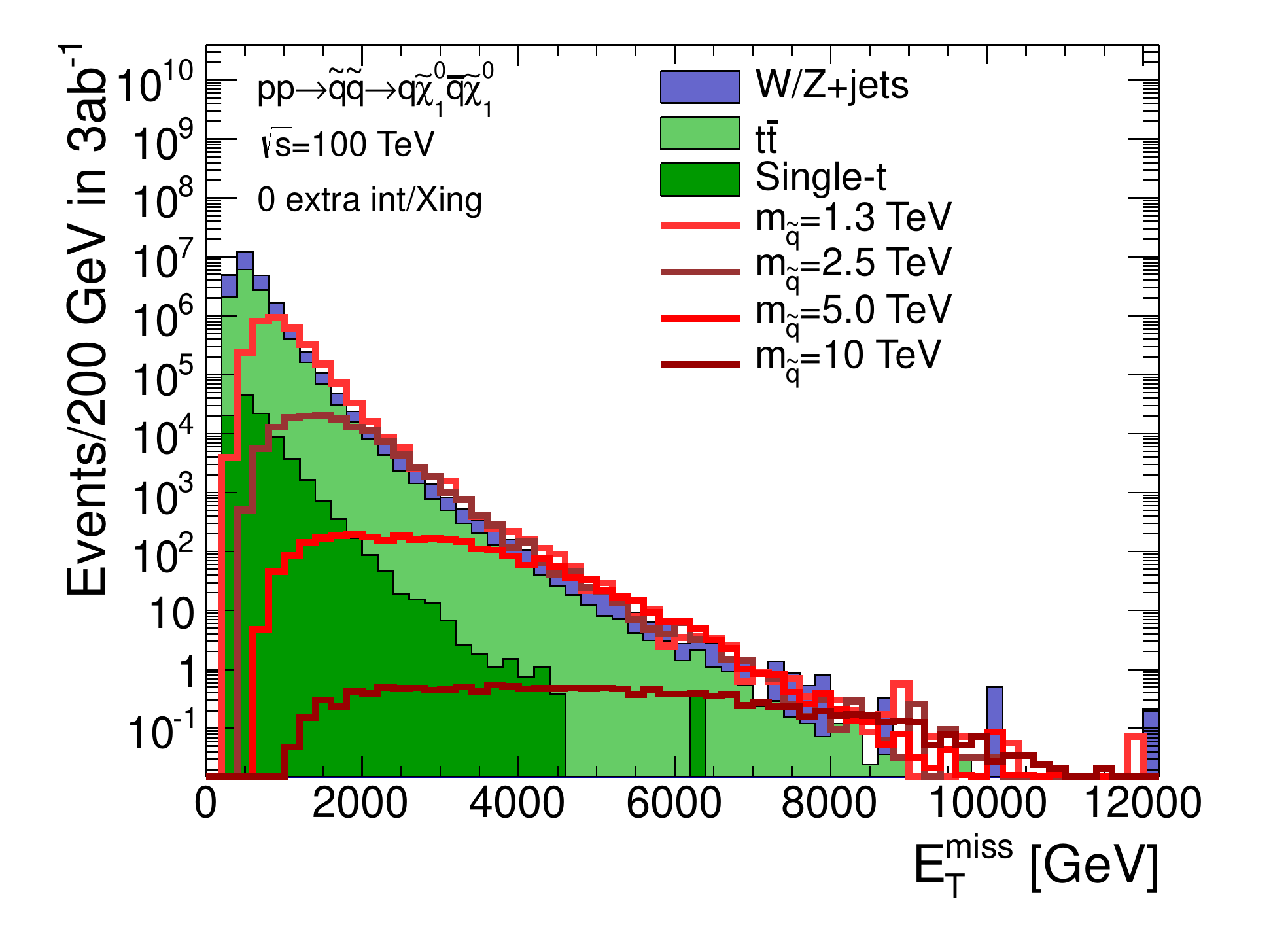}
    \includegraphics[width=0.45\textwidth]{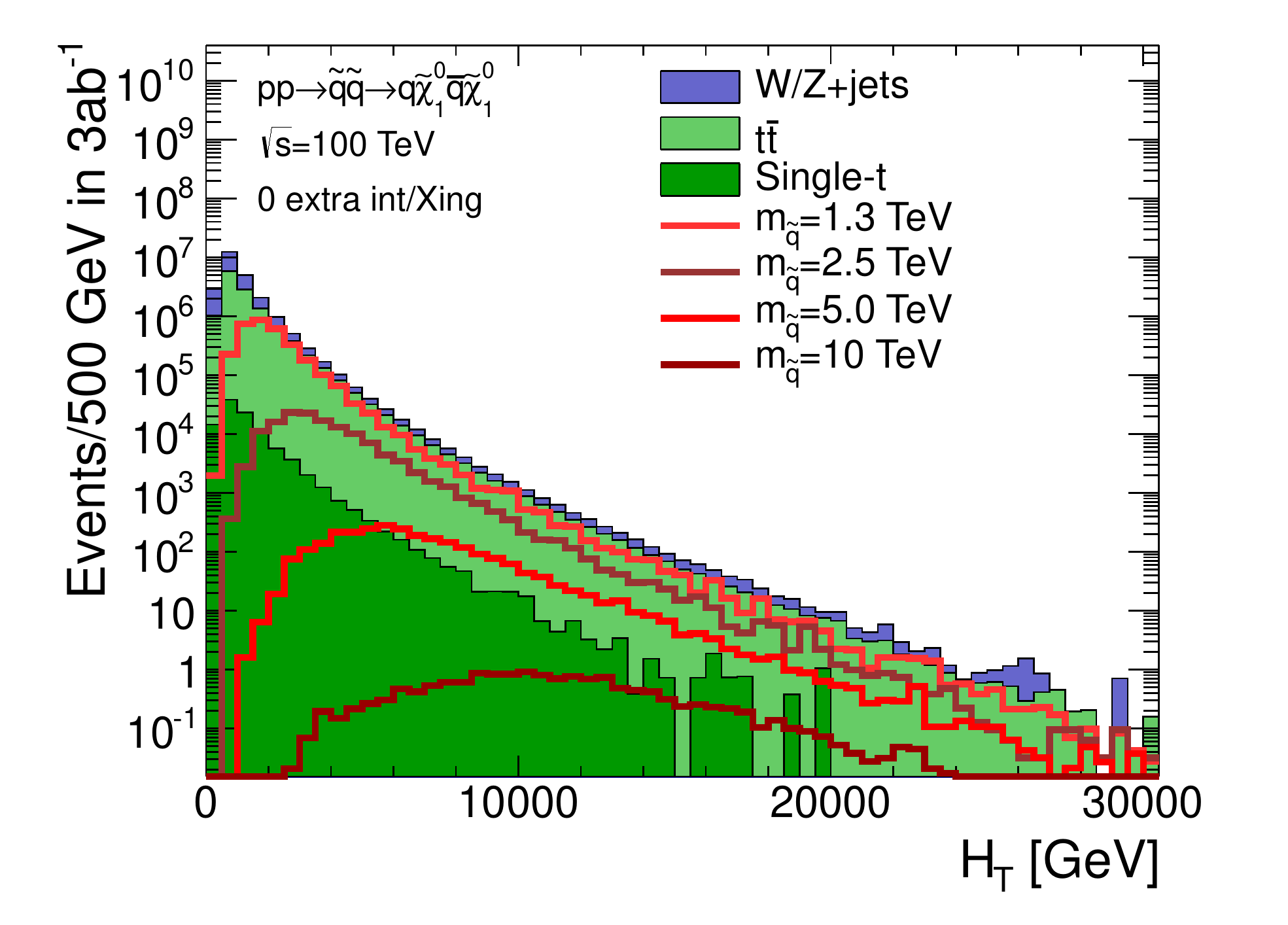}
    \caption{Histogram of \MET~[left] and \HT~[right] after preselection cuts for background and a range of squark-neutralino models at a $100$ TeV proton collider.  The neutralino mass is $1\GeV$ for all signal models.}
    \label{fig:alljet_presel_distributions_SQSQB_100TeV}
  \end{center}
\end{figure}

\hiddensubsection{Analysis: 100 TeV}
Figure~\ref{fig:alljet_presel_distributions_SQSQB_100TeV} shows the background and three signal distributions for a $100$ TeV proton collider in the two kinematic variables which are scanned in this analysis: \MET~[left] and \HT~[right].  In Table~\ref{tab:SNbulkCounts_100TeV} we give the number of events after each stage of cuts for the dominant backgrounds and a two signal models.   From this table, it is clear that a $100$ TeV proton collider with $3000 \text{ fb}^{-1}$ would be able to exclude (but not discover) squarks with mass of $8 \TeV$.

\begin{table}[h!]
\renewcommand{\arraystretch}{1.4}
\setlength{\tabcolsep}{4pt}
\footnotesize
\vskip 10pt
  \begin{centering}
    \begin{tabular}{| r | r r | r | r  r r |}
\hline
&&&&\multicolumn{3}{c|}{$m_{\tilde{q}}$ [GeV]}\\
                                               Cut&                 $V$+jets&               $t\overline{t}$&                 Total BG&                      
$3162$&                    $ 5012$ &                     $7944$\\
\hline
\hline
$                             \mathrm{Preselection}$&{$1.64 \times 10^{9}$}&{$3.33 \times 10^{9}$}&{$4.97 \times 10^{9}$}&$2.01 \times 10^{5}$&$1.44 \times 10^{4}$&                    $  668$\\
\hline
$\MET/\sqrt{\HT }> 15 \GeV^{1/2}$&{$3.59 \times 10^{7}$}&{$3.31 \times 10^{7}$}&{$6.90 \times 10^{7}$}&$1.62 \times 10^{5}$&$1.26 \times 10^{4}$&                     $ 614$\\
$p_T^{\mathrm{leading}} < 0.4\times \HT $&{$1.19 \times 10^{7}$}&{$1.25 \times 10^{7}$}&{$2.44 \times 10^{7}$}&$4.14 \times 10^{4}$&$2.63 \times 10^{3}$&                    $ 96.9$\\
\hline
\hline
$         \MET > 5550 \GeV$&\multirow{2}{*}{                       $24$}&\multirow{2}{*}{                    $ 21.2$}&\multirow{2}{*}{                     $45.1$}&\multirow{2}{*}{                \color{red}   $  73.3$}&\multirow{2}{*}{                    $ 60.1$}&\multirow{2}{*}{                    $ 19.9$}\\
$                          \HT > 900 \GeV$&&&&&&\\
\hline
$         \MET > 4900 \GeV$&\multirow{2}{*}{                     $61.2$}&\multirow{2}{*}{                   $  53.3$}&\multirow{2}{*}{                    $  114$}&\multirow{2}{*}{                    $  119$}&\multirow{2}{*}{         \color{red}              $143$}&\multirow{2}{*}{                     29.5}\\
$                         \HT > 5450 \GeV$&&&&&&\\
\hline
$        \MET > 6150 \GeV$&\multirow{2}{*}{                      $9.2$}&\multirow{2}{*}{                      $8.4$}&\multirow{2}{*}{                     $17.6$}&\multirow{2}{*}{                     $11.2$}&\multirow{2}{*}{                     $  17$}&\multirow{2}{*}{              \color{red}       $ 11.5$}\\
$                         \HT > 8200 \GeV$&&&&&&\\
\hline
    \end{tabular}
    \caption{Number of expected events for $\sqrt{s} = 100$ TeV with 3000 fb$^{-1}$ integrated luminosity for the background processes and selected squark masses for the squark-neutralino model.  The neutralino mass is $1 \GeV$.  Three choices of cuts on \MET~and \HT~are provided, and for each squark mass column the entry in the row corresponding to the ``optimal" cuts is marked in red.
    }
    \label{tab:SNbulkCounts_100TeV}
  \end{centering}
\end{table}

\begin{figure}[bp]
  \centering
  \includegraphics[width=.48\columnwidth]{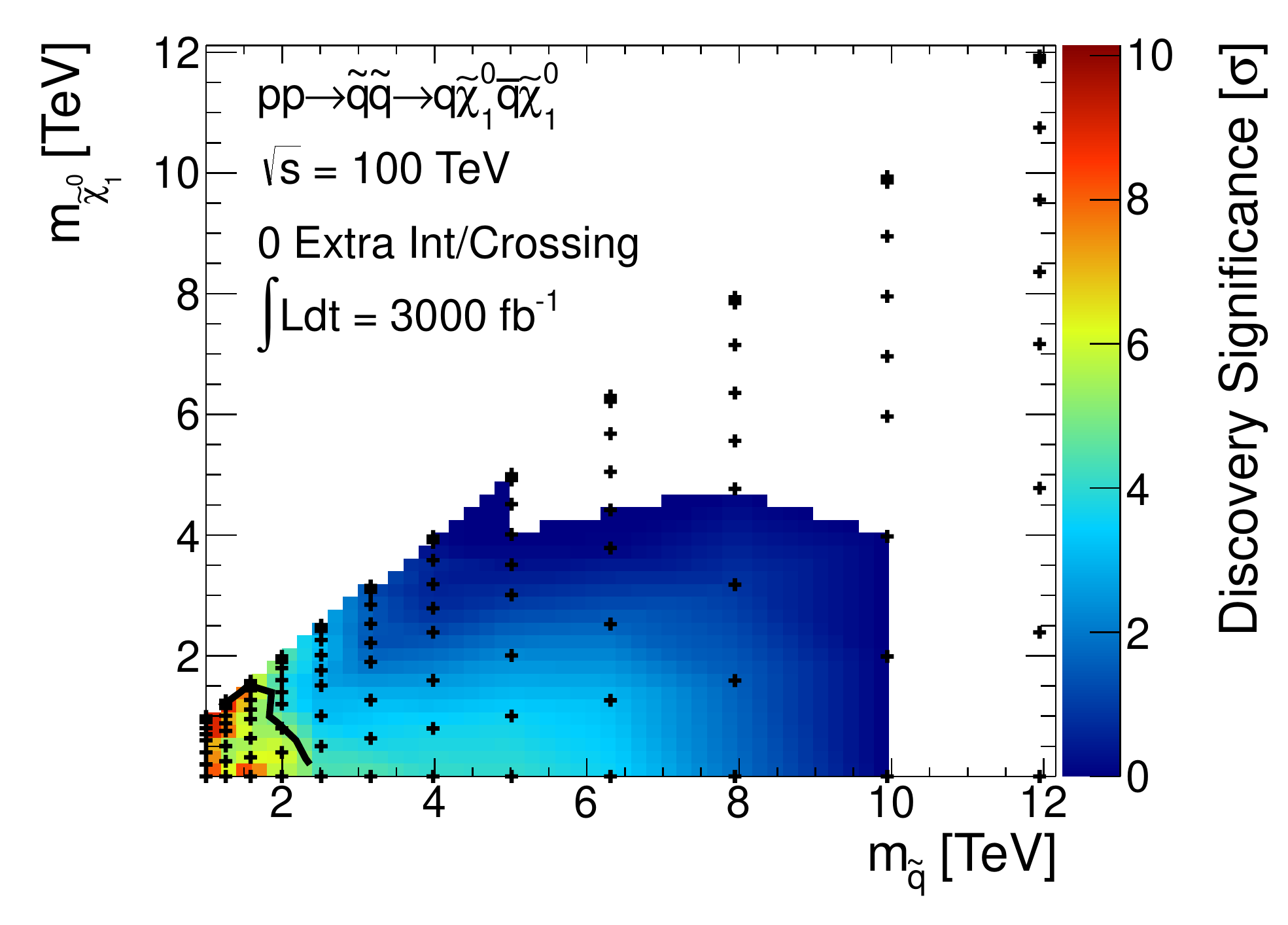}
  \includegraphics[width=.48\columnwidth]{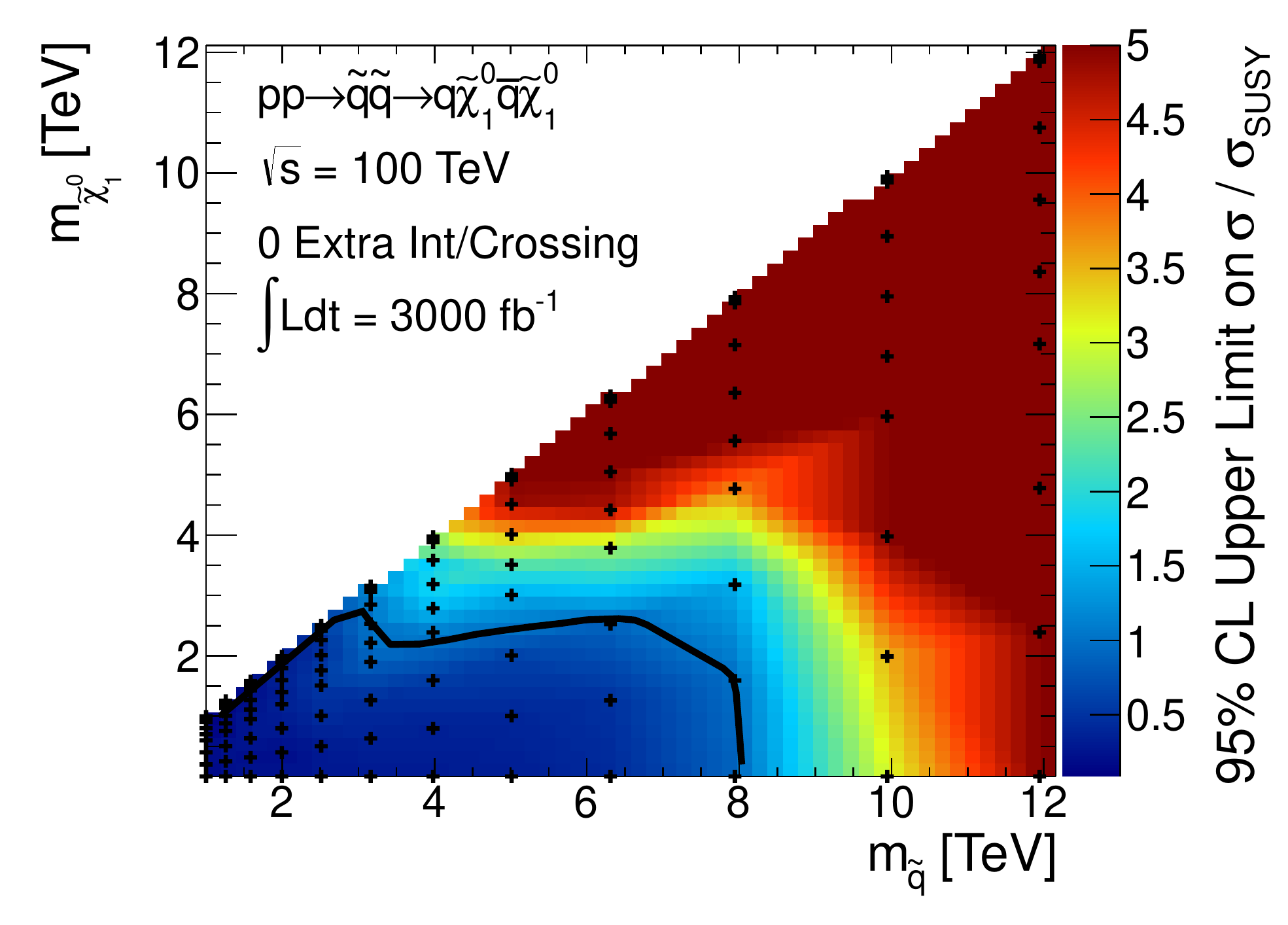}
  \caption{Results for the squark-neutralino model with light flavor decays are given in the $m_{\widetilde{\chi}_1^0}$ versus $m_{\widetilde{q}}$ plane.  The left [right] panels shows the expected $5\sigma$ discovery reach [95\% confidence level upper limits] for squark-anti-squark production at a $100$ TeV proton collider.  Mass points to the left/below the contours are expected to be probed with $3000$ fb$^{-1}$ of data [right].  A 20\% systematic uncertainty is assumed for the backgrounds. Pileup is not included.}
    \label{fig:SQSQ_100_NoPileUp_results}
\end{figure}

\hiddensubsection{Results: 100 TeV}
The results for the squark-neutralino model are shown in Fig.~\ref{fig:SQSQ_100_NoPileUp_results} for a $100$ TeV proton collider.  Discovery significance [$95\%$ CL exclusion] contours in the $m_{\widetilde{\chi}_1^0}$ versus $m_{\widetilde{q}}$ plane are shown on the left [right].  As expected, the reach is significantly smaller than for the gluino-neutralino model with light flavor decays.  

Using the NLO gluino pair production cross section one can make a very naive estimate for the reach of a given collider.  For example, we find that the choice of squark mass which would yield $10$ events at $3000 \text{ fb}^{-1}$ is $14.8 \TeV$.  This roughly corresponds to the maximal possible reach one could expect for a given luminosity using $100 \TeV$ proton collisions.  

Using a realistic simulation framework along with the search strategy employed here the $100 \TeV$ $3000 \text{ fb}^{-1}$ limit with massless neutralinos is projected to be $8.0 \TeV$ (corresponding to 849 events).  Given this huge number of events, it is possible that a different (or more sophisticated) search strategy would allow for greater sensitivity to these models --- this is outside the purview of the current study.  Finally, we note that the $100 \TeV$ proton collider with $3000 \text{ fb}^{-1}$ could discover a squark as heavy as $2.4\TeV$ if the neutralino is massless.  Compared to the $14$ and $33$ TeV searches, the squark reach degrades less rapidly as the neutralino mass is increased from the massless limit. The next section provides a comparison of the impact that the four collider scenarios studied here can have on the parameter space of this model.

\begin{figure}[h!]
  \centering
  \includegraphics[width=.48\columnwidth]{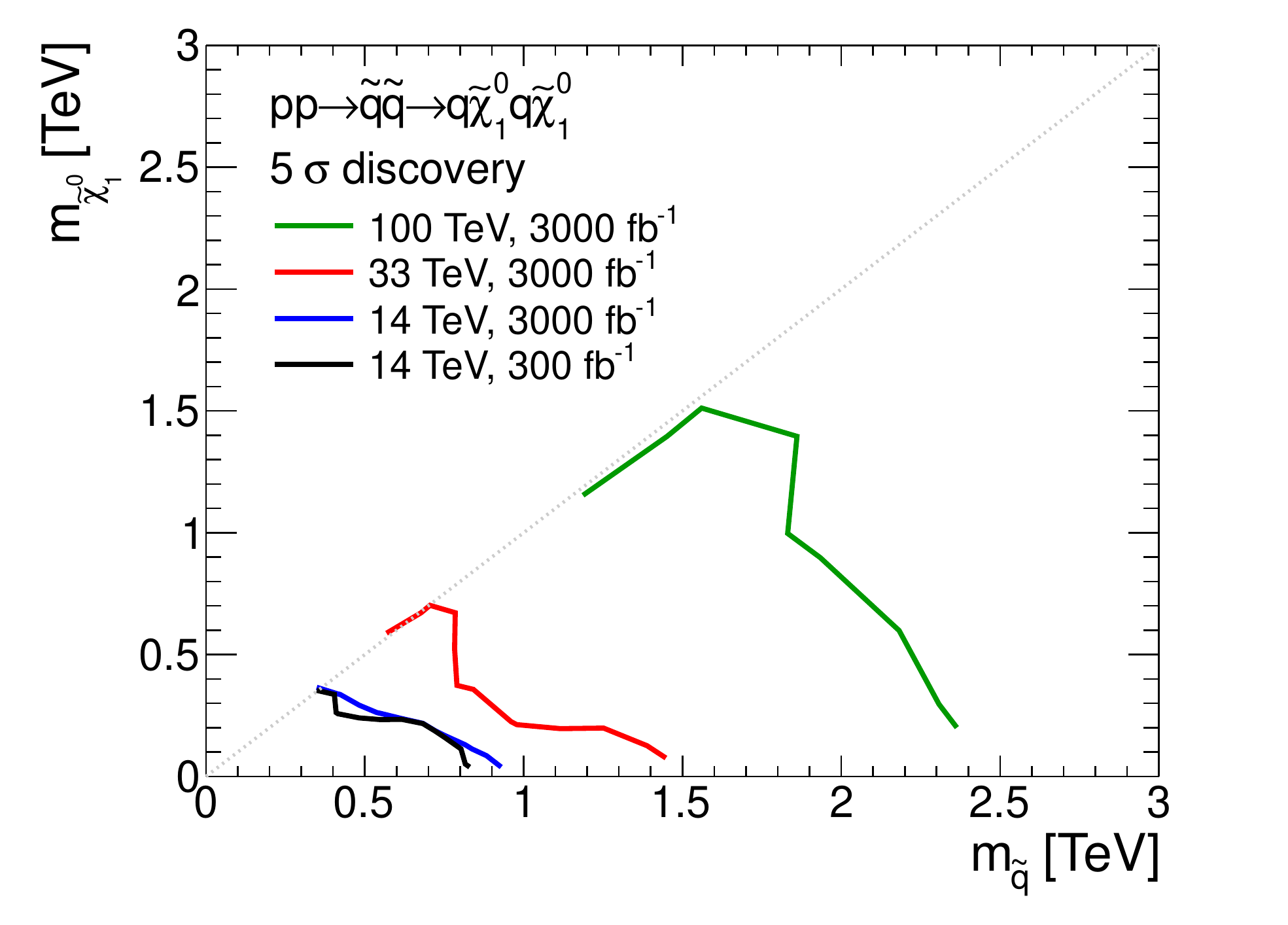}
  \includegraphics[width=.48\columnwidth]{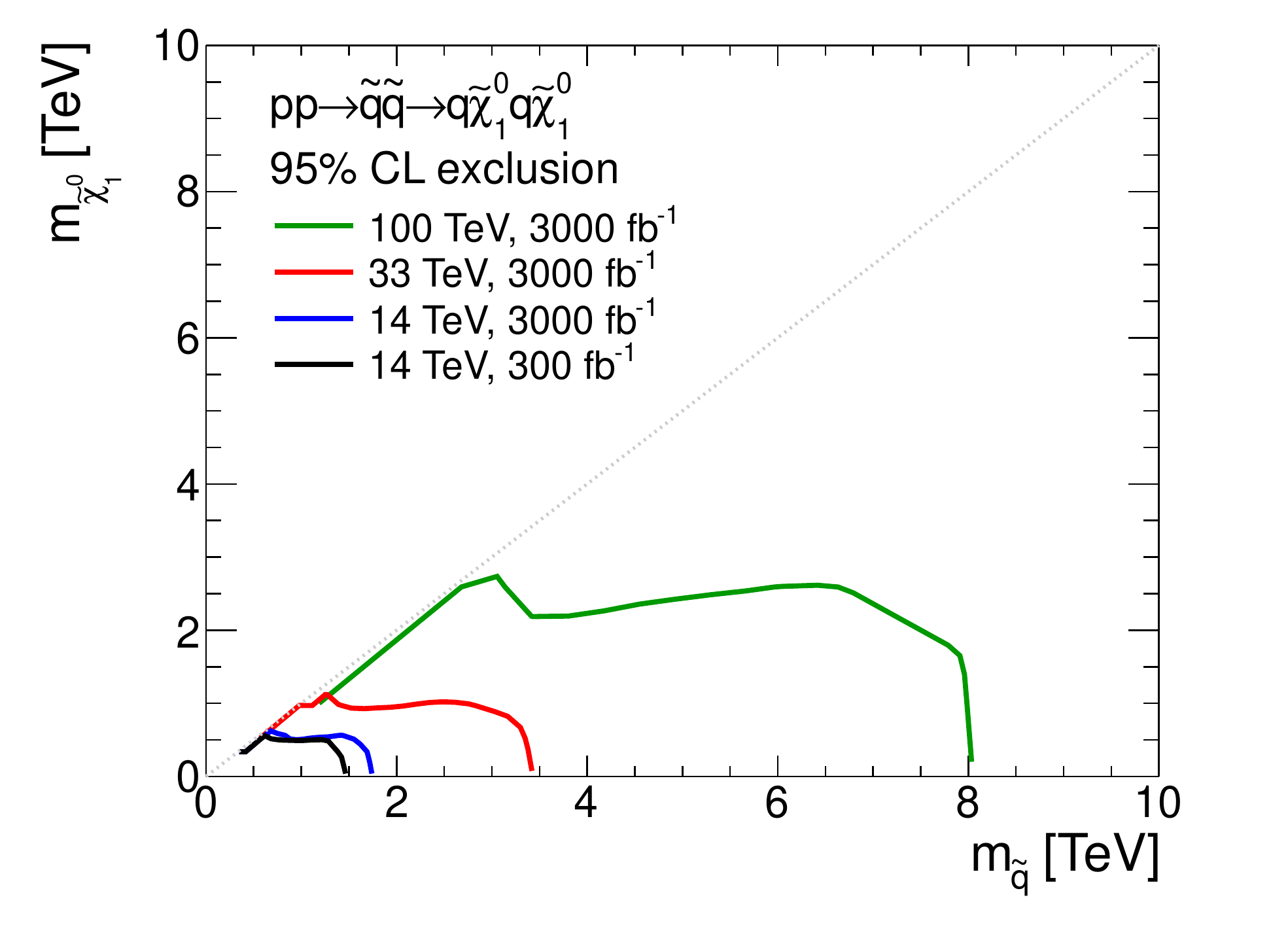}
  \caption{Results for the squark-neutralino model.  The left [right] panel shows the $5\,\sigma$ discovery reach [$95\%$ CL exclusion] for the four collider scenarios studied here.  A $20\%$ systematic uncertainty is assumed and pileup is not included.}
    \label{fig:SQSQ_Comparison}
\end{figure}

\pagebreak
\hiddensubsection{Comparing Colliders}
The squark-neutralino model has a similar multi-jet plus \MET~signature to the gluino-neutralino model with light flavor decays.  However, the squark-neutralino model is more difficult to probe due to the smaller number of hard jets in the final state coupled with the substantially smaller production cross section.  Since this model provides a more challenging scenario, it is interesting to understand the impact that can be made on exploring the parameter space with different collider scenarios.  Figure \ref{fig:SQSQ_Comparison} shows the $5\sigma$ discovery reach [$95\%$ CL exclusion] for two choices of integrated luminosity at $14$ TeV, along with the reach using the full data set assumed for $33$ and $100$ TeV.  

In general, we find that due to the small cross sections, it is very difficult to distinguish this model from background with discovery level significance\footnote{It is worth noting that this search, which was devised originally to target gluinos, has not been extensively optimized for the signature of squark pair production.  It is possible that a search exactly tailored to this signal could improve the reach beyond what is found here.}. Consequentially, the discovery reach does not appear to significantly improve with the $14$ TeV luminosity upgrade. The discovery reach in the massless neutralino limit also scales slowly with the CM energy, increasing only by a factor of $3$ from $14$ TeV to $100$ TeV, compared to a factor of $5$ for the gluino-neutralino model. 

The exclusion reach for the squark-neutralino models is much more favorable in comparison. At this level of significance the background systematics are less difficult to overcome, and the limits scale much more favorably with luminosity and CM energy, as in the gluino-neutralino model. Figure \ref{fig:GOGO_Comparison} makes a compelling case for investing in future proton colliders which can operate at these high energies.

\section{The Compressed Squark-Neutralino Model}
\label{sec:QNCompressed}
The results presented in the previous section were derived using a search which targeted the bulk of the  squark-neutralino Simplified Model parameter space.  In the compressed region where
\be
m_{\widetilde{q}} - m_{\widetilde{\chi}_1^0} \equiv \Delta m \ll m_{\widetilde{q}}
\ee
a different search strategy is required.  For parameters in this range, the jets which result from the direct decays of the squark will be very soft and one has to rely on ISR jets to discriminate these models from background.  These signatures will be very similar to those produced by the compressed gluino-neutralino model with light flavor decays, and therefore the backgrounds will be identical to those described above in Sec.~\ref{sec:Backgrounds_CompressedGOGO}.  Therefore, we will use the same search strategies described above in Sec.~\ref{sec:FourAnalysisStrategies}.  

\begin{figure}[!h]
\begin{center}
\includegraphics[width=0.48\textwidth]{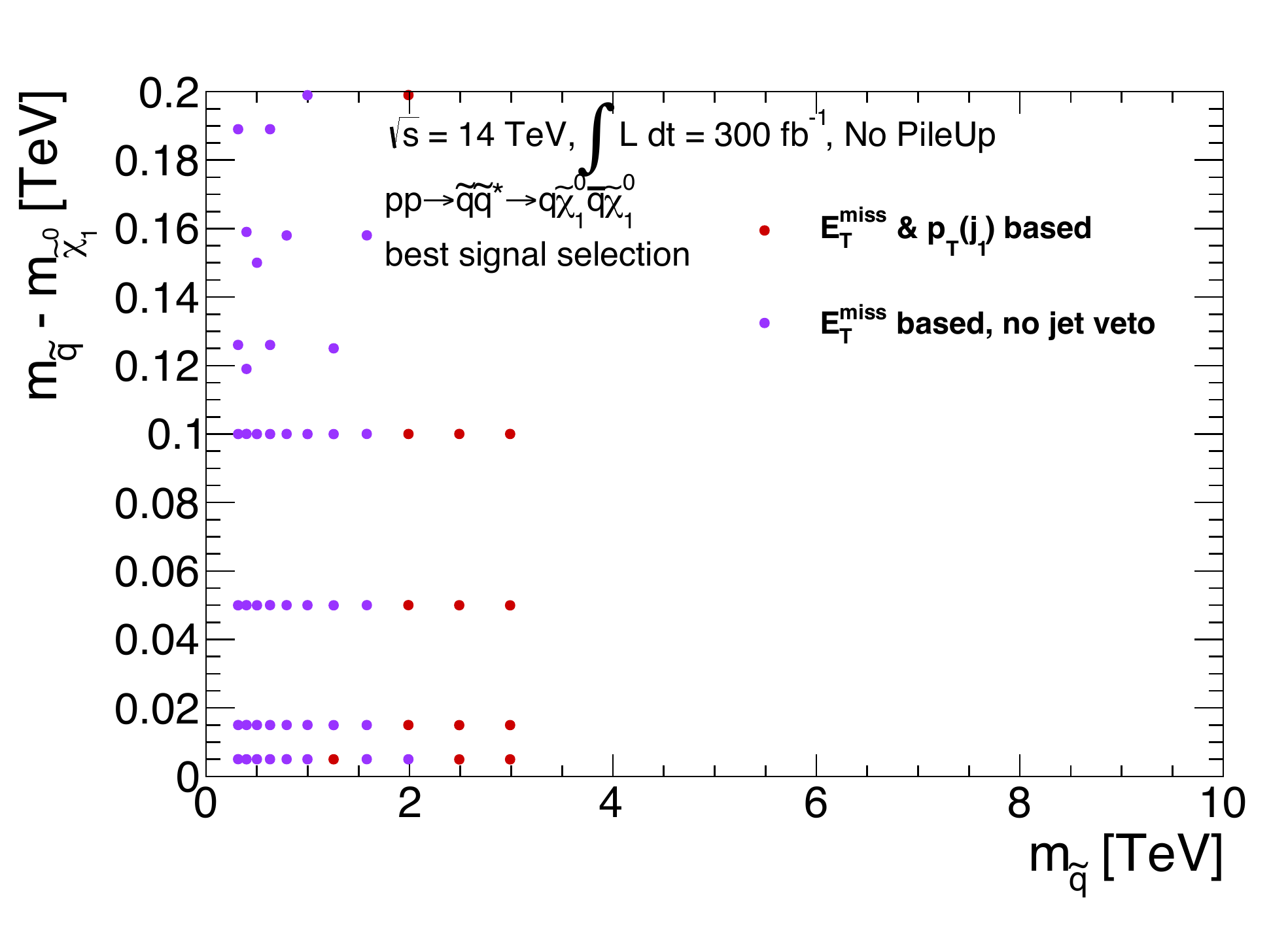}
\includegraphics[width=0.48\textwidth]{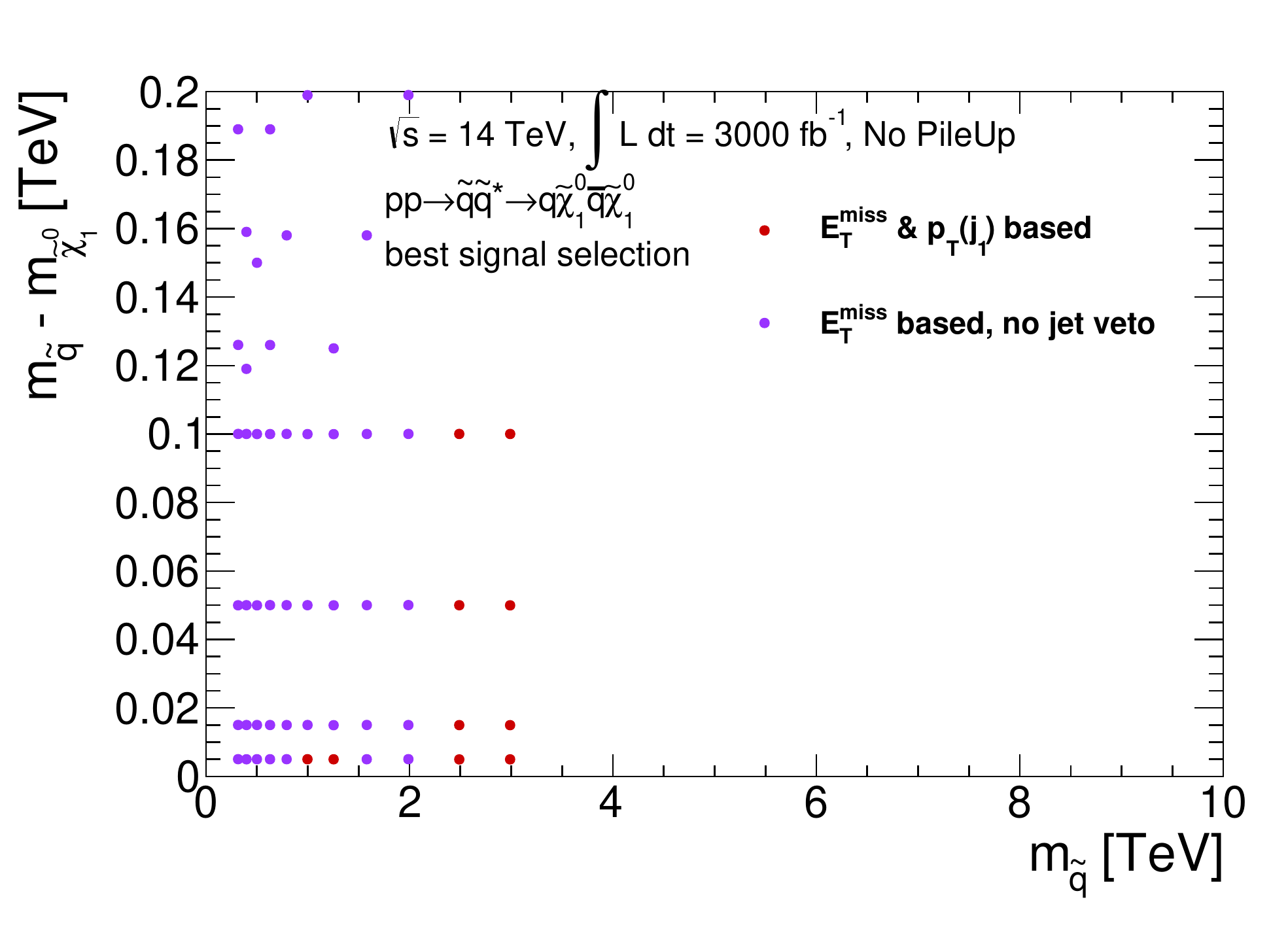}
\caption{The choice of analysis strategy that leads to the best discovery reach for a given point in parameter space for an integrated luminosity of 300 fb$^{-1}$ [left] and 3000 fb$^{-1}$ [right] for the compressed region of the squark-neutralino model. The colors refer to the analyses as presented above: red circle = leading jet based, purple circle = $\MET$-based.}
\label{fig:SQSQCompressedBestSearch}
\end{center}
\end{figure}

\begin{figure}[!h]
\begin{center}
\includegraphics[width=0.6\textwidth]{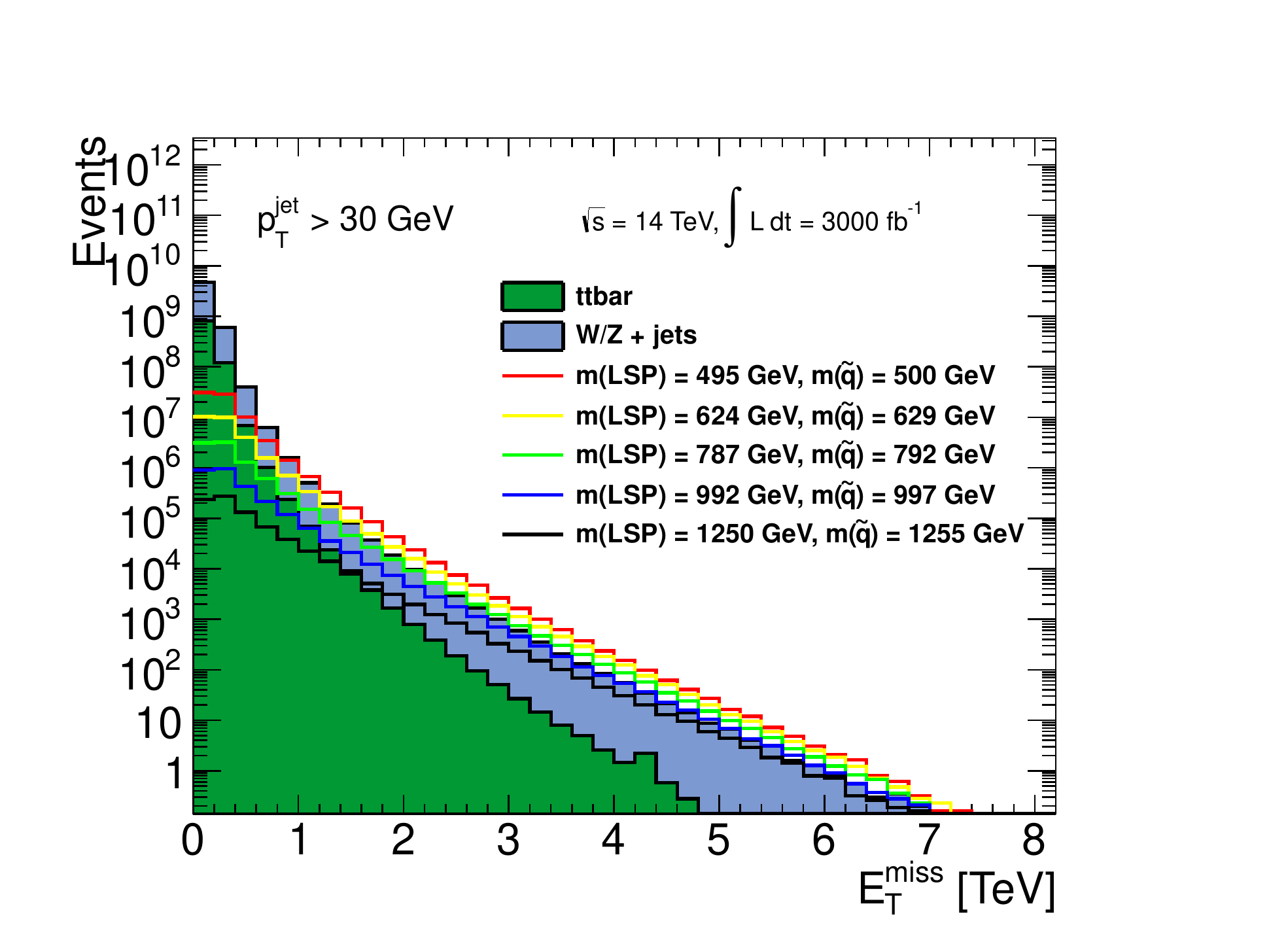}
\caption{Histogram of \MET~for signal and background after the preselection for a range of squark and neutralino masses in the compressed region.}
\label{fig:lj_pt_presel_distributions_SQSQ}
\end{center}
\end{figure}

\hiddensubsection{Analysis: 14 TeV}
As can be seen in Fig~\ref{fig:SQSQCompressedBestSearch}, for the very small squark masses excludable in the compressed region the only relevant strategy is the \MET~based search.  A histogram of the discriminating variable relevant for this search is shown in Fig.~\ref{fig:lj_pt_presel_distributions_SQSQ}.  We also give the number of events after cuts for this strategy in Table~\ref{tab:SQSQ_met_counts_14TeV}.  It is clear that for low mass squarks, it is possible that the signal could be distinguished over background.

\begin{table}[h!]
\begin{centering}
\renewcommand{\arraystretch}{1.6}
\setlength{\tabcolsep}{5pt}
\footnotesize
\vskip 10pt
\begin{tabular}{|r|rr|r|rrr|}
\hline
&&&&\multicolumn{3}{c|}{$\big(m_{\widetilde{q}},\,m_{\widetilde{\chi}^0_1}\big)\quad$ [GeV]} \\
Cut & $V$+jets & $t\,\overline{t}$ & Total BG & $(500,\,495)$ & $(792,\,787)$ & $(997,\,992)$\\
\hline
\hline
$\text{Preselection}$ & $1.3 \times 10^{9}$ & $1.2 \times 10^{8}$ & $1.4 \times 10^{9}$ & $5.8 \times 10^{6}$ & $4.3 \times 10^{5}$ & $9.8 \times 10^{4}$\\
\hline
$p_{T}^{\text{leadjet}} > 110 \text{ GeV}, |\eta^{\text{leadjet}}| < 2.4$ & $7.9 \times 10^{8}$ & $8.1 \times 10^{7}$ & $8.7 \times 10^{8}$ & $4.5 \times 10^{6}$ & $3.5 \times 10^{5}$ & $8.0 \times 10^{4}$\\
\hline
\hline
$\MET > 500 \text{ GeV}$ & $2.1 \times 10^{6}$ & $1.5 \times 10^{5}$ & $2.3 \times 10^{6}$ & $4.5 \times 10^{5}$ & $5.2 \times 10^{4}$ & $1.4 \times 10^{4}$\\
\hline
$\MET > 1 \text{ TeV}$ & $4.9 \times 10^{4}$ & $2.1 \times 10^{3}$ & $5.2 \times 10^{4}$ & $3.0 \times 10^{4}$ & $5.0 \times 10^{3}$ & $1.6 \times 10^{3}$\\
 \hline
$\MET > 2 \text{ TeV}$ & $278$ & $3$ & $282$ & $237$ & $64$ & $24$\\
\hline 
\end{tabular}
\caption{Number of expected events for $\sqrt{s} = 14$ TeV and $3000$ fb$^{-1}$ for the background processes and selected signal processes. The selection without a veto on additional jets with cuts on \MET~is applied. Three choices of cuts are provided for illustration.}
\label{tab:SQSQ_met_counts_14TeV}
\end{centering}
\end{table}

\begin{figure}[!htb]
\begin{center}
\includegraphics[width=0.48\textwidth]{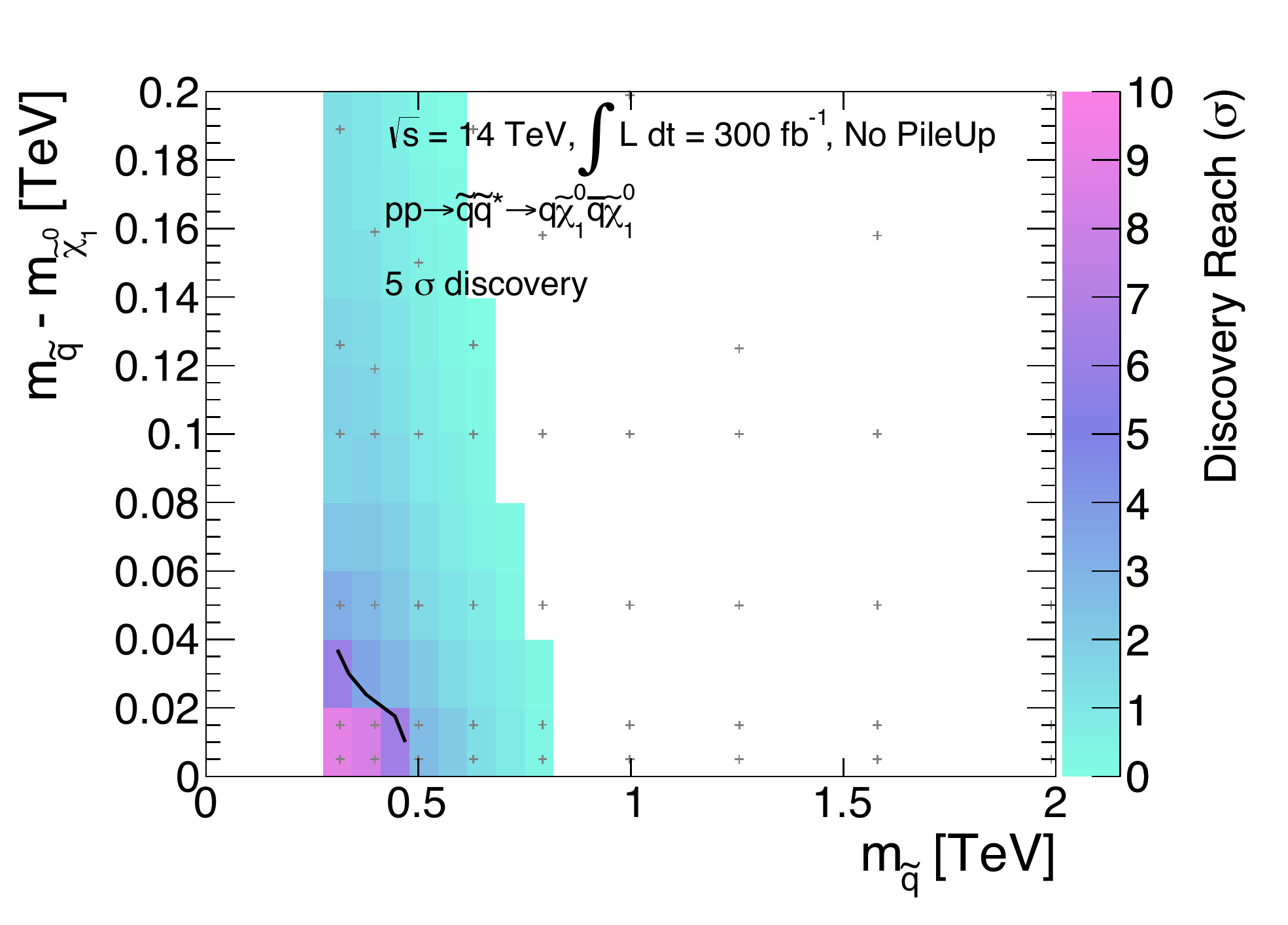}
\includegraphics[width=0.48\textwidth]{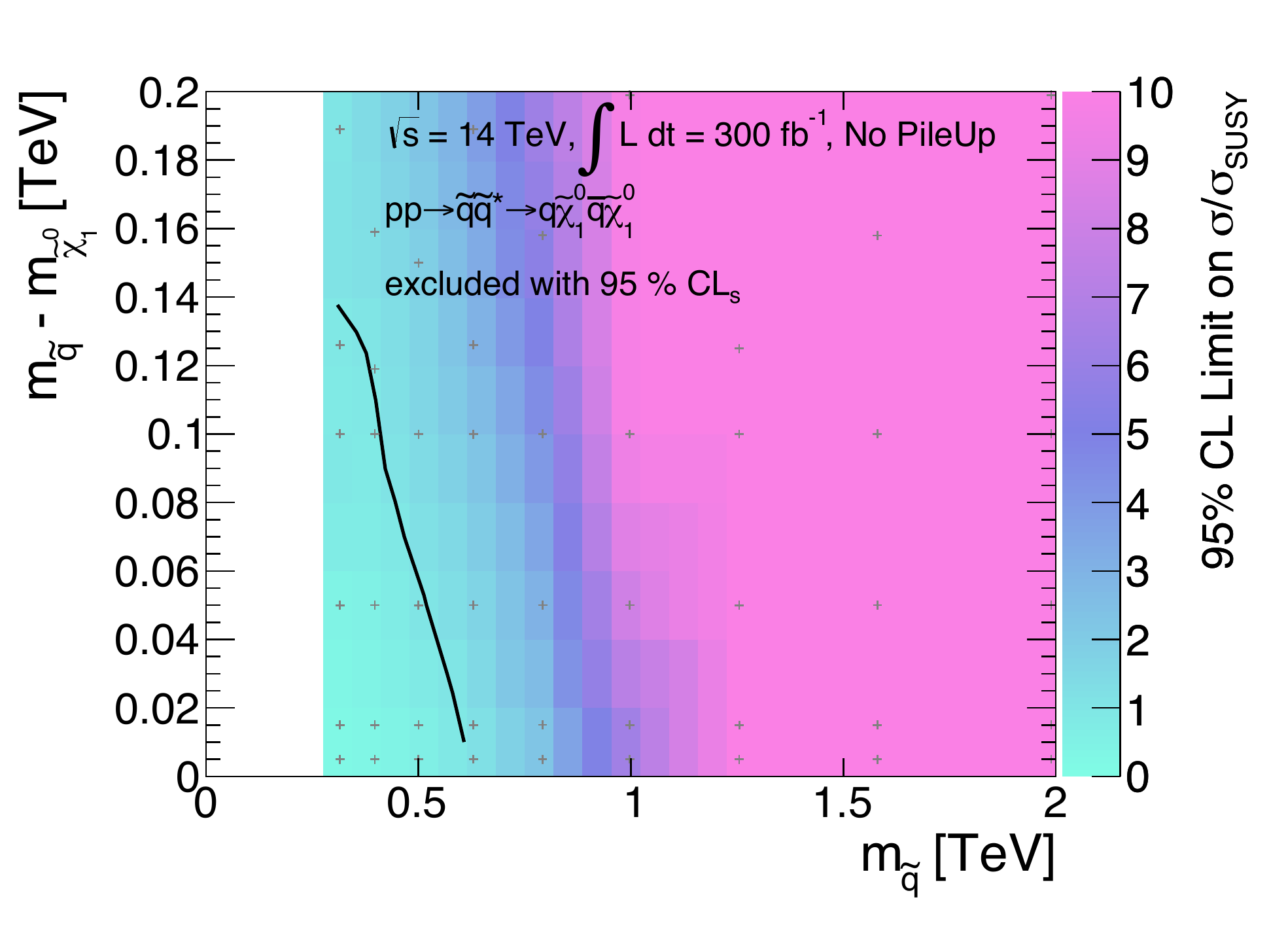}
\includegraphics[width=0.48\textwidth]{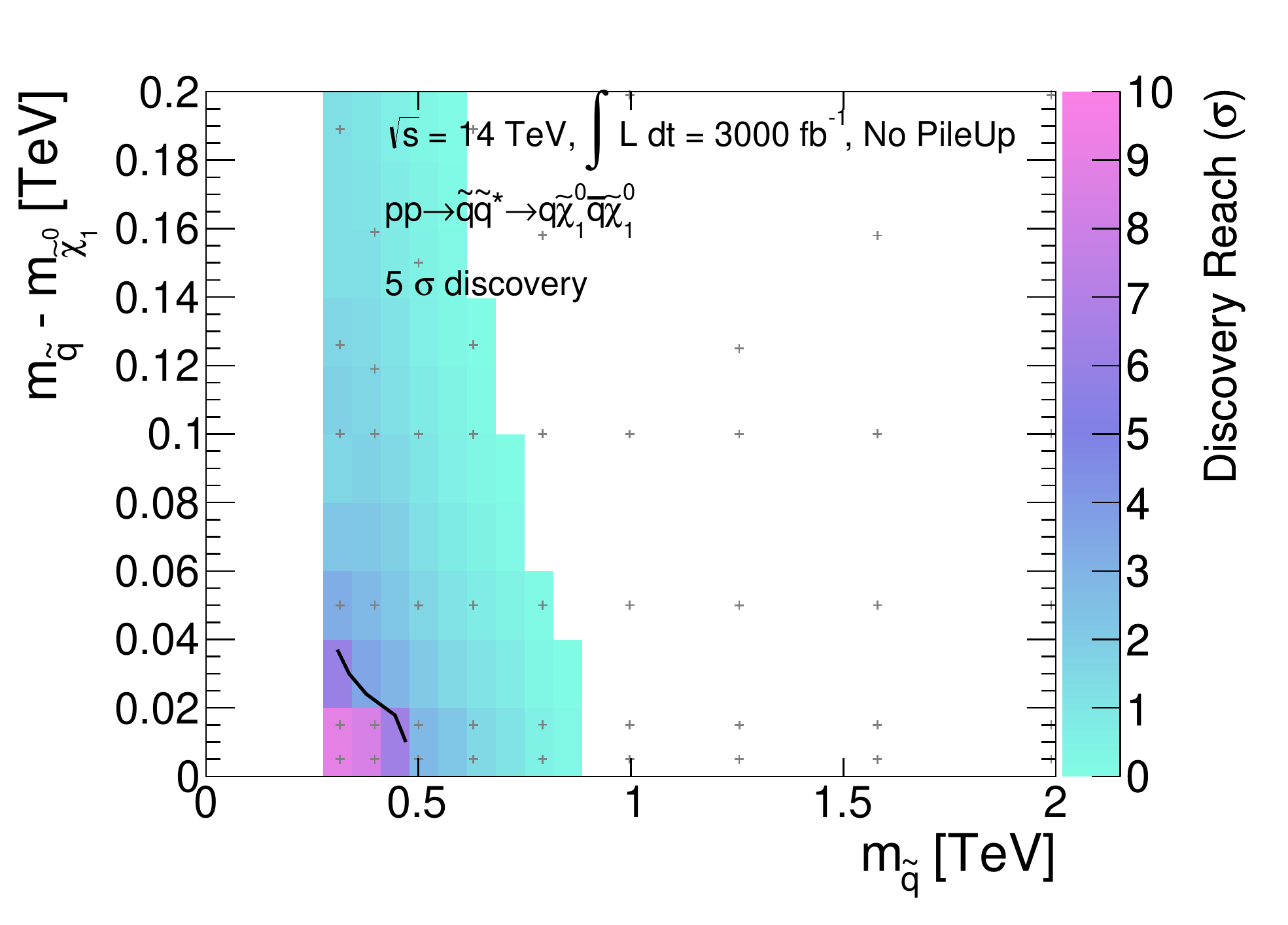}
\includegraphics[width=0.48\textwidth]{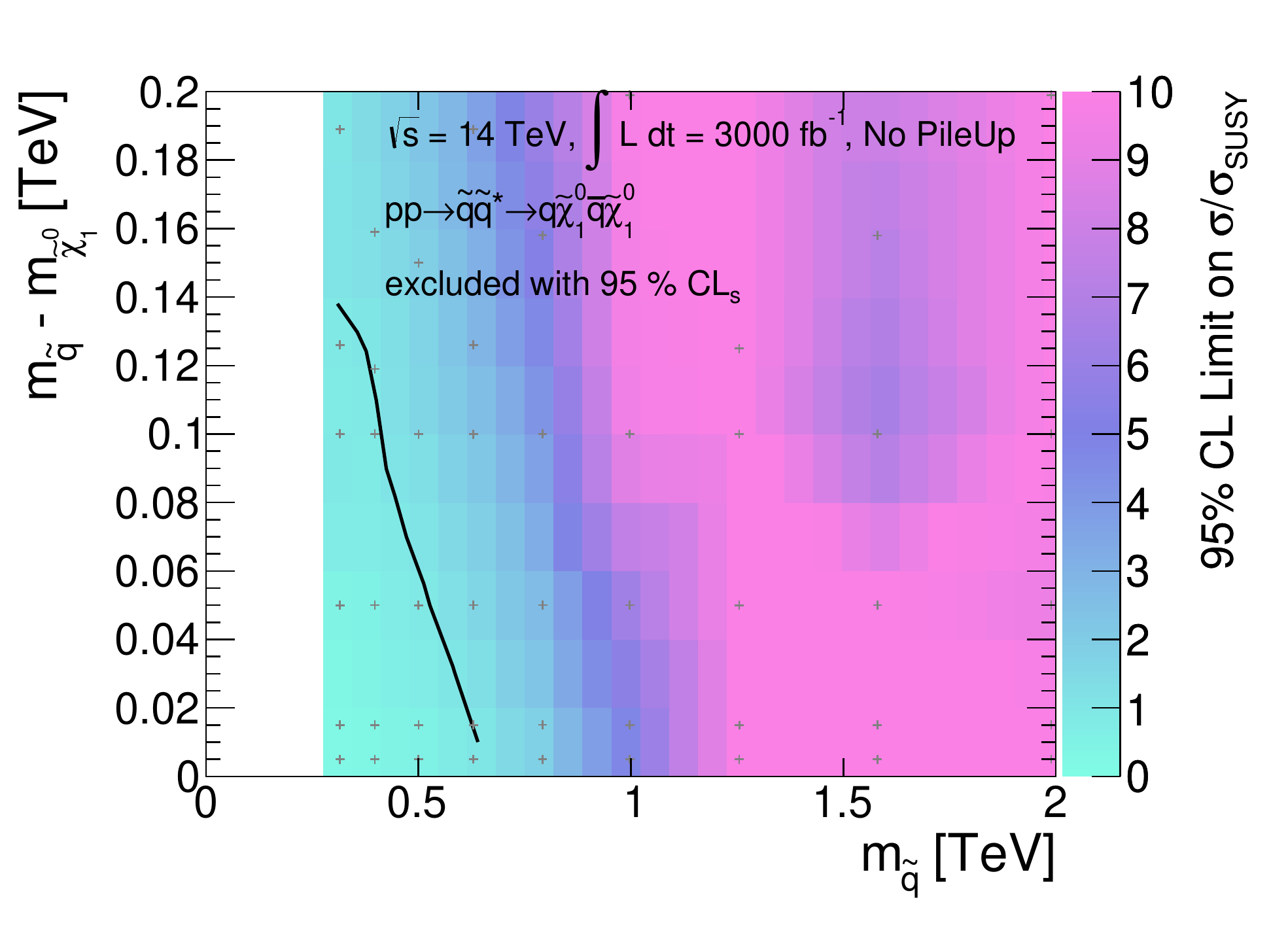}
\caption{Results for the compressed squark-neutralino model with light flavor decays are given in the $m_{\widetilde{q}} - m_{\widetilde{\chi}_1^0}$ versus $m_{\widetilde{q}}$ plane.  The top [bottom] row shows the expected $5\sigma$ discovery reach [95\% confidence level upper limits] for squark-anti-squark production.  Mass points to the left/below the contours are expected to be probed at 300 fb$^{-1}$ [left] and 3000 fb$^{-1}$ [right].  A 20\% systematic uncertainty is assumed for the background.  Pileup is not included.}
\label{fig:SNLightFlavorCompressedResults}
\end{center}
\end{figure}

\pagebreak
\hiddensubsection{Results: 14 TeV}
The results for the squark neutralino model in the compressed region of parameter space are given in Fig.~\ref{fig:SNLightFlavorCompressedResults}.  As discussed above, only the \MET~based strategy (see Sec.~\ref{sec:FourAnalysisStrategies}) is relevant for this model at the $14$ TeV LHC.   It is possible to exclude  (discover) squarks in the degenerate limit with mass less than $\sim 650 \GeV (500\GeV)$ with $300 \text{ fb}^{-1}$ of data. Increasing the integrated luminosity by a factor of 10  has a minimal impact on the discovery reach for compressed squark models. This search improves the exclusion (discovery) reach near the degenerate limit by roughly $300\GeV (150\GeV)$ compared to the $\HT$-based analysis described in Sec.~\ref{sec:QN}; the $\HT$-based searches do not begin to set stronger limits until $\Delta\gtrsim50\GeV$.  Finally, we note that given our results for the compressed gluino-neutralino study in Sec.~\ref{sec:GOGOpileup_compressed} above, pileup is not expected to have a significant impact on these conclusions.  The compressed region of this model will be difficult to probe at the $14 \TeV$ LHC, but will still represent a significant improvement over current bounds.

\begin{figure}[!htb]
\begin{center}
\includegraphics[width=0.48\textwidth]{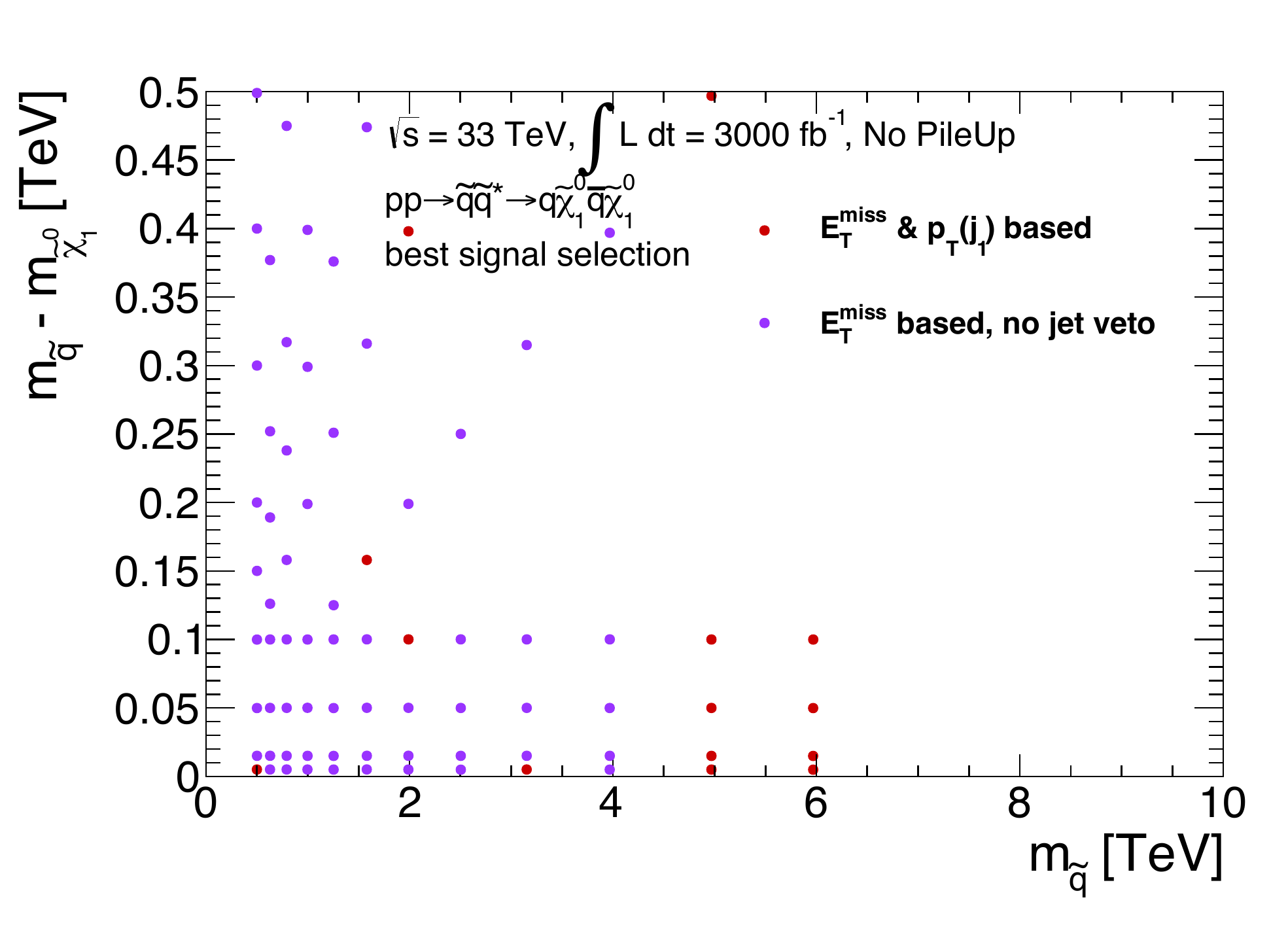}
\caption{The choice of analysis that lead to the best discovery reach for a given point in parameter space for an integrated luminosity of 3000 fb$^{-1}$ at a $33$ TeV proton collider for the compressed region of the squark-neutralino Simplified Model with light flavor decays. The colors refer to the analyses as presented above: red circle = leading jet based, purple circle = $\MET$-based.}
\label{fig:SNLightFlavorCompressedBestSearch_33TeV}
\end{center}
\end{figure}

\begin{figure}[!htb]
\begin{center}
\includegraphics[width=0.6\textwidth]{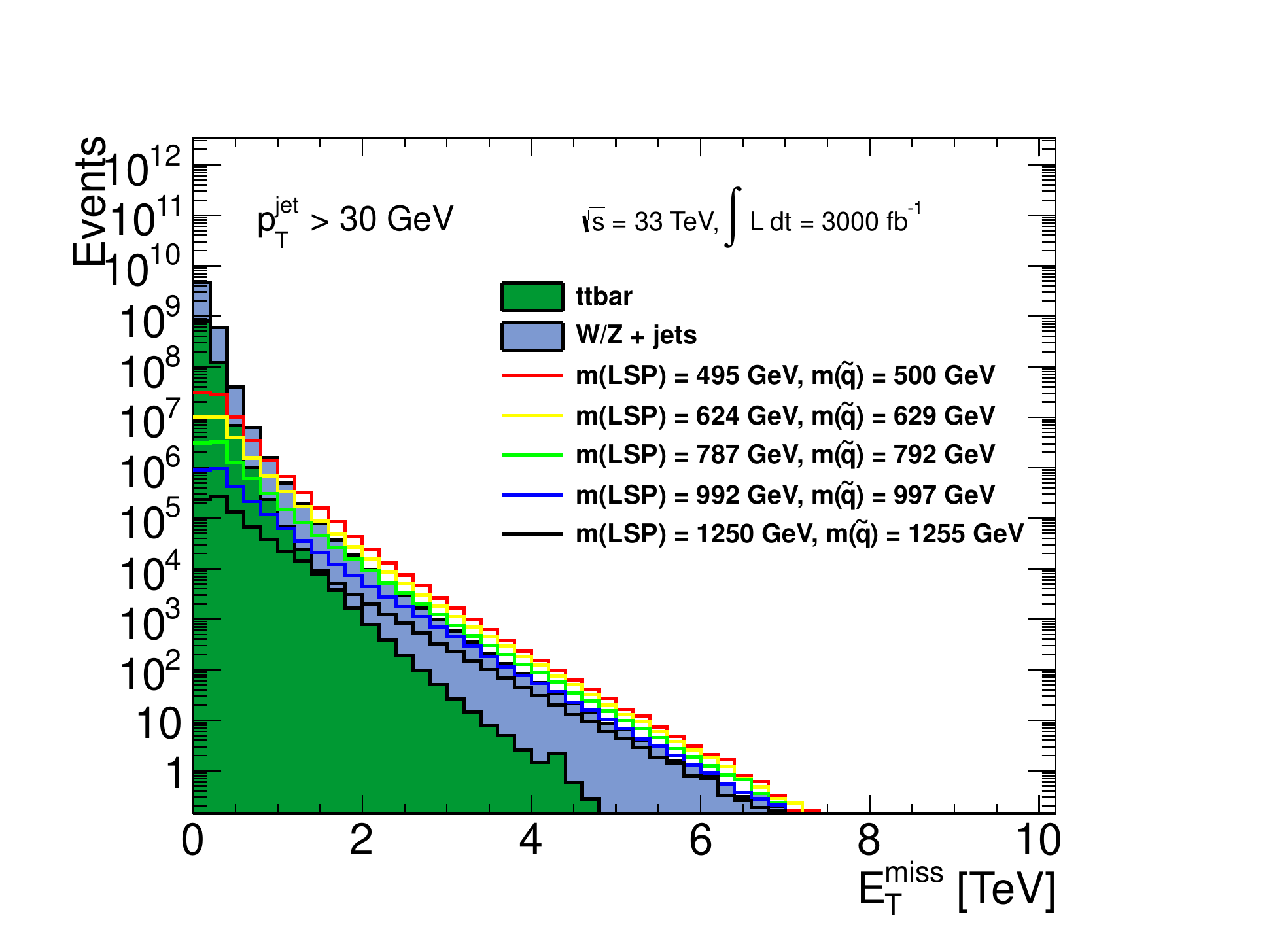}
\caption{Histogram of $\MET$ for signal and background at a $33$ TeV proton collider after the preselection for a range of squark and neutralino masses in the compressed region.}
\label{fig:SQSQ_met_presel_distributions_33TeV}
\end{center}
\end{figure}

\pagebreak
\hiddensubsection{Analysis: 33 TeV}
As discussed above in the context of the compressed region of the gluino-neutralino model, the \MET~based search tends to be more powerful at the higher energy colliders since the probability of having multiple ISR jets increases.  From Fig.~\ref{fig:SNLightFlavorCompressedBestSearch_33TeV}, it is clear that the relevant strategy in the region that can be probed by this machine is the \MET~based search.  A histogram of the discriminating variable relevant for this search is shown in  Fig.~\ref{fig:SQSQ_met_presel_distributions_33TeV}.  We also give the number of events after cuts for this strategy in Table~\ref{tab:SQSQ_met_counts_33TeV}.  It is clear that for low mass squarks, it is possible that the signal could be distinguished over background.

\begin{table}[h!]
\begin{centering}
\renewcommand{\arraystretch}{1.6}
\setlength{\tabcolsep}{5pt}
\footnotesize
\vskip 10pt
\begin{tabular}{|r|rr|r|rrr|}
\hline
&&&&\multicolumn{3}{c|}{$\big(m_{\widetilde{q}},\,m_{\widetilde{\chi}^0_1}\big)\quad$ [GeV]} \\
Cut & $V$+jets & $t\,\overline{t}$ & Total BG & $(797,\,787)$ & $(997,\,992)$ & $(1580,\,1575)$\\
\hline
\hline
Preselection & $4.2 \times 10^{9}$ & $8.6 \times 10^{8}$ & $5.1 \times 10^{9}$ & $8.4 \times 10^{6}$ & $ 2.6 \times 10^{6}$ & $2.0 \times 10^{5}$\\
\hline
$p_{T}^{\text{leadjet}} > 110 \text{ GeV}, |\eta^{\text{leadjet}}| < 2.4$ & $2.6 \times 10^{9}$ & $6.5 \times 10^{8}$ & $3.3 \times 10^{9}$ & $7.2 \times 10^{6}$ & $2.3 \times 10^{6}$ & $1.8 \times 10^{5}$\\
\hline
\hline
$\MET > 1 \text{ TeV}$ & $7.5 \times 10^{5}$ & $1.1 \times 10^{5}$ & $8.6 \times 10^{5}$ & $3.5 \times 10^{5}$ & $\color{black}1.5 \times 10^{5}$ & $2.0 \times 10^{4}$\\
\hline
$\MET > 3 \text{ TeV}$ & $1.5 \times 10^{3}$ & $62$ & $1.5 \times 10^{3}$ & $2.1 \times 10^{3}$ & $1.3 \times 10^{3}$ & $315$ \\
\hline
$\MET > 5 \text{ TeV}$ & $19$ & $0$ & $19$ & $30$ & $20$ & $7$ \\
\hline
\end{tabular}
\caption{Number of expected events for $\sqrt{s} = 33$ TeV and $3000$ fb$^{-1}$ for the background processes and selected signal processes. The selection without a veto on additional jets with cuts on \MET~is applied. Three choices of cuts are provided for illustration.}
\label{tab:SQSQ_met_counts_33TeV}
\end{centering}
\end{table}

\begin{figure}[!htb]
\begin{center}
\includegraphics[width=0.48\textwidth]{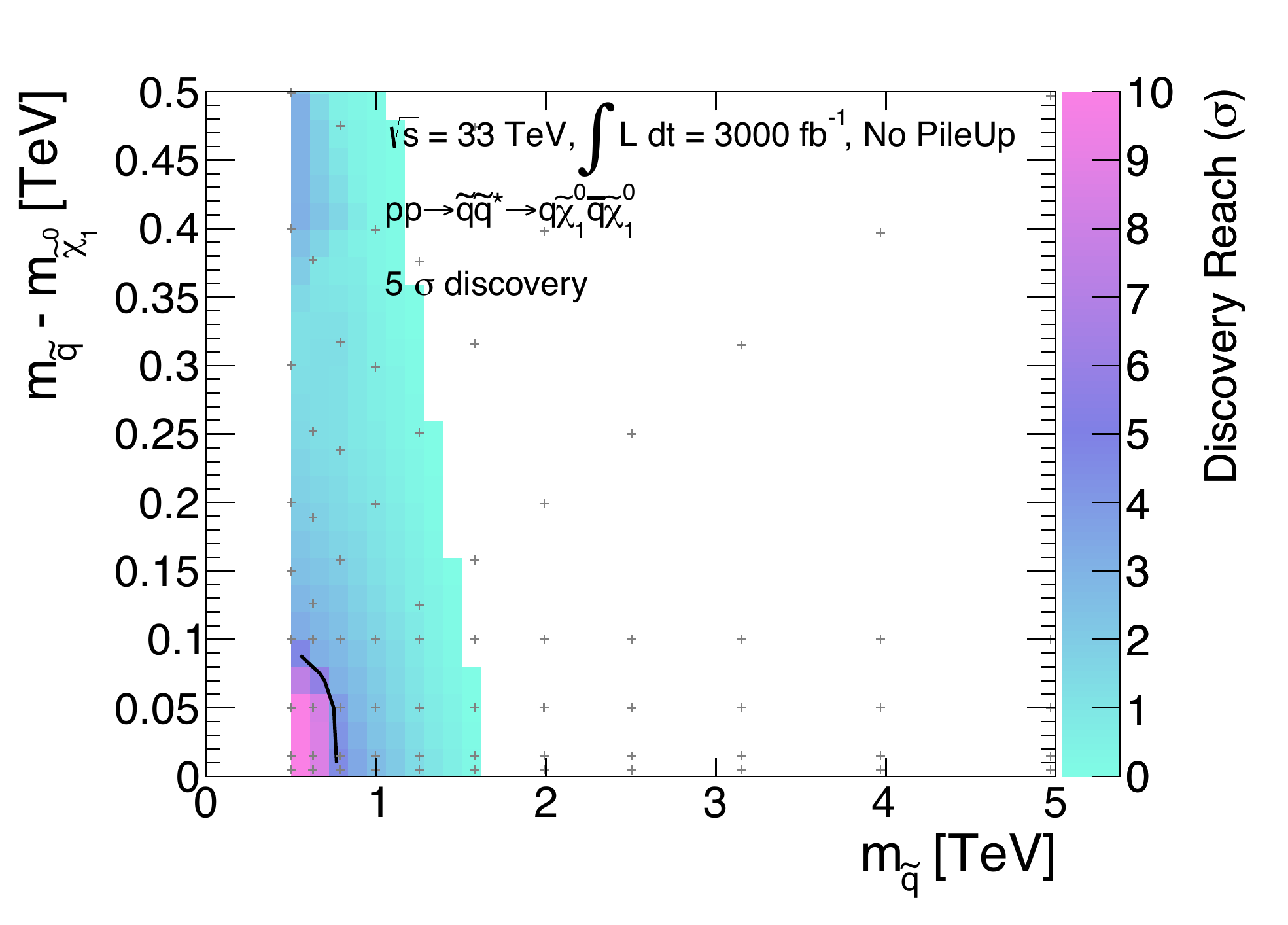}
\includegraphics[width=0.48\textwidth]{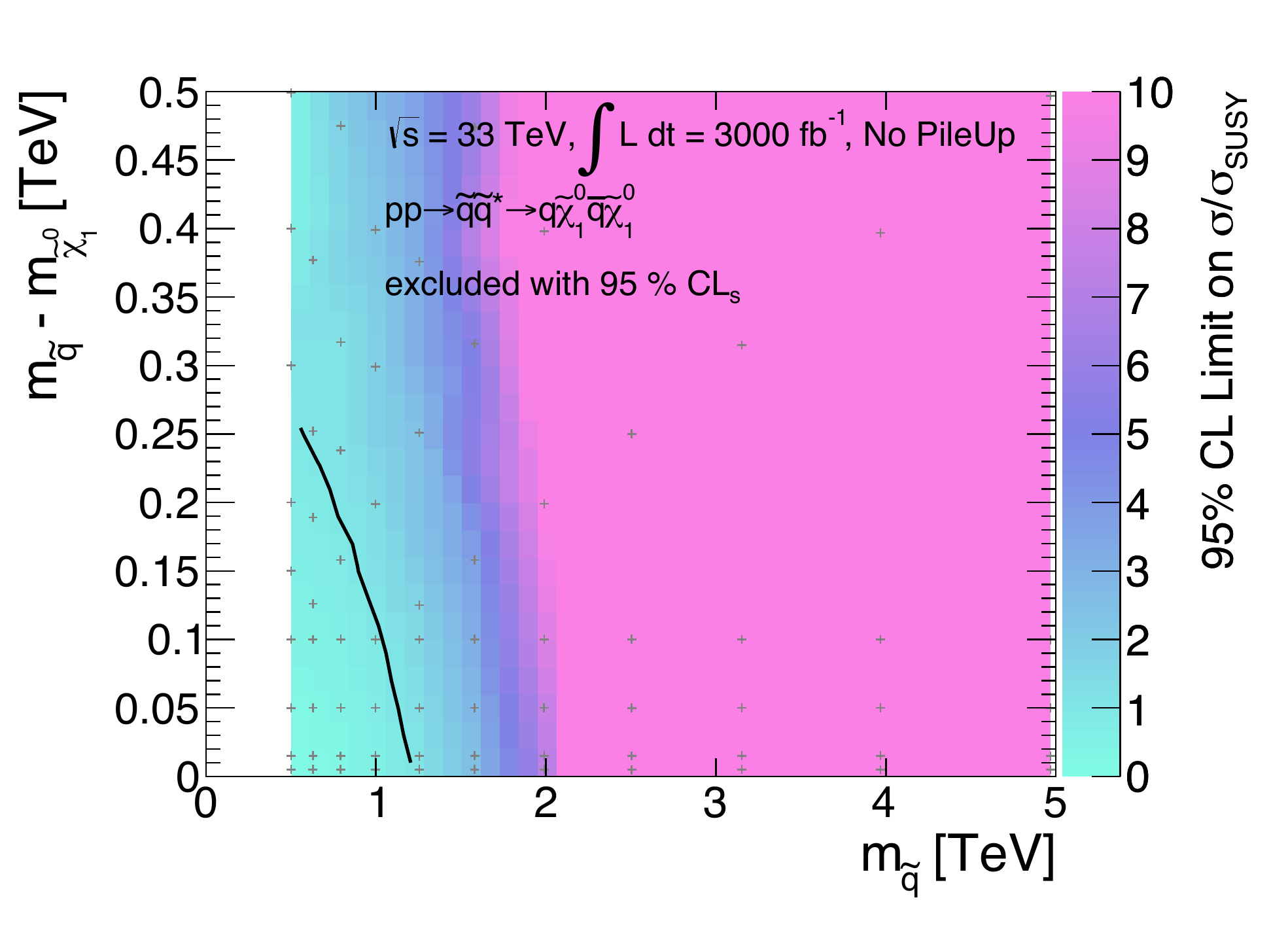}
\caption{Results for the compressed squark-neutralino model with light flavor decays at a $33$ TeV proton collider are given in the $m_{\widetilde{q}} - m_{\widetilde{\chi}_1^0}$ versus $m_{\widetilde{g}}$ plane.  The left [right] plot shows the expected $5\sigma$ discovery reach [95\% confidence level upper limits] for squark-anti-squark pair production.  Mass points to the left/below the contours are expected to be probed at $3000$ fb$^{-1}$ [right].  A 20\% systematic uncertainty is assumed for the background and pileup is not included.}
\label{fig:SNLightFlavorCompressedResults_33TeV}
\end{center}
\end{figure}

\pagebreak
\hiddensubsection{Results: 33 TeV}
The results for the squark neutralino model in the compressed region of parameter space are given in Fig.~\ref{fig:SNLightFlavorCompressedResults_33TeV}.  As discussed above, only the \MET~based strategy (see Sec.~\ref{sec:FourAnalysisStrategies}) is relevant for this model at a $33$ TeV proton collider.   It is possible to exclude (discover) squarks in the degenerate limit with mass less than $\sim1.2 (0.7) \TeV$ with $3000 \text{ fb}^{-1}$ of data.  This does not substantially improves the discovery reach near the degenerate limit compared to the $\HT$-based analysis described in Sec.~\ref{sec:QN}, but does improve the exclusion reach by roughly $200\GeV$ for $\Delta \lesssim 100 \GeV$.  Note that given our results for the compressed gluino-neutralino study in Sec.~\ref{sec:GOGOpileup_compressed} above, pileup is not expected to have a significant impact on these conclusions.  This search demonstrates that a $33$ TeV machine will be relevant to our understanding of the difficult to probe compressed region of this model.

\pagebreak
\hiddensubsection{Analysis: 100 TeV}

\begin{figure}[!htb]
\begin{center}
\includegraphics[width=0.48\textwidth]{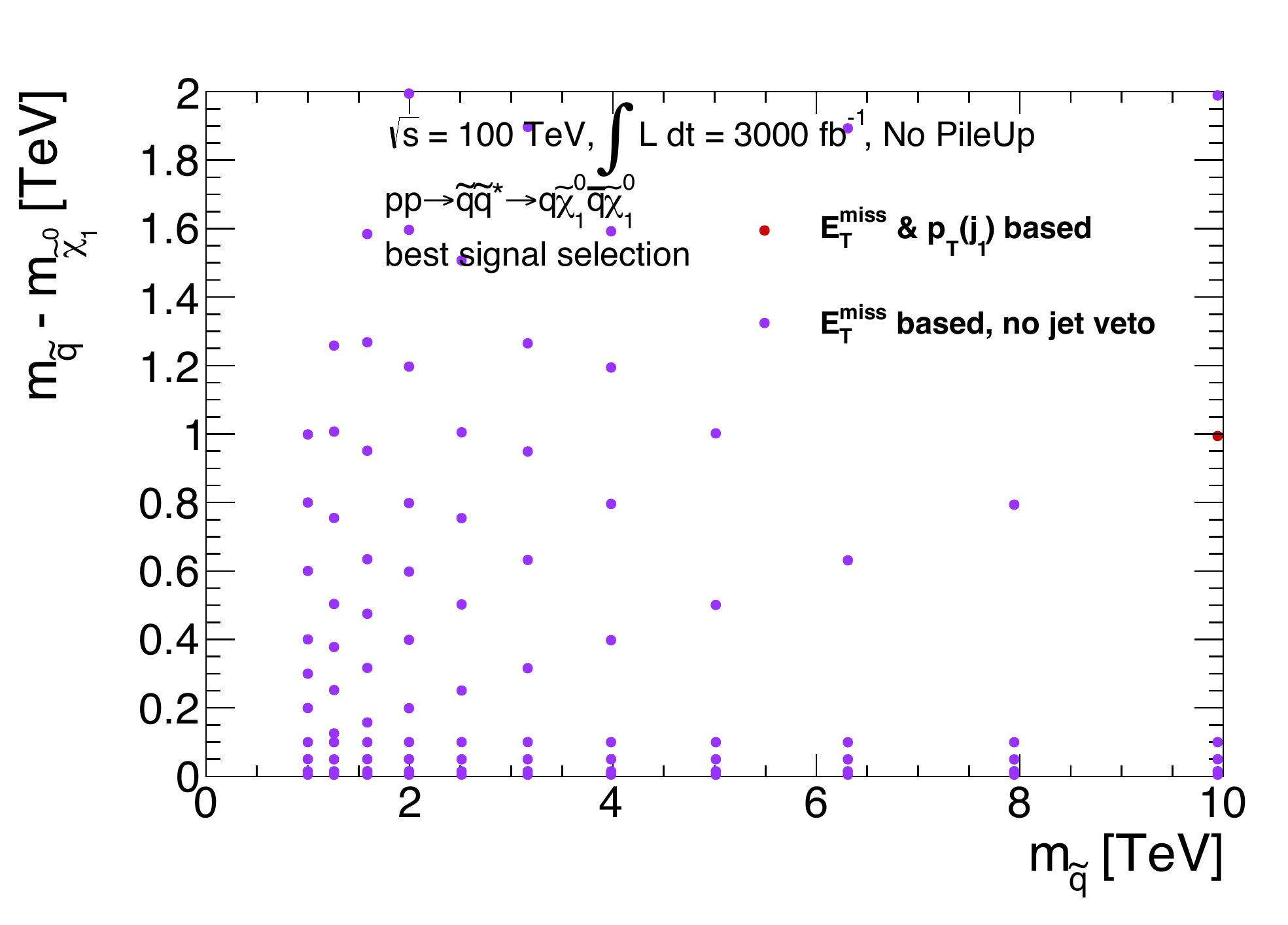}
\caption{The choice of analysis that lead to the best discovery reach for a given point in parameter space for an integrated luminosity of 3000 fb$^{-1}$ at a $100$ TeV proton collider for the compressed region of the squark-neutralino Simplified Model with light flavor decays. The colors refer to the analyses as presented above: red circle = leading jet based, purple circle = $\MET$-based.}
\label{fig:SNLightFlavorCompressedBestSearch_100TeV}
\end{center}
\end{figure}

\begin{figure}[!htb]
\begin{center}
\includegraphics[width=0.6\textwidth]{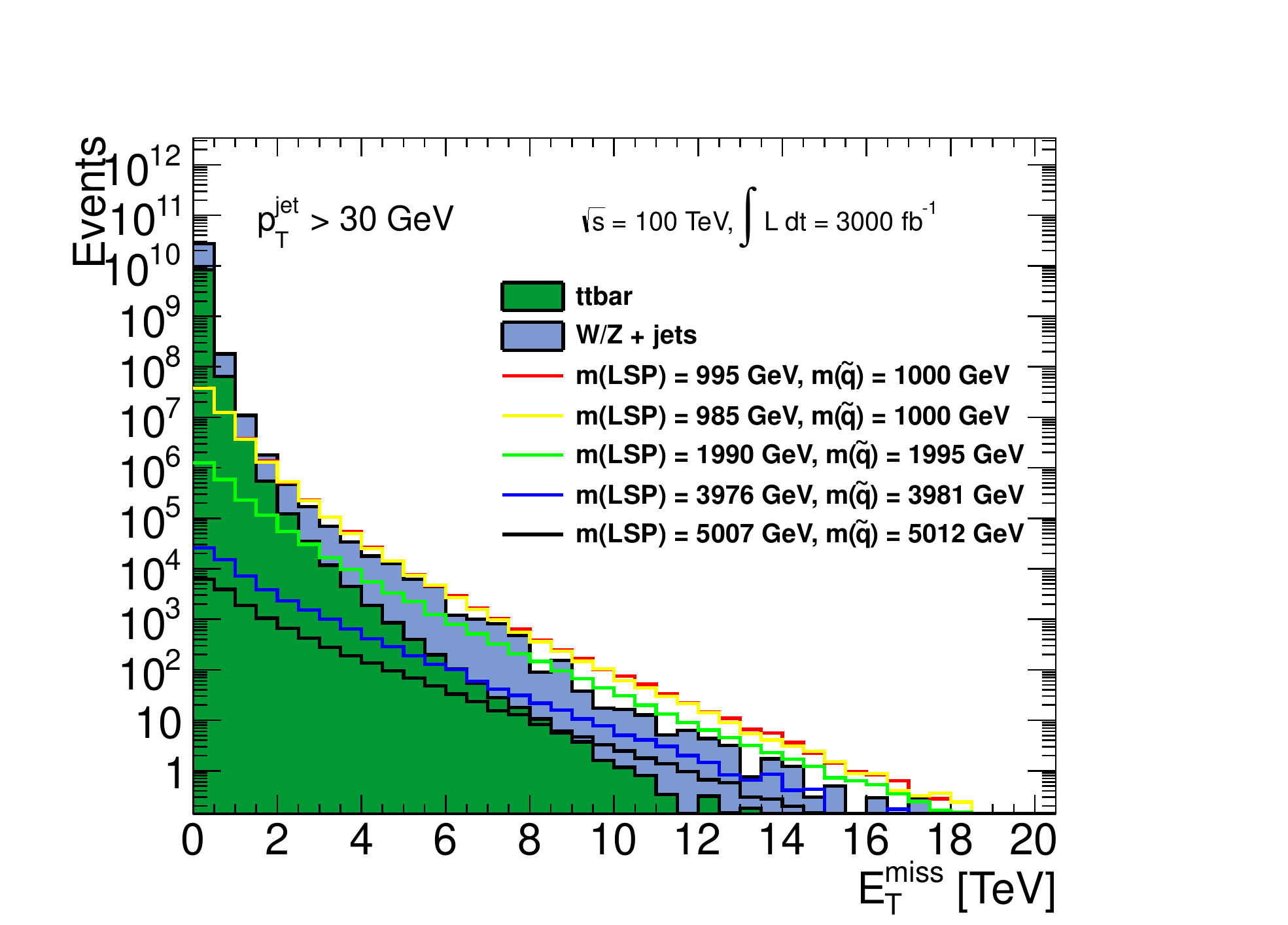}
\caption{Histogram of $\MET$ for signal and background at a $100$ TeV proton collider after the preselection for a range of squark and neutralino masses in the compressed region.}
\label{fig:SQSQ_met_presel_distributions_100TeV}
\end{center}
\end{figure}

 As discussed above in the context of the compressed region of the gluino-neutralino model, the \MET~based search tends to be more powerful at the higher energy colliders since the probability of having multiple ISR jets increases.  From Fig.~\ref{fig:SNLightFlavorCompressedBestSearch_100TeV}, it is clear that the relevant strategy in the region that can be probed by this machine is the \MET~based search.  A histogram of the discriminating variable relevant for this search is shown in Fig.~\ref{fig:SQSQ_met_presel_distributions_100TeV}.  We also give the number of events after cuts for this strategy in Table~\ref{tab:SQSQ_met_counts_100TeV}.  It is clear that for low mass squarks, it is possible that the signal could be distinguished over background.

\begin{table}[h!]
\begin{centering}
\renewcommand{\arraystretch}{1.6}
\setlength{\tabcolsep}{5pt}
\footnotesize
\vskip 10pt
\begin{tabular}{|r|rr|r|rrr|}
\hline
&&&&\multicolumn{3}{c|}{$\big(m_{\widetilde{q}},\,m_{\widetilde{\chi}^0_1}\big)\quad$ [GeV]} \\
Cut & $V$+jets & $t\,\overline{t}$ & Total BG & $(1995,\,1990)$ & $(2512,\,2507)$ & $(5012,\,5007)$\\
\hline
\hline
$\text{Preselection}$ & $1.7 \times 10^{10}$ & $7.0 \times 10^{9}$ & $2.4 \times 10^{10}$ & $2.0 \times 10^{6}$ & $6.4 \times 10^{5}$ & $1.4 \times 10^{4}$\\
\hline
$p_{T}^{\text{leadjet}} > 110 \text{ GeV}, |\eta^{\text{leadjet}}| < 2.4$ & $1.2 \times 10^{10}$ & $6.1 \times 10^{9}$ & $1.9 \times 10^{10}$ & $2.0 \times 10^{6}$ & $6.4 \times 10^{5}$ & $1.4 \times 10^{4}$\\
\hline
\hline
$\MET > 3 \text{ TeV}$ & $1.3 \times 10^{5}$ & $2.0 \times 10^{4}$ & $1.5 \times 10^{5}$ & $4.1 \times 10^{4}$ & $1.9 \times 10^{4}$ & $935$ \\
\hline
$\MET > 6 \text{ TeV}$ & $3.6 \times 10^{3}$ & 229 & $3.8 \times 10^{3}$ & $2.3 \times 10^{3}$ & $1. \times 10^{3}$ & $116$ \\
\hline
$\MET > 9 \text{ TeV}$ & $100$ & $9$ & $109$ & $206$ & $130$ & $17$ \\
\hline
\end{tabular}
\caption{Event yields for background and selected signal points for $\sqrt{s} = 100$ TeV and $3000$ fb$^{-1}$ in the event selection with cuts on $\MET$.}
\label{tab:SQSQ_met_counts_100TeV}
\end{centering}
\end{table}

\begin{figure}[!htb]
\begin{center}
\includegraphics[width=0.48\textwidth]{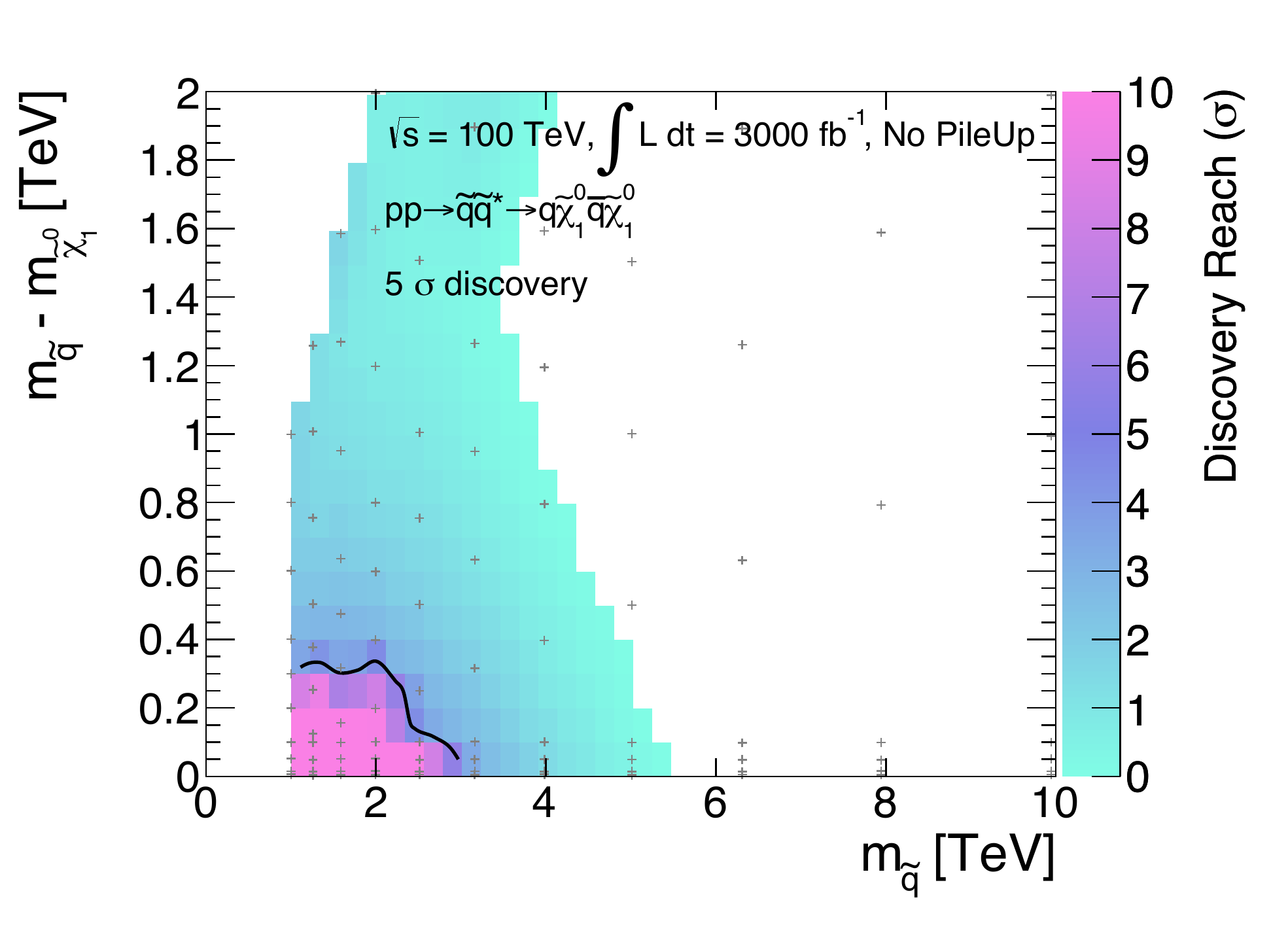}
\includegraphics[width=0.48\textwidth]{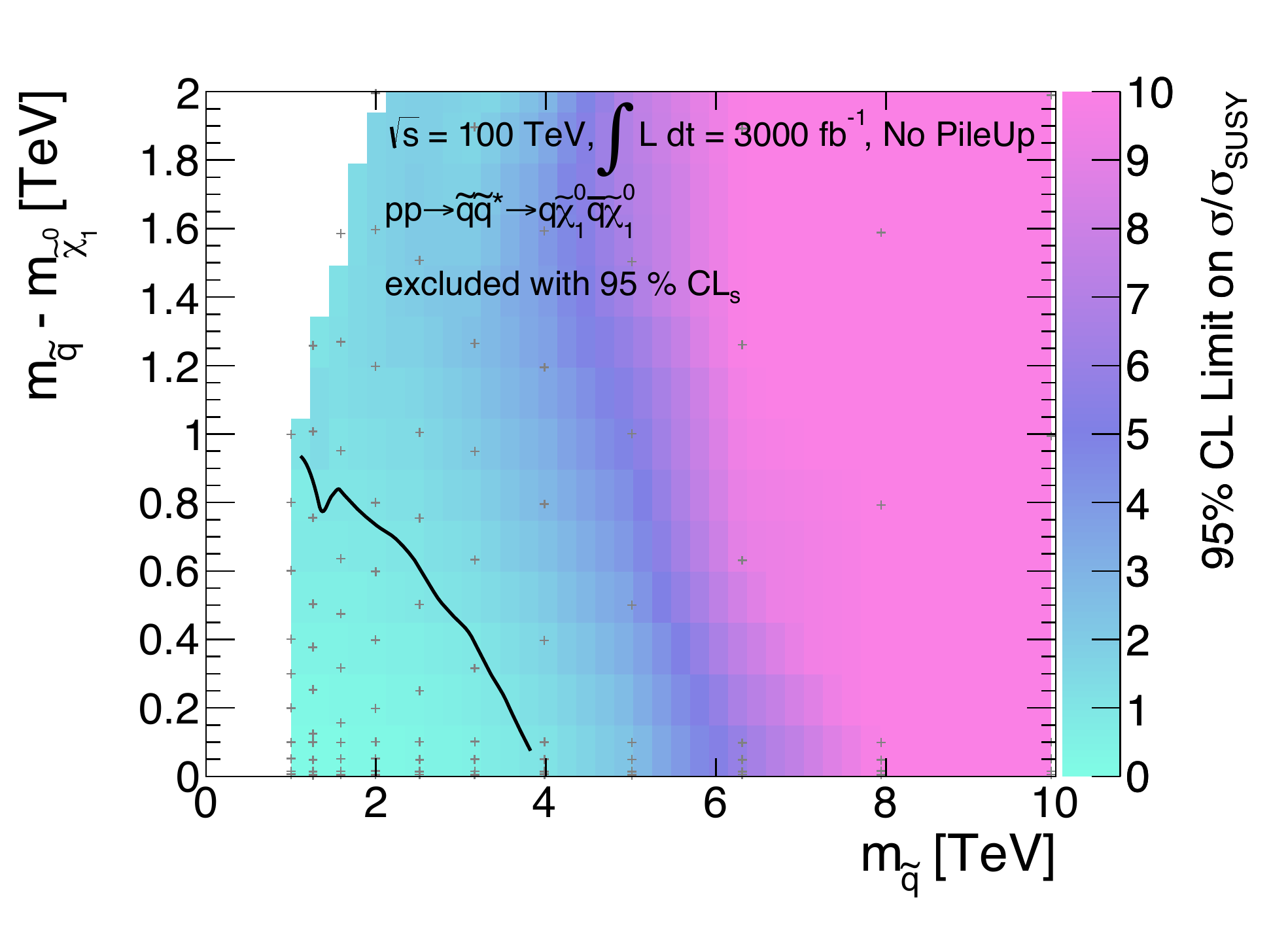}
\caption{Results for the compressed squark-neutralino model with light flavor decays at a $100$ TeV proton collider are given in the $m_{\widetilde{q}} - m_{\widetilde{\chi}_1^0}$ versus $m_{\widetilde{g}}$ plane.  The left [right] plot shows the expected $5\sigma$ discovery reach [95\% confidence level upper limits] for squark-anti-squark pair production.  Mass points to the left/below the contours are expected to be probed at $3000$ fb$^{-1}$ [right].  A 20\% systematic uncertainty is assumed for the background and pileup is not included.}
\label{fig:SNLightFlavorCompressedResults_100TeV}
\end{center}
\end{figure}

\hiddensubsection{Results: 100 TeV}
The results for the squark neutralino model in the compressed region of parameter space are given in Fig.~\ref{fig:SNLightFlavorCompressedResults_100TeV}.  As discussed above, only the \MET~based strategy (see Sec.~\ref{sec:FourAnalysisStrategies}) is relevant for this model at a $100$ TeV proton collider.   It is possible to exclude (discover) squarks in the degenerate limit with mass less than $\sim 4 \TeV (3 \TeV)$ with $3000 \text{ fb}^{-1}$ of data. This improves the exclusion (discovery) reach near the degenerate limit compared to the $\HT$-based analysis described in Sec.~\ref{sec:QN} by roughly $1.5\TeV (1.8\TeV)$ for $\Delta \lesssim 200 \GeV$. Note that given our results for the compressed gluino-neutralino study in Sec.~\ref{sec:GOGOpileup_compressed} above, pileup is not expected to have a significant impact on these conclusions.  This search demonstrates that a $100$ TeV machine will be relevant to our understanding of the difficult to probe compressed region of this model.

\begin{figure}[h!]
  \centering
  \includegraphics[width=.48\columnwidth]{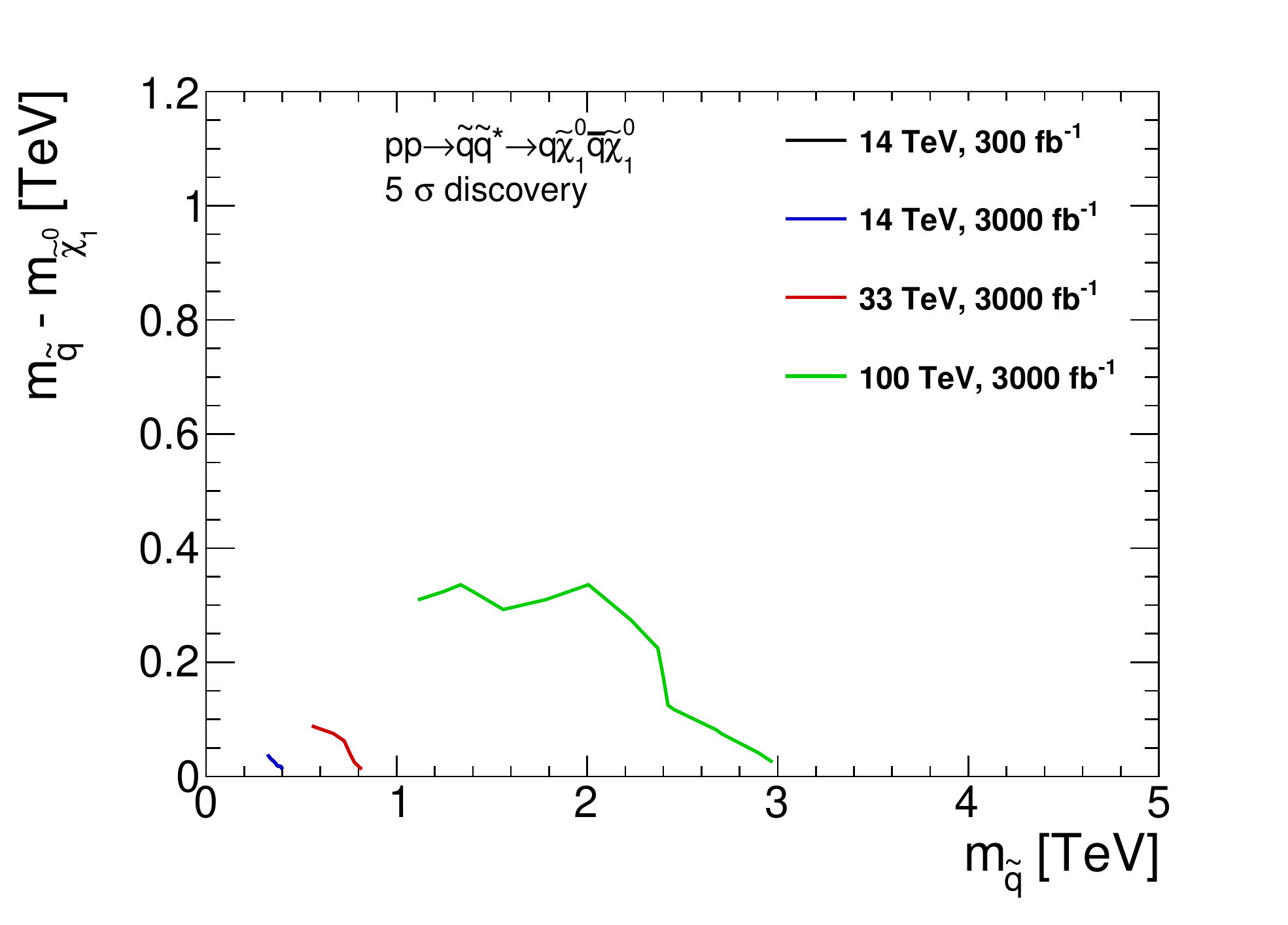}
  \includegraphics[width=.48\columnwidth]{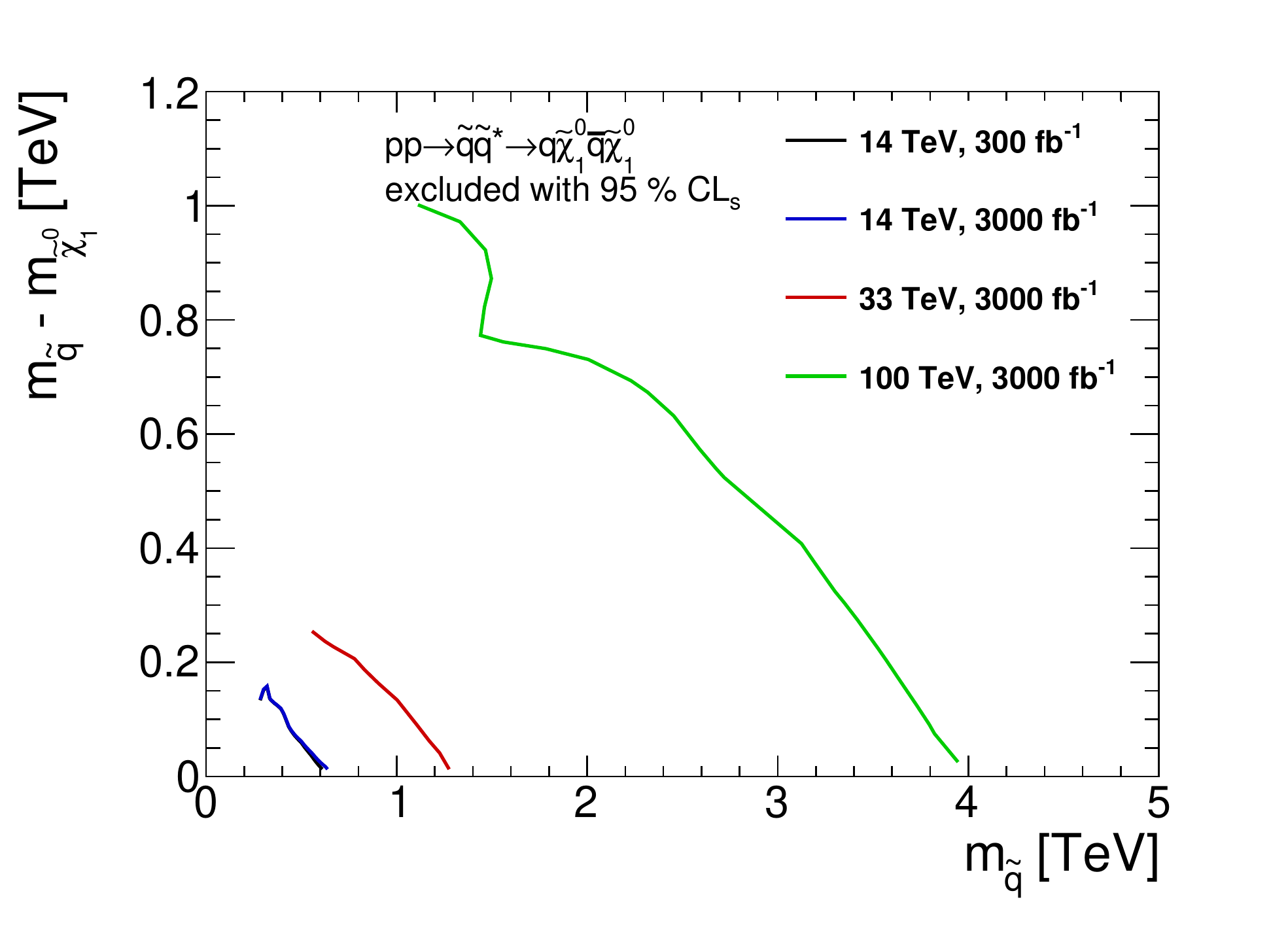}
  \caption{Results for the squark-neutralino model with light flavor decays for the analyses that target the compressed region of parameter space.  The left [right] panel shows the $5\,\sigma$ discovery reach [$95\%$ CL exclusion] for the four collider scenarios studied here.  A $20\%$ systematic uncertainty is assumed and pileup is not included.}
    \label{fig:SQSQ_Compressed_Comparison}
\end{figure}

\hiddensubsection{Comparing Colliders}
The compressed region of the squark-neutralino model has a similar signature to the compressed gluino-neutralino model with light flavor decays.  However, the squark-neutralino model is more difficult to probe due to the substantially smaller production cross section.  Since this model provides a more challenging scenario, it is interesting to understand the impact that can be made on exploring the parameter space with different collider scenarios.  Figure \ref{fig:SQSQ_Compressed_Comparison} shows the $5\sigma$ discovery reach [$95\%$ CL exclusion] for two choices of integrated luminosity at $14$ TeV, along with the reach using the full data set assumed for $33$ and $100$ TeV.  

In general, we find that due to the small cross sections, it is very difficult to distinguish this model from background with discovery level significance. Consequentially, the discovery reach does not appear to significantly improve with the $14$ TeV luminosity upgrade. The discovery reach increases by a factor of $\sim 6$ from $14$ TeV to $100$ TeV, but in absolute terms remains small.  The exclusion reach for the compressed squark-neutralino model is more favorable in comparison. At this level of significance the background systematics are less difficult to overcome, and the limits scale much more favorably with luminosity and CM energy. For higher center-of-mass-energy, these searches specially targeted at the compressed region also become more and more important to fill in the gap in the reach of the untargeted search described in Sec.~\ref{sec:QN}. Figure \ref{fig:SQSQ_Compressed_Comparison} makes a compelling case for investing in future proton colliders which can operate at these high energies.

\pagebreak
\section{The Gluino-Squark-Neutralino Model}
\label{sec:GluinoSquarkNeutralino}
In the ``gluino-squark-neutralino model", the gluino $\widetilde{g}$ and the first and second generation squarks $\widetilde{q}$ are all allowed to be kinematically accessible. The only relevant parameters are the squark mass $m_{\widetilde{q}}$, which is taken to be universal for the first two generations, the gluino mass $m_{\widetilde{g}}$, and the neutralino mass $m_{\widetilde{\chi}^0}$. For this study we fix the neutralino mass $m_{\widetilde{\chi}^0}=1{\rm~GeV}$, which captures the relevant kinematics for $m_{\widetilde{g}},m_{\widetilde{q}}\gg m_{\widetilde{\chi}^0}$. The decay mode is chosen depending on the mass hierarchy.  The model is summarized as:
\NegSpace
\begin{center}
\renewcommand{\arraystretch}{1.6}
\setlength{\tabcolsep}{12pt}
\begin{tabular}{c|c|c}
BSM particles & production & decay \\
\hline
\rule{0pt}{5ex}
\multirow{6}{*}{$\widetilde{g},\,\widetilde{q},\,\widetilde{\chi}^0_1$} & $p\,p \rightarrow \widetilde{g}\,\,\widetilde{g}$ & \multirow{3}{*}{$\widetilde{g} \rightarrow \begin{cases}
\widetilde{q}\,\,\overline{q} & \text{for}\,\,\, m_{\widetilde{g}} > m_{\widetilde{q}} \\
q\,\overline{q}\,\widetilde{\chi}^0_1 & \text{for}\,\,\, m_{\widetilde{g}} \simeq m_{\widetilde{q}} \\
q\,\overline{q}\,\widetilde{\chi}^0_1 & \text{for}\,\,\,  m_{\widetilde{g}} < m_{\widetilde{q}}
\end{cases}
$} \\
 & $p\,p \rightarrow \widetilde{g}\,\,\widetilde{q}$ & \\
 & $p\,p \rightarrow \widetilde{g}\,\,\widetilde{q}^*$ & \\
 & $p\,p \rightarrow \widetilde{q}\,\,\widetilde{q}^*$ & \multirow{3}{*}{$\widetilde{q} \rightarrow \begin{cases}
q\,\widetilde{\chi}^0_1  & \text{for}\,\,\, m_{\widetilde{g}} > m_{\widetilde{q}} \\
q\,\widetilde{\chi}^0_1 & \text{for}\,\,\, m_{\widetilde{g}} \simeq m_{\widetilde{q}} \\
q\,\widetilde{g}& \text{for}\,\,\,  m_{\widetilde{g}} < m_{\widetilde{q}}
\end{cases}
$} \\
  & $p\,p \rightarrow \widetilde{q}\,\,\widetilde{q}$ &\\
  & & 
\end{tabular}
\end{center}

For a full MSSM model, which in particular would imply a specific neutralino composition, there will in general be a non-zero branching ratio for the squark to decay to a neutralino and a quark when kinematically allowed. If the decay directly to a gluino is kinematically allowed however it will tend to dominate, and in this study for simplicity we assume that the squark is weakly coupled to the neutralino and decays to the gluino proceed with 100\% branching ratio when kinematically allowed. Likewise for $m_{\widetilde{g}} > m_{\widetilde{q}}$, the branching ratio of the gluino to 3-body versus 2-body decays depends on the masses and coupling of the squarks to the neutralino, and we take the 2-body branching ratio to be 100\% in this region of parameter space.  To capture the transition region where $m_{\widetilde{g}} \simeq m_{\widetilde{q}}$, parameter choices along the line $m_{\widetilde{g}} = m_{\widetilde{q}}$ are included; the gluino decay is taken to be 3-body and the squarks are assumed to decay directly to the neutralino.

This model is a good proxy for comparing the power of searches which rely on the traditional jets and $\MET$ style hadron collider search strategy to discriminate against background.  The final state ranges from two to four (or more) hard jets from the decay (depending on the production channel) and missing energy.   The current preliminary limits on this model using $20$ fb$^{-1}$ of $8$ TeV data are $m_{\widetilde{g}} = 1750 \text{ GeV}$ and $m_{\widetilde{q}} = 1600 \text{ GeV}$ (ATLAS \cite{ATLAS-CONF-2013-047}) assuming a massless neutralino.

We simulated matched \texttt{MadGraph} samples for $\big(\widetilde{g}\,\widetilde{g}\big),\,\big(\widetilde{q}\,\widetilde{q}^*\big),\,\big(\widetilde{q}\,\widetilde{q}\big),\,\big(\widetilde{g}\,\widetilde{q}\big),\,\big(\widetilde{g}\,\widetilde{q}^*\big)$ production with up to 2 additional generator level jets for the following points in parameter space:  
\begin{center}
\renewcommand{\arraystretch}{1.3}
\setlength{\tabcolsep}{12pt}
\begin{tabular}{c|l}
BSM particles & masses \\
\hline
\hline
\multirow{2}{*}{$m_{\widetilde{g}}$ $\big[14 \TeV\big]$}  & $(315, 397, 500, 629, 792, 997, 1255, 1580, 1989, 2489, 2989, $ \\
				&  $ \,3489, 3989, 4489) \GeV$ \\
\multirow{2}{*}{$m_{\widetilde{g}}$ $\big[33 \TeV\big]$}  & $(1000, 1259, 1585, 1995, 2512, 3162, 3981, 4981, 5981, 6981, $ \\
				&  $7981, 8981, 9981  \,) \GeV$ \\	
\multirow{2}{*}{$m_{\widetilde{g}}$ $\big[100 \TeV\big]$}  & $(1000, 1259, 1585, 1995, 2512, 3162, 3981, 5012, 6310, 7944,$ \\
				&  $9944, 11944, 13944, 15944, 17944, 19944, 21944, 23944  \,) \GeV$ \\								
\hline
\multirow{2}{*}{$m_{\widetilde{q}}$ $\big[14 \TeV\big]$}  & $(315, 397, 500, 629, 792, 997, 1255, 1580, 1989, 2489, 2989, $ \\
				&  $ \,3489, 3989, 4489) \GeV$ \\
\multirow{2}{*}{$m_{\widetilde{q}}$ $\big[33 \TeV\big]$}  & $(1000, 1259, 1585, 1995, 2512, 3162, 3981, 4981, 5981, 6981,$ \\
				&  $7981, 8981, 9981  \,) \GeV$ \\	
\multirow{2}{*}{$m_{\widetilde{q}}$ $\big[100 \TeV\big]$}  & $(1000, 1259, 1585, 1995, 2512, 3162, 3981, 5012, 6310, 7944, $ \\
				&  $9944, 11944, 13944, 15944, 17944, 19944, 21944, 23944 \,) \GeV$ \\					
\hline
$m_{\widetilde{\chi}^0_1}$ & $1 \GeV$
\end{tabular}
\end{center}

The signatures of this model are essentially a mixture of the gluino-neutralino and squark-neutralino Simplified Models except for slight variations in the kinematics due to the presence of on-shell states in the decays.   Therefore, the dominant backgrounds will be the same as described in Sec.~\ref{sec:GOGO_Backgrounds}, and we use the same search strategy described in detail in Sec.~\ref{sec:GOGO_Strategy}. Again, based on the results of studying the effect of pile-up on this search strategy in Sec.~\ref{sec:pileupGOGO}, we present results only for the no pile-up scenario and expect that that pileup will not have a significant impact on the results.

When both the gluino and squarks are kinematically accessible, the total cross section for this Simplified Model is significantly enhanced with respect to the limit where either particle is decoupled due to the presence of the associated production channel $\widetilde{g}\,\widetilde{q}$ and also due to $t$-channel diagrams which open the important $\widetilde{q}\,\widetilde{q}$ channel and tend to dominate the $\widetilde{q}\,\widetilde{q}^*$ cross sections.    It is important to note that even when the squarks or gluinos are kinematically inaccessible these t-channel processes still can dominate the cross section. For this reason the limits we obtain within the scanned range of gluino and squark masses do not reach the asymptotic values that can be inferred from the gluino-neutralino and squark-neutralino Simplified Models.

\begin{figure}[h!]
  \begin{center}
    \includegraphics[width=0.45\textwidth]{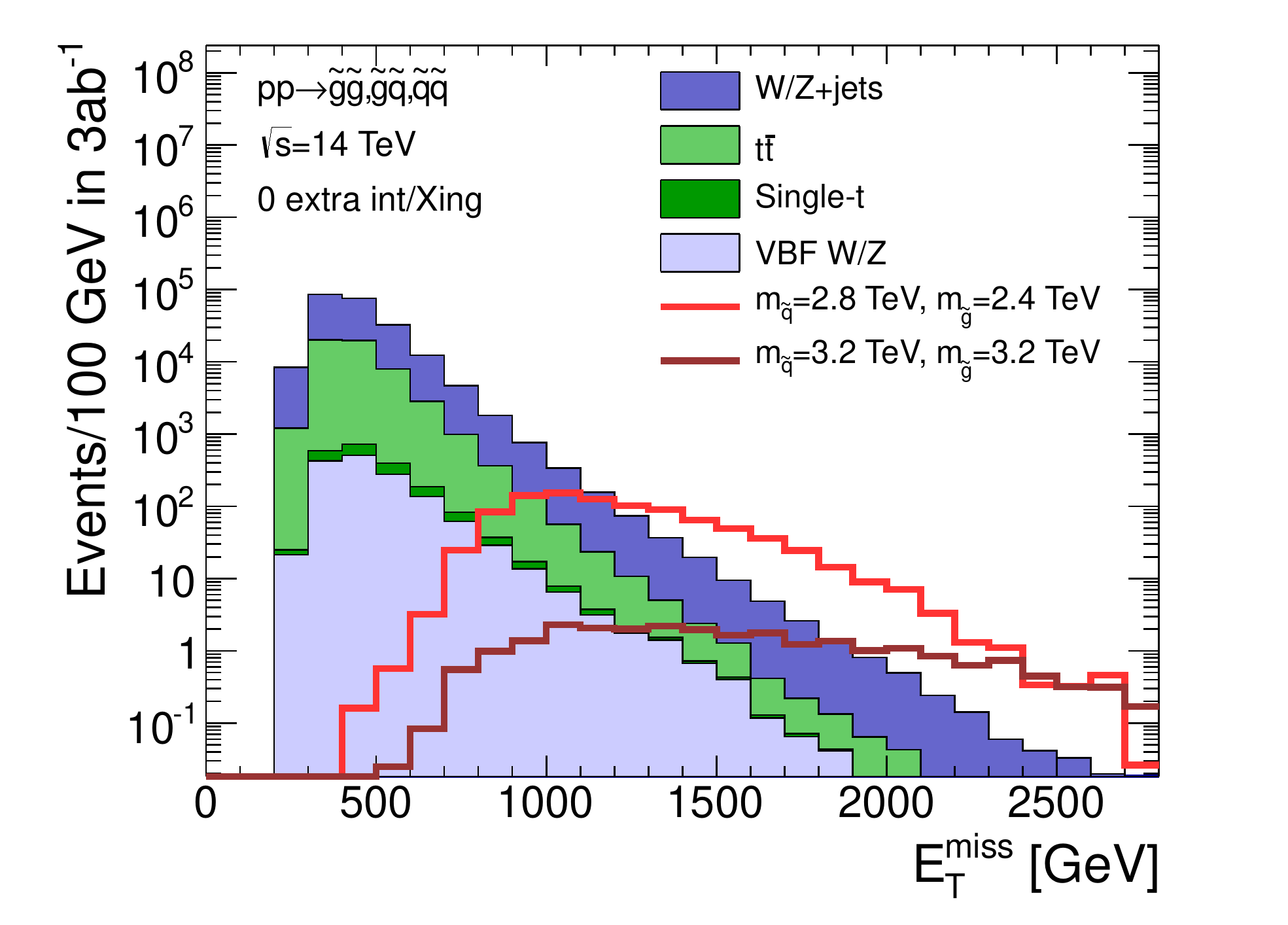}
    \includegraphics[width=0.45\textwidth]{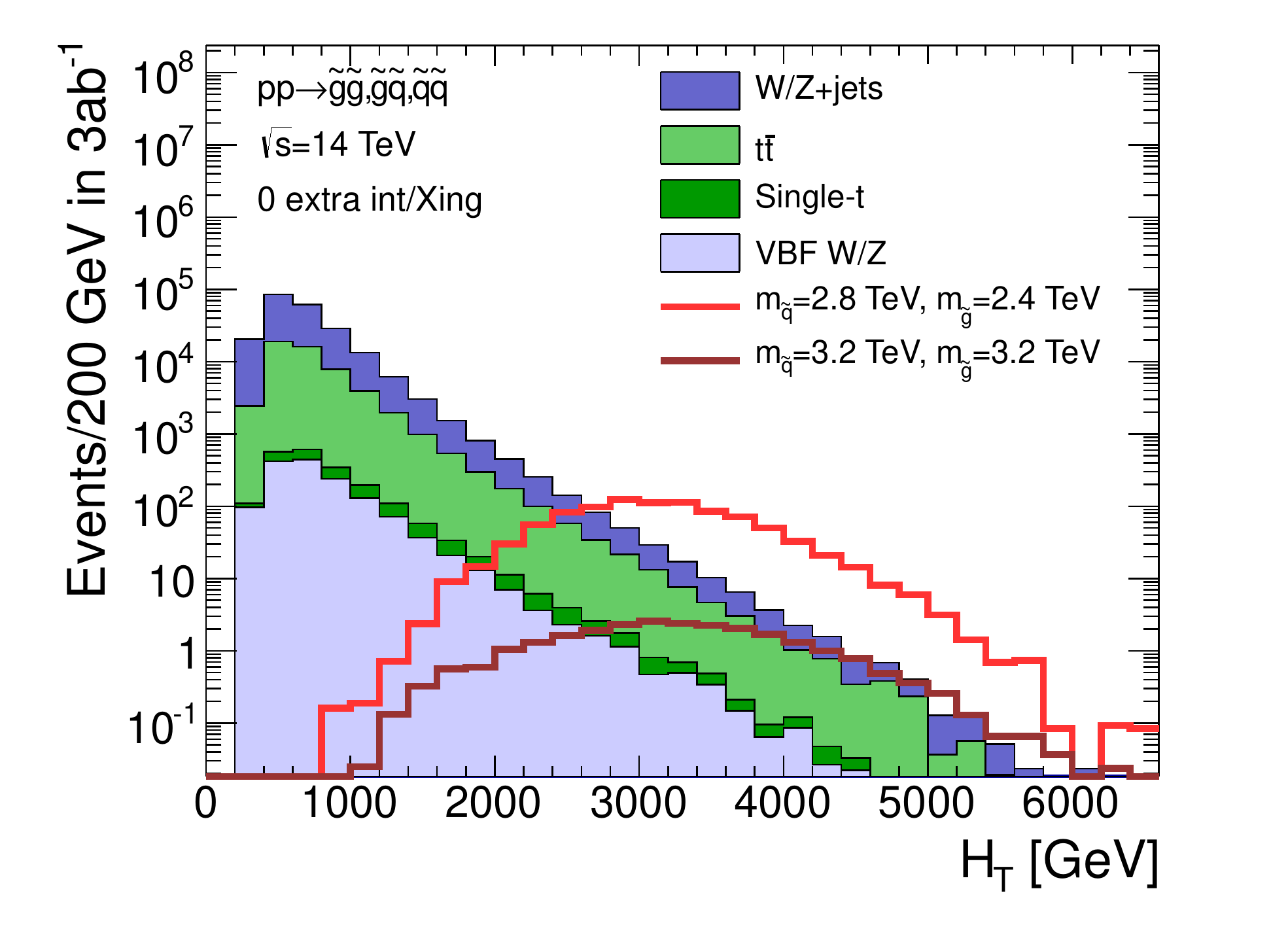}
    \caption{
    Histogram of \MET~[left] and \HT~[right] after preselection cuts for background and a range of gluino-squark models at a $14$ TeV proton collider.  The neutralino mass is $1\GeV$ for all signal models.}
    \label{fig:alljet_presel_distributions_GOSQ}
  \end{center}
\end{figure}

\hiddensubsection{Analysis: 14 TeV}
Figure~\ref{fig:alljet_presel_distributions_GOSQ} gives histograms for $\MET$ [left] and $\HT$ [right] for two example models at the $14$ TeV LHC.  Comparing with the analogous gluino-neutralino (Fig.~\ref{fig:GOGO_alljet_presel_distributions}) and squark-neutralino (Fig.~\ref{fig:alljet_presel_distributions_SQSQB}) distributions, it is clear that the total BSM cross section in this model is enhanced.  This will lead to a significant improvement in mass reach with respect to the previous results.  

The number of events that result from the cut flow used in this search are shown in Table~\ref{tab:GSbulkCounts}.  By comparing the optimal cuts which result from this analysis to the to the cuts employed for the gluino-neutralino model in Table~\ref{tab:GNbulkCounts} and the squark-neutralino model in Table~\ref{tab:SNbulkCounts}, it is clear that the optimization procedure can take advantage of the the larger cross sections by utilizing significantly harder cuts.

\begin{table}[h!]
\renewcommand{\arraystretch}{1.4}
\setlength{\tabcolsep}{6pt}
\footnotesize
\vskip 10pt
  \begin{centering}
    \begin{tabular}{| r | r r | r | r  r |}
\hline
&&&&\multicolumn{2}{c|}{$\left(m_{\widetilde{g}},m_{\widetilde{q}}\right)\,\,\,\,\,\,\,\,\,$ [TeV]}\\
                                               Cut&                 $V$+jets&               $t\bar{t}$&                 Total BG&                 $  (2.4,2.8)$&                   $ (3.2,3.2)$\\
\hline
\hline
$                             \text{Preselection}$&{$2.07 \times 10^{7}$}&{$2.47 \times 10^{7}$}&{$4.54 \times 10^{7}$}&$2.75 \times 10^{3}$&                      136\\
$\MET/\sqrt{\HT} > 15 \text{ GeV}^{1/2}$&{$4.45 \times 10^{5}$}&{$1.20 \times 10^{5}$}&{$5.65 \times 10^{5}$}&$1.34 \times 10^{3}$&                      109\\
$p_T^{\text{leading}} < 0.4\times \HT$&{$1.69 \times 10^{5}$}&{$5.16 \times 10^{4}$}&{$2.21 \times 10^{5}$}&                      937&                     25.4\\
\hline
\hline
$\MET > 1250$ GeV&\multirow{2}{*}{                     10.8}&\multirow{2}{*}{                        5}&\multirow{2}{*}{                     15.8}&\multirow{2}{*}{              \color{red} 264}&\multirow{2}{*}{                     12.8}\\
$\HT > 3000$ GeV          &&&&&\\
\hline
$\MET > 1850$ GeV&\multirow{2}{*}{                        1}&\multirow{2}{*}{                      0.2}&\multirow{2}{*}{                      1.2}&\multirow{2}{*}{                     30.5}&\multirow{2}{*}{              \color{red}  6.1}\\
$\HT > 2850$ GeV          &&&&&\\
\hline
    \end{tabular}
    \caption{Number of expected events for $\sqrt{s} = 14$ TeV with 3000 fb$^{-1}$ integrated luminosity for the background processes and selected gluino and squark masses for the gluino-squark model with light flavor decays.  The neutralino mass is $1 \GeV$.  Two choices of cuts on \MET~and \HT~are provided, and for each  mass column the entry in the row corresponding to the ``optimal" cuts is marked in red.
    }
    \label{tab:GSbulkCounts}
  \end{centering}
\end{table}

\begin{figure}[h!]
  \centering
  \includegraphics[width=.48\columnwidth]{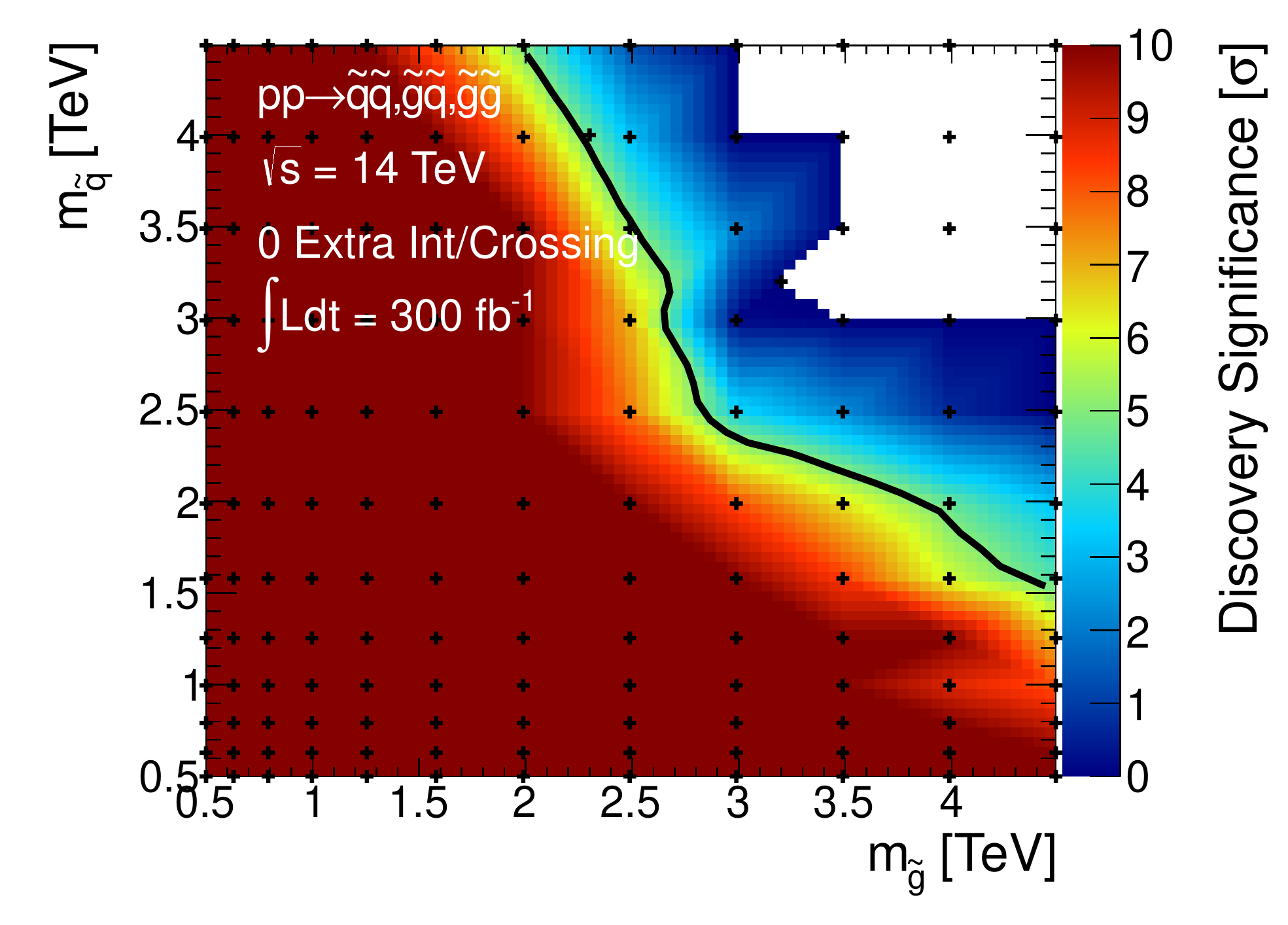}
  \includegraphics[width=.48\columnwidth]{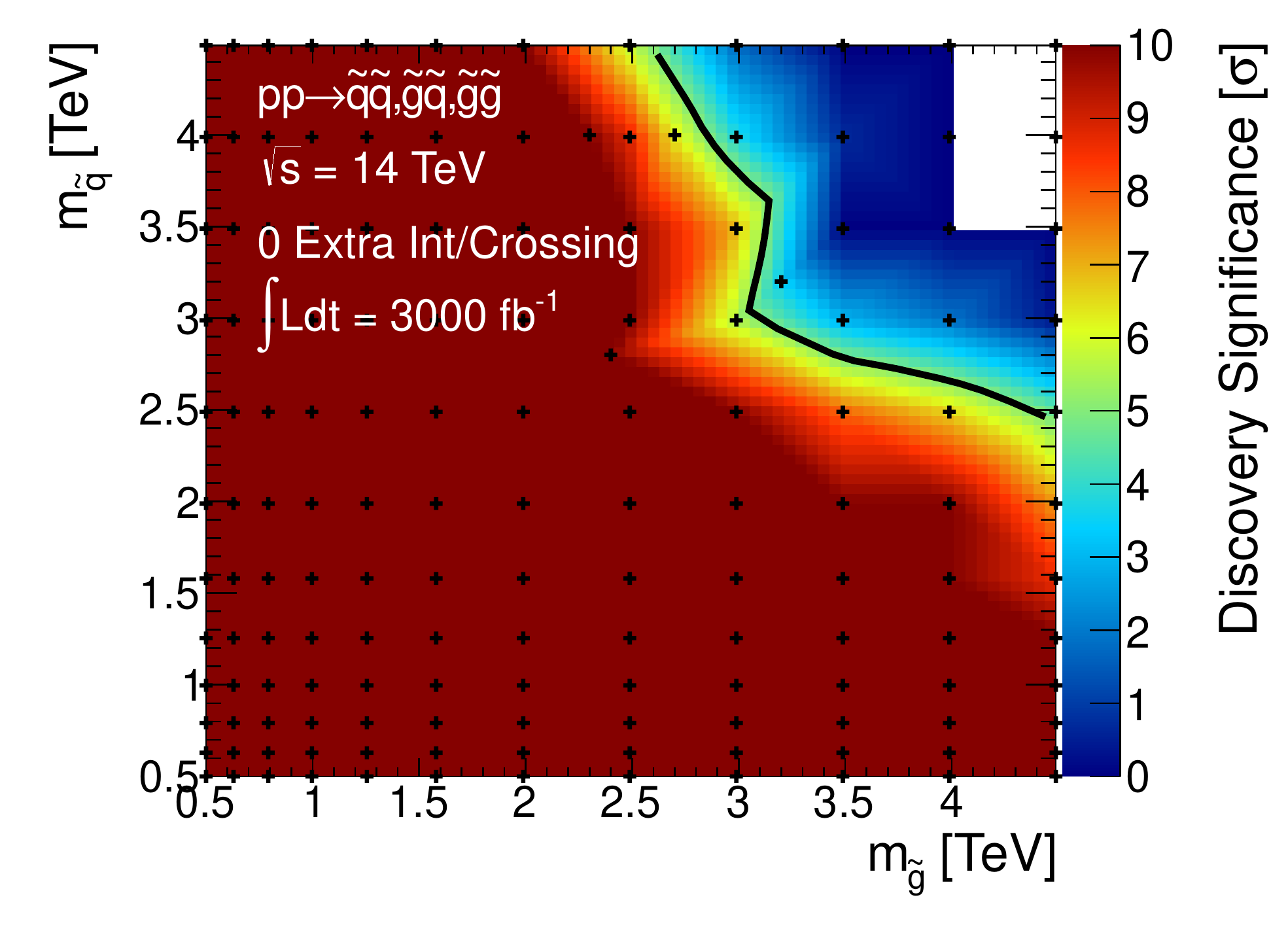}
  \includegraphics[width=.48\columnwidth]{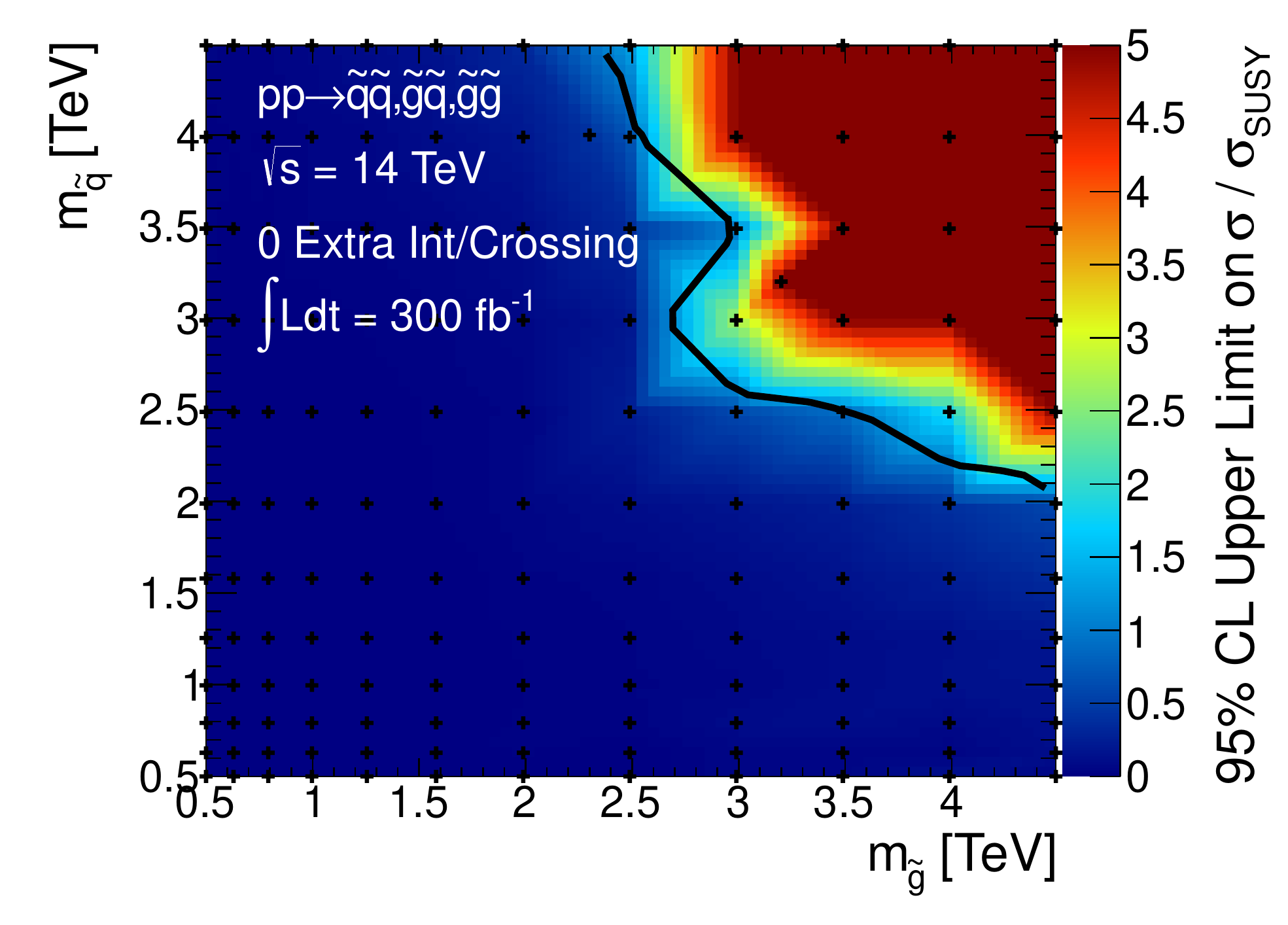}
  \includegraphics[width=.48\columnwidth]{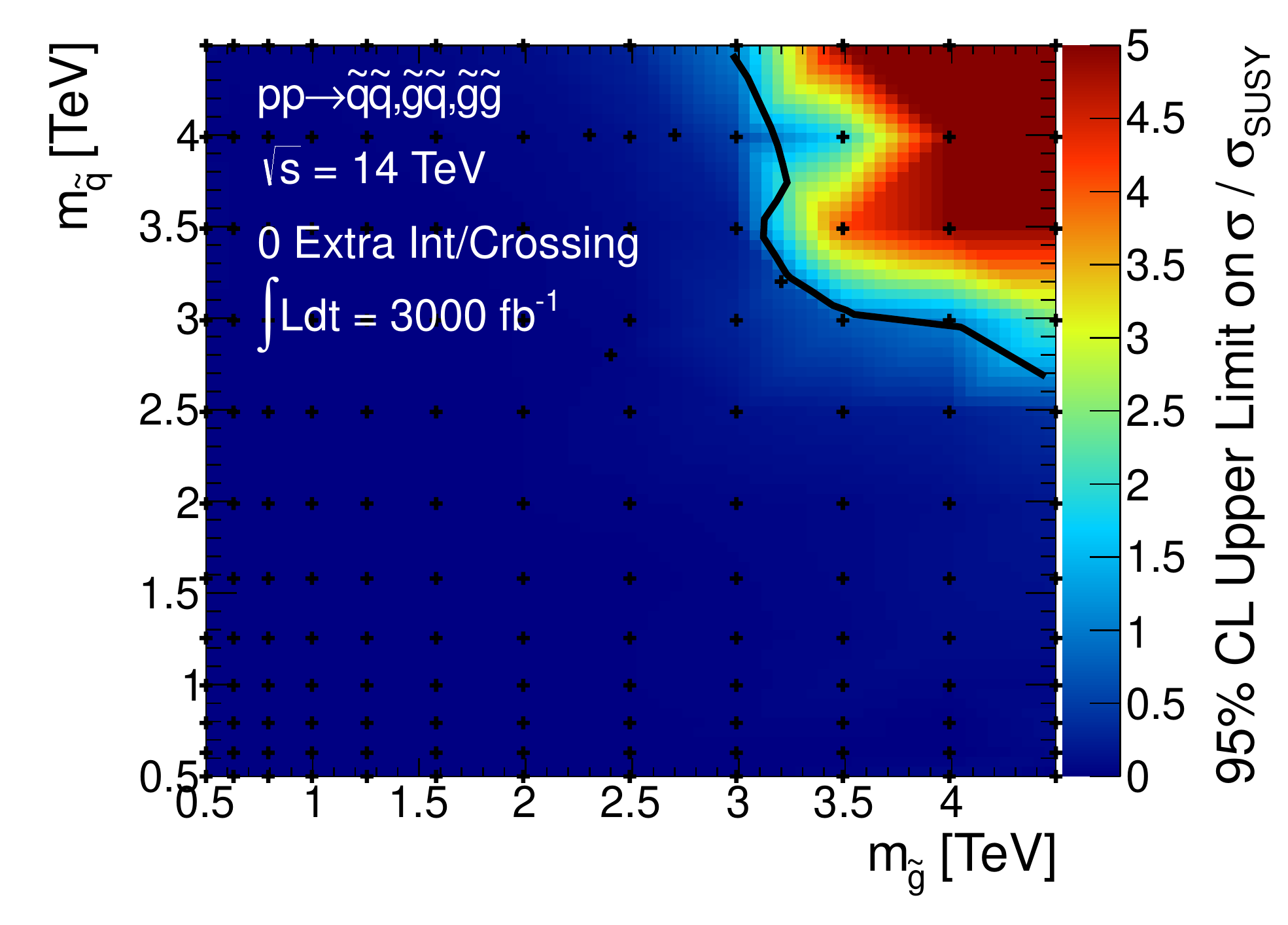}
  \caption{Results for the gluino-squark model with a massless neutralino are given in the $m_{\widetilde{g}}$ versus $m_{\widetilde{q}}$ plane.  The top [bottom] row shows the expected $5\sigma$ discovery reach [95\% confidence level upper limits] for the combined production channels at the $14$ TeV LHC.  Mass points to the left/below the contours are expected to be probed at 300 fb$^{-1}$ [left] and 3000 fb$^{-1}$ [right].  A 20\% systematic uncertainty is assumed and pileup is not included.}
\label{fig:GOSQ_14_NoPileUp_results}
\end{figure}

\pagebreak
\hiddensubsection{Results: 14 TeV}
The results for the gluino-squark-neutralino model are shown in Fig.~\ref{fig:GOSQ_14_NoPileUp_results}.  In the bulk of the parameter space, the $14 \TeV$ LHC with $3000 \text{ fb}^{-1}$ could discover a model with $m_{\widetilde{g}} \simeq 3 \TeV$ and $m_{\widetilde{q}} \simeq 3 \TeV$.  When compared to the maximal reach for the gluino-neutralino model of $m_{\widetilde{g}} \simeq 2.3 \TeV$ as shown in Fig.~\ref{fig:GOGO_14_NoPileUp_results}, the large cross section for the additional channels explains this $\sim30\%$ improvement. 

Using the NLO gluino pair production cross section one can make a very naive estimate for the reach of a given collider.  We found the choice of gluino and squark masses which would yield a fixed number of events at $14 \TeV$ and $300 \text{ fb}^{-1}$ $\big(3000 \text{ fb}^{-1}\big)$ for three choices in the $m_{\widetilde{q}}-m_{\widetilde{g}}$ plane: there would be $10$ events when $m_{\widetilde{g}} = m_{\widetilde{q}} = 3.5 \TeV\, (3.9 \TeV)$; when $m_{\widetilde{g}} = 2.7 \TeV\,(3.2 \TeV)$ and the squark mass is at the edge of the region simulated; and when $m_{\widetilde{q}} = 3.3 \TeV\,(3.9 \TeV)$ and the gluino mass is at the edge of the region simulated.  This roughly corresponds to the maximal possible reach one could expect for a given luminosity using $14 \TeV$ proton collisions.  

Using a realistic simulation framework along with the search strategy employed here the $14 \TeV$ $300 \text{ fb}^{-1}$ limits are projected to be $m_{\widetilde{g}} = m_{\widetilde{q}} = 2.8$ (corresponding to 155 events); $m_{\widetilde{g}} = 2.4 \TeV$ (corresponding to 43 events) and the squark mass is at the edge of the region simulated;  $m_{\widetilde{q}} = 2.1 \TeV$ (corresponding to 774 events) and the gluino mass is at the edge of the region simulated.  The $14 \TeV$ $3000 \text{ fb}^{-1}$ limits are projected to be $m_{\widetilde{g}} = m_{\widetilde{q}} = 3.2$ (corresponding to 293 events); $m_{\widetilde{g}} = 3.0 \TeV$ (corresponding to 23 events) and the squark mass is at the edge of the region simulated;  $m_{\widetilde{q}} = 2.7 \TeV$ (corresponding to 953 events) and the gluino mass is at the edge of the region simulated.  Clearly the search does better with light gluinos.  This is likely related to the four jet preselection requirement.  It would be investigating to understand what additional search regions could be used to push the mass reach even further; this is beyond the scope of this work.

\hiddensubsection{Analysis: 33 TeV}

\begin{figure}[h!]
  \begin{center}
    \includegraphics[width=0.45\textwidth]{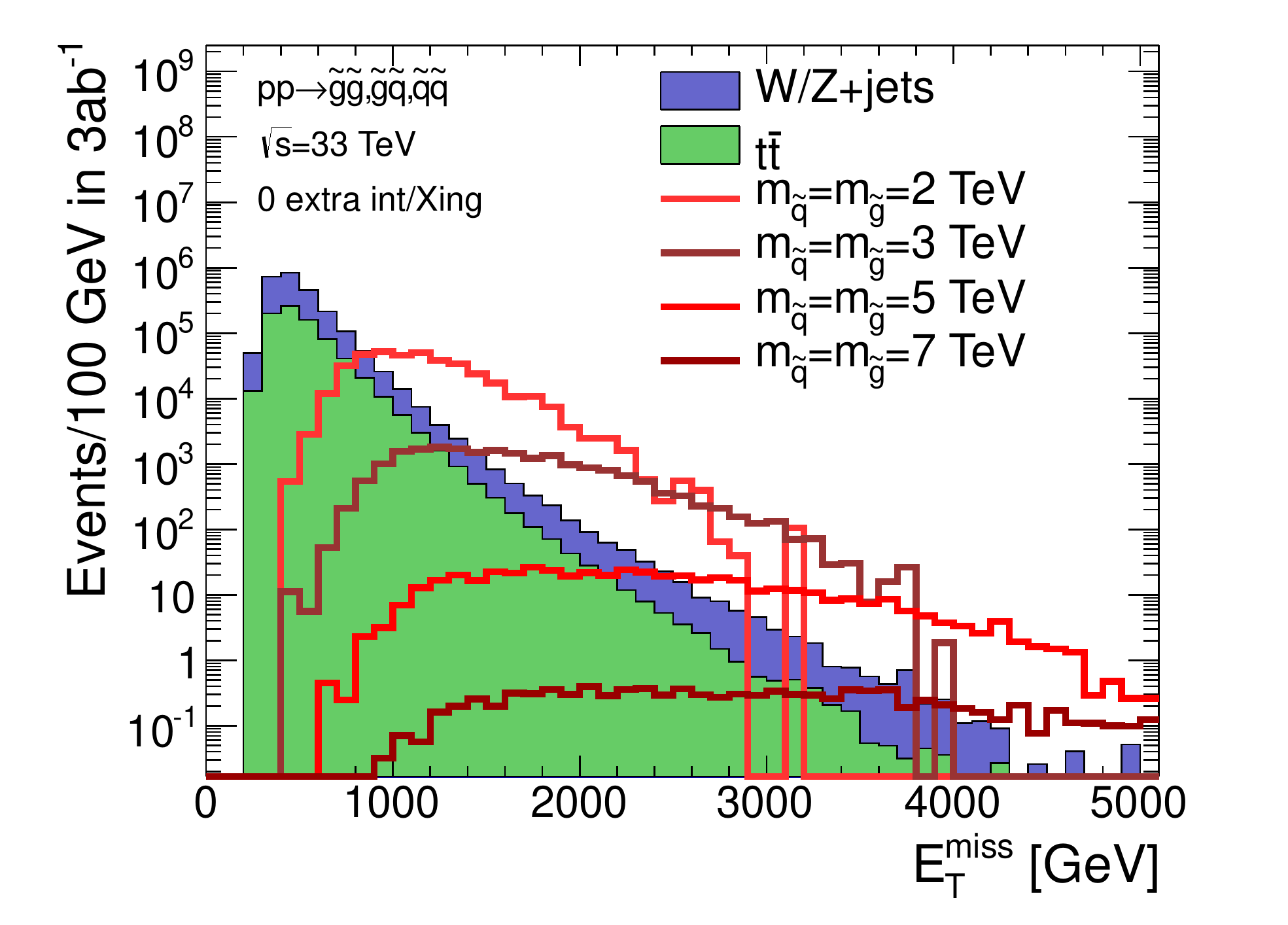}
    \includegraphics[width=0.45\textwidth]{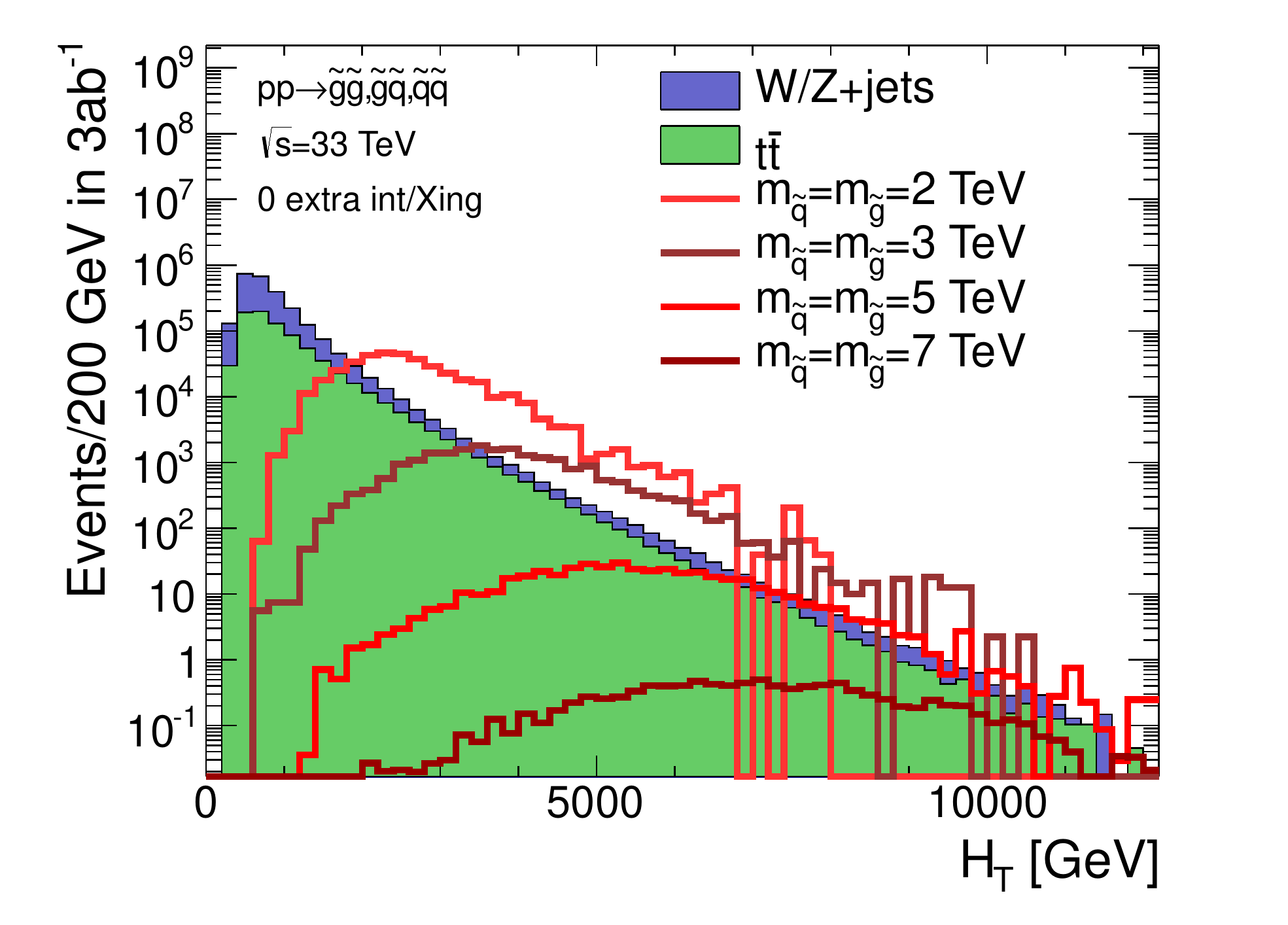}
    \caption{
    Histogram of \MET~[left] and \HT~[right] after preselection cuts for background and a range of gluino-squark models at a $33$ TeV proton collider.  The neutralino mass is $1\GeV$ for all signal models.}
    \label{fig:alljet_presel_distributions_GOSQ_33TeV}
  \end{center}
\end{figure}

Figure~\ref{fig:alljet_presel_distributions_GOSQ_33TeV} gives histograms for $\MET$ [left] and $\HT$ [right] for two example models at the $33$ TeV LHC.  Comparing with the analogous gluino-neutralino (Fig.~\ref{fig:GOGO_alljet_presel_distributions_33TeV}) and squark-neutralino (Fig.~\ref{fig:alljet_presel_distributions_SQSQB_33TeV}) distributions, it is clear that the total BSM cross section in this model is enhanced.  This will lead to a significant improvement in mass reach with respect to the previous results.  

The number of events that result from the cut flow used in this search are shown in Table~\ref{tab:GSbulkCounts_33TeV}.  By comparing the optimal cuts which result from this analysis to the cuts employed for the gluino-neutralino model in Table~\ref{tab:GNbulkCounts33TeV} and the squark-neutralino model in Table~\ref{tab:SNbulkCounts_33TeV}, it is clear that the optimization procedure can take advantage of the the larger cross sections by utilizing significantly harder cuts.

\begin{table}[h!]
\renewcommand{\arraystretch}{1.4}
\setlength{\tabcolsep}{6pt}
\footnotesize
\vskip 10pt
  \begin{centering}
    \begin{tabular}{| r | r r | r | r  r  |}
\hline
&&&&\multicolumn{2}{c|}{$\left(m_{\widetilde{g}},m_{\widetilde{q}}\right)\,\,\,\,\,\,\,\,\,$ [TeV]}\\
                                               Cut&                 $V$+jets&               $t\overline{t}$&                 Total BG&                 $( 3.2, 3.2)$&                 $( 6.0,6.0)$                \\
\hline
$                             \mathrm{Preselection}$&{$1.55 \times 10^{8}$}&{$2.86 \times 10^{8}$}&{$4.42 \times 10^{8}$}&$8.64 \times 10^{4}$&                      $361$ \\
\hline
\hline
$\MET/\sqrt{\HT} > 15 \GeV^{1/2}$&{$4.50 \times 10^{6}$}&{$1.93 \times 10^{6}$}&{$6.44 \times 10^{6}$}&$6.55 \times 10^{4}$&                      $321$\\
$p_{\mathrm{T}}^{\mathrm{leading}} < 0.4\HT$&{$1.70 \times 10^{6}$}&{$8.02 \times 10^{5}$}&{$2.50 \times 10^{6}$}&$2.14 \times 10^{4}$&                     $69.8$ \\
\hline
\hline
$         \MET > 2650 \GeV$&\multirow{2}{*}{                     $60.4$}&\multirow{2}{*}{                     $16.4$}&\multirow{2}{*}{                     $76.8$}&\multirow{2}{*}{\color{red} $1.79 \times 10^{3}$}&\multirow{2}{*}{                     40.1}\\
$                         \HT > 2700 \GeV$&&&&&\\
\hline
$         \MET > 3700 \GeV$&\multirow{2}{*}{                      $1.1$}&\multirow{2}{*}{                      $0.3$}&\multirow{2}{*}{                      $1.3$}&\multirow{2}{*}{                     $30.4$}&\multirow{2}{*}{                   \color{red}  $14.2$}\\
$                         \HT > 5350 \GeV$&&&&&\\
\hline
    \end{tabular}
    \caption{Number of expected events for $\sqrt{s} = 33$ TeV with 3000 fb$^{-1}$ integrated luminosity for the background processes and selected gluino and squark masses for the gluino-squark model with light flavor decays.  The neutralino mass is $1 \GeV$.  Three choices of cuts on \MET~and \HT~are provided, and for each  mass column the entry in the row corresponding to the ``optimal" cuts is marked in red.
    }
    \label{tab:GSbulkCounts_33TeV}
  \end{centering}
\end{table}

\begin{figure}[h!]
  \centering
  \includegraphics[width=.48\columnwidth]{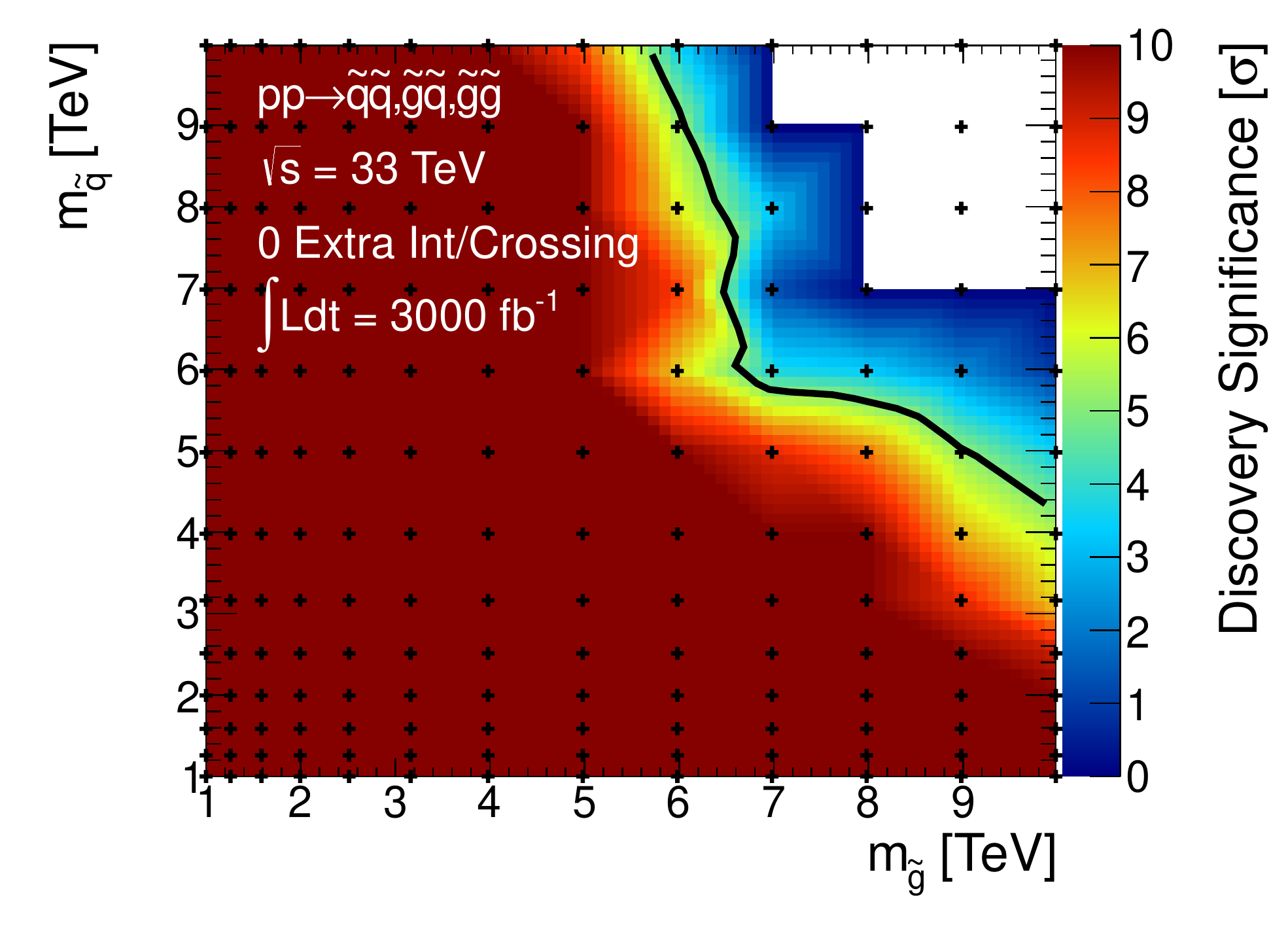}
  \includegraphics[width=.48\columnwidth]{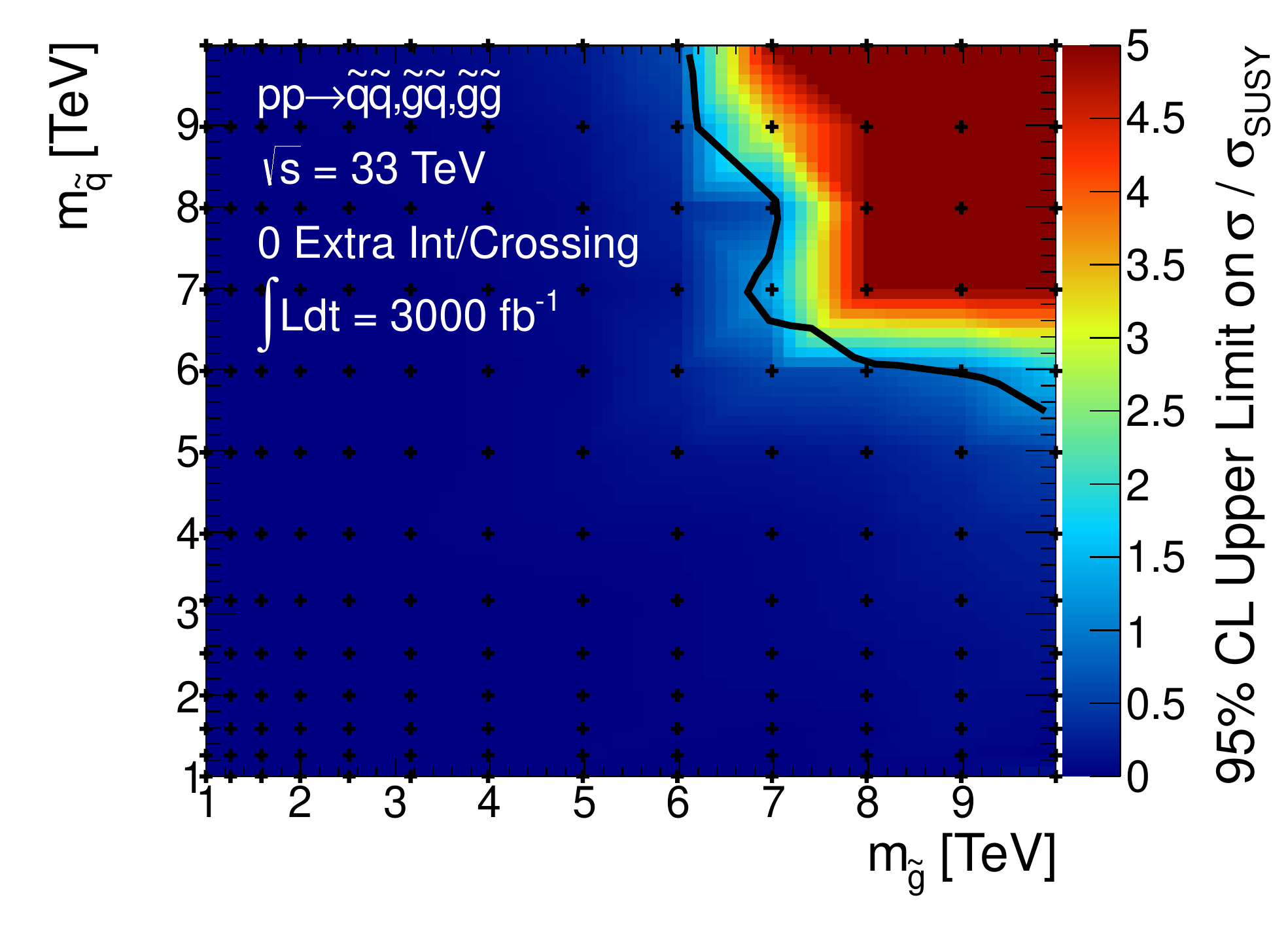}
  \caption{Results for the gluino-squark model with a massless neutralino are given in the $m_{\widetilde{g}}$ versus $m_{\widetilde{q}}$ plane.  The left [right] panel shows the expected $5\sigma$ discovery reach [95\% confidence level upper limits] for the combined production channels at a $100$ TeV proton collider.  Mass points to the left/below the contours are expected to be probed with 3000 fb$^{-1}$ of data.  A 20\% systematic uncertainty is assumed and pileup is not included.}
\label{fig:GOSQ_33_NoPileUp_results}
\end{figure}

\hiddensubsection{Results: 33 TeV}
The results for the gluino-squark-neutralino model are shown in Fig.~\ref{fig:GOSQ_33_NoPileUp_results}.  In the bulk of the parameter space, a $33 \TeV$ proton collider with $3000 \text{ fb}^{-1}$ could discover a model with $m_{\widetilde{g}} \simeq 6.5 \TeV$ and $m_{\widetilde{q}} \simeq 6 \TeV$.  When compared to the maximal reach for the gluino-neutralino model of $m_{\widetilde{g}} \simeq 4.8 \TeV$ as shown in Fig.~\ref{fig:GOGO_33_NoPileUp_results}, the large cross section for the additional channels explains this $\sim30\%$ improvement.  

Using the NLO gluino pair production cross section one can make a very naive estimate for the reach of a given collider.  We found the choice of gluino and squark masses which would yield a fixed number of events at $33 \TeV$ and $3000 \text{ fb}^{-1}$ for three choices in the $m_{\widetilde{q}}-m_{\widetilde{g}}$ plane: there would be $10$ events when $m_{\widetilde{g}} = m_{\widetilde{q}} = 8.2 \TeV$; when $m_{\widetilde{g}} = 6.1 \TeV$ and the squark mass is at the edge of the region simulated; and when $m_{\widetilde{q}} = 5.5 \TeV$ and the gluino mass is at the edge of the region simulated.  This roughly corresponds to the maximal possible reach one could expect for a given luminosity using $33 \TeV$ proton collisions.  

Using a realistic simulation framework along with the search strategy employed here the $33 \TeV$ $3000 \text{ fb}^{-1}$ limits are projected to be $m_{\widetilde{g}} = m_{\widetilde{q}} = 6.8$ (corresponding to 132 events); $m_{\widetilde{g}} = 6.1 \TeV$ (corresponding to 21 events) and the squark mass is at the edge of the region simulated;  $m_{\widetilde{q}} = 5.5 \TeV$ (corresponding to 473 events) and the gluino mass is at the edge of the region simulated.  Clearly the search does better with light gluinos.  Furthermore, we find that we are closer to the ideal limit than in the $14 \TeV$ case.  Both of these facts are likely related to the four jet preselection requirement.  While investigating the reach that could be extracted using additional search regions is beyond the scope of this work, it would be interesting to understand what it takes to push the mass reach even further.

\hiddensubsection{Analysis: 100 TeV}

\begin{figure}[tbp]
  \begin{center}
    \includegraphics[width=0.45\textwidth]{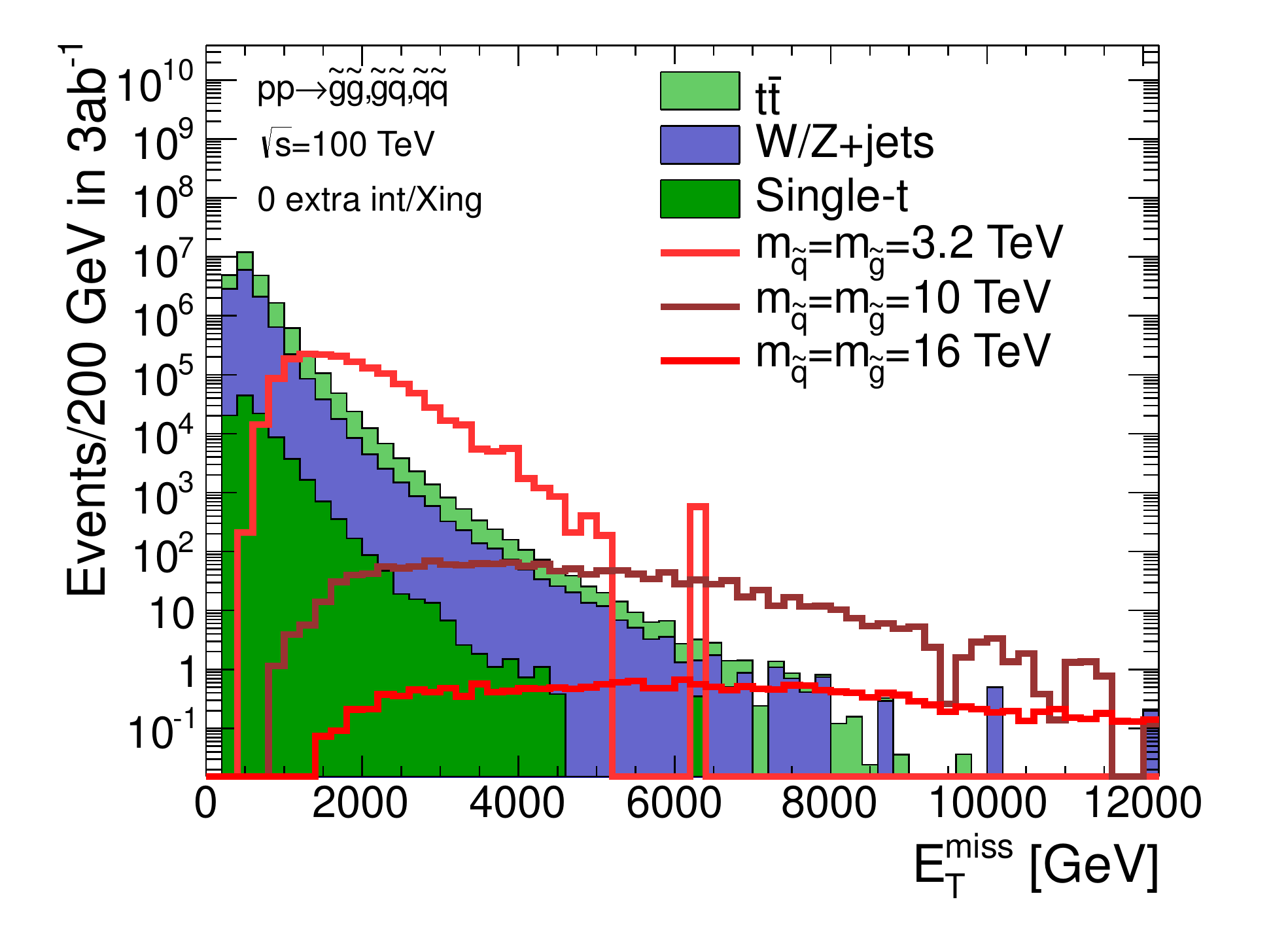}
    \includegraphics[width=0.45\textwidth]{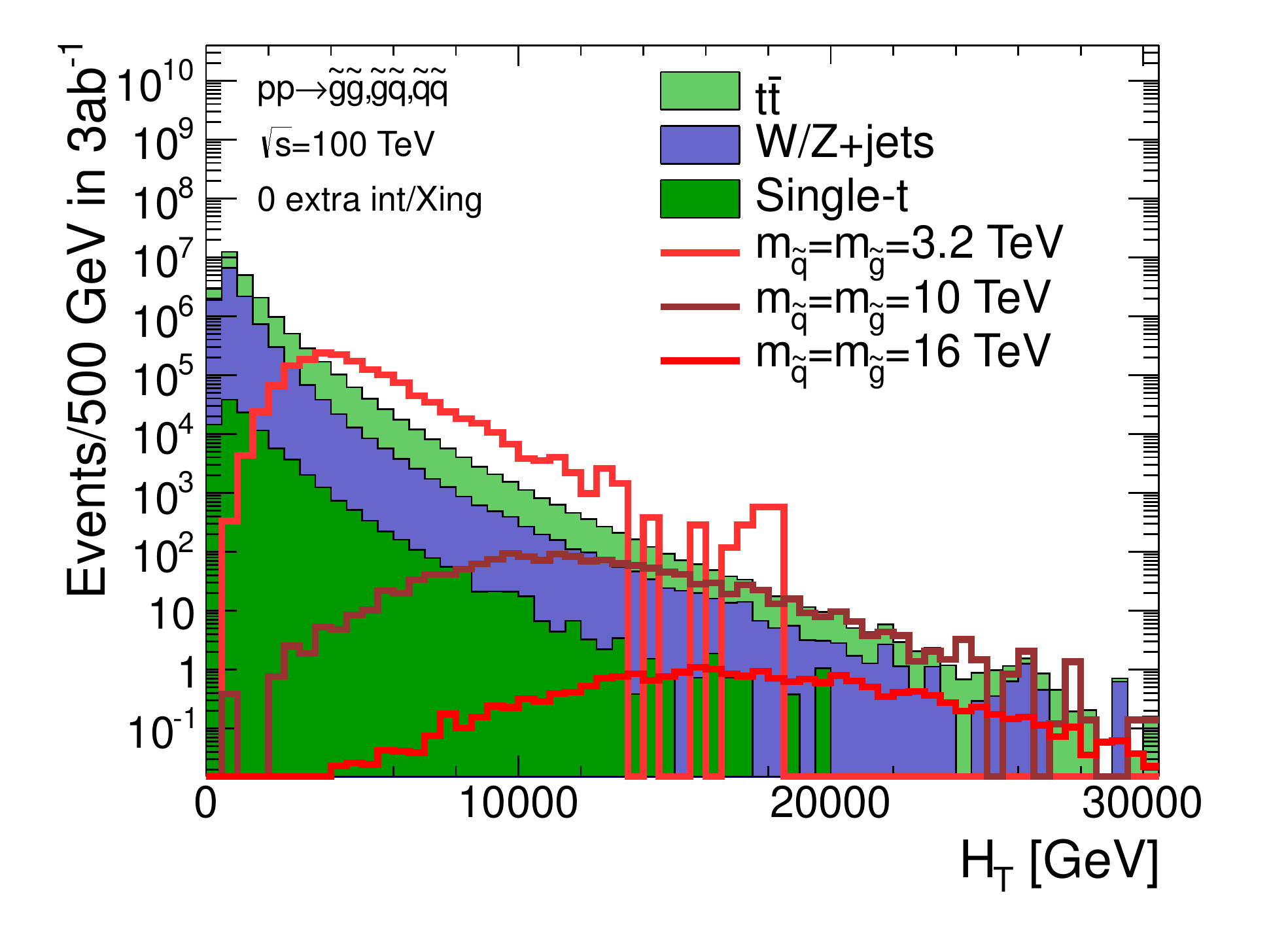}
    \caption{
    Histogram of \MET~[left] and \HT~[right] after preselection cuts for background and a range of gluino-squark models at a $100$ TeV proton collider.  The neutralino mass is $1\GeV$ for all signal models.}
    \label{fig:alljet_presel_distributions_GOSQ_100TeV}
  \end{center}
\end{figure}

Figure~\ref{fig:alljet_presel_distributions_GOSQ_100TeV} gives histograms for $\MET$ [left] and $\HT$ [right] for two example models at a $100$ TeV proton collider.  Comparing with the analogous gluino-neutralino (Fig.~\ref{fig:GOGO_alljet_presel_distributions_100TeV}) and squark-neutralino (Fig.~\ref{fig:alljet_presel_distributions_SQSQB_100TeV}) distributions, it is clear that the total BSM cross section in this model is enhanced.  This will lead to a significant improvement in mass reach with respect to the previous results.  

The number of events that result from the cut flow used in this search are shown in Table~\ref{tab:GSbulkCounts_100TeV}. Comparing the optimal cuts which result for this model to the cuts employed for the gluino-neutralino model in Table~\ref{tab:GNbulkCounts100TeV} and the squark-neutralino model in Table~\ref{tab:SNbulkCounts_100TeV}, it is clear that the optimization procedure can take advantage of the the larger cross sections by utilizing significantly harder cuts.

\begin{table}[tbp]
\renewcommand{\arraystretch}{1.4}
\setlength{\tabcolsep}{6pt}
\footnotesize
\vskip 10pt
  \begin{centering}
    \begin{tabular}{| r | r r | r | r  r |}
\hline
&&&&\multicolumn{2}{c|}{$\left(m_{\widetilde{g}},m_{\widetilde{q}}\right)\,\,\,\,\,\,\,\,\,$ [TeV]}\\
                                               Cut&                 $V$+jets&               $t\overline{t}$&                 Total BG&                 
$(8, 8)$&                    $(16,16)$\\
\hline
\hline
$                             \mathrm{Preselection}$&{$1.64 \times 10^{9}$}&{$3.33 \times 10^{9}$}&{$4.97 \times 10^{9}$}&$2.69 \times 10^{4}$&                      111\\
\hline
$\MET/\sqrt{\HT} > 15 \GeV^{1/2}$&{$3.59 \times 10^{7}$}&{$3.31 \times 10^{7}$}&{$6.90 \times 10^{7}$}&$2.41 \times 10^{4}$&                      $107$\\
$p_T^{\mathrm{leading}} < 0.4\times \HT $&{$1.19 \times 10^{7}$}&{$1.25 \times 10^{7}$}&{$2.44 \times 10^{7}$}&$7.34 \times 10^{3}$&                     $20.5$\\
\hline
\hline
$         \MET > 5700 \GeV$&\multirow{2}{*}{                       $13$}&\multirow{2}{*}{                     $14.9$}&\multirow{2}{*}{                    $ 27.8$}&\multirow{2}{*}{                 \color{red}    $ 771$}&\multirow{2}{*}{                      $ 12$}\\
$                         \HT > 8000 \GeV$&&&&&\\
\hline
$        \MET > 5800 \GeV$&\multirow{2}{*}{                      $0.4$}&\multirow{2}{*}{                      $2.5$}&\multirow{2}{*}{                     $ 2.9$}&\multirow{2}{*}{                     $41.7$}&\multirow{2}{*}{                 \color{red}     $5.3$}\\
$                        \HT > 17800 \GeV$&&&&&\\                                               
\hline
    \end{tabular}
    \caption{Number of expected events for $\sqrt{s} = 100$ TeV with 3000 fb$^{-1}$ integrated luminosity for the background processes and selected gluino and squark masses for the gluino-squark model with light flavor decays.  The neutralino mass is $1 \GeV$.  Two choices of cuts on \MET~and \HT~are provided, and for each  mass column the entry in the row corresponding to the ``optimal" cuts is marked in red.
    }
    \label{tab:GSbulkCounts_100TeV}
  \end{centering}
\end{table}

\begin{figure}[h!]
  \centering
  \includegraphics[width=.48\columnwidth]{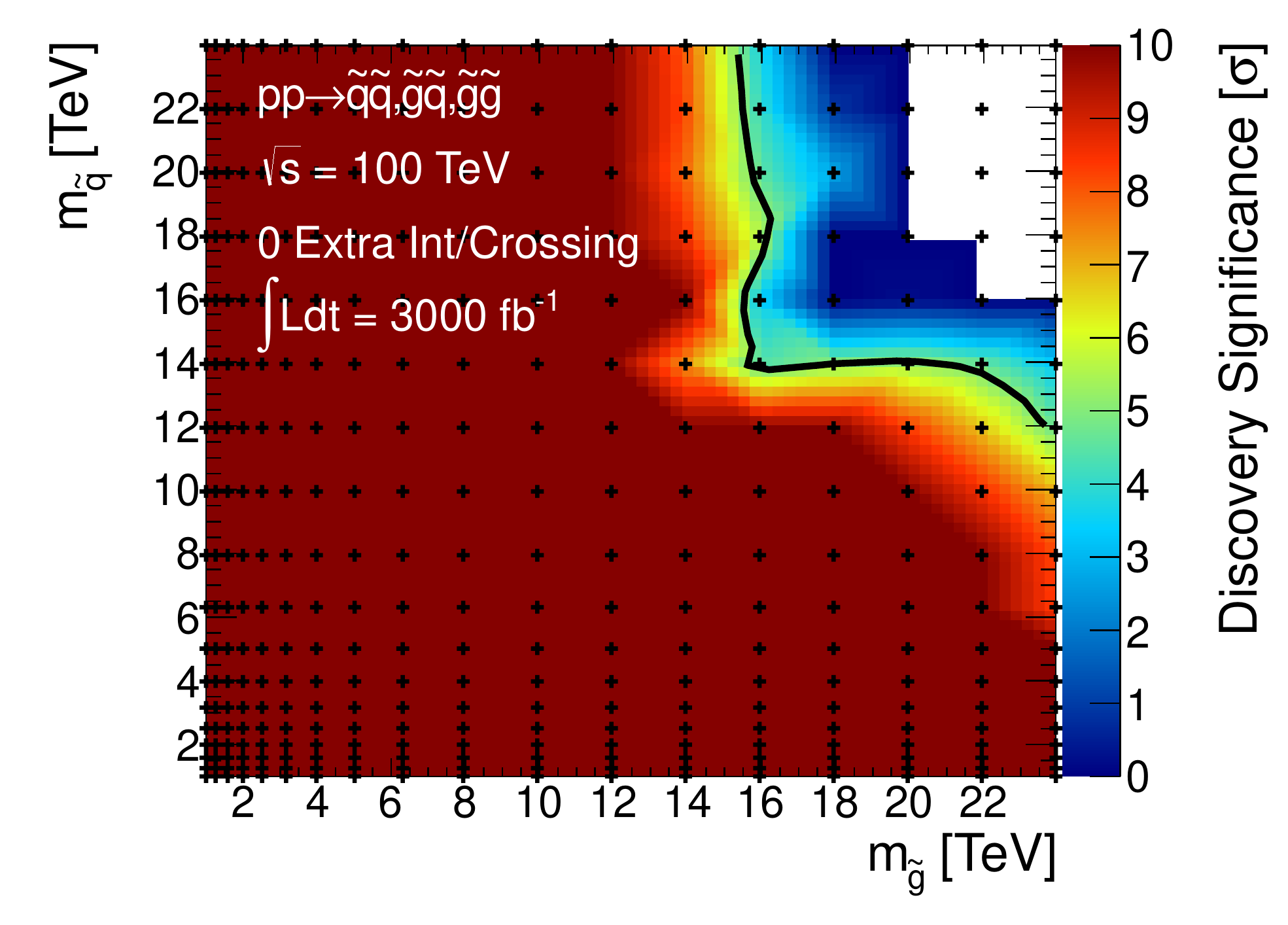}
  \includegraphics[width=.48\columnwidth]{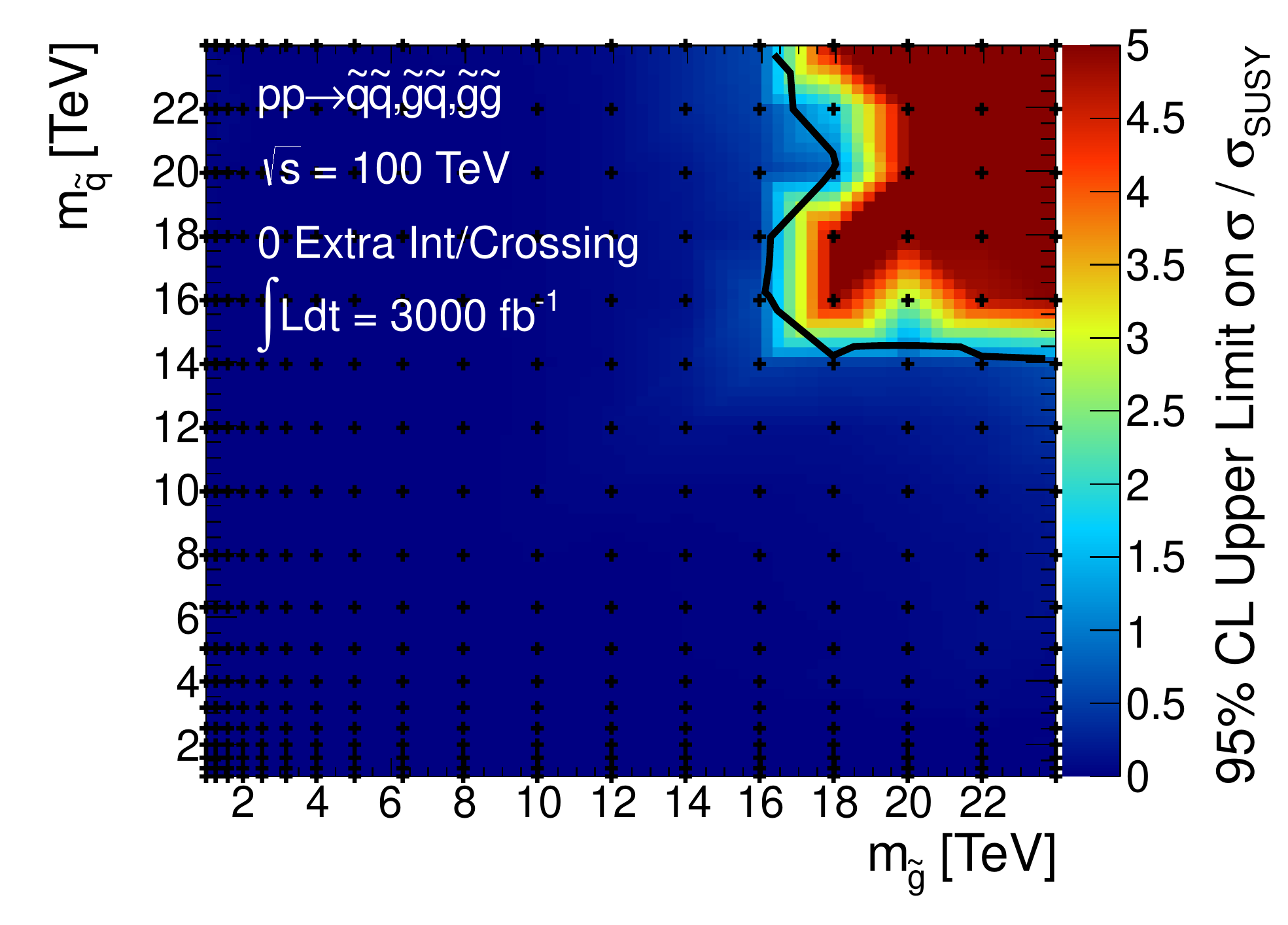}
  \caption{Results for the gluino-squark model with a massless neutralino are given in the $m_{\widetilde{g}}$ versus $m_{\widetilde{q}}$ plane.  The left [right] panel shows the expected $5\sigma$ discovery reach [95\% confidence level upper limits] for the combined production channels at a $100$ TeV proton collider.  Mass points to the left/below the contours are expected to be probed with 3000 fb$^{-1}$ of data.  A 20\% systematic uncertainty is assumed and pileup is not included.}
\label{fig:GOSQ_100_NoPileUp_results}
\end{figure}

\pagebreak
\hiddensubsection{Results: 100 TeV}
The results for the gluino-squark-neutralino model are shown in Fig.~\ref{fig:GOSQ_100_NoPileUp_results}.  In the bulk of the parameter space, a $100 \TeV$ proton collider with $3000 \text{ fb}^{-1}$ could discover a model with $m_{\widetilde{g}} \simeq 16 \TeV$ and $m_{\widetilde{q}} \simeq 14 \TeV$.  When compared to the maximal reach for the gluino-neutralino model of $m_{\widetilde{g}} \simeq 11 \TeV$ as shown in Fig.~\ref{fig:GOGO_100_NoPileUp_results}, the large cross section for the additional channels explains this $\sim30\%$ improvement.  

Using the NLO gluino pair production cross section one can make a very naive estimate for the reach of a given collider.  The choice of gluino and squark masses which would yield a fixed number of events at $100 \TeV$ and $3000 \text{ fb}^{-1}$ for three choices in the $m_{\widetilde{q}}-m_{\widetilde{g}}$ plane: there would be $10$ events when $m_{\widetilde{g}} = m_{\widetilde{q}} = 20.4 \TeV$; when $m_{\widetilde{g}} = 16.5 \TeV$ and the squark mass is at the edge of the region simulated; and when $m_{\widetilde{q}} = 19.6 \TeV$ and the gluino mass is at the edge of the region simulated.  This roughly corresponds to the maximal possible reach one could expect for a given luminosity using $100 \TeV$ proton collisions.  

Using a realistic simulation framework along with the search strategy employed here the $100 \TeV$ $3000 \text{ fb}^{-1}$ limits are projected to be $m_{\widetilde{g}} = m_{\widetilde{q}} = 16$ (corresponding to 136 events); $m_{\widetilde{g}} = 16 \TeV$ (corresponding to 13 events) and the squark mass is at the edge of the region simulated;  $m_{\widetilde{q}} = 14 \TeV$ (corresponding to 169 events) and the gluino mass is at the edge of the region simulated.  Clearly the search does better with light gluinos. Furthermore, we find that we are closer to the ideal limit than in the $14 \TeV$ and $33 \TeV$ cases.  Both of these facts are likely related to the four jet preselection requirement.  While investigating the reach that could be extracted using additional search regions is beyond the scope of this work, it would be interesting to understand what it takes to push the mass reach even further.

The next section provides a comparison of the impact that the four collider scenarios studied here can have on the parameter space of this model.

\begin{figure}[h!]
  \centering
  \includegraphics[width=.48\columnwidth]{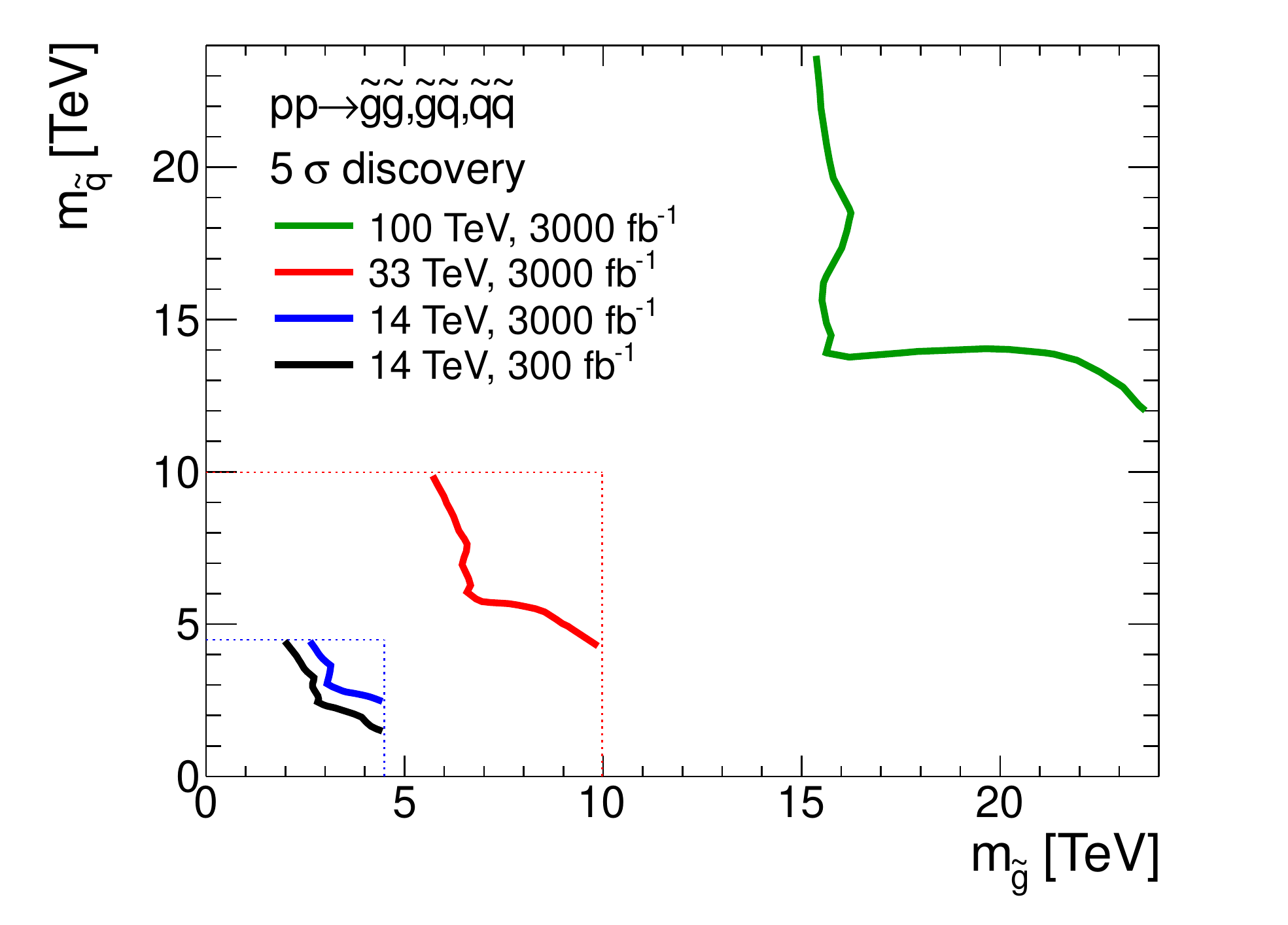}
  \includegraphics[width=.48\columnwidth]{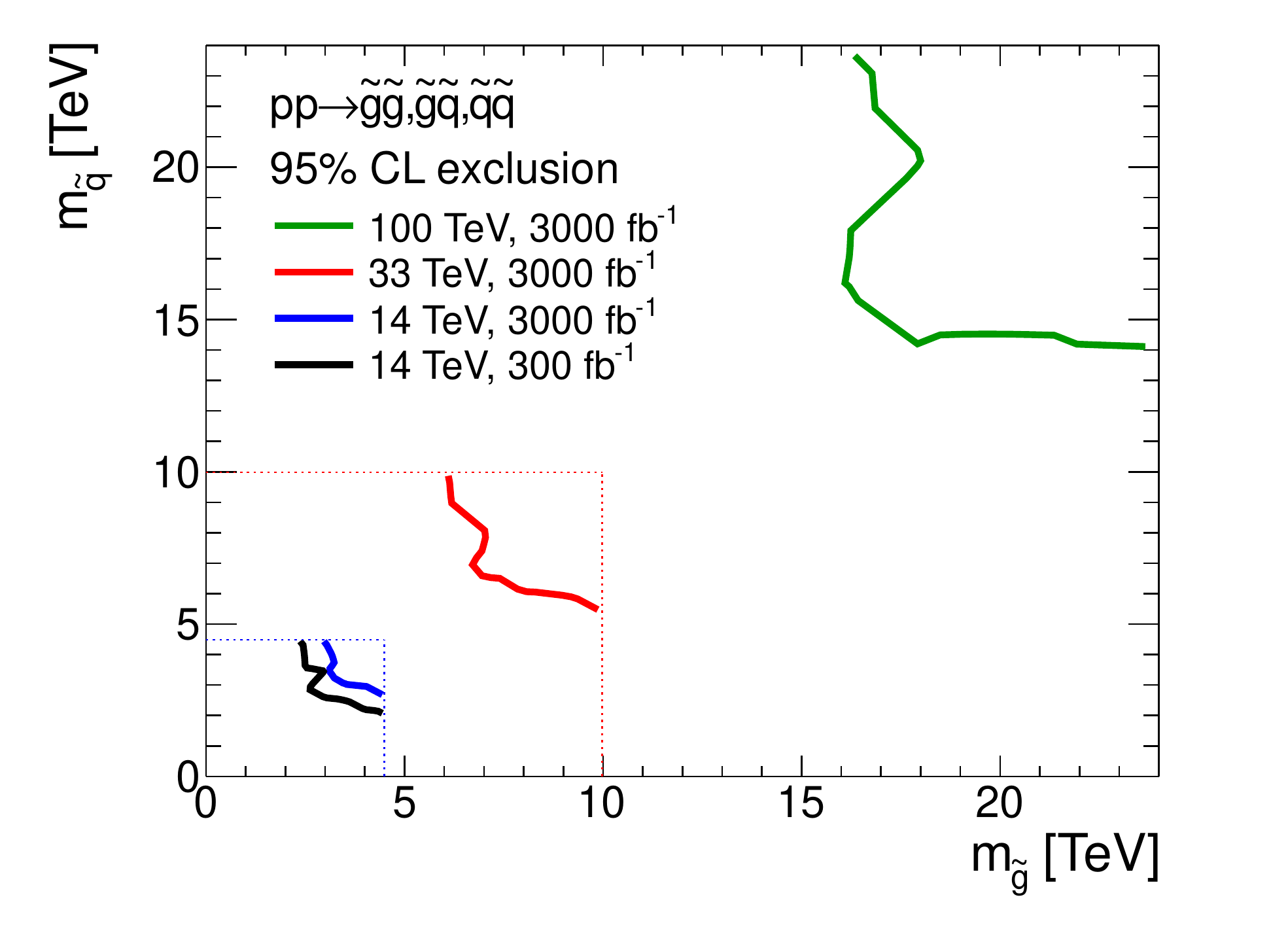}
  \caption{Results for the gluino-squark-neutralino model.  The neutralino mass is taken to be $1 \GeV$.  The left [right] panel shows the $5\,\sigma$ discovery reach [$95\%$ CL exclusion] for the four collider scenarios studied here.  A $20\%$ systematic uncertainty is assumed and pileup is not included.}
    \label{fig:GOSQ_Comparison}
\end{figure}

\pagebreak
\hiddensubsection{Comparing Colliders}
The multi-jet plus \MET~signature of the gluino-squark-neutralino model with light flavor decays provides a useful case study with which to compare the potential impact of different proton colliders. Due to the large production cross sections, this model is also interesting as one of the most striking possible cases of accessible new physics. Figure \ref{fig:GOSQ_Comparison} shows the $5\sigma$ discovery reach [$95\%$ CL exclusion] for two choices of integrated luminosity at $14$ TeV, along with the full data set assumed for $33$ and $100$ TeV.  Because the high mass signal regions are relatively low-background for this model, a factor of $10$ increase in luminosity leads to roughly a factor of 10 increase in cross-section reach at edges of the limits. In terms of mass reach, this corresponds roughly to a respectable $500$ GeV improvement. Again increasing the center-of-mass energy has a tremendous impact on the experimentally available parameter space, since now much heavier gluinos can be produced without relying on the tails of parton distributions to supply the necessary energy.  Figure \ref{fig:GOSQ_Comparison} makes a compelling case for investing in future proton colliders which can operate at these high energies.

\pagebreak
\section{The Gluino-Neutralino Model with Heavy Flavor Decays}
\label{sec:GNHeavyFlavor}
In the ``gluino-neutralino model with heavy flavor decays", the gluino $\widetilde{g}$ is the only kinematically accessible colored particle. The squarks are completely decoupled and do not contribute to gluino production diagrams. The gluino undergoes a prompt three-body decay through off-shell stops,  $\widetilde{g} \rightarrow t\,\overline{t}\,\widetilde{\chi}^0_1$, where $t$ is the top quark and $\widetilde{\chi}^0_1$ is a neutralino LSP. The only two relevant parameters are the gluino mass $m_{\widetilde{g}}$ and the neutralino mass $m_{\widetilde{\chi}^0_1}$.  This model can be summarized by:
\begin{center}
\NegSpace
\renewcommand{\arraystretch}{1.3}
\setlength{\tabcolsep}{12pt}
\begin{tabular}{c|c|c}
BSM particles & production & decays \\
\hline
$\widetilde{g},\,\widetilde{\chi}^0_1$ & $p\,p\rightarrow \widetilde{g}\, \widetilde{g}$ & $\widetilde{g} \rightarrow t\,\overline{t}\,\widetilde{\chi}^0_1 $ 
\end{tabular}
\end{center}

This model has a variety of motivations.  Perhaps the most compelling are ``natural" SUSY scenarios \cite{Dimopoulos:1995mi, Cohen:1996vb, Papucci:2011wy, Brust:2011tb, Essig:2011qg}, where the stop mass is assumed to be below the (stronger) bounds on first and second generation squark masses; for some examples of explicit constructions, see \cite{ArkaniHamed:1997fq, Craig:2012hc, Craig:2012di, Csaki:2012fh, Larsen:2012rq, Cohen:2012rm, Randall:2012dm}.  If both the stop and gluino are kinematically accessible for a given center-of-mass energy, the gluino would be visible above background before that of the stop; this Simplified Model reproduces the first signature of this paradigm.  Note that in these models, the gluino decays involving on-shell stops.  However, the final state are identical and the kinematics are similar enough that the reach is qualitatively reproduced by the results presented below.  The current preliminary limits on this model using $20$ fb$^{-1}$ of $8$ TeV data are $m_{\widetilde{g}} = 1400 \text{ GeV}$ (ATLAS \cite{ATLAS-CONF-2013-061}) and $m_{\widetilde{g}}  = 1310 \text{ GeV}$ (CMS \cite{Chatrchyan:2013iqa}) assuming a massless neutralino.

There is also a class of split-SUSY models where the inaccessible stops are somewhat lighter than the other squarks --- this Simplified Model acts as an excellent proxy for the first signatures of these scenarios.  There are compelling reasons to believe this is a ``preferred" spectrum.  Renormalization group evolution tends to reduce the stop mass with respect to the first/second generation squarks (due to the large top Yukawa coupling) \cite{Martin:1997ns}.  Also, assuming the MSSM, avoiding flavor and/or CP violation bounds would imply that the squarks have masses $\gtrsim 1000 $ \cite{Altmannshofer:2013lfa}, while for $\tan \beta \gtrsim 2$ the stops would be lighter than $\OO(100 \TeV)$ \cite{Giudice:2011cg} in order to yield a $125 \GeV$ Higgs boson.

Finally, we note that this model is interesting from an experimental perspective.  The model produces two $t\,\overline{t}$ pairs along with considerable \MET~(away from the compressed region of parameter space), and therefore provides an interesting benchmark scenario for searches involving  a combination of hadronic activity, leptonic signatures and b-tagging.  As described in detail below, a search which requires same-sign di-leptons (SSDL) is one viable approach to eliminating the SM background since this final state is highly suppressed in the SM.  We note that this was the only channel explored in this scenario; it would be interesting to investigate how an all hadronic final state would perform at the higher energy machines.

We simulated matched \texttt{MadGraph} samples for $\widetilde{g}\,\widetilde{g}$ with up to 2 additional generator level jets for the following points in parameter space:\footnote{We include $1\GeV$ for an example where the neutralino is effectively massless; the second line of neutralino masses is chosen to cover the bulk of the gluino-neutralino plane; the final line is chosen to ensure coverage in the ``compressed" region.}  

\begin{center}
\NegSpace
\renewcommand{\arraystretch}{1.3}
\setlength{\tabcolsep}{12pt}
\begin{tabular}{c|l}
BSM particles & masses \\
\hline
\hline
$m_{\widetilde{g}}$ $\big[14 \TeV\big]$ & $ (315, 397, 500, 629, 792, 997, 1255, 1580, 1989, 2489, 2989, 3489) \GeV$   \\
$m_{\widetilde{g}}$ $\big[33 \TeV\big]$ & $ (500, 629, 792, 997, 1255, 1580, 1989, 2504, 3152, $ \\
 & $3968, 4968, 5968, 6968) \GeV $\\
$m_{\widetilde{g}}$ $\big[100 \TeV\big]$ & $ (1000, 1259, 1585, 1995, 2512, 3162, 3981, 5012, 6310,$   \\
 &  $7944, 9944, 11944, 13944, 15944) \GeV$ \\
\hline
					    & $1\GeV$ \\
$m_{\widetilde{\chi}^0_1}$  & $(0.5, 0.7, 0.9)\times (m_{\widetilde{g}} - 2m_t)$ \\
					    & $m_{\widetilde{g}} - 2m_t - 10\GeV$
\end{tabular}
\end{center}

\subsection{Dominant Backgrounds}
\label{sec:GOGO_HeavyFlavor_Backgrounds}
The analysis used to derive the results below requires an SSDL pair, which is very efficient at eliminating backgrounds. The dominant background is top pair production, where both tops decay leptonically (the di-leptonic channel).  There are subdominant backgrounds from $W\,b\,b$, which are accounted for by including the $BJ$ Snowmass particle container \cite{Avetisyan:2013onh}.  All backgrounds simulated for Snowmass are included and their rates are found to be negligible.

\subsection{Analysis Strategy}
\label{sec:GOGO_HeavyFlavor_Strategy}
The gluino-neutralino model with heavy flavor decays can be probed with an analysis that is inspired by the CMS collaboration in~\cite{Chatrchyan:2012paa}.  A SSDL pair is required and any remaining leptons are not allowed to form a $Z$-boson.  Since the SSDL requirement is very effective at suppressing backgrounds, only mild cut on \MET~is necessary to observe this model.  This implies that this search will also be very effective in the compressed regions of parameter space where $m_{\widetilde{g}} \simeq m_{\widetilde{\chi}_1^0}$.

In detail, our analysis strategy proceeds as follows:

\textbf{\textsc{Preselection}}
\NegSpace
\begin{itemize}
\item At least one SSDL pair, where the leptons are required to have $p_T > 20 \GeV$ and $|\eta| <2.5$
\item At least two $b$-tagged jets
\item The invariant mass of the SSDL pair  $>12 \GeV$ to suppress low mass resonances
\item Veto an event where a third lepton $\ell_i$ reconstructs a $Z$-boson with either of the leptons from the SSDL pair $\ell_j$: $76 \GeV < m_{\ell_i\,\ell_j} < 106 \GeV$ is vetoed, where the third lepton is required to have $p_T > 10 \GeV$ and $|\eta| < 2.5$
\item $(\HT)_\text{jets} > 80 \GeV$
\item $\MET > 50 \GeV$
\end{itemize}

\textbf{\textsc{Search Strategy}: Define 8 signal regions}

After preselection, the following are used as discriminating variables.  Eight model points, three with very low LSP mass, three with medium LSP mass, and two with high LSP mass are used to define eight signal regions, which rely on some combination of the following cuts.
\NegSpace
\begin{itemize}
\item symmetric $M_{T2} > \big(\text{symmetric } M_{T2}\big)_{\text{optimal}}$
\item $p_T > \big(p_T\big)_\text{optimal}$ for the hardest lepton
\item $\MET > \big(\MET \big)_\text{optimal}$
\item $N_\text{jets} > \big(N_\text{jets} \big)_\text{optimal}$
\item $N_{b\text{-jets}} > \big(N_{b\text{-jets}} \big)_\text{optimal}$
\item $\meff > \big(\meff \big)_\text{optimal}$
\item $(\HT)_{\text{jets}}  >  \big( (\HT)_{\text{jets}} \big)_\text{optimal}$
\end{itemize}
\NegSpace
Symmetric $M_{T2}$ is defined in the canonical way~\cite{Lester:1999tx, Barr:2003rg, Burns:2008va}, where the SSDL pair is used for the visible signal and the invisible particle test mass is assumed to be zero; \meff~is defined as the scalar sum of the $p_T$ of all visible objects and \MET.

The goal is to attempt to provide as much total coverage in the Simplified Model plane as possible.  Therefore, the cuts range from very stringent (for the light gluino mass/zero neutralino mass points) to very inclusive (for the compressed and heavy spectra).  Approximate signal regions will be defined for each center-of-mass energy below.  These provide a sense of how the cuts scale with luminosity and energy.

\hiddensubsection{Analysis: 14 TeV}
\addcontentsline{toc}{subsection}{8.3-8.9 $\,\,\,\,$ Analysis and Results Including the Impact of Pileup: $14$, $33$, and $100$ TeV}
In this section, the signal regions are are presented.  The choice of cuts depends on the assumed integrated luminosity.   Table \ref{tab:SSDLCuts14TeV300ifb} [\ref{tab:SSDLCuts14TeV3000ifb}] provides the values that are relevant for the $300$ fb$^{-1}$ $\big[3000$ fb$^{-1}\big]$ results.  Also shown in these tables are number of events from the two dominant contributions to the background along with the total number of background events after cuts (including all contributions).   When evaluating the gluino reach, we compute the signal efficiency after cuts for each signal region and use the one with the highest significance.

\topskip0pt
\vspace*{\fill}
\begin{table}[]
\renewcommand{\arraystretch}{1.55}
\setlength{\tabcolsep}{6pt}
\footnotesize
\centering
\begin{tabular}{c|c|c|c}
signal region & BSM masses & cuts & backgrounds\\
\hline
\hline
\multirow{3}{*}{SR1} &\multirow{3}{*}{\minitab[c]{ $m_{\widetilde{g}} = 997 \GeV$  \\ $m_{\widetilde{\chi}_1^0} = 1 \GeV$  }}&  \multirow{3}{*}{ \minitab[c]{ $ \MET \gtrsim 250 \GeV$ \\ $ N_{\text{jets}} > 5; \quad N_{b} > 2 $ }} & \multirow{3}{*}{\minitab[c|c]{ $t\,\overline{t}$ & 0.025   \\ di-boson & 0.37\\ \hline  total & 0.40  }}  \\
                                 &                                                      & & \\
                                 &&&\\
\hline
\multirow{3}{*}{SR2} &\multirow{3}{*}{\minitab[c]{ $m_{\widetilde{g}} = 997 \GeV$ \\ $m_{\widetilde{\chi}_1^0} = 641 \GeV$  }}& $ \MET \gtrsim 250 \GeV; \quad \HT \gtrsim 700 \GeV $ &  \multirow{3}{*}{\minitab[c|c]{ $t\,\overline{t}$ &  0.025  \\ di-boson & 0.37 \\ \hline  total & 0.40  }} \\
                                 &                                                      & $ \meff \gtrsim 1000 \GeV $ & \\                                 
                                &                                                       & $ N_{\text{jets}} > 5; \quad N_b > 2  $ & \\
                               
\hline
\multirow{3}{*}{SR3} &\multirow{3}{*}{\minitab[c]{ $m_{\widetilde{g}} = 1580 \GeV$ \\ $m_{\widetilde{\chi}_1^0} = 1 \GeV$  }}& $ \MET \gtrsim 300 \GeV; \quad \HT \gtrsim 800 \GeV $ &  \multirow{3}{*}{\minitab[c|c]{ $t\,\overline{t}$ &  0.020  \\ di-boson & 0.0064 \\ \hline  total & 0.031  }}  \\
                                 &                                                      & $ \meff \gtrsim 1500 \GeV $ & \\  
                                 &                                                      & $  N_{\text{jets}} > 5; \quad N_b > 2 $ & \\
\hline
\multirow{3}{*}{SR4} &\multirow{3}{*}{\minitab[c]{ $m_{\widetilde{g}} = 1580 \GeV$  \\ $m_{\widetilde{\chi}_1^0} = 1224 \GeV$  }}&  \multirow{3}{*}{ \minitab[c]{$ \MET \gtrsim 600 \GeV; \quad \meff \gtrsim 1500 \GeV  $ \\ $  N_{\text{jets}} > 5; \quad N_b > 2  $ }}&  \multirow{3}{*}{\minitab[c|c]{di-boson &  0.0064  \\ tri-boson & 0.0003 \\ \hline  total & 0.0067  }}  \\
                                 &                                                      & & \\
                                 &&&\\
\hline
\multirow{3}{*}{SR5} &\multirow{3}{*}{\minitab[c]{ $m_{\widetilde{g}} = 2489 \GeV$  \\ $m_{\widetilde{\chi}_1^0} = 1 \GeV$  }}& \multirow{3}{*}{ \minitab[c]{$ \HT \gtrsim 2000 \TeV; \quad  \meff \gtrsim 2500 \GeV  $ \\ $  N_{\text{jets}} > 7 $}} &  \multirow{3}{*}{\minitab[c|c]{ $t\,\overline{t}$ &  0.0072  \\ di-boson & 0.013 \\ \hline  total & 0.022  }} \\
                                 &                                                      & & \\
                                 &&&\\
\hline
\multirow{3}{*}{SR6} & \multirow{3}{*}{\minitab[c]{$m_{\widetilde{g}} = 2489 \GeV$ \\  $m_{\widetilde{\chi}_1^0} = 2133 \GeV$}} &  \multirow{3}{*}{$N_{\text{jets}} > 7; \quad N_b > 2 $} &  \multirow{3}{*}{\minitab[c|c]{$t\,W$ & 0.96  \\ di-boson & 0.77 \\ \hline  total &  2.1 }}  \\
& & & \\
 &&&\\
\hline
\multirow{2}{*}{SR7} &\multirow{2}{*}{\minitab[c]{ $m_{\widetilde{g}} = 3489 \GeV$  \\ $m_{\widetilde{\chi}_1^0} = 1 \GeV$  }}& $ \text{ sym-}M_{T2} \gtrsim 400 \GeV $ &  \multirow{2}{*}{\minitab[c|c]{$t\,\overline{t}$ & 0.021  \\  \hline  total &  0.021 }}\\
                                &                                                       & $ N_{\text{jets}} > 7  $ & \\
\hline                    
\multirow{3}{*}{SR8} & \multirow{3}{*}{\minitab{$m_{\widetilde{g}} = 3489 \GeV$ \\ $m_{\widetilde{\chi}_1^0} = 3133 \GeV$}} &  \multirow{3}{*}{$ N_{\text{jets}} > 7  $} & \multirow{3}{*}{\minitab[c|c]{$t\,\overline{t}$ & 0.39  \\ di-boson & 1.1 \\ \hline  total &  1.8 }} \\
&  & \\
&&&\\
\hline            
\end{tabular}
\caption{The eight signal regions defined for the SSDL $14$ TeV $300 \text{ fb}^{-1}$ search.  Also shown are the dominant two contributions to the background along with the total background after cuts, which also includes all subdominant backgrounds.}
\label{tab:SSDLCuts14TeV300ifb}
\end{table}
\vspace*{\fill}

\topskip0pt
\vspace*{\fill}
\begin{table}[]
\renewcommand{\arraystretch}{1.55}
\setlength{\tabcolsep}{6pt}
\footnotesize
\centering
\begin{tabular}{c|c|c|c}
signal region & BSM masses & cuts & backgrounds\\
\hline
\hline
\multirow{3}{*}{SR1} &\multirow{3}{*}{\minitab[c]{ $m_{\widetilde{g}} = 997 \GeV$  \\ $m_{\widetilde{\chi}_1^0} = 1 \GeV$  }}& $ \MET \gtrsim 250 \GeV; \quad \HT \gtrsim 500 \GeV $ &  \multirow{3}{*}{\minitab[c|c]{$t\,\overline{t}$ & 0.061  \\ $t\,W$ & 0.019 \\ \hline  total &  0.086 }}  \\
                                 &                                                      & $ \meff \gtrsim 1000 \GeV; \quad \text{sym-}M_{T2} \gtrsim 60 \GeV $ &\\      
                                 &                                                      & $ N_{\text{jets}} > 5;\quad N_{b} > 2 $ &\\ 
\hline
\multirow{3}{*}{SR2} &\multirow{3}{*}{\minitab[c]{ $m_{\widetilde{g}} = 997 \GeV$  \\ $m_{\widetilde{\chi}_1^0} = 641 \GeV$  }}&  \multirow{3}{*}{ \minitab[c]{ $ \MET \gtrsim 150 \GeV; \quad \text{sym-}M_{T2} \gtrsim 60 \GeV $ \\ $ N_{\text{jets}} > 4; \quad N_{b} > 2 $ }}& \multirow{3}{*}{\minitab[c|c]{$t\,\overline{t}$ & 0.87  \\ di-boson & 0.15 \\ \hline  total &  1.04 }}  \\    
                                 &                                                      & &  \\
                                 &&& \\
\hline
\multirow{3}{*}{SR3} & \multirow{3}{*}{\minitab[c]{ $m_{\widetilde{g}} = 1580 \GeV$  \\ $m_{\widetilde{\chi}_1^0} = 1 \GeV$  }}& $ \MET \gtrsim 250 \GeV; \quad  \HT \gtrsim 700 \GeV $ & \multirow{3}{*}{\minitab[c|c]{$t\,\overline{t}$ & 0.061  \\ tri-boson & 0.0053 \\ \hline  total &  0.069 }}   \\
                                 &                                                      & $ \meff \gtrsim 1500 \GeV; \quad \text{sym-}M_{T2} \gtrsim 50 \GeV $ & \\  
                                 &                                                      & $  N_{\text{jets}} > 5; \quad N_b > 2 $ & \\
\hline
\multirow{3}{*}{SR4} &\multirow{3}{*}{\minitab[c]{ $m_{\widetilde{g}} = 1580 \GeV$  \\ $m_{\widetilde{\chi}_1^0} = 1224 \GeV$  }}& $ \MET \gtrsim 150 \GeV; \quad  \meff \gtrsim 500 \GeV  $ & \multirow{3}{*}{\minitab[c|c]{$t\,\overline{t}$ & 0.061  \\ $t\,W$ & 0.019 \\ \hline  total &  0.088 }}   \\
                                 &                                                      & $ \text{sym-}M_{T2} \gtrsim 60 \GeV $ & \\                                   
                                 &                                                      & $  N_{\text{jets}} > 5; \quad  N_b > 2 $ & \\
\hline
\multirow{3}{*}{SR5} &\multirow{3}{*}{\minitab[c]{ $m_{\widetilde{g}} = 2489 \GeV$  \\ $m_{\widetilde{\chi}_1^0} = 1 \GeV$  }}& $ \MET \gtrsim 250 \GeV; \quad \HT \gtrsim 1000 \GeV $ & \multirow{3}{*}{\minitab[c|c]{$t\,W$ & 0.013 \\ \hline  total &  0.013 }}   \\
                                 &                                                      & $ \meff \gtrsim 2000 \GeV; \quad \text{sym-}M_{T2} \gtrsim 180 \GeV  $ & \\
                                &                                                       & $  N_{\text{jets}} > 5; \quad N_b > 2 $ & \\                     
\hline
\multirow{3}{*}{SR6} &\multirow{3}{*}{\minitab[c]{ $m_{\widetilde{g}} = 2489 \GeV$  \\ $m_{\widetilde{\chi}_1^0} = 2133 \GeV$  }}& $ \MET \gtrsim 150 \GeV; \quad  \HT \gtrsim 300 \GeV  $ & \multirow{3}{*}{\minitab[c|c]{$t\,\overline{t}$ & 0.14  \\ tri-boson & 0.15 \\ \hline  total & 0.35 }}   \\
                                 &                                                      & $ \meff \gtrsim 500 \GeV $ & \\
                                &                                                       & $  N_{\text{jets}} > 5; \quad N_b > 2 $ & \\    
\hline
\multirow{3}{*}{SR7} &\multirow{3}{*}{\minitab[c]{ $m_{\widetilde{g}} = 3489 \GeV$  \\ $m_{\widetilde{\chi}_1^0} = 1 \GeV$  }}& $ \MET \gtrsim 100 \GeV; \quad \HT \gtrsim 1000 \GeV $ & \multirow{3}{*}{\minitab[c|c]{tri-boson & 0.0027  \\ \hline  total &  0.0027 }}   \\
                                &                                                       & $ \meff \gtrsim 2000 \GeV; \quad \text{sym-}M_{T2} \gtrsim 100 \GeV $ & \\
                                &                                                       & $  N_{\text{jets}} > 5; \quad N_b > 2 $ & \\                      
\hline                    
\multirow{3}{*}{SR8} &\multirow{3}{*}{\minitab[c]{ $m_{\widetilde{g}} = 3489 \GeV$  \\ $m_{\widetilde{\chi}_1^0} = 3133 \GeV$  }}& \multirow{3}{*}{\minitab[c]{$ \HT \gtrsim 200 \GeV; \quad  \meff \gtrsim 400 \GeV $ \\  $  N_{\text{jets}} > 5; \quad N_b > 2 $}}& \multirow{3}{*}{\minitab[c|c]{$t\,\overline{t}$ & 1.1  \\ $t\,W$ & 4.7 \\ \hline  total &  6.3 }}   \\
                                &                                                       & & \\    
                                &&&\\
\hline            
\end{tabular}
\caption{The eight signal regions defined for the SSDL $14$ TeV $3000 \text{ fb}^{-1}$ search.  Also shown are the dominant two contributions to the background along with the total background after cuts, which also includes all subdominant backgrounds.}
\label{tab:SSDLCuts14TeV3000ifb}
\end{table}

\vspace*{\fill}

\pagebreak
$ $

\begin{figure}[h!]
  \centering
  \includegraphics[width=.48\columnwidth]{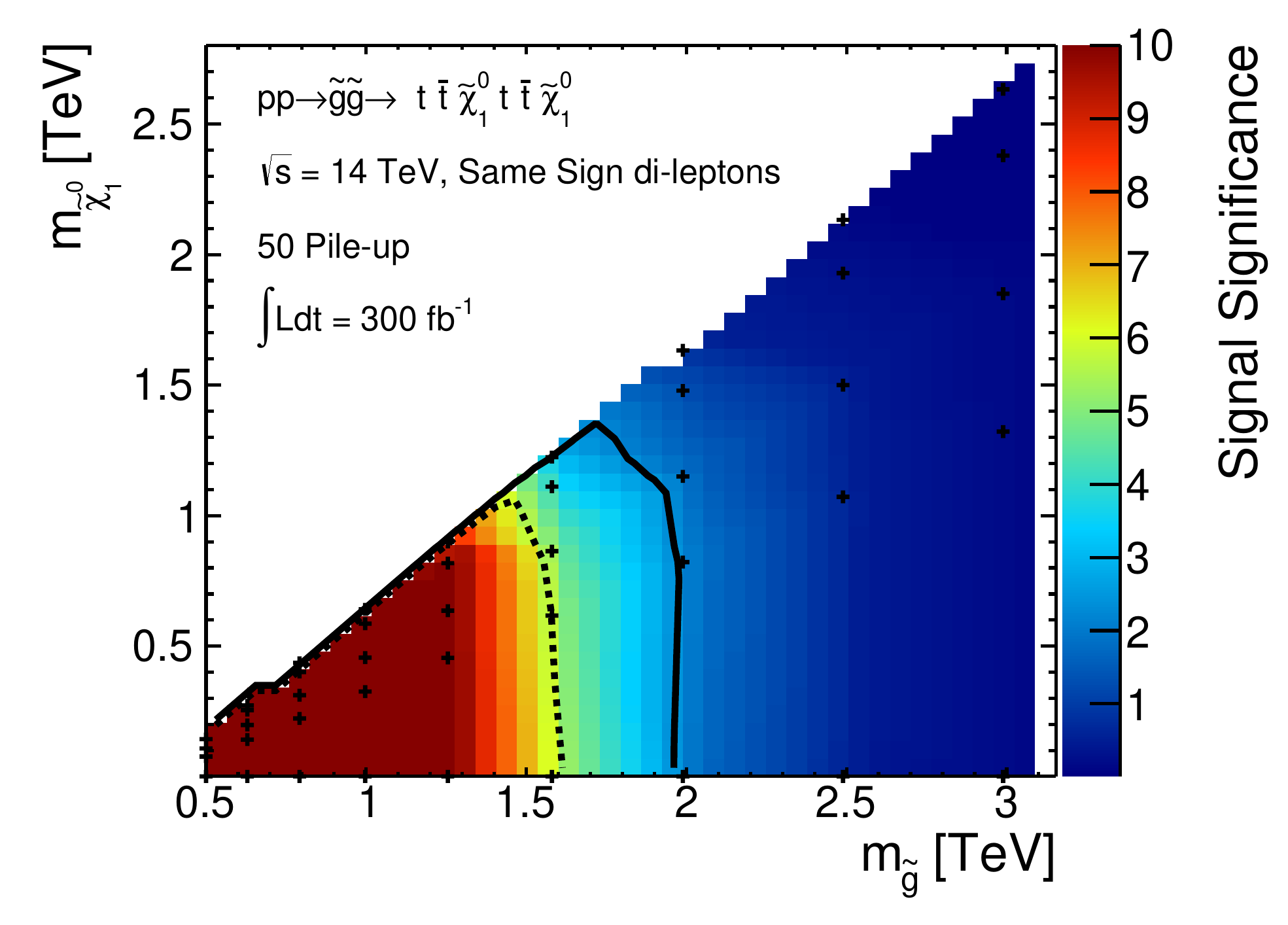}
  \includegraphics[width=.48\columnwidth]{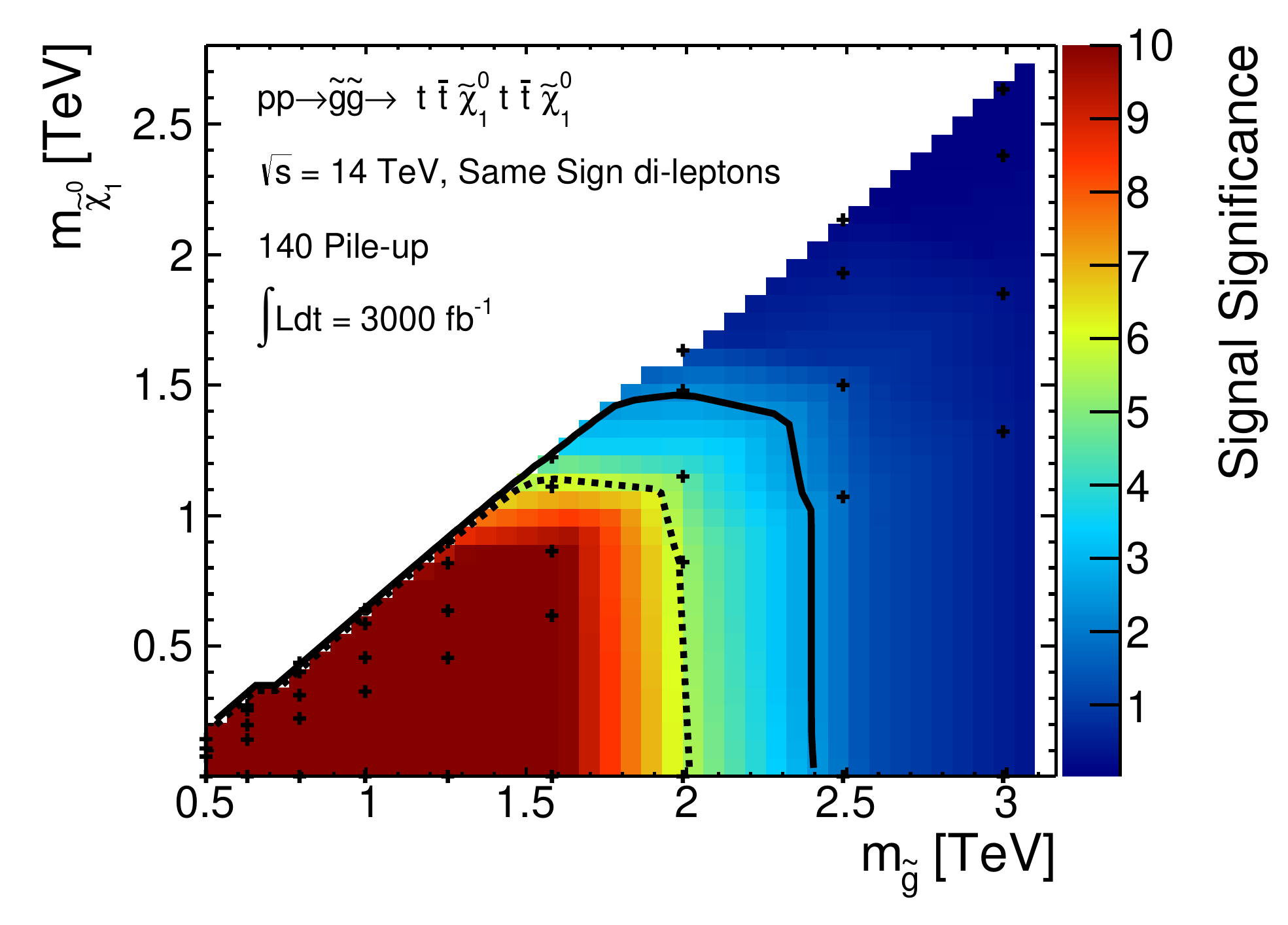}
    \includegraphics[width=.48\columnwidth]{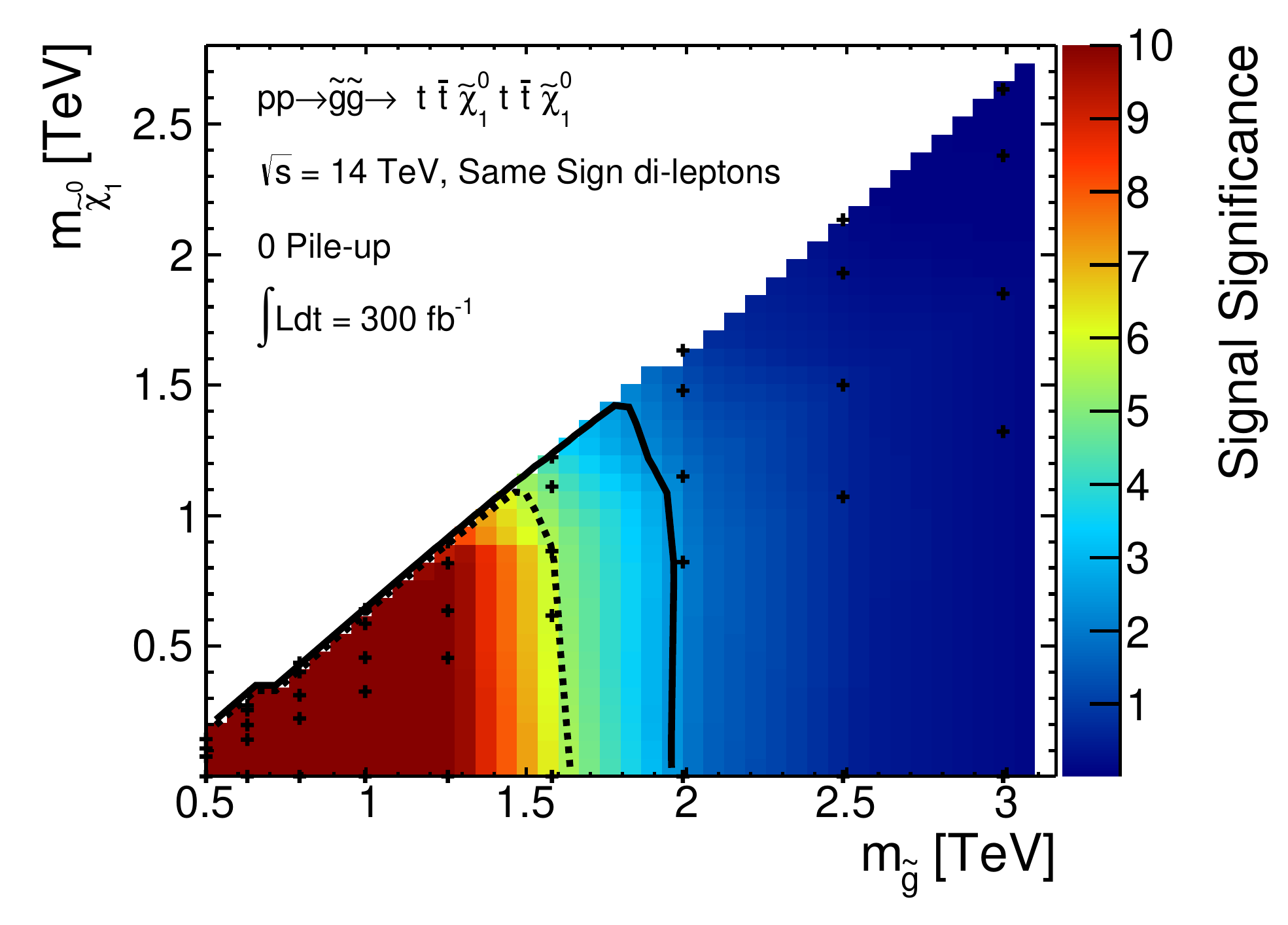}
  \includegraphics[width=.48\columnwidth]{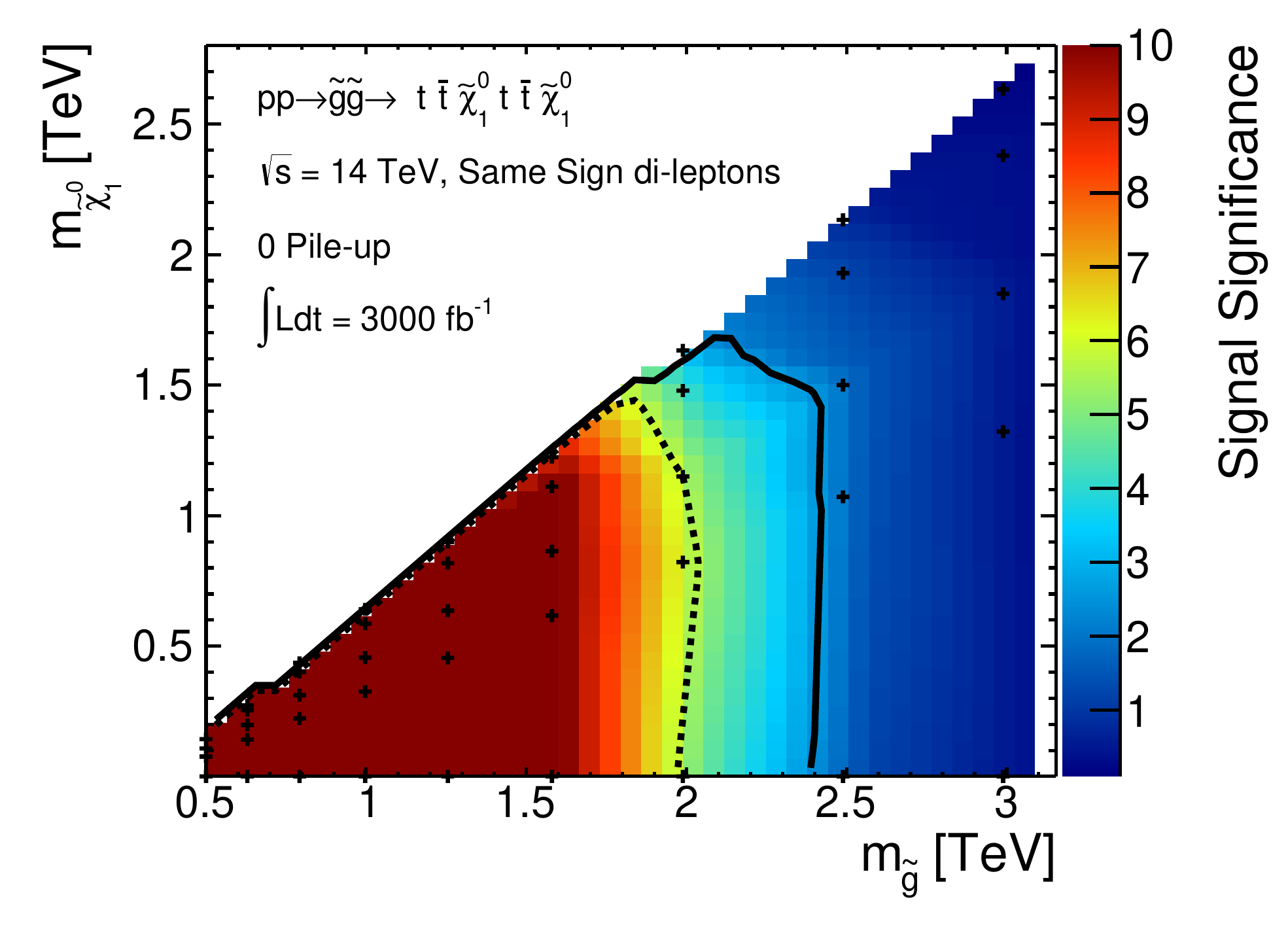}
  \caption{Results for the gluino-neutralino model with heavy flavor decays are given in the $m_{\widetilde{\chi}_1^0}$ versus $m_{\widetilde{g}}$ plane.  The solid [dotted] lines shows the expected $5\sigma$ discovery reach [95\% confidence level upper limits] for gluino pair production.  Mass points to the left/below the contours are expected to be probed at the $14\TeV$ LHC using 300 fb$^{-1}$ [left] and 3000 fb$^{-1}$ [right] of integrated luminosity.  A 20\% systematic uncertainty is assumed.  The $300 \text{ fb}^{-1}$ result on the left includes an average of $50$ pileup events; the$3000 \text{ fb}^{-1}$ result on the right includes an average of $140$ pileup events.  The results on the bottom do not include the effects of pileup.}
\label{fig:GOGO_HeavyFlavor_14_PileUp_results}
\end{figure}

\hiddensubsection{Results: 14 TeV}
Using the signal regions outlined in the previous section, we determine the ability of the $14$ TeV LHC to probe this model in the SSDL channel.  These results are presented in Fig.~\ref{fig:GOGO_HeavyFlavor_14_PileUp_results}, where we show the $95\%$ CL exclusion [solid line] and the $5\sigma$ discovery contours in the $m_{\widetilde{\chi}_1^0}$ versus $m_{\widetilde{g}}$ plane.  A $20\%$ systematic uncertainty has been assumed for the background.  Also shown on the bottom row of this figure are the results without including pileup.  The dominant effect of pileup on this analysis is that it can contaminate the lepton isolation cones, thereby reducing the signal strength.  At $14$ TeV this effect is significant only in the high mass compressed region, where it slightly weakens the limits. 

Using the NLO gluino pair production cross section one can make a very naive estimate for the reach of a given collider.  For example, we find that the choice of gluino mass which would yield $10$ SSDL events (accounting for the leptonic branching ratios) at $300 \text{ fb}^{-1}$  $\big(3000 \text{ fb}^{-1}\big)$ is $2.3 \TeV$ ($2.8 \TeV$).  This roughly corresponds to the maximal possible reach one could expect for a given luminosity using $14 \TeV$ proton collisions.  

Using a realistic simulation framework along with the search strategy employed here the $14 \TeV$ $300 \text{ fb}^{-1}$ limit is projected to be $1.9 \TeV$ (corresponding to 73 events), and the $3000 \text{ fb}^{-1}$ limit is projected to be $2.4 \TeV$ (corresponding to 67 events).   Finally, we note that the $14 \TeV$ LHC with $3000 \text{ fb}^{-1}$ could discover a gluino (with $\widetilde{g}\rightarrow t\,\overline{t}\,\widetilde{\chi}_1^0$) as heavy as $2.0\TeV$ if the neutralino is massless.  Note that due to the relatively weak cuts that can be placed on \MET, the SSDL signal is robust against models with almost degenerate gluino and neutralino. 

\pagebreak
\hiddensubsection{Analysis: 33 TeV}
In this section, the signal regions are presented for the SSDL search at $33 \TeV$ assuming $3000 \text{ fb}^{-1}$.  Table \ref{tab:SSDLCuts33TeV} provides the values from an optimization for $3000$ fb$^{-1}$ of data.  Also shown in these tables are number of events from the two dominant contributions to the background along with the total number of background events after cuts (including all contributions).  When evaluating the gluino reach, we compute the signal efficiency after cuts for each signal region and use the one with the highest significance. 

\vspace{20pt}
\begin{table}[h!]
\renewcommand{\arraystretch}{1.4}
\setlength{\tabcolsep}{6pt}
\centering
\footnotesize
\begin{tabular}{c|c|c|c}
signal region & BSM masses & cuts & backgrounds\\
\hline
\hline
\multirow{3}{*}{SR1} &\multirow{3}{*}{\minitab[c]{ $m_{\widetilde{g}} = 997 \GeV$  \\ $m_{\widetilde{\chi}_1^0} = 1 \GeV$  }}& \multirow{3}{*}{\minitab[c]{ $\MET \gtrsim 100 \GeV; \quad \meff > 1000 \GeV$ \\ $\HT \gtrsim 500 \GeV; \quad \text{sym-}M_{T2} \gtrsim 40 \GeV$ \\ $N_{\text{jets}} > 5; \quad N_b > 2$ }} & \multirow{3}{*}{\minitab[c|c]{$t\,\overline{t}$ & 19   \\ di-boson & 0.17 \\ \hline  total & 19.2  }}  \\
                                 &                                                      & $  $ \\
                                &                                                       & $  $ \\                            
\hline          
\multirow{3}{*}{SR2} &\multirow{3}{*}{\minitab[c]{ $m_{\widetilde{g}} = 997 \GeV$  \\ $m_{\widetilde{\chi}_1^0} = 641 \GeV$  }}&  \multirow{3}{*}{\minitab[c]{  $ \MET \gtrsim 50 \GeV; \quad \HT \gtrsim 400 \GeV $ \\ $ \meff \gtrsim 700 \GeV $ \\ $N_{\text{jets}} > 5; \quad N_b > 2 $  }} & \multirow{3}{*}{\minitab[c|c]{$t\,\overline{t}$ & 460   \\ di-boson & 9.2 \\ \hline  total & 470  }}  \\
                                 &                                                      & $  $ \\
                                &                                                       & $  $ \\
\hline   
\multirow{3}{*}{SR3} &\multirow{3}{*}{\minitab[c]{ $m_{\widetilde{g}} = 1989 \GeV$  \\ $m_{\widetilde{\chi}_1^0} = 1 \GeV$  }}&  \multirow{3}{*}{\minitab[c]{ $ \MET \gtrsim 500 \GeV; \quad \HT \gtrsim 1500 \GeV $ \\ $ \meff \gtrsim 2200 \GeV; \quad \text{sym-}M_{T2} \gtrsim 200 \GeV $ \\  $N_{\text{jets}} > 7; \quad N_b > 2$   }} & \multirow{3}{*}{\minitab[c|c]{$t\,\overline{t}$ & 0.081  \\ tri-boson & 0.0062 \\ \hline  total &  0.087 }}  \\
                                 &                                                      & $  $ \\
                                &                                                       & $  $ \\
\hline   
\multirow{3}{*}{SR4} &\multirow{3}{*}{\minitab[c]{ $m_{\widetilde{g}} = 1989 \GeV$  \\ $m_{\widetilde{\chi}_1^0} = 1633 \GeV$  }}&  \multirow{3}{*}{\minitab[c]{  $ \MET \gtrsim 600 \GeV; \quad \meff \gtrsim 3000 \GeV $ \\ $ N_{\text{jets}} > 7; \quad N_b > 2 $  }} & \multirow{3}{*}{\minitab[c|c]{$t\,\overline{t}$ & 0.20  \\ $t\,j$ & 0.035 \\ \hline  total &  0.26 }}  \\
                                 &                                                      & $  $ \\
                                &                                                       & $  $ \\
\hline   
\multirow{3}{*}{SR5} &\multirow{3}{*}{\minitab[c]{ $m_{\widetilde{g}} = 3152 \GeV$  \\ $m_{\widetilde{\chi}_1^0} = 1 \GeV$  }}&  \multirow{3}{*}{\minitab[c]{  $ \MET \gtrsim 1000 \GeV; \quad \meff \gtrsim 2000 \GeV $ \\ $ \text{sym-}M_{T2} \gtrsim 300 \GeV $ \\ $ N_{\text{jets}} > 7; \quad N_b > 2 $  }} & \multirow{3}{*}{\minitab[c|c]{tri-boson & 0.0062  \\ \hline  total &  0.0062 }}  \\
                                 &                                                      & $  $ \\
                                &                                                       & $  $ \\
\hline   
\multirow{3}{*}{SR6} &\multirow{3}{*}{\minitab[c]{ $m_{\widetilde{g}} = 3152 \GeV$  \\ $m_{\widetilde{\chi}_1^0} = 2796 \GeV$  }}&  \multirow{3}{*}{\minitab[c]{ $ N_{\text{jets}} > 7; \quad N_b > 3 $ }} & \multirow{3}{*}{\minitab[c|c]{$t\,\overline{t}$ &  2.9 \\ $t\,j$ & 0.12 \\ \hline  total &  3.0 }}  \\
                                 &                                                      & $  $ \\
                                &                                                       & $  $ \\
\hline   
\multirow{3}{*}{SR7} &\multirow{3}{*}{\minitab[c]{ $m_{\widetilde{g}} = 4968 \GeV$  \\ $m_{\widetilde{\chi}_1^0} = 1 \GeV$  }}&  \multirow{3}{*}{\minitab[c]{ $ \MET \gtrsim 100 \GeV; \quad \HT \gtrsim 800 \GeV $ \\ $ \meff \gtrsim 1500 \GeV $ \\ $ N_{\text{jets}} > 7; \quad N_b > 1 $   }} & \multirow{3}{*}{\minitab[c|c]{$t\,\overline{t}$ &  390 \\ di-boson & 0.75 \\ \hline  total &  400 }}  \\
                                 &                                                      & $  $ \\
                                &                                                       & $  $ \\
\hline   
\multirow{3}{*}{SR8} &\multirow{3}{*}{\minitab[c]{ $m_{\widetilde{g}} = 4968 \GeV$  \\ $m_{\widetilde{\chi}_1^0} = 4612 \GeV$  }}&  \multirow{3}{*}{\minitab[c]{ $ \meff \gtrsim 400 \GeV; \quad \HT \gtrsim 150 \GeV $ \\ $ \text{sym-}M_{T2} \gtrsim 100 \GeV $ \\ $ N_{\text{jets}} > 7; \quad N_b > 2 $   }}  & \multirow{3}{*}{\minitab[c|c]{$t\,\overline{t}$ & 1.6  \\ $t\,j$ & 0.12 \\ \hline  total &  1.8 }}  \\
                                 &                                                      & $  $ \\
                                &                                                       & $  $ \\
\hline   
\end{tabular}
\caption{The eight signal regions defined for the SSDL $33$ TeV $3000 \text{ fb}^{-1}$ search.  Also shown are the dominant two contributions to the background along with the total background after cuts, which also includes all subdominant backgrounds.}
\label{tab:SSDLCuts33TeV}
\end{table}

\begin{figure}[h!]
  \centering
  \includegraphics[width=.48\columnwidth]{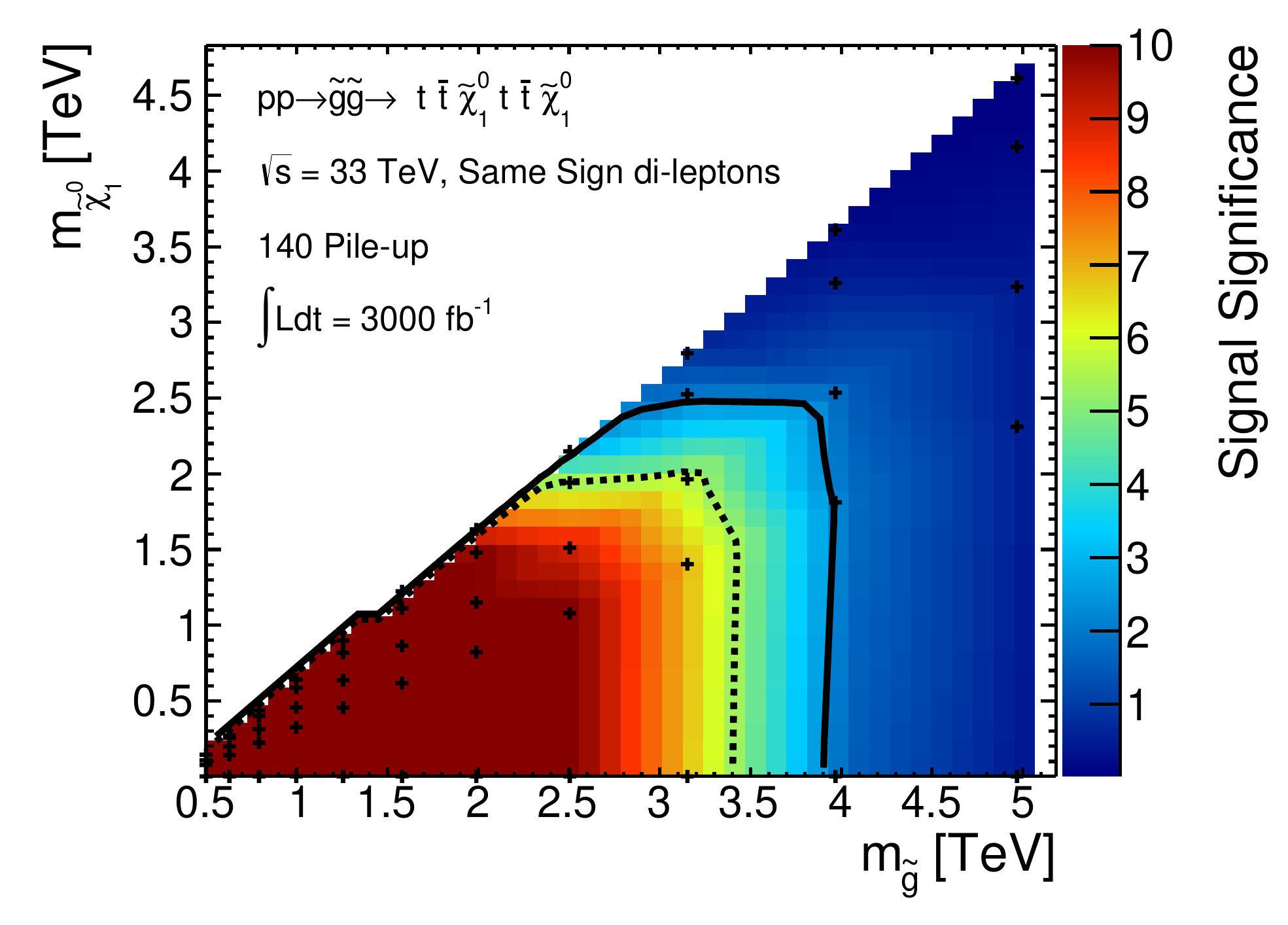}
    \includegraphics[width=.48\columnwidth]{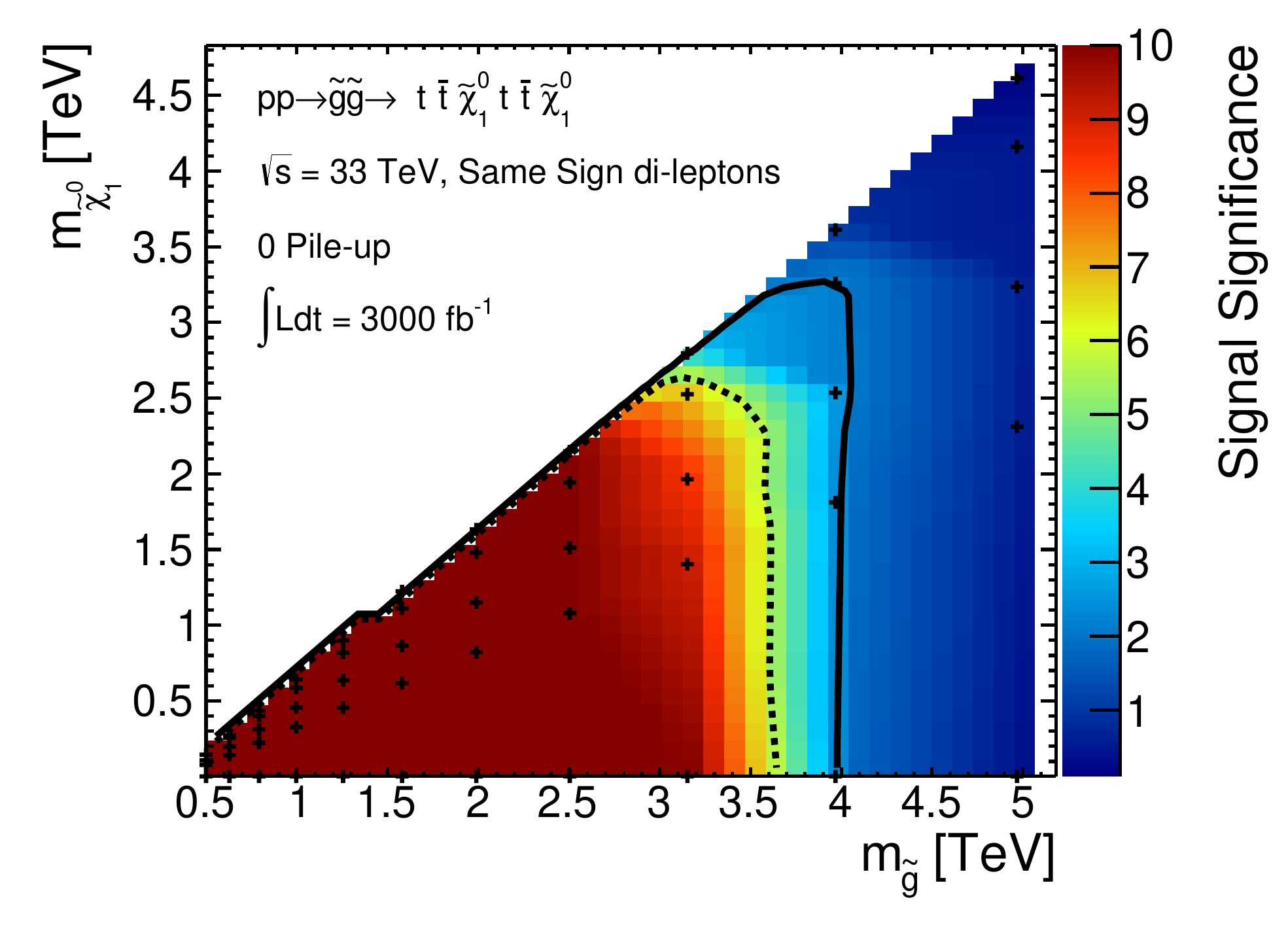}
\caption{Results for the gluino-neutralino model with heavy flavor decays are given in the $m_{\widetilde{\chi}_1^0}$ versus $m_{\widetilde{g}}$ plane.  The solid [dotted] lines shows the expected $5\sigma$ discovery reach [95\% CL upper limits] for gluino pair production.  Mass points to the left/below the contours are expected to be probed at a $33\TeV$ proton collider using 3000 fb$^{-1}$ of integrated luminosity.  A 20\% systematic uncertainty is assumed.   For the figure on the left [right], an average of $140$ $[0]$ pileup events are included.}
\label{fig:GOGO_HeavyFlavor_33_PileUp_results}
\end{figure}

\pagebreak
\hiddensubsection{Results: 33 TeV}
Using the signal regions outlined in the previous section, we determine the ability of the $33$ TeV LHC to probe this model in the SSDL channel.   These results are presented in Fig.~\ref{fig:GOGO_HeavyFlavor_33_PileUp_results}, where we show the $95\%$ CL exclusion [solid line] and the $5\sigma$ discovery contours in the $m_{\widetilde{\chi}_1^0}$ versus $m_{\widetilde{g}}$ plane.  A $20\%$ systematic uncertainty has been assumed for the background.  Also shown in the right panel of this figure are the results without including pileup.  The dominant effect of pileup on this analysis is that it can contaminate the lepton isolation cones, thereby reducing the signal strength.  As we go to higher CM energy colliders the min bias events have higher $p_T$ and begin to affect the lepton isolation more significantly.  At $33$ TeV, in contrast with $14$ TeV, we find a small change in the overall reach and a substantial change in the reach for the high mass compressed region.  

\vspace{20pt}

Using the NLO squark pair production cross section one can make a very naive estimate for the reach of a given collider.  For example, we find that the choice of gluino mass which would yield $10$ SSDL events, \emph{i.e.}, appropriately accounting for the branching ratio, at $3000 \text{ fb}^{-1}$ is $5.5 \TeV$.  This roughly corresponds to the maximal possible reach one could expect for a given luminosity using $33 \TeV$ proton collisions.  

Using a realistic simulation framework along with the search strategy employed here the $33 \TeV$ $3000 \text{ fb}^{-1}$ limit is projected to be $4.0 \TeV$ (corresponding to 243 events).   Finally, we note that a $33 \TeV$ proton collider with $3000 \text{ fb}^{-1}$ could discover a gluino (with $\widetilde{g}\rightarrow t\,\overline{t}\,\widetilde{\chi}_1^0$) as heavy as $3.4\TeV$ if the neutralino is massless.  Note that due to the relatively weak cuts that can be placed on \MET, the SSDL signal is robust against models with almost degenerate gluino and neutralino. 

\pagebreak
\hiddensubsection{Analysis: 100 TeV}
In this section, the signal regions are are presented for the SSDL search at $100 \TeV$ assuming $3000 \text{ fb}^{-1}$.  Table \ref{tab:SSDLCuts100TeV} provides the values from an optimization for $3000$ fb$^{-1}$ of data.  Also shown in these tables are number of events from the two dominant contributions to the background along with the total number of background events after cuts (including all contributions).   When evaluating the gluino reach, we compute the signal efficiency after cuts for each signal region and use the one with the highest significance.

\vspace{20pt}

\begin{figure}[h!]
  \centering
  \includegraphics[width=.48\columnwidth]{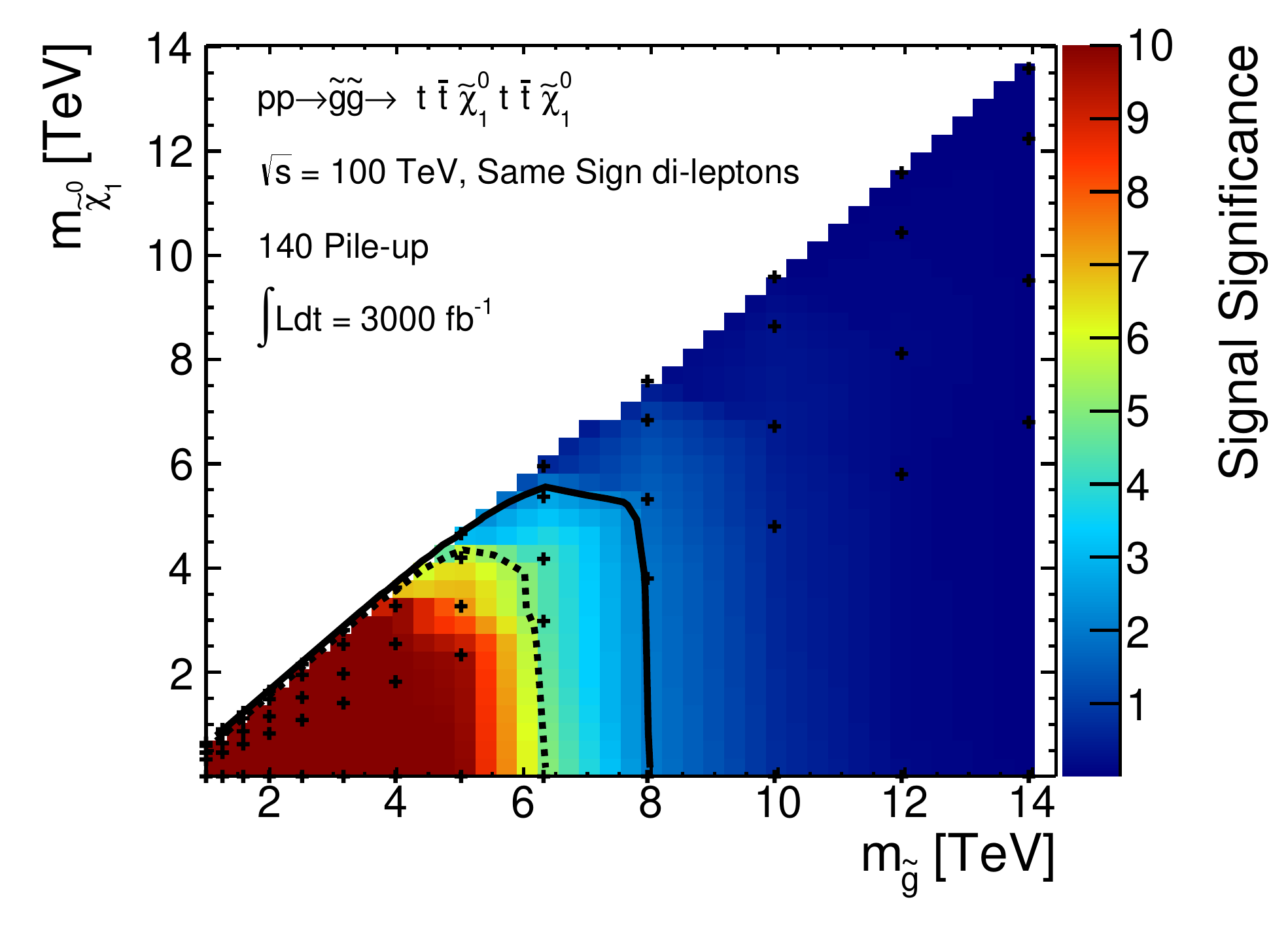}
    \includegraphics[width=.48\columnwidth]{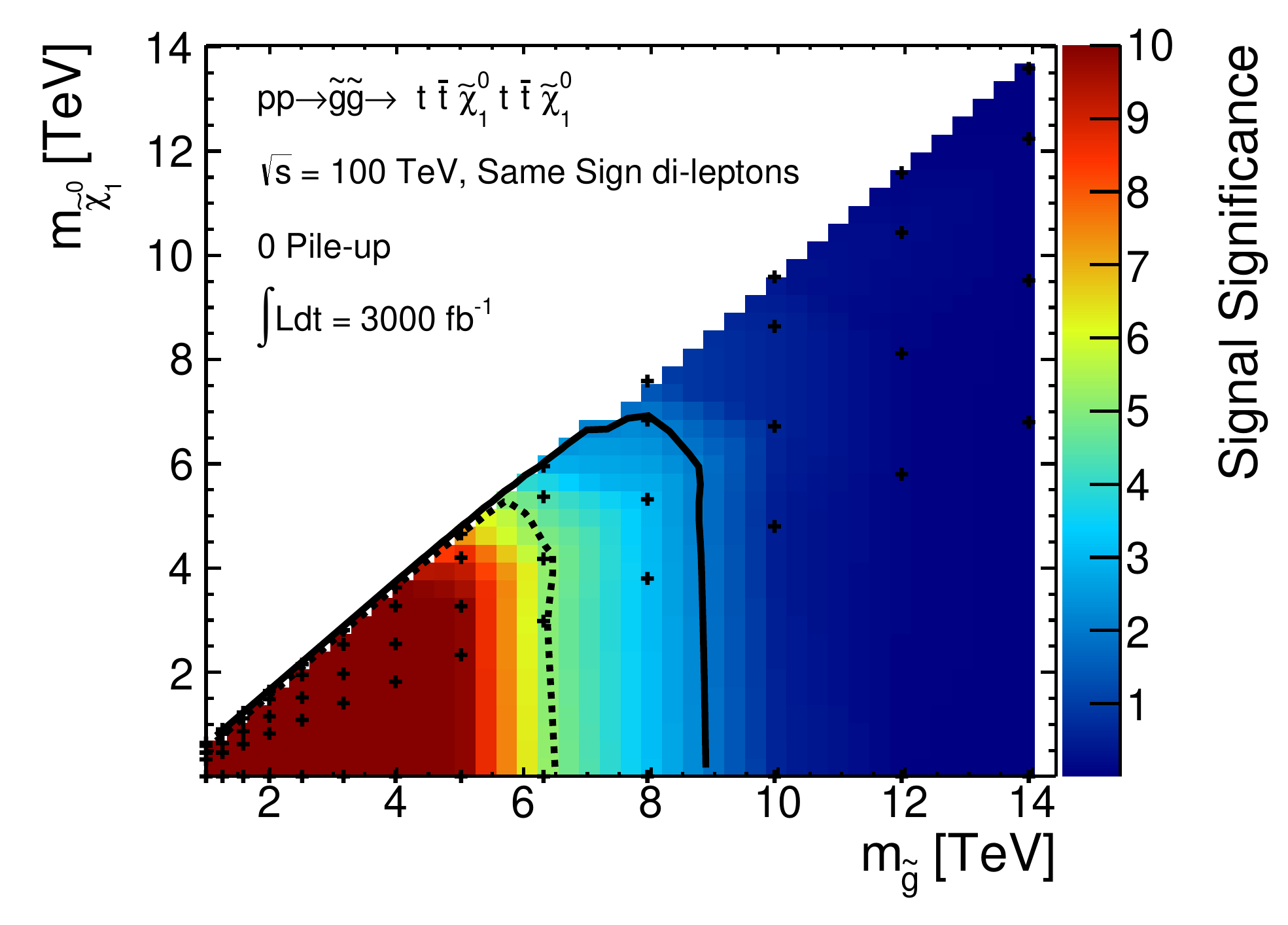}
\caption{Results for the gluino-neutralino model with heavy flavor decays are given in the $m_{\widetilde{\chi}_1^0}$ versus $m_{\widetilde{g}}$ plane.  The solid [dotted] lines shows the expected $5\sigma$ discovery reach [95\% confidence level upper limits] for gluino pair production.  Mass points to the left/below the contours are expected to be probed at a $100\TeV$ proton collider using 3000 fb$^{-1}$ of integrated luminosity.  A 20\% systematic uncertainty is assumed.  For the figure on the left [right], an average of $140$ $[0]$ pileup events are included.}
\label{fig:GOGO_HeavyFlavor_100_PileUp_results}
\end{figure}

\begin{table}[h!]
\renewcommand{\arraystretch}{1.4}
\setlength{\tabcolsep}{6pt}
\centering
\footnotesize
\begin{tabular}{c|c|c|c}
signal region & BSM masses & cuts & backgrounds\\
\hline
\hline
\multirow{3}{*}{SR1} &\multirow{3}{*}{\minitab[c]{ $m_{\widetilde{g}} = 1995 \GeV$  \\ $m_{\widetilde{\chi}_1^0} = 1 \GeV$  }}& \multirow{3}{*}{\minitab[c]{ $ \MET \gtrsim 400 \GeV; \quad \HT \gtrsim 1500 \GeV $ \\ $ \meff \gtrsim 2000 \GeV; \quad \text{sym-}M_{T2} \gtrsim 100 \GeV $ \\ $ N_{\text{jets}} > 7; \quad N_b > 3 $   }} & \multirow{3}{*}{\minitab[c|c]{$t\,\overline{t}$ & 0.53  \\ \hline  total & 0.53  }}  \\
                                 &                                                      & $  $ \\
                                &                                                       & $  $ \\
\hline          
\multirow{3}{*}{SR2} &\multirow{3}{*}{\minitab[c]{ $m_{\widetilde{g}} = 1995 \GeV$  \\ $m_{\widetilde{\chi}_1^0} = 1639 \GeV$  }}&  \multirow{3}{*}{\minitab[c]{ $ \MET \gtrsim 400 \GeV; \quad \HT \gtrsim 1200 \GeV $ \\ $ \meff \gtrsim 2000 \GeV; \quad \text{sym-}M_{T2} \gtrsim 100 \GeV $ \\ $ N_{\text{jets}} > 7; \quad N_b > 3 $   }} & \multirow{3}{*}{\minitab[c|c]{$t\,\overline{t}$ &  0.53 \\  \hline  total &  0.53 }}  \\
                                 &                                                      & $  $ \\
                                &                                                       & $  $ \\
\hline   
\multirow{3}{*}{SR3} &\multirow{3}{*}{\minitab[c]{ $m_{\widetilde{g}} = 3981 \GeV$  \\ $m_{\widetilde{\chi}_1^0} = 1 \GeV$  }}&  \multirow{3}{*}{\minitab[c]{ $ \MET \gtrsim 1200 \GeV; \quad \HT \gtrsim 3500 \GeV $ \\ $ \meff \gtrsim 4500 \GeV; \quad \text{sym-}M_{T2} \gtrsim 350 \GeV $ \\ $ N_{\text{jets}} > 7; \quad N_b > 2 $   }} & \multirow{3}{*}{\minitab[c|c]{$t\,\overline{t}$ & 0.60  \\ \hline  total & 0.60  }}  \\
                                 &                                                      & $  $ \\
                                &                                                       & $  $ \\
\hline   
\multirow{3}{*}{SR4} &\multirow{3}{*}{\minitab[c]{ $m_{\widetilde{g}} = 3981 \GeV$  \\ $m_{\widetilde{\chi}_1^0} = 3625 \GeV$  }}&  \multirow{3}{*}{\minitab[c]{ $ \MET \gtrsim 1200 \GeV; \quad \HT \gtrsim 4000 \GeV $ \\ $ \meff \gtrsim 5000 \GeV $ \\ $ N_{\text{jets}} > 7; \quad N_b > 2 $   }} & \multirow{3}{*}{\minitab[c|c]{$t\,\overline{t}$ & 0.85  \\ $t\,W$ & 0.14 \\ \hline  total & 1.0  }}  \\
                                 &                                                      & $  $ \\
                                &                                                       & $  $ \\
\hline   
\multirow{3}{*}{SR5} &\multirow{3}{*}{\minitab[c]{ $m_{\widetilde{g}} = 7944 \GeV$  \\ $m_{\widetilde{\chi}_1^0} = 1 \GeV$  }}&  \multirow{3}{*}{\minitab[c]{ $ \MET \gtrsim 800 \GeV; \quad \HT \gtrsim 3000 \GeV $ \\ $ \meff \gtrsim 5000 \GeV $ \\ $ N_{\text{jets}} > 7; \quad N_b > 2 $   }} & \multirow{3}{*}{\minitab[c|c]{$t\,\overline{t}$ & 3.6  \\ $t\,W$ & 0.20 \\ \hline  total &  3.8 }}  \\
                                 &                                                      & $  $ \\
                                &                                                       & $  $ \\
\hline   
\multirow{3}{*}{SR6} &\multirow{3}{*}{\minitab[c]{ $m_{\widetilde{g}} = 7944 \GeV$  \\ $m_{\widetilde{\chi}_1^0} = 7588 \GeV$  }}&  \multirow{3}{*}{\minitab[c]{ $ \HT \gtrsim 400 \GeV; \quad \meff \gtrsim 800 \GeV $ \\ $ N_{\text{jets}} > 7; \quad N_b > 2 $   }} & \multirow{3}{*}{\minitab[c|c]{$t\,\overline{t}$ & 8100  \\ $t\,W$ & 190 \\ \hline  total & 8300  }}  \\
                                 &                                                      & $  $ \\
                                &                                                       & $  $ \\
\hline   
\multirow{3}{*}{SR7} &\multirow{3}{*}{\minitab[c]{ $m_{\widetilde{g}} = 15944 \GeV$  \\ $m_{\widetilde{\chi}_1^0} = 1 \GeV$  }}&  \multirow{3}{*}{\minitab[c]{  $ \MET \gtrsim 500 \GeV; \quad \HT \gtrsim 4000 \GeV $ \\ $ \meff \gtrsim 6000 \GeV $ \\ $ N_{\text{jets}} > 5; \quad N_b > 1 $  }} & \multirow{3}{*}{\minitab[c|c]{$t\,\overline{t}$ & 45  \\ single boson & 1.8 \\ \hline  total &  50 }}  \\
                                 &                                                      & $  $ \\
                                &                                                       & $  $ \\
\hline   
\multirow{3}{*}{SR8} &\multirow{3}{*}{\minitab[c]{ $m_{\widetilde{g}} = 15944 \GeV$  \\ $m_{\widetilde{\chi}_1^0} = 15588 \GeV$  }}&  \multirow{3}{*}{\minitab[c]{ $ \HT \gtrsim 200 \GeV; \quad \meff \gtrsim 400 \GeV $ \\ $ \text{sym-}M_{T2} \gtrsim 100 \GeV $ \\ $ N_{\text{jets}} > 5; \quad N_b > 2 $   }}  & \multirow{3}{*}{\minitab[c|c]{$t\,\overline{t}$ & 2200  \\ $t\,W$ & 16 \\ \hline  total &   2200 }}  \\
                                 &                                                      & $  $ \\
                                &                                                       & $  $ \\
\hline   
\end{tabular}
\caption{The eight signal regions defined for the SSDL $100$ TeV $3000 \text{ fb}^{-1}$ search.  Also shown are the dominant two contributions to the background along with the total background after cuts, which also includes all subdominant backgrounds.}
\label{tab:SSDLCuts100TeV}
\end{table}
\hiddensubsection{Results: 100 TeV}
Using the signal regions outlined in the previous section, we determine the ability of the $100$ TeV LHC to probe this model in the SSDL channel.  These results are presented in Fig.~\ref{fig:GOGO_HeavyFlavor_100_PileUp_results}, where we show the $95\%$ CL exclusion [solid line] and the $5\sigma$ discovery contours in the $m_{\widetilde{\chi}_1^0}$ versus $m_{\widetilde{g}}$ plane.  A $20\%$ systematic uncertainty has been assumed for the background.  Also shown in the right panel of this figure are the results without including pileup.  The dominant effect of pileup on this analysis is that it can contaminate the lepton isolation cones, thereby reducing the signal strength.   As we go to higher CM energy colliders the min bias events have higher $p_T$ and begin to affect the lepton isolation more significantly.  At $100$ TeV, this effect is significant enough to decrease the limits on the gluino mass in this analysis by almost $1 \TeV$.  Note that the lepton isolation cuts were not optimized for the higher pile-up and CM energy environments in this study; an interesting direction for future work would be to study how this issue can be ameliorated.

\vspace{20pt}

Using the NLO gluino pair production cross section one can make a very naive estimate for the reach of a given collider.  For example, we find that the choice of gluino mass which would yield $10$ SSDL events at $3000 \text{ fb}^{-1}$ is $12.7 \TeV$.  This roughly corresponds to the maximal possible reach one could expect for a given luminosity using $100 \TeV$ proton collisions.  

Using a realistic simulation framework along with the search strategy employed here the $100 \TeV$ $3000 \text{ fb}^{-1}$ limit is projected to be $8.8 \TeV$ (corresponding to 224 events).   Finally, we note that a $100 \TeV$ proton collider with $3000 \text{ fb}^{-1}$ could discover a gluino (with $\widetilde{g}\rightarrow t\,\overline{t}\,\widetilde{\chi}_1^0$) as heavy as $6.4\TeV$ if the neutralino is massless.  Note that due to the relatively weak cuts that can be placed on \MET, the SSDL signal is robust against models with almost degenerate gluino and neutralino. 

\begin{figure}[h!]
  \centering
  \includegraphics[width=.48\columnwidth]{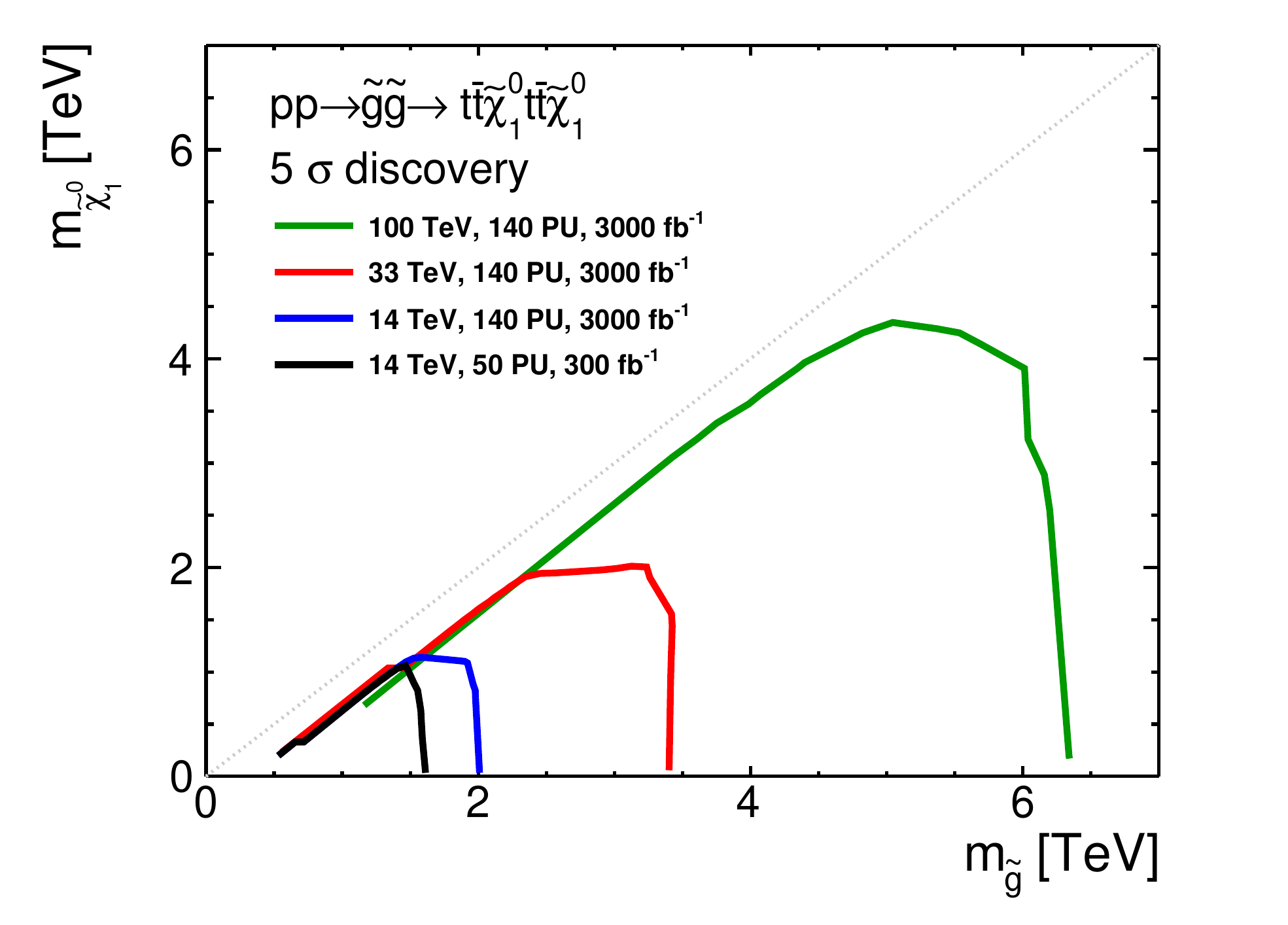}
    \includegraphics[width=.48\columnwidth]{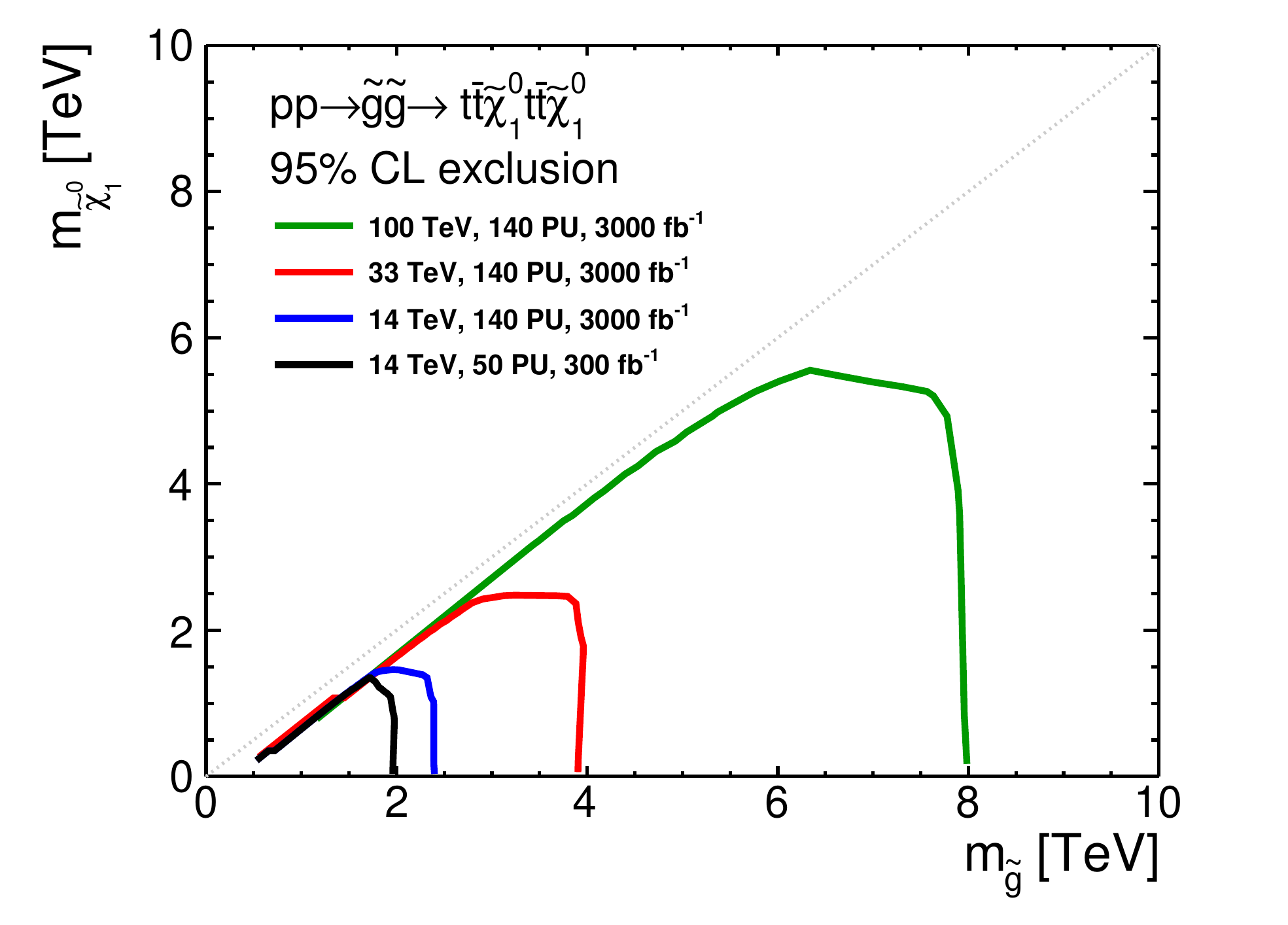}
  \caption{Results for the gluino-squark-neutralino model.  The neutralino mass is taken to be $1 \GeV$.  The left [right] panel shows the $5\,\sigma$ discovery reach [$95\%$ CL exclusion] for the four collider scenarios studied here.  A $20\%$ systematic uncertainty is assumed and pileup is included.}
\label{fig:GOGO_HeavyFlavor_Comparison}
\end{figure}

\hiddensubsection{Comparing Colliders}
The same-sign di-lepton signature of the gluino-neutralino model with heavy flavor decays provides a useful case study with which to compare the potential impact of different proton colliders.  Due to theoretical motivation in the context of both natural SUSY and split SUSY models, this final state is a very important signature of new physics to consider. Figure \ref{fig:GOGO_HeavyFlavor_Comparison} shows the $5\sigma$ discovery reach [$95\%$ CL exclusion] for two choices of integrated luminosity at $14$ TeV, along with the full data set assumed for $33$ and $100$ TeV.  At the LHC, a factor of $10$ increase in luminosity leads to an improved reach of roughly $500$ GeV.  Increasing the center-of-mass energy has a tremendous impact on the experimentally available parameter space, since now much heavier gluinos can be produced without relying on the tails of parton distributions to supply the necessary energy.  Figure \ref{fig:GOGO_HeavyFlavor_Comparison} makes a compelling case for investing in future proton colliders which can operate at these high energies.

Note that studying other final states for this decay channel was outside the scope of this project.  In light of these results though, it would be interesting to see if an all hadronic search would lead to improvements in the projected limits, especially since lepton efficiencies are significantly affected at high CM energies by the pile-up conditions and the highly boosted top quarks, and similarly to veto $\tau$-tagged jets to further reduce $W/Z$+jets. In particular, when considering searches at a $100 \TeV$ collider, it would be interesting to investigate the fat top jet signatures of this model with very heavy gluinos.
\vspace{20pt}

\pagebreak
\section{Outlook}
Particle accelerators are one of the primary tools for experimentally investigating questions related to the microscopic properties of our Universe.  Given the $20+$ year time scale required to build one of these machines, it is important to think carefully about their physics capabilities.  A wide variety of studies are required for making informed decisions on the machine requirements such as center-of-mass energy, instantaneous and integrated luminosity; issues related to detailed detector design specifications must also be addressed.  

This paper presents some of the first comprehensive comparisons between the upcoming $14 \TeV$ run of the CERN Large Hadron Collider and future possible experiments (for other recent studies see \cite{Agashe:2013fda,Avetisyan:2013rca,Bhattacharya:2013iea,Varnes:2013pxa,Andeen:2013zca,Apanasevich:2013cta,Degrande:2013yda,Stolarski:2013msa,Yu:2013wta,Zhou:2013raa}).  This includes a high-luminosity program at $14 \TeV$, and experiments that will collide protons at energies between $\sim 33$ and $\sim 100 \TeV$.  Our goal was to obtain the best limits possible with generic, signature-based searches that are not overly tuned to specific models.

We assessed the physics potential of these collider scenarios by performing analyses using Monte Carlo samples for signals and backgrounds with a realistic detector model.  Results obtained with a fast detector simulation provide a complimentary estimate to a simple rescaling of existing limits, where the $8$ TeV search strategies could be more sophisticated or involve more signal regions than the approaches taken here.  Performing searches on Monte Carlo also provides insight into the impact of effects such as systematic uncertainties and pileup as a function of the center-of-mass energy.

In particular, we studied the reach for four SUSY Simplified Models whose experimental signatures are driven by the production of colored states.  Three analysis strategies were employed in deriving these projections: a jets + \MET~analysis which optimized over \HT~and \MET, a mono-jet analysis with either an inclusive or exclusive jet requirement, and a same-sign di-lepton search.  Table \ref{tab:summary} shows the discovery potential [exclusion reach] for the different collider scenarios.

\begin{table}[h!]
\vspace{-10pt}
\small
\renewcommand{\arraystretch}{1.5}
\setlength{\tabcolsep}{5pt}
\begin{center}
\begin{tabular}{c||c|c|c|c}
 Simplified Model & $14 \TeV$ $300 \text{ fb}^{-1}$ &  $14 \TeV$ $3000 \text{ fb}^{-1}$ &  $33 \TeV$ &  $100 \TeV$ \\
\hline
\hline
$\widetilde{g}$ - $\widetilde{\chi}_1^0$ &  \multirow{3}{*}{\minitab[c]{ $1.9 \TeV$ \\ $\big[2.3 \TeV\big]$}} &   \multirow{3}{*}{\minitab[c]{ $2.2 \TeV$ \\ $\big[2.7 \TeV\big]$}} &   \multirow{3}{*}{\minitab[c]{ $5.0 \TeV$ \\ $\big[5.8 \TeV\big]$}}  &   \multirow{3}{*}{\minitab[c]{ $11 \TeV$ \\ $\big[13.5 \TeV\big]$}} \\
$\text{light flavor decays}$  & & & & \\
$m_{\widetilde{\chi}_1^0} \simeq 0$  & & & & \\
 \vspace{-13pt} &&&&\\  
\hline
$\widetilde{g}$ - $\widetilde{\chi}_1^0$  &  \multirow{3}{*}{\minitab[c]{ $0.75 \TeV$ \\ $\big[0.9 \TeV\big]$}} &   \multirow{3}{*}{\minitab[c]{ $0.9 \TeV$ \\ $\big[1.0 \TeV\big]$}} &   \multirow{3}{*}{\minitab[c]{ $1.5 \TeV$ \\ $\big[1.8 \TeV\big]$}}  &   \multirow{3}{*}{\minitab[c]{ $4.6 \TeV$ \\ $\big[5.5 \TeV\big]$}} \\
$\text{light flavor decays}$  & & & & \\
$m_{\widetilde{g}} \simeq m_{\widetilde{\chi}_1^0}$  & & & & \\
 \vspace{-13pt} &&&&\\  
\hline
$\widetilde{q}$ - $\widetilde{\chi}_1^0$   &  \multirow{3}{*}{\minitab[c]{ $0.80 \TeV$ \\ $\big[1.5 \TeV\big]$}} &   \multirow{3}{*}{\minitab[c]{ $0.9 \TeV$ \\ $\big[1.7 \TeV\big]$}} &   \multirow{3}{*}{\minitab[c]{ $1.4 \TeV$ \\ $\big[3.4 \TeV\big]$}} &   \multirow{3}{*}{\minitab[c]{ $2.4 \TeV$ \\ $\big[8.0 \TeV\big]$}} \\
$\text{light flavor decays}$ & & & &\\
$m_{\widetilde{\chi}_1^0} \simeq 0$  & & & & \\
 \vspace{-13pt} &&&&\\  
\hline
$\widetilde{q}$ - $\widetilde{\chi}_1^0$ &  \multirow{3}{*}{\minitab[c]{ $0.45 \TeV$ \\ $\big[0.65 \TeV\big]$}} &   \multirow{3}{*}{\minitab[c]{ $0.45 \TeV$ \\ $\big[0.70 \TeV\big]$}} &   \multirow{3}{*}{\minitab[c]{ $0.80 \TeV$ \\ $\big[1.3 \TeV\big]$}} &   \multirow{3}{*}{\minitab[c]{ $3.0 \TeV$ \\ $\big[3.9 \TeV\big]$}} \\
$\text{light flavor decays}$ & & & &\\
$m_{\widetilde{q}} \simeq m_{\widetilde{\chi}_1^0}$  & & & &\\
 \vspace{-13pt} &&&&\\  
\hline
$\widetilde{g}$ - $\widetilde{q}$- $\widetilde{\chi}_1^0$ &  \multirow{3}{*}{\minitab[c]{ $2.7 \TeV$ \\ $\big[2.8 \TeV\big]$}} &   \multirow{3}{*}{\minitab[c]{ $3.0 \TeV$ \\ $\big[3.2 \TeV\big]$}} &   \multirow{3}{*}{\minitab[c]{ $6.6 \TeV$ \\ $\big[6.8 \TeV\big]$}} &   \multirow{3}{*}{\minitab[c]{ $15.5 \TeV$ \\ $\big[16 \TeV\big]$}} \\
$\text{light flavor decays}$ & & & &\\
$m_{\widetilde{g}}\simeq m_{\widetilde{q}}$ and $m_{\widetilde{\chi}_1^0} \simeq 0$ & & & &\\
 \vspace{-13pt} &&&&\\  
\hline
$\widetilde{g}$ - $\widetilde{\chi}_1^0$  &  \multirow{3}{*}{\minitab[c]{ $1.6 \TeV$ \\ $\big[1.9 \TeV\big]$}} &   \multirow{3}{*}{\minitab[c]{ $2.0 \TeV$ \\ $\big[2.4 \TeV\big]$}} &   \multirow{3}{*}{\minitab[c]{ $3.4 \TeV$ \\ $\big[3.9 \TeV\big]$}} &   \multirow{3}{*}{\minitab[c]{ $6.3 \TeV$ \\ $\big[8.8 \TeV\big]$}} \\
$\text{heavy flavor decays}$ & & & &\\
$m_{\widetilde{\chi}_1^0} \simeq 0$  & & & & \\
\end{tabular}
\end{center}
\caption{This table summarizes the expected discovery reach [$95\%$ CL limits] as computed using the search strategies employed in this study.}
\label{tab:summary}
\end{table}

\normalsize

The results clearly demonstrate that these machines can have a substantial impact on our understanding of the parameter space of these models.  They also address several big-picture questions when comparing colliders.  In particular, it is possible to understand ``how do analyses scale between these different machines?" by studying this work.  

One example of how collider physics evolves as one moves to higher $\sqrt{s}$ is seen in the composition of the jets + \MET~backgrounds.  This was most obvious in the $\widetilde{g} \rightarrow q\,\overline{q} \,\chi_1^0$ study of Sec.~\ref{sec:GNLightFlavor}, where the dominant background was $W/Z + \text{jets}$ at $14 \TeV$, but became $t\,\overline{t}$ at $100 \TeV$.  Another important lesson was illustrated in using the same-sign di-lepton approach to the $\widetilde{g} \rightarrow t\,\overline{t} \,\chi_1^0$ final state, where it was clear that the impact of pileup changed significantly between $14 \TeV$ and $100 \TeV$.  
In particular, it is likely that lepton isolation requirements will need to evolve to cope with higher \pt\ objects and harder pileup interactions at high $\sqrt{s}$.

\pagebreak
There are many exciting opportunities for progress.
This paper provides a concrete starting point for understanding the new physics potential of experiments that would collide protons at energies approaching the boundary of what humans can hope to achieve.  
By providing a quantitative analysis of several SUSY Simplified Models, these results help define many challenges and opportunities for future hadron colliders.

\pagebreak
\ack
We thank Michael Peskin and Gavin Salam for helpful discussions.  TC is supported by the US DoE under contract number DE-AC02-76SF00515 and was supported in part by the NSF under Grant No. PHYS-1066293 while enjoying the hospitality of the Aspen Center for Physics.  AH is supported by the Alexander von Humboldt Foundation, Germany.  KH is supported by an NSF Graduate Research Fellowship under Grant number DGE-0645962 and by the US DoE under contract number DE-AC02-76SF00515, and was partially supported by ERC grant BSMOXFORD no. 228169.  SP was supported in part by the Department of Energy grant DE-FG02-90ER40546 and the FNAL LPC Fellowship.  JGW is supported by the US DoE under contract number DE-AC02-76SF00515. The authors are supported by grants from the Department of Energy Office of Science, the Alfred P. Sloan Foundation and the Research Corporation for Science Advancement.

\appendix
\section*{Appendix: Simulation Framework}
\addcontentsline{toc}{section}{Appendix: Simulation Framework}
This appendix is devoted to the details of our simulation procedure for the signal events.  The publicly available Snowmass backgrounds \cite{Avetisyan:2013onh} were used for all Standard Model Monte Carlo events. 

Unless otherwise specified we applied the following systematic uncertainties to all analyses:
\begin{itemize}
\vspace{-5pt}
\item luminosity: 2.8\%
\item PDF uncertainty: 5\%
\item signal acceptance: 15\%
\item background normalization: 20\%
\end{itemize}
\vspace{-10pt}

Parton level events were generated using \texttt{Madgraph5}  v1.5.10 \cite{Alwall:2011uj}.  All signals involve the pair production of SUSY particles and are matched using MLM matching up to 2 additional jets.  The $k_T$-ordered shower scheme with a matching scale of \texttt{qcut}=\texttt{xqcut}=$100\GeV$ was used.  Note that we do not account for any possible inadequacies inherent in the current Monte Carlo technology, \emph{e.g.} electroweak gauge bosons are not included in the shower.

The gluinos and squarks were treated as stable at the parton level.  These events were subsequently decayed and showered using \texttt{Pythia6} \cite{Sjostrand:2006za} and passed through the \texttt{Delphes} detector simulation \cite{deFavereau:2013fsa} using the ``Snowmass" detector parameter card \cite{Anderson:2013kxz}.  Total production cross sections were computed at NLO using a modified version of \texttt{Prospino} v2.1 \cite{Beenakker:1996ed, Beenakker:1996ch, Beenakker:1997ut}.

It proved to be advantageous to use a weighted event procedure.  In particular, it was our goal to accurately model the tails of the distributions in the compressed region which is notoriously difficult to simulate.  To this end we used a variation of the procedure developed for the Snowmass Standard Model background generation \cite{Avetisyan:2013onh}.  Since the only jets at the parton level were due to ISR, we could use generator level $H_T$ variable which is built into \texttt{Madgraph5}.  We will 
refer to cuts on this variable as \texttt{htmin}, \texttt{htmax}.  This allowed us to bin events in ``recoil" due to the presence of ISR --- these are exactly the types of events which contribute in the compressed region.

In detail, the process for each parameter point is:
\begin{enumerate}
\item \emph{Compute the approximate differential cross section with respect to $H_{T}$}. We run \texttt{Madgraph} in ``survey" mode (using the command \texttt{bin/madevent survey}) while incrementing the \texttt{htmin} cut to determine the cross section as a function of this cut, $\sigma_i \equiv \sigma(H_{T} > \texttt{htmin}_i)$ with
\begin{equation*}
\texttt{htmin}_{i=0\dots n} = \{0 \mbox{ GeV},\,100\mbox{ GeV},\,200\mbox{ GeV},\dots\}
\end{equation*}
We find that subsequent steps of 100 GeV provide an accurate enough characterization of the cross section for our purposes. We increase the cut until $\sigma_i < 1/\mathcal{L}$ where $\mathcal{L}$ is the luminosity for which good statistics are desired. The differential cross section is calculated from the differences:
\begin{equation*} 
\text{d}\sigma_i = \sigma_{i+1}-\sigma_{i}.
\end{equation*}

\item \emph{Determine bins of $H_{T}$ for event generation.} We define ${\rm bin}_{\alpha=1\dots m}$ by $\texttt{htmin}_\alpha\le H_{T} < \texttt{htmax}_\alpha$. We choose bin edges based on a ``weight fraction" $x$ with $0<x\leq1$ as follows:

\begin{enumerate}
\item The lower edge of the first bin is $\texttt{htmin}_1 = 0{\rm~GeV}$.
\item The upper edge of the first bin $\texttt{htmax}_1$ is chosen to be the smallest value such that $\sigma_1 \ge x\times \sigma_{\rm tot}$.
\item The remaining upper bin edges $\texttt{htmax}_{\alpha=2\dots m}$ are chosen similarly with each bin as small as possible such that  
\begin{equation}
\sigma_\alpha \equiv  \sigma(\texttt{htmax}_\alpha > H_{T} > \texttt{htmin}_\alpha) > x \times\sigma(H_{T} > \texttt{htmin}_\alpha),
\end{equation}
with $\texttt{htmin}_{\alpha}=\texttt{htmax}_{\alpha-1}$, where $\sigma({\rm bin}_k)$ is the sum over $\text{d}\sigma_i$ for the range associated with bin$_k$.
\item The final bin is inclusive and determined by $\sigma({\rm bin}_m) \times \mathcal{L} < N/10$, where $N$ is the total number of events to be generated in the final bin.
\end{enumerate}

Note that $x = 0.9$ was used for this study.

\item \emph{Generate weighted events.} We generate $N\simeq5\times10^4$ generator-level events in each of the $m$ bins.  For each bin separately, the events are showered, decayed, and matched in $\texttt{Pythia}$ and reconstructed in $\texttt{Delphes}$. After matching, each bin has $n_k \le N$ events and an associated matched cross section  $\sigma_{{\rm LO-matched}}$.
\end{enumerate}

\pagebreak
\addcontentsline{toc}{section}{References}
\footnotesize
\setstretch{1.0}
\bibliography{SimplifiedModels}
\end{document}